\documentclass[11pt]{article}
\usepackage{amsmath,amssymb,amsthm,color,graphicx,verbatim}
\usepackage{euscript}
\usepackage{times}
\usepackage{epic,eepic}

\newtheorem{theorem}{Theorem}[section]
\newtheorem{proposition}[theorem]{Proposition}

\newtheorem{lemma}[theorem]{Lemma}
\newtheorem{claim}[theorem]{Claim}

\graphicspath{{Figs/}}

\def\A{\mathcal{A}}
\def\I{\mathcal{I}}
\def\J{\mathcal{J}}
\def\TF {{\Sigma}}
\def\Q{\sigma}
\def\SQ{\chi}
\def\TQ {{\varrho}}
\def\reals{{\mathbb R}}

\def\G{{\mathcal G}}
\def\eps{{\varepsilon}}

\def\CL{{\sf H}}

\def\F{\mathcal{F}}
\def\FQ{\mathcal{Q}}
\def\t{\lambda}
\def\sp{\vartheta}
\def\k{k}

\def\T{\mathcal{T}}

\def\L{L}

\def\conn#1#2{[#1,#2]}

\def\H{{\cal H}}

\def\DT{\mathop{\mathrm{DT}}}
\def\VD{\mathop{\mathrm{VD}}}

\newcommand{\ignore}[1]{}

\def\marrow{{\marginpar[\hfill$\longrightarrow$]{$\longleftarrow$}}}

\def\natan#1{{\sc Natan says: }{\marrow\sf #1}}

\begin{document}
\begin{titlepage}
\title{On Kinetic Delaunay Triangulations: A Near Quadratic Bound for Unit Speed Motions}
\author{Natan Rubin\thanks{%
Universit\'{e} Pierre \& Marie Curie, Institut de Math\'{e}matiques de Jussieu (UMR 7586), 4 Place Jussieu, 75252 Paris Cedex, France.
Email: {\tt rubinnat.ac@gmail.com}. Work on this paper was partly done while the author was a Minerva Postdoctoral Fellow at the Department of Mathematics and Computer Science of the Free University of Berlin. Work on this paper was partly supported by Minerva Fellowship Program of the Max Planck Society, and by Fondation Sciences Math\'{e}matiques de Paris.}}

\maketitle
\begin{abstract}

Let $P$ be a collection of $n$ points in the plane, each moving along some straight line at unit speed.
We obtain an almost tight upper bound of $O(n^{2+\eps})$, for any $\eps>0$, on the maximum number of discrete changes that the Delaunay triangulation $\DT(P)$ of $P$ experiences during this motion.
Our analysis is cast in a purely topological setting, where we only assume that
(i) any four points can be co-circular at most three times, and (ii) no triple of points can be collinear more than twice; these assumptions hold for unit speed motions.
\end{abstract}
\end{titlepage}
\section{Introduction}\label{Sec:Intro}
\noindent{\bf Delaunay triangulations.} 
Let $P$ be a finite set of points in the plane. 
Let $\VD(P)$ and $\DT(P)$ denote the Euclidean Voronoi diagram and Delaunay
triangulation of $P$, respectively. 
The Delaunay triangulation consists of all 
triangles spanned by $P$ whose circumcircles do not contain points of $P$ in their
interior. A pair of points $p,q\in P$ are connected by a Delaunay edge if and only if
there is a circle passing through $p$
and $q$ that does not contain any point of $P$ in its interior.
%

Delaunay triangulations and their duals, Voronoi diagrams, are among the most extensively and longest studied constructs in computational geometry, with a wide range of applications.
For a {\it static} point set $P$, both $\DT(P)$ and $\VD(P)$ have linear complexity and can be computed in optimal $O(n\log n)$ time.
See \cite{AK,Ed2,Fortune} for surveys and a
textbook on these structures. 
The problem has also been studied in the {\it dynamic} setting, where one seeks to maintain $\DT(P)$ and $\VD(P)$ under updates of $P$ (insertion and deletion of points); see, e.g., \cite{Chan}. 


\smallskip
\noindent{\bf The kinetic setting: Previous work.}
In many applications of Delaunay/Voronoi methods (e.g., mesh generation and kinetic collision detection) the points of the input set $P$ are moving continuously, so
these diagrams need to be efficiently updated during the motion.
Even though the motion of the points is continuous, the combinatorial structure of the Voronoi and
Delaunay diagrams changes only at
discrete times when certain critical events occur. 
Interest in efficient maintenance of geometric
structures under simple motion\footnote{While there are several ways to define this notion, the simplest would be to assume that each coordinate of each point $p=p(t)$ in $P$ is a is fixed-degree polynomial in $t$.} of the underlying point set goes back at least to 
Atallah \cite{a-dcg-83,a-sdcgp-85}. 

For the purpose of kinetic maintenance, Delaunay triangulations are 
nice structures, because, as mentioned above, they admit local 
certifications associated with individual triangles (namely, that their circumcircles be $P$-empty).  This makes 
it simple to maintain $\DT(P)$ under point motion: an update is 
necessary only when one of these empty circumcircle conditions 
fails---this (typically) corresponds to co-circularities of certain subsets of
four points, where the relevant circumcircle is $P$-empty. Whenever such an event, referred to as a {\it Delaunay co-circularity} in this paper, happens, 
a single edge flip easily restores Delaunayhood.\footnote{We assume that the motion of the points is generic, so that no more than four points can become co-circular at any given time.} 
In addition, the Delaunay triangulation changes when some triple of points of $P$ become collinear on the boundary of the convex hull of $P$; see below for details.
Hence, the performance of any Voronoi- or Delaunay-based kinetic algorithm depends on the maximum possible number of {\it discrete changes}, that is, Delaunay co-circularities and convex hull collinearities, which $\DT(P)$ experiences during the motion of its points.

This paper studies the best-known formulation of the problem, in which each point of $P$ moves along a straight line with unit speed; see \cite{TOPP,Fortune}. In this case, the (previously) best-known upper bound on the number of discrete changes in $\DT(P)$ is $O(n^3)$. 
In the more general (and even more difficult) version of the problem, each point of $P$
moves with so-called pseudo-algebraic motion of constant description complexity. This implies (in particular) that any four points are co-circular at most $s$ times, and any triple of points can are collinear at most $s'$ times, for some constants $s,s'>0$.  
Given these (purely topological) restrictions on the continuous motion of $P$,
Fu and Lee \cite{FuLee}, and Guibas et al.~\cite{gmr-vdmpp-92} established a roughly cubic upper bound of
$O(n^2 \lambda_{s+2}(n))$, where $\lambda_s(n)$ is the (almost linear) maximum length
of an $(n,s)$-Davenport-Schinzel sequence~\cite{SA95}. 
A substantial gap exists between these near-cubic upper bounds
and the best known quadratic lower bound~\cite{SA95}. Closing this gap has been in the 
computational geometry lore for many years, and is considered as one of the major open problems in the field. It is listed as Problem 2 in the TOPP project; see \cite{TOPP}.
A recent work \cite{Natan} by the author provides an almost tight bound of $O(n^{2+\eps})$, for any $\eps>0$, for a more restricted version of the problem, in which any four points can be co-circular at most {\it twice}.


In view of the very slow progress on the above general problem, several alternative structures were studied. For example, Chew \cite{Chew} proved that $\VD(P)$ undergoes a near-quadratic number of discrete changes if it is defined with respect to a ``polygonal" distance function.
More recent studies \cite{AWY,KRS} show how to maintain a (non-Delaunay) triangulation of $P$ so that it undergoes only a near-quadratic number of changes.  Agarwal et al.~\cite{Stable}
show how to efficiently maintain a so called {\it $\alpha$-stable} subgraph of the Euclidean $\DT(P)$, which experiences only a near-quadratic number of changes, and whose edges are robust with respect to small changes in the underlying norm. 

\smallskip
\noindent{\bf Our result.}
We study the problem in a purely topological setup, where we assume that (i)
any four points of $P$ are co-circular at most three times during their (continuous) motion, and (ii) any three points of $P$ can be collinear at most twice. For any point set $P$ whose motion satisfies these two axioms, we derive a nearly tight upper bound of $O(n^{2+\eps})$, for any $\eps>0$, on the overall number of discrete changes experienced by $\DT(P)$. As is well known (and briefly discussed in Appendix \ref{AppSec:UnitSpeeds}), these properties hold for points that move along straight lines with a common (unit) speed, so our near-quadratic bound holds in this case.

\smallskip
\noindent{\bf Proof ingredients.} 
The majority of the discrete changes in $\DT(P)$ occur at moments $t_0$ when some four points $p,q,a,b\in P$ are co-circular, and the corresponding circumdisc contains no other points of $P$. We refer to these events as {\it Delaunay co-circularities}. Suppose that $p,a,q,b$ appear along their common circumcircle in this order, so $ab$ and $pq$ form the chords of the convex quadrilateral spanned by these points. Right before $t_0$, one of the chords, say $pq$, is Delaunay and thus admits a $P$-empty disc whose boundary contains $p$ and $q$. 
Right after time $t_0$, the edge $pq$ is replaced in $\DT(P)$ by $ab$, an operation known as an {\it edge-flip}.  
Informally, this happens because the Delaunayhood of $pq$ is violated by $a$ and $b$: Any disc whose boundary contains $p$ and $q$ contains at least one of the points $a,b$. 
If $pq$ does not re-enter $\DT(P)$ after time $t_0$, we can charge the event at time $t_0$ to the edge $pq$, for a total of $O(n^2)$ such events. We thus assume that $pq$ is again Delaunay at some moment $t_1>t_0$. 

One of the major observations used in our analysis is that one of the following always holds: either the Delaunayhood of $pq$ is interrupted during $(t_0,t_1)$ by at least $k^2$ pairs $u,v\in P$, or this edge can be made Delaunay throughout $(t_0,t_1)$ by removal of at most $\Theta(k)$ points of $P$. In the former case, each violating pair $u,v$ contributes during $(t_0,t_1)$ either a co-circularity of $p,q,u,v$, or a collinearity in which one of the points $u$ or $v$ crosses $pq$.
This fairly simple observation lies at the heart of our charging strategy.

\smallskip
\noindent {\bf Combinatorial charging.}
Our goal is to derive a recurrence formula for the maximum number $N(n)$ of such Delaunay co-circularities induced by any set $P$ of $n$ points (whose motion satisfies the above conditions). Notice that the number of {\it all} co-circularities, each defined by some four points of $P$, can be as large as $\Theta(n^4)$. The challenge is thus to show that the vast majority of co-circularity events are not Delaunay (i.e., their corresponding circumdiscs are penetrated by additional points of $P$). 

In Section \ref{Sec:Prelim} we study the set of all co-circularities that involve some disappearing Delaunay edge $pq$ and some other pair of points of $P\setminus \{p,q\}$, and occur during the period $(t_0,t_1)$ when $pq$ is absent\footnote{In fact, the analysis in Section \ref{Sec:Prelim} is more general, and applies to any interval $(t_0,t_1)$ with the property that $pq$ is Delaunay at one of its endpoints $t_0,t_1$.} from $\DT(P)$. A co-circularity is called {\it $k$-shallow} if its circumdisc contains at most $k$ points of $P$.
If we find at least $\Omega(k^2)$ such $k$-shallow co-circularities\footnote{Each of them would become a Delaunay co-circularity after removal of at most $k$ points of $P$.}, involving $p,q,$ and another pair of points, we can charge them for the disappearance of $pq$. We use the routine probabilistic argument of Clarkson and Shor \cite{CS} to show that the number of Delaunay co-circularities, for which this simple charging works, is $O\left(k^2N(n/k)\right)$.
Informally, this term that such Delaunay co-circularities contribute to the overall recurrence formula (see, e.g., \cite{ASS} and \cite{ConstantLines}), yields a near-quadratic bound for $N(n)$. Similarly, if we find a ``shallow" collinearity of $p,q$ and another point (one halfplane bounded by the line of collinearity contains at most $k$ points), we can charge the disappearance of $pq$ to this collinearity. A combination of the Clarkson-Shor technique with the known near-quadratic bound on the number of topological changes in the convex hull of $P$ (see \cite[Section 8.6.1]{SA95}) yields an additional near-quadratic term in the recurrence.

\smallskip
\noindent{\bf Probabilistic refinement.} It thus remains to bound the number of the above Delaunay co-circularities, for which $p$ and $q$ participate in fewer shallow co-circularities and in no shallow collinearity during $(t_0,t_1)$. In this case, we show, in what follows we refer to as the {\it Red-Blue Theorem} (or Theorem \ref{Thm:RedBlue}), that one can restore the Delaunayhood of $pq$ throughout $(t_0,t_1)$ by removal of some subset $A$ of at most $3k$ points of $P$. 
To bound the maximum number of such ``non-chargeable" events, we incorporate them into more structured topological configurations (or, more precisely, processes), which are likely to show up (in the style of the Clarkson-Shor argument) in a reduced Delaunay triangulation $\DT(R)$, defined over a random sample $R\subset P$ of $\Theta(n/k)$ points.

For example, suppose that the above co-circularity at time $t_0$, is the {\it last} co-circularity of $p,q,a,b$. 
Then (at least) one of the points $a$ or $b$ must hit the edge $pq$ before it re-enters $\DT(P)$ at time $t_1$.
Clearly, the point which crosses $pq$, let it be $a$, must belong to $A$.
Notice that the following two events occur simultaneously, with probability $\Omega\left(1/k^3\right)$: (1) the random sample $R$ contains the crossing triple $p,a,q$, and (2) none of the points of $A\setminus \{a\}$ belong to $R$. 
In such case, we say that the edge $pq$ undergoes 
a {\it Delaunay crossing by} $a$ in the {\it refined} triangulation $\DT(R)$, which takes place during a certain subinterval $I\subset [t_0,t_1]$ (such that (i) $a$ hits $pq$ during $I$, (ii) $pq\in \DT(R)$ at the beginning and the end of $I$, and (iii) $pq\not\in \DT(R)$ in the interior of $I$, but belongs to $\DT(R\setminus \{a\})$ throughout $I$). 
A symmetric (time-reversed) argument applies if we encounter the {\it first} co-circularity of $p,q,a,b$.

As argued in the predecessor paper \cite{Natan}, Delaunay crossings are especially nice objects due to their strict structural properties. In particular, as shown in \cite{Natan}: (i) The edges $pa$ and $aq$ belong to $\DT(R)$ throughout the above interval $I$, and (ii) Assuming $a$ hits $pq$ exactly once during $I$, every other point $w\in R\setminus \{p,q,a\}$ is involved during this interval in a co-circularity with $p,q,a$.

\smallskip
\noindent {\bf The roadmap.}
In Section \ref{Sec:ReduceToCrossings} we show that the number of Delaunay co-circularities is dominated by the maximum possible number of Delaunay crossings. Notice the previously sketched argument (which appears in \cite{Natan}) works only for the first and the last Delaunay co-circularities of the quadruple. 

To extend the above reduction to the remaining, ``middle" Delaunay co-circularities, we resort in Section \ref{Sec:ReduceToCrossings} to a fairly simple argument, expressing the maximum possible number of such co-circularities in
terms of the numbers of extremal Delaunay co-circularities and Delaunay crossings that arise in smaller-size subsets of $P$.

In Section \ref{Sec:CrossOnce}, we recall (or re-establish) several structural properties of Delaunay crossings, which will be used throughout the rest of the analysis. 
Informally, our goal is to show that, for an average pair $(p,r)$, the point $r$ is involved in ``few" crossings of $p$-incident edges.
To do so, we express the number of Delaunay crossings in terms of the maximum number of certain {\it quadruples} in $P$. Each such quadruple $\Q=(p,q,a,r)$ is composed of a pair of ``consecutive" Delaunay crossings of $p$-adjacent edges $pq$ and $pa$, by the same point $r$. 


In Section \ref{Sec:CountQuad} we apply the routine ``charge-or-refine" strategy (via our Red-Blue Theorem) 
to analyze the maximum number of the above quadruples. 
This is done in several steps. At each stage we first try to dispose of as many quadruples as possible by charging each of them either to sufficiently many ``shallow" co-circularities (or collinearities), or to one of the several kinds of ``terminal" triples, for which we provide back in Section \ref{Sec:CrossOnce} a direct quadratic bound on their number.

There are two main types of such terminal triples $(p,q,a)$. In one of them, we have a {\it double Delaunay crossing}---the point $a$ crosses $pq$ twice during the interval $I$. In the other the same triple performs two distinct ``single" Delaunay crossings, where, say, $a$ crosses $pq$ during one crossing, and $q$ crosses $pa$ during the second one. In both cases the number of such triples is only $O(n^2)$.

Each step of the analysis enforces additional constraints on the surviving quadruples. There are two main types of such constraints. The first is to enforce more Delaunay crossings involving sub-triples of the points of the quadruple. The other is to enforce ``almost-Delaunayhood" of various pairs of points in the quadruple, for progressively larger time intervals. By this we mean that the corresponding edge is Delaunay if we remove from $P$ a small subset of points. The ultimate goal is to enforce sufficiently many Delaunay crossings, so that some triple of points undergoes {\it two} distinct Delaunay crossings. As mentioned above, this is the main type of the ``terminal" configurations, for which we have a quadratic bound on their number.

Each step of the analysis yileds a recurrence formula that involves ``near-quadratic" terms (of the kind mentioned earlier) plus terms involving further-constrained configurations, until we finally bottom out (in Section \ref{Sec:Terminal}) by reaching the terminal triples mentioned above. In each of the recurrences we make use of the Clarkson-Shor probabilistic argument \cite{CS}, in order to get rid of the small ``obstruction" subset of $P$ that we need to remove; this is done by passing to a random sample of $P$, the standard style of \cite{CS}. The overall collection of recurrences solves to a near-quadratic bound, in a manner similar to many earlier works involving such recurrences (see, e.g., \cite{ASS,Envelopes3D,ConstantLines,Overlay4D,MichaHigher} and \cite[Section 7.3.2]{SA95}).

Unfortunately, the analysis is fairly involved and consists of many steps. 
In addition to the aforementioned type of quadruples (formed by pairs of Delaunay crossings), we use two additional classes of quadruples which are studied in Sections \ref{Sec:Special} and \ref{Sec:Terminal}, respectively. Note that only the last kind of configurations, referred to as {\it terminal quadruples}, can always be traced to some of the above ``terminal" triples. 

We postpone the rest of this discussion until Section \ref{Subsec:Roadmap}, where we provide a more detailed summary of the three classes of quadruples, and of the connections between these classes, and the Delaunay crossings.
 
Finally, we emphasize that the contribution of the paper, and its main ideas, are delivered already in Sections \ref{Sec:Intro} through \ref{Sec:CrossOnce}.



\smallskip
\paragraph{Acknowledgements.} I would like to thank my former Ph.D. advisor Micha Sharir 
whose dedicated support made this work possible.
In particular, I would like to thank him for the insightful discussions, and, especially, for his invaluable help in the preparation and careful reading of this paper. 

\section{Geometric Preliminaries}\label{Sec:Prelim}
\noindent{\bf Delaunay co-circularities.} Let $P$ be a collection of $n$ points moving along pseudo-algebraic trajectories in the plane, so that any four points are co-circular at most {\it three} times, and any three points can be collinear at most {\it twice} during the motion. 
In addition, we assume, without loss of generality, that the trajectories of the points of $P$ satisfy all the standard general position assumptions; see Appendix \ref{AppSec:GenPos} for more details. 

\begin{figure}[htb]
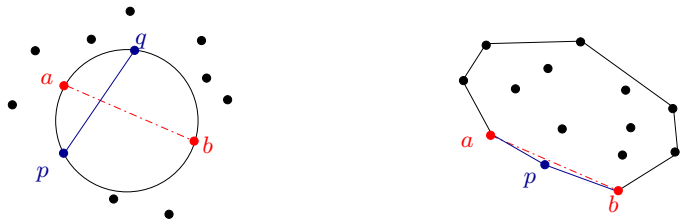

\begin{center}
\input{DelaunayCocirc.pstex_t}\hspace{3cm}\input{HullEvent.pstex_t}
\caption{\small{Left: A Delaunay co-circularity of $a,b,p,q$. An old Delaunay edge $pq$ is replaced by the new edge $ab$. Right: A collinearity of $a,p,b$ right before $p$ ceases being a vertex on the boundary of the convex hull. 
}}
\label{Fig:DelaunayEvents}
\end{center}

\end{figure}

The Delaunay triangulation $\DT(P)$ changes at discrete time moments $t_0$ when one of the following two types of events occurs.

(i) Some four points $a,b,p,q$  of $P$ become co-circular, so that the cicrumdisc of $p,q,a,b$ is {\it empty}, i.e., does not contain any point of $P$ in its interior. We refer to such events as {\it Delaunay co-circularities}. See Figure \ref{Fig:DelaunayEvents} (left). At each such co-circularity $\DT(P)$ undergoes an {\it edge-flip}, where an old Delaunay edge $pq$ is replaced by the ``opposite" edge $ab$.

(ii) Some three points $a,b,p$ of $P$ become collinear on the boundary of the convex hull of $P$. Assume that $p$ lies between $a$ and $b$. In this case, if $p$ moves into the interior of the hull then the triangle $abp$ becomes a new Delaunay triangle, and if $p$ moves outside and becomes a new vertex, the old Delaunay triangle $abp$ shrinks to a segment and disappears. See Figure \ref{Fig:DelaunayEvents} (right).
The number of such collinearities on the convex hull boundary is known to be at most nearly quadratic; see, e.g., \cite[Section 8.6.1]{SA95} and below. 

In view of the above, it suffices to obtain a near-quadratic bound on the number of Delaunay co-circularities. Hence, the rest of this paper is devoted to proving the following main result:

\begin{theorem}\label{Thm:OverallBound}
Let $P$ be a collection of $n$ points moving along pseudo-algebraic trajectories in the plane, so that (i) any four points of $P$ are co-circular at most three times, and (ii) no triple of points can be collinear more than twice. Then $P$ admits at most $O(n^{2+\eps})$ Delaunay co-circularities, for any $\eps>0$.
\end{theorem}

In what follows, we use $N(n)$ to denote the maximum possible number of Delaunay co-circularities that can arise in a set of $n$ points whose motion satisfies the above assumptions.

\paragraph{Shallow co-circularities.}
We say that a co-circularity event, where four points of $P$ become co-circular, has {\it level} $k$ if its corresponding circumdisc contains exactly $k$ points of $P$ in its interior. In particular, the Delaunay co-circularities have level $0$. The co-circularities having level at most $k$ are called {\it $k$-shallow}.

We can bound the maximum possible number of $k$-shallow co-circularities (for $k\geq 1$) in terms of the maximum number of Delaunay co-circularities in smaller-size point sets using the following fairly general argument of Clarkson and Shor \cite{CS}.
Consider a random sample $R$ of $\Theta(n/k)(<n/2)$ points of $P$ and observe that any $k$-shallow co-circularity (with respect to $P$) becomes a Delaunay co-circularity (with respect to $R$) with probability $\Theta(1/k^4)$. (For this to happen, the four points of the co-circularity have to be chosen in $R$, and the at most $k$ points of $P$ inside the circumdisc must not be chosen; see \cite{CS} for further details.) 
Hence, the overall number of $k$-shallow co-circularities is $O(k^4 N(n/k))$.

\paragraph{Shallow collinearities.}
Similar notations apply to collinearities of triples of points $p,q,r$. A collinearity of $p,q,r$ is called {\it $k$-shallow} if the number of points of $P$ to the left, or to the right, of the line through $p,q,r$ is at most $k$. The above probabilistic argument of Clarkson and Shor implies, in a similar manner, that the number of such events, for $k\geq 1$, is $O(k^3 \CL(n/k))$, where $\CL(m)$ denotes the maximum number of discrete changes of the convex hull of an $m$-point subset of $P$. 
As shown, e.g., in \cite[Section 8.6.1]{SA95},  $\CL(m)=O(m^2\beta(m))$, where $\beta(\cdot)$ is an extremely slowly growing function.\footnote{Specifically,
$\beta(n)=\frac{\lambda_{s+2}(n)}{n}$, where $s$ is the maximum number of collinearities of any fixed triple of points, and where $\lambda_{s+2}(n)$ is the maximum length of $(n,s+2)$-Davenport-Schinzel sequences \cite{SA95}.} We thus get that the number of $k$-shallow collinearities is $O(kn^2\beta(n/k))=O(kn^2\beta(n))$.

\begin{figure}[htb]
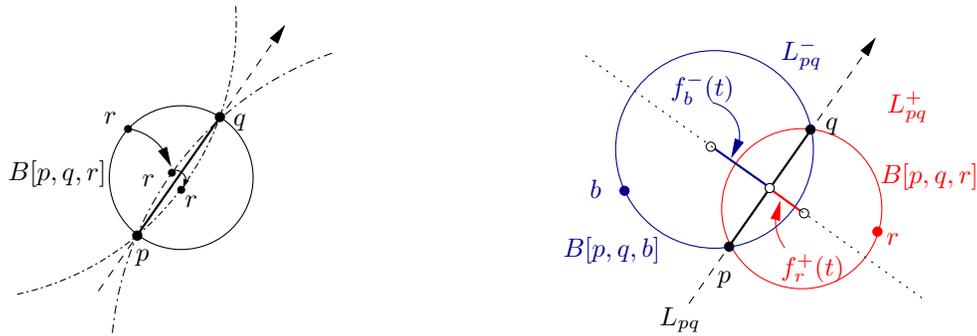

\begin{center}
\input{Circumdisc.pstex_t}\hspace{3cm}\input{RedBlueFuncs.pstex_t}
\caption{\small Left: The circumdisc $B[p,q,r]$ of $p,q$ and $r$ moves continuously as long as these three points are not collinear, and then flips over to the other side of the line of collinearity after the collinearity. 
Right: A snapshot at moment $t$. 
In the depicted configuration we have $f_b^-(t)<0<f_r^+(t)$.}
\label{Fig:RedBlueFuncs}
\end{center}
\end{figure}

For every ordered pair $(p,q)$ of points of $P$, denote by $\L_{pq}$ the line passing through $p$ and $q$ and oriented from $p$ to $q$. Define $\L_{pq}^-$ (resp., $\L_{pq}^+$) to be the halfplane to the left (resp., right) of $\L_{pq}$.
Notice that $\L_{pq}$ moves continuously with $p$ and $q$ (since, by assumption, $p$ and $q$ never coincide during the motion). Note also that $\L_{pq}$ and $\L_{qp}$ are oppositely oriented and that $\L_{pq}^+=\L_{qp}^-$ and $\L_{pq}^-=\L_{qp}^+$.
We also orient the edge $pq$ connecting $p$ and $q$ from $p$ to $q$, so that the edges $pq$ and $qp$ have opposite orientations.

Any three points $p,q,r$ span a circumdisc $B[p,q,r]$ which moves continuously with $p,q,r$ as long as $p,q,r$ are not collinear. See Figure \ref{Fig:RedBlueFuncs} (left). When $p,q,r$ become collinear, say, when $r$ crosses $pq$ from $\L_{pq}^-$ to $\L_{pq}^+$, the circumdisc $B[p,q,r]$ changes instantly from being all of $\L_{pq}^+$ to all of $\L_{pq}^-$.
Similarly, when $r$ crosses $\L_{pq}$ from $\L_{pq}^-$ to $\L_{pq}^+$ {\it outside} $pq$, the circumdisc changes instantly from $\L_{pq}^-$ to $\L_{pq}^+$. Symmetric changes occur when $r$ crosses $\L_{pq}$ from $\L_{pq}^+$ to $\L_{pq}^-$.

\paragraph{\bf The red-blue arrangement.} 
As in \cite{gmr-vdmpp-92,Natan}, we use the so called red-blue arrangement to facilitate the analysis of co-circularities whose corresponding discs touch the same two points $p,q\in P$. For the sake of completeness, we provide  below a formal definition of this arrangement.

For a fixed ordered pair $p,q\in P$, we call a point $a$ of $P\setminus\{p,q\}$ {\it red} (with respect to the oriented edge $pq$) if $a\in \L_{pq}^+$; otherwise it is {\it blue}.

We define, for each $r\in P\setminus\{p,q\}$, a pair of partial functions $f_r^+,f_r^-$ over the time axis as follows.
If $r\in \L_{pq}^+$ at time $t$ then $f_r^-(t)$ is undefined, and $f^+_r(t)$ is the signed distance of the center $c$ of $B[p,q,r]$ from $\L_{pq}$; it is positive (resp., negative) if $c$ lies in $\L_{pq}^+$ (resp., in $\L_{pq}^-$). A symmetric definition applies when $r\in \L_{pq}^-$. Here too $f^-_r(t)$ is positive (resp., negative) if the center of $B[p,q,r]$ lies in $\L_{pq}^+$ (resp., in $\L_{pq}^-$). We refer to $f_r^+$ as the {\it red function} of $r$ (with respect to $pq$) and to $f_r^-$ as the {\it blue function} of $r$. Note that at all times when $p,q,r$ are not collinear, exactly one of $f_r^+,f_r^-$ is defined. See Figure \ref{Fig:RedBlueFuncs} (right).
The common points of discontinuity of $f_r^+,f_r^-$ occur at moments when $r$ crosses $\L_{pq}$. Specifically, $f_r^+$ tends to $+\infty$ before $r$ crosses $\L_{pq}$ from $\L_{pq}^+$ to $\L_{pq}^-$ outside the segment $pq$, and it tends to $-\infty$ when $r$ does so within $pq$; the behavior of $f_r^-$ is fully symmetric.

\begin{figure}[htb]
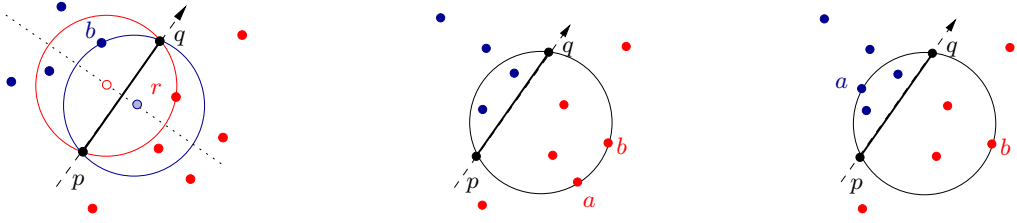

\begin{center}
\input{Envelopes.pstex_t}\hspace{2.5cm}\input{RedRedCocirc.pstex_t}\hspace{2.5cm}\input{RedBlueCocirc.pstex_t}
\caption{\small Left: A snapshot at fixed time $t$. The red and blue envelopes $E^+,E^-$ coincide with the functions $f_r^+,f_b^-$, respectively. The edge $pq$ is not a Delaunay edge because $E^+(t)$ (the hollow center) is smaller than $E^-(t)$ (the shaded center). Center and right: Red-red and red-blue co-circularities.}
\label{Fig:Envelopes}
\end{center}
\vspace{-0.2cm}
\end{figure}

Let $E^+$ denote the lower envelope of the red functions, and let $E^-$ denote the upper envelope of the blue functions. The edge $pq$ is a Delaunay edge at time $t$ if and only if $E^-(t)<E^+(t)$. Any disc whose bounding circle passes through $p$ and $q$ which is centered anywhere in the interval $(E^-(t),E^+(t))$ along the perpendicular bisector of $pq$ (with the origin on this line lying at the midpoint of $pq$) is empty at time $t$, and thus serves as a witness to $pq$ being Delaunay.
If $pq$ is not Delaunay at time $t$, there is a pair of a red function $f_r^+(t)$ and a blue function $f_b^-(t)$ such that $f_r^+(t)<f_b^-(t)$.
For example, we can take $f_r^+$ (resp., $f_b^-$) to be the function attaining $E^+$ (resp., $E^-$) at time $t$; see Figure \ref{Fig:Envelopes} (left). In such a case, we say that the Delaunayhood of $pq$ is {\it violated} by the pair of points $r,b\in P$ that define $f_r^+,f_b^-$. Note that in general there can be many pairs $(r,b)$ that violate $pq$ (quadratically many in the worst case).

Hence, at any time when the edge $pq$ joins or leaves $\DT(P)$, via a Delaunay co-circularity involving $p$, $q$, and two other points of $P$, we have $E^-(t)=E^+(t)$. In this case the two other points, $a,b$, are such that one of them, say $a$, lies in $\L_{pq}^+$ and $b$ lies in $\L_{pq}^-$, and $E^+(t)=f_a^+(t), E^-(t)=f_b^-(t)$.

Let $\A=\A_{pq}$ denote the arrangement of the $2n-4$ functions $f_r^+(t),f_r^-(t)$, for $r\in P\setminus\{p,q\}$, drawn in the parametric $(t,\rho)$-plane, where $t$ is the time and $\rho$ measures signed distance to the midpoint of $pq$ along the perpendicular bisector of $pq$. We label each vertex of $\A$ as red-red, blue-blue, or red-blue, according to the colors of the two functions meeting at the vertex. Note that our general position assumptions imply that $\A$ is also in general position, so that no three functions pass through a common vertex, and no pair of functions are tangent to each other. As discussed above, the functions forming $\A$ have in general discontinuities, at the corresponding collinearities. At the time $t_0$ of each such collinearity, a red function $f_r^+$ tends to $\infty$ or $-\infty$ on one side of $t_0$, and is replaced on the other side of $t_0$ by the corresponding blue function $f_r^-$ which tends to $-\infty$ or $\infty$, respectively.

An intersection between two red functions $f_a^+,f_b^+$ corresponds to a co-circularity event which involves $p,q,a$ and $b$, occurring when both $a$ and $b$ lie in $\L_{pq}^+$.
Similarly, an intersection of two blue functions $f_a^-,f_b^-$ corresponds to a co-circularity event
involving $p,q,a,b$ where both $a$ and $b$ lie in $\L_{pq}^-$. Also, an intersection of a red fuction $f_a^+$ and a blue function $f_b^-$ represents a co-circularity of $p,q,a,b$, where $a\in \L_{pq}^+$ and $b\in \L_{pq}^-$. We label these co-circularities, as we labeled the vertices of $\A$, as red-red, blue-blue, and red-blue (all with respect to $pq$), depending on the respective colors of $a$ and $b$. See Figure \ref{Fig:Envelopes} (center and right).

It is instructive to note that in any co-circularity of four points of $P$ there are exactly two pairs (the opposite pairs in the co-circularity)
with respect to which the co-circularity is red-blue, and four pairs (the adjacent pairs) with respect to which the co-circularity is ``monochromatic". When the co-circularity is Delaunay, the two pairs for which the co-circularity is red-blue are those that enter or leave the Delaunay triangulation $\DT(P)$ (one pair enters and one leaves). The Delaunayhood of pairs for which the co-circularity is monochromatic is not affected by the co-circularity, which appears in the corresponding arrangement as a {\it breakpoint} of either $E^+(t)$ or $E^-(t)$.

The following useful result on $\A_{pq}$, which is one of the major tools in our analysis, was established in \cite{Natan} by applying routine techniques for analyzing planar arrangements. For the sake of completeness, we provide its proof in Appendix \ref{Append:RedBlue}.

\begin{theorem}[Red-blue Theorem]\label{Thm:RedBlue}
Let $P$ be a collection of $n$ points moving in the plane as described above. Suppose that an edge $pq$ belongs to $\DT(P)$ at (at least) one of the two moments $t_0$ and $t_1$, for $t_0<t_1$.
Let $k>12$ be some sufficiently large constant.\footnote{The constants is the $O(\cdot)$ and $\Omega(\cdot)$ notations do not depend on $k$.}
Then one of the following conditions holds:\\
\indent (i) There is a $k$-shallow collinearity which takes place during $(t_0,t_1)$, and involves $p$, $q$ and another point $r$.\\
\indent (ii) There are $\Omega(k^2)$ $k$-shallow red-red, red-blue, or blue-blue co-circularities (with respect to $pq$) which occur during $(t_0,t_1)$.\\
\indent (iii) There is a subset $A\subset P$ of at most $3k$ points whose removal guarantees that $pq$ belongs to $\DT(P\setminus A)$ throughout $(t_0,t_1)$.
\end{theorem}

Notice that we do not assume that $pq$ leaves $\DT(P)$ at any moment during $(t_0,t_1)$ (in that case, case (iii) holds, with $A=\emptyset$). Note also that, although we do not need this property, the theorem continues to hold in the more general setting of pseudo-algebraic motions of constant description complexity.


\section{From Delaunay Co-Circularities to Delaunay Crossings}\label{Sec:ReduceToCrossings}
Let $P$ be a set of $n$ points moving in the plane, so that any four points can be co-circular at most three times, and any triple of points can be collinear more than twice. For the sake of brevity, we will often take these topological restrictions for granted. As before, $N(n)$ denotes the maximum possible number of Delaunay co-circularities that can arise in such a set $P$.

In this section we introduce the notion of a Delaunay crossing, which plays a central role both in this paper and in its predecessor \cite{Natan}, and express the above quantity $N(n)$ in terms of the maximum numbers of Delaunay crossings that can arise in smaller sets of moving points. 

\smallskip
\noindent{\bf Delaunay crossings.}
A {\it Delaunay crossing} is a triple $(pq,r,I=[t_0,t_1])$, where $p,q,r\in P$ and $I$ is a time interval, such that 
\begin{enumerate}
\item $pq$ 
leaves $\DT(P)$ at time $t_0$, and returns at time $t_1$ (and $pq$ does not belong to $\DT(P)$ during $(t_0,t_1)$),
\item $r$ crosses the segment $pq$ {\it at least} once\footnote{And at most twice, by assumption.} during $I$, and
\item $pq$ is an edge of $\DT(P\setminus \{r\})$ during $I$ (i.e., removing $r$ restores the Delaunayhood of $pq$ during the entire time interval $I$).
\end{enumerate}

\begin{figure}[htb]
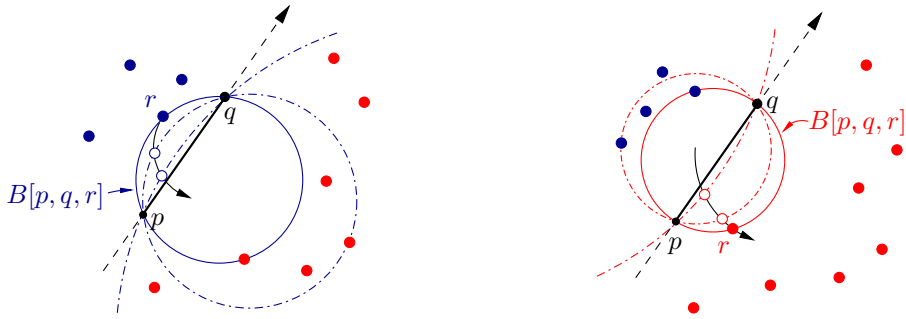

\begin{center}
\input{DelaunayCrossing.pstex_t}\hspace{3cm}\input{DelaunayCrossing1.pstex_t}
\caption{\small A Delaunay crossing of $pq$ by $r$ from $\L_{pq}^-$ to $\L_{pq}^+$. Several snapshots of the continuous motion of $B[p,q,r]$ before and after $r$ crosses $pq$ are depicted (in the left and right figures, respectively). Hollow points specify the positions of $r$ when $pq\not \in \DT(P)$. The solid circle in the left (resp., right) figure is the Delaunay co-circularity that destroys (resp., restores) the Delaunayhood of $pq$.}
\label{Fig:DelaunayCrossing}
\end{center}
\vspace{-0.5cm}
\end{figure} 

Note that each of the Delaunay co-circularities that destroys the Delaunayhood of $pq$ at time $t_0$ and restores it at time $t_1$ must involve $r$.

Note that we also allow Delaunay crossings, where the point $r$ hits $pq$ at one (or both) of the times $t_0,t_1$. In this case, the crossed edge $pq$ leaves the convex hull of $P$ at time $t_0$, or enters it at time $t_1$, so the overall number of such ``degenerate" crossings does not exceed $O(n^2\beta(n))$, and we may ignore them in what follows. 


Assuming $n\geq 5$, it is easy to see that the third condition is equivalent to the following condition, expressed in terms of the red-blue arrangement $\A_{pq}$ associated with $pq$: The point $r$ participates only in red-blue co-circularites during the interval $I$, and these are the only red-blue co-circularities that occur during $I$.\footnote{If $n=4$, then, in order for (3) to hold, we also need that the remaining point of $P$ does not cross $pq$ during $I$.} More specifically, note that $r$ is red during some portion of $I$ and is blue during the complementary portion (both portions are not necessarily connected). During the former portion the graph of $f_r^+$ coincides with the red lower envelope $E^+$ (otherwise $E^+(t)<E^-(t)$ would hold sometime during $I$ even after removal of $r$), so it can only meet the graphs of blue functions. Similarly, during the latter portion $f_r^-$ coincides with the blue upper envelope $E^-$, so it can only meet the graphs of red functions. When passing from the former portion to the latter, $f_r^+$ goes down to $-\infty$, meeting all blue functions below it, and then it is replaced by $f_r^-$, which goes down from $\infty$.
See Figure \ref{Fig:DelaunayCrossing} for a schematic illustration of this behavior.

Notice that no points, other than $r$, cross $pq$ during $I$ (any such crossing would clearly contradict the third condition at the very moment when it occurs).
Moreover, $r$ does not cross $\L_{pq}$ outside $pq$ during $I$; otherwise $pq$ would belong to $\DT(P)$ when $r$ belongs to $\L_{pq}\setminus pq$.

\medskip
\noindent{\bf Types of Delaunay co-circularities.} We say that a co-circularity event at time $t_0$ involving $a,b,p,q$ has {\it index} $1,2$, or $3$ if this is, respectively, the first,  the second, or the third co-circularity involving $a,b,p,q$. 
A co-circularity is {\it extremal} if its index is $1$ or $3$, and the co-circularities with index $2$ are referred to as {\it middle} co-circularities. 

Let $C(n)$ denote the maximum possible number of Delaunay crossings that can arise in a set of $n$ moving points $\reals^2$.
To bound $N(n)$ in terms of $C(n)$ (or, more precisely, in terms of $C(m)$, for some $m\leq n$), we first develop a recurrence which expresses the maximum possible number $N_E(n)$ of extremal Delaunay co-circularities in $P$ in terms of $C(n/k)$.
(In \cite{Natan}, there were no middle co-circularities, so the same argument worked for {\it all} Delaunay co-circularities.)
We then express the maximum possible number $N_M(n)$ of middle Delaunay co-circularities in $P$ in terms of  $C(n/k)$ and $N_E(n/k)$. (Here $k$ is an arbitrary sufficiently large parameter.)


\smallskip
\noindent{\bf The number of extremal co-circularities.} 
Consider a Delaunay co-circularity event at time $t_0$ at which an edge $pq$ of $\DT(P)$ is replaced by another edge $ab$, because of an extremal red-blue co-circularity (with respect to $pq$, and, for that matter, also with respect to $ab$) of level $0$ (that is, a co-circularity that is Delaunay).
Without loss of generality, assume that the co-circularity of $p,q,a,b$ has index $3$ (the case of index $1$ is handled fully symmetrically, by reversing the direction of the time axis).
 
There are at most $O(n^2)$ such events for which the vanishing edge $pq$ never reappears in $\DT(P)$, so we focus on the Delaunay co-circularities (of index $3$) whose corresponding edge $pq$ rejoins $\DT(P)$ at some future moment $t_1>t_0$. (As reviewed in Section \ref{Sec:Prelim}, $\DT(P)$ experiences then either a red-blue Delaunay co-circularity with respect to $pq$, or a hull event, when $pq$ is crossed by a point of $P\setminus \{p,q\}$. In the latter case, $pq$ is not strictly Delaunay at time $t_1$, and joins $\DT(P)$ right after $t_1$.) Note that in this case, at least one of the two other points $a,b$ involved in the co-circularity at time $t_0$ must cross $pq$ at some time between $t_0$ and $t_1$. Indeed, otherwise $p,q,a$ and $b$ would have to become co-circular again, in order to ``free" $pq$ from its non-Delaunayhood, which is impossible since our co-circularity has index $3$. 
More generally, we have the following lemma: 

\begin{lemma}\label{Lemma:MustCross}
Assume that the Delaunayhood of $pq$ is violated at time $t_0$ (or rather right after it) by the points $a\in \L_{pq}^-$ and $b\in \L_{pq}^+$. 
Furthermore, suppose that $pq$ re-enters $\DT(P)$ at some future time $t_1>t_0$.
Then at least one of the followings occurs during $(t_0,t_1]$: 

\medskip
(1) The point $a$ crosses $pq$ from $\L_{pq}^-$ to $\L_{pq}^+$.\\
\indent(2) The point $b$ crosses $pq$ from $\L_{pq}^+$ to $\L_{pq}^-$.\\
\indent(3) The four points $p,q,a,b$ are involved in a red-blue co-circularity.

\smallskip
Furthermore, the Delaunayhood of $pq$ is violated by $a$ and $b$ (so, in particular, the segments $pq$ and $ab$ intersect) after time $t_0$ and until the first time in $(t_0,t_1]$ when at least one of the events in (1)--(3) occurs. 
\end{lemma}

Clearly, the third scenario is not possible if the co-circularity at time $t_0$ has index $3$. A symmetric version of Lemma \ref{Lemma:MustCross} applies if the Delaunayhood of $pq$ is violated right {\it before} time $t_0$ by $a$ and $b$, and this edge is Delaunay at an {\it earlier} time $t_1<t_0$.

\begin{figure}[htb]
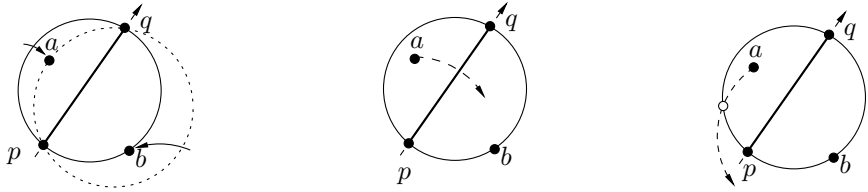

\begin{center}
\input{AfterLastCocirc.pstex_t}\hspace{2.5cm}\input{CrossWithin.pstex_t}\hspace{2.5cm}\input{NotOutside.pstex_t}
\caption{\small Proof of Lemma \ref{Lemma:MustCross}. Left: The setup right after time $t_0$. Center and right: the point $a$ can leave $B[p,q,b]\cap \L_{pq}^-$ (before $b$ leaves the symmetric cap $B[p,q,a]\cap \L_{pq}^+$) in two possible ways, corresponding to cases (1) and (3) of the lemma.}
\label{Fig:StaysViolated}
\end{center}
\vspace{-0.3cm}
\end{figure}

\begin{proof}
Refer to Figure \ref{Fig:StaysViolated}.
Clearly, the Delaunayhood of $pq$ remains violated by $a$ and $b$ after time $t_0$ as long as
$a$ remains within the cap $B[p,q,b]\cap \L_{pq}^-$, and $b$ remains within the cap $B[p,q,a]\cap \L_{pq}^+$ (as depicted in the left figure). 

Consider the first time $t^*\in (t_0,t_1]$ when the above state of affairs ceases to hold. Notice that the Delaunayhood of $pq$ is violated by $a$ and $b$ (so, in particular, $pq$ is intersected by $ab$) throughout the interval $(t_0,t^*)$.
Assume without loss of generality that $a$ leaves the the cap $B[p,q,b]\cap \L_{pq}^-$. If $a$ crosses $pq$, then the first scenario holds. Otherwise, $a$ can leave the above cap only through the boundary of $B[p,q,b]$ (as depicted in the right figure), so the third scenario occurs.
\end{proof}

Notice, however, that the points of $P$ can define $\Omega(n^3)$ collinearities, so a naive charging of extremal Delaunay co-circularities to collinearities of type (1) or (2) in Lemma \ref{Lemma:MustCross} will not lead to a near-quadratic upper bound. Before we get to this (major) issue in our analysis, we begin by laying down the infrastructure of our charging scheme, similar to the one used in \cite{Natan}.

We fix some sufficiently large constant parameter $k>12$ and 
apply Theorem \ref{Thm:RedBlue} to the edge $pq$ over the interval $(t_0,t_1)$ of its absense from $\DT(P)$.
Assume first that one of the conditions (i) or (ii) of the theorem holds, so we can charge the co-circularity of $p,q,a,$ and $b$ either to $\Omega(k^2)$ $k$-shallow
co-circularities (each involving $p,q,$ and some two other points of $P$), or to a $k$-shallow collinearity (involving $p,q,$ and some third point of $P$). As argued in Section \ref{Sec:Prelim}, the overall number of $k$-shallow co-circularities is $O(k^4 N(n/k))$. Each $k$-shallow co-circularity is charged by only $O(1)$ Delaunay co-circularities in this manner,\footnote{Indeed, there are at most $O(1)$ ways to guess $p$ and $q$ among the four points of the charged co-circularity, and then the charging co-circularity corresponds to the latest previous disappearance of $pq$ from $\DT(P)$.} and it has to ``pay" only $O(1/k^2)$ units every time it is charged. Similarly, as already argued, the number of $k$-shallow collinearities is $O(kn^2\beta(n))$, and each such collinearity is charged by at most $O(1)$ Delaunay co-circularities. Hence, there are at most  $O(k^2N(n/k)+kn^2\beta(n))$ Delaunay co-circularities for which one of the conditions (i) or (ii) holds. 

Assume then that condition (iii) holds for our co-circularity. By assumption, there is a set $A$ of at most $3k$ points (necessarily including at least one of $a$ or $b$) whose removal ensures the Delaunayhood of $pq$ throughout $(t_0,t_1)$. By Lemma \ref{Lemma:MustCross}, at least one the two points $a,b$, let it be $a$, crosses $pq$ during $(t_0,t_1)$. As we will shortly show, in the reduced triangulation\footnote{To simplify the ongoing discourse, we apply slight abuse of notation, where we refer to certain non-Delaunay events as occurring {\it in} a suitable triangulation. These events are closely related to the changes that the triangulation undergoes, even though they themselves are not part of the Delaunay triangulation.} $\DT(P\setminus A\cup\{a\})$, the collinearity of $p,q$ and $a$ can be turned into one or several Delaunay crossings.


We can now express the number of remaining Delaunay co-circularities of index $3$ in terms of the maximum possible number of Delaunay crossings. Recall that for each such co-circularity there is a set $A$ of at most $3k$ points whose removal restores the Delaunayhood of $pq$ throughout $[t_0,t_1]$. In addition, we assume that $a$ hits $pq$ during $(t_0,t_1]$, and then $a\in A$.

We sample at random (and without replacement) a subset $R\subset P$ of $n/k$ points, and notice that the following two events occur simultaneously with probability at least $\Omega(1/k^3)$: (1) the points $p,q,a$ belong to $R$, and (2) none of the points of $A\setminus\{a\}$ belong to $R$. 
Since $a$ crosses $pq$ during $[t_0,t_1]$, and $pq$ is Delaunay at time $t_0$ and (right after) time $t_1$, the sample $R$ induces a Delaunay crossing $(pq,a,I)$, for some time interval $I\subset [t_0,t_1]$. (If $a$ crosses $pq$ twice, we have either two separate Delaunay crossings, which occur at disjoint sub-intervals of $(t_0,t_1)$, or only one Delaunay crossing, during which $a$ crosses $pq$ twice. This depends on whether $pq$ manages to become Delaunay in $\DT(R)$ in between these crossings.)
We charge the disappearance of $pq$ from $\DT(P)$ to this crossing (or to the first such crossing if there are two) and note that the charging is unique (i.e., every Delaunay crossing $(pq,a,I)$ in $\DT(R)$ is charged by at most one disappearance $t_0$ of the respective edge $pq$ from $\DT(P)$, which is {\it last} such disappearance of $pq$ before $a$ hits $pq$ in $I$).
Hence, the number of Delaunay co-circularities of this kind is bounded by $O(k^3 C(n/k))$, where $C(n)$ denotes, as above, the maximum number of Delaunay crossings induced by any collection $P$ of $n$ points whose motion satisfies the above assumptions.


If the Delaunay co-circularity of $p,q,a,b$ has index $1$, we reverse the direction of the time axis and argue as above for the edge $ab$ instead of $pq$. We thus obtain the following recurrence for the maximum possible number $N_E(n)$ of extremal Delaunay co-circularities:

\begin{equation}\label{Eq:FirstRecurrence}
N_E(n)=O\left(k^3C(n/k)+k^2N(n/k)+kn^2\beta(n)\right).
\end{equation}

\paragraph{Remark.} Our analysis will generate many recurrences of similar nature. Informally, each recurrence will have ``quadratic" terms (such as the second and the third terms in (\ref{Eq:FirstRecurrence})), which, in themselves, lead to a near-quadratic bound, and ``non-quadratic" terms (such as the first one in (\ref{Eq:FirstRecurrence})), which delegate the charging to new quantities. These quantities will generate recurrences of their own, of a similar nature, and the process will bottom out, in Section \ref{Sec:Terminal}, with recurrences that have only ``quadratic" terms. Using known techniques, such as in \cite{Envelopes3D} and \cite[Section 7.3.2]{SA95}, the whole system of recurrences will yield a near quadratic bound (for all the involved quantities).

\medskip
\noindent{\bf The number of middle Delaunay co-circularities.} We now develop a recurrence that expresses the number of middle Delaunay co-circularities in terms of $C(n/k)$, $N_E(n/k)$, and $N(n/k)$, for an appropriate constant parameter $k$.

Consider such a middle co-circularity event at time $t_0$, when an edge $pq$ of $\DT(P)$ is replaced by another edge $ab$.
As in the previous case, there are at most $O(n^2)$ such events for which the vanishing edge $pq$ never reappears in $\DT(P)$, so we focus on middle Delaunay co-circularities  whose corresponding edge $pq$ rejoins $\DT(P)$ at some future moment $t_1>t_0$.

Once again, we fix a sufficiently large constant $k>12$ and apply Theorem \ref{Thm:RedBlue} to the red-blue arrangement of $pq$ over the interval $(t_0,t_1)$. Assume first that one of the Conditions (i) and (ii) is satisfied, or that one of the points $a,b$ hits $pq$ during $(t_0,t_1]$. Then the preceding analysis (used for extremal Delaunay co-circularities) can be applied, essentially verbatim, in this case too, and it implies that the number of such middle co-circularities is 
$O\left(k^3C(n/k)+k^2N(n/k)+kn^2\beta(n)\right)$.

Assuming that the above scenario does not occur,
the four points $p,q,a,b$ are involved in an additional red-blue co-circularity during $(t_0,t_1]$, which ``frees" $pq$ from its violation by $a$ and $b$. Moreover, there is a set $A$ of at most $3k$ points whose removal restores the Delaunayhood of $pq$ throughout $[t_0,t_1]$. 
Let $t_0\leq t^*\leq t_1$ be the time of the additional (third) co-circularity of $p,q,a,b$, and
let $B^*$ be the corresponding circumdisc of $p,q,a,b$ at time $t^*$.

If $B^*$ contains at most $14k$ points, we can charge the disappearance of $pq$ to the resulting $14k$-shallow extremal co-circularity. Clearly, any such co-circularity of index $3$ is charged for at most one middle Delaunay co-circularity. Moreover, the number of $14k$-shallow extremal co-circularities is bounded by $O\left(k^4N_E(n/k)\right)$ using the standard probabilistic argument of Clarkson and Shor \cite{CS}.
Hence, this scenario arises for at most $O\left(k^4N_E(n/k)\right)$ middle Delaunay co-circularities.

\begin{figure}[htb]
\begin{center}
\input{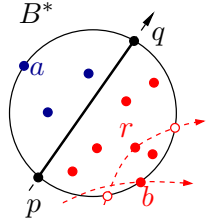}
\caption{\small Analysis of middle Delaunay co-circularities. The four points $p,q,a,b$ are involved, during $[t_0,t_1]$, in their third co-circularity, whose respective circumdisc $B^*$ contains at least $7k$ red points. At least $k$ red points $r$, whose red level ranges between $3k$ and $7k$, do not hit $pq$ during $[t_0,t_1]$.}
\label{Fig:Middle}
\end{center}
\vspace{-0.3cm}
\end{figure}

Now assume that $B^*$ contains at least $14k$ points of $P$. Without loss of generality, assume that the cap $B\cap \L_{pq}^+$ contains at least $7k$ points of $P$. That is, the corresponding red function, say $f_b^+$, has level at least $7k$ in the red arrangement at time $t^*$. 
Refer to Figure \ref{Fig:Middle}.
Let $r$ be a red point whose respective function $f_r^+$ lies, at time $t^*$, at red level between $3k$ and $7k-1$. That is, the number of red points in the circumdisc $B[p,q,r]$ ranges from $3k$ to $7k-1$. Then the number of blue points in $B[p,q,r]$ is at most $3k$. Indeed, if there were more that $3k$ blue points in $B[p,q,r]$ then after removing $A$ this disc would still contain at least one blue point and at least one red point (possibly $r$ itself), so $pq$ could not be Delaunay at time $t^*$. 
Since $f_r^+<f_b^+$, this disc also contains $a$ (which is still a blue point on the boundary of $B[p,q,b]$), so the Delaunayhood of $pq$ is violated at time $t^*$ by $r$ and $a$. Before $pq$ re-enters $\DT(P)$ at time $t_1$, one of the following must happen, according to Lemma \ref{Lemma:MustCross}: Either $r$ hits\footnote{Recall that, by assumption, $a$ does not hit $pq$ in the present case.} $pq$ or the points $p,q,r,a$ are involved in a red-blue co-circularity (when $a$ leaves $B[p,q,r]$ and before $r$ hits $\L_{pq}$). A fairly symmetric argument shows that either $r$ hits $pq$, or $p,q,r,a$ are involved in a red-blue co-circularity during $(t_0,t^*)$ (when $a$ enters $B[p,q,r]$). 
Note, however, that $pq$ is hit by at most $3k$ points during $(t_0,t_1]$, all of them in $A$. Thus, at least $k$ such points $r$
do not hit $pq$ during $(t_0,t_1]$, so each of them is
involved in two co-circularities with $p,q,a$ during $(t_0,t_1]$: one before $t^*$, and another afterwards. 

Fix a point $r$, as above, which does not cross $pq$. 
Notice that at least one of the two promised co-circularities of $p,q,r,a$ is extremal.
If the above extremal co-circularity of $p,q,r,a$, occuring at some $t^{**}\in (t_0,t_1)$, is $(11k)$-shallow, we charge it for the disappearance of $pq$. As before, this charging is unique, and the number of charged co-circularities is $O(k^4N_E(n/k))$. 
Otherwise, the boundary of $B[p,q,r]$ is crossed during the interval $(t^{*},t^{**})$ (or $(t^{**},t^{*})$) by at least $k$ points, so the triple $p,q,r$ defines $\Omega(k)$ $(11k)$-shallow co-circularities involving $p,q$ during $(t_0,t_1)$. 

Repeating the same argument for the (at least) $k$ possible choices of $r$, we obtain $\Omega(k^2)$ $(11k)$-shallow co-circularities, each involving $p,q$ and some other pair of points and occurring during $(t_0,t_1]$. As in Case (ii) of Theorem \ref{Thm:RedBlue}, we charge these co-circularities for the disappearance of $pq$.

We have thus established the following recurrence for the maximum possible number $N_M(n)$ of middle Delaunay co-circularities for a set of $n$ moving points:

\begin{equation}\label{Eq:MiddleRecurrence}
N_M(n)=O\left(k^4N_E(n/k)+k^2N(n/k)+kn^2\beta(n)+k^3C(n/k)\right).
\end{equation}

Informally, and as will be argued rigorously later on, the combination of (\ref{Eq:FirstRecurrence}) and (\ref{Eq:MiddleRecurrence}) implies that the maximum number of extremal Delaunay co-circularities is asymptotically dominated by the maximum number of Delaunay crossings (assuming it is at least quadratic).

\section{The Number of Delaunay crossings}\label{Sec:CrossOnce}
The remainder of the paper is devoted to deriving a recurrence relation for the maximum number $C(n)$ of Delaunay crossings induced by any set $P$ of $n$ moving points as above.
In this section we establish several basic properties of Delaunay crossings, and outline the forthcoming stages of their analysis.
The eventual system of recurrences that we will derive will express $C(n)$ in terms of the maximum number of Delaunay co-circularities of smaller-size sets, plus a nearly quadratic additive term.
Plugging that relation into (\ref{Eq:FirstRecurrence}) will yield the near-quadratic bound on $N(n)$ that was asserted in Theorem \ref{Thm:OverallBound}.

\subsection{Delaunay crossings: the key properties}\label{Subsec:SingleProp}
Consider a Delaunay crossing $(pq,r,I)$. Recall that $p,q,r$ can be collinear at most twice.
Moreover, both collinearities can (but do not have to) occur during the interval $I$ of the same Delaunay crossing of $pq$ by $r$. Clearly, $r$ cannot hit $\L_{pq}$ {\it outside $pq$} during $I$ because, at such an ``outer" collinearity, $pq$, which is Delaunay when $r$ is removed, would also be Delaunay in the presence of $r$.

The Delaunay crossing of $pq$ by $r$ is called {\it single} (resp., {\it double}) if $r$ hits $pq$ exactly once (resp., twice) during the corresponding interval $I$ of $pq$'s absence from $\DT(P)$. 

\begin{figure}[htbp]
\begin{center}
\input{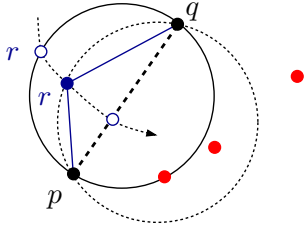}
\caption{\small Lemma \ref{Lemma:Crossing}. If $(pq,r,I)$ is a Delaunay crossing, then each of $pr,rq$ belongs to $\DT(P)$ throughout $I$.}
\label{Fig:StayDelaunay1}
\end{center}
\end{figure}

The following lemma holds for both types of Delaunay crossings (see Figure \ref{Fig:StayDelaunay1}).

\begin{lemma}\label{Lemma:Crossing}
If $(pq,r,I=[t_0,t_1])$ is a Delaunay crossing then each of the edges $pr,rq$ belongs to $\DT(P)$ throughout $I$.
\end{lemma}

Lemma \ref{Lemma:Crossing}, whose explicit proof appears in the predecessor paper \cite{Natan}, is a direct corollary of the following well-known result on {\it static} Delaunay triangulations:

\begin{lemma}\label{Lemma:Incremental}
Let $Q$ be a finite set of points in $\reals^2$, and let $r$ be a point not in $Q$. Let $pq$ be an edge that is Delaunay in $Q$, but not in $Q\cup \{r\}$. Then the triangulation $\DT(Q\cup \{r\})$ includes the two edges $pr$ and $qr$.
\end{lemma}

For the sake of completeness, we prove Lemma \ref{Lemma:Incremental} in Appendix \ref{AppendSec:Incremental}.


In the full version of the predecessor paper \cite{Natan}, we obtain an upper bound of $O(n^2)$ on the number of double Delaunay crossings. Since the argument from \cite{Natan} holds (as is) also in the setting studied by this paper, we have the following theorem.

\begin{theorem}\label{Thm:SpecialCrossings}
Any set $P$ of $n$ moving points, as above, induces at most $O(n^2)$ double Delaunay crossings.
\end{theorem}

For the sake of completeness, we supply the complete analysis of double Delaunay crossings in Appendix \ref{Subsec:Double}.

It therefore 
suffices to establish a suitable recurrence for the maximum possible number of single Delaunay crossings, and this is what is undertaken in the the remainder of the paper is devoted to the study of the latter crossings. 
For the sake of brevity, we shall often refer to single Delaunay crossings simply as Delaunay crossings, and use $C(n)$ to denote the maximum number of single Delaunay crossings.

We next establish several topological properties of (single) Delaunay crossings.


\paragraph{Single Delaunay crossings: notational conventions.}
Recall from Section \ref{Sec:Prelim} that every edge $pq$ is oriented from $p$ to $q$, and its corresponding line $\L_{pq}$ splits the plane into the left halfplane $\L_{pq}^-$ and the right halfplane $\L_{pq}^+$.

Without loss of generality, we assume in what follows that, for any single Delaunay crossing $(pq,r,I=[t_0,t_1])$, the point $r$ crosses $pq$ from $\L_{pq}^-$ to $\L_{pq}^+$ during $I$. Recall that $r$ cannot cross $\L_{pq}$ outside $pq$ during $I$, so this is the {\it only} collinearity of $p,q,r$ in $I$.
If $r$ crosses $pq$ in the opposite direction, we denote this crossing as $(qp,r,I=[t_0,t_1])$.

Note that every such Delaunay crossing $(pq,r,I)$ is uniquely determined by the respective ordered triple $(p,q,r)$, because there can be at most one collinearity\footnote{If $r$ hits $pq$ twice, then the other crossing of $pq$ by $r$ is from $\L_{pq}^+$ back to $\L_{pq}^-$.} where $r$ crosses the line $\L_{pq}$ within $pq$ from $\L_{pq}^-$ to $\L_{pq}^+$.

For convenience of reference, we label each such crossing $(pq,r,I)$ as {\it a clockwise $(p,r)$-crossing}, and as {\it a counterclockwise $(q,r)$-crossing}, with an obvious meaning of these labels.



The following lemma lies at the heart of our analysis.

\begin{figure}[htbp]
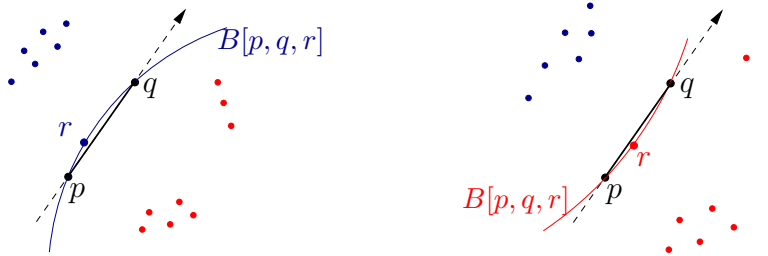

\begin{center}
\input{BeforeCrossing.pstex_t}\hspace{3cm}\input{AfterCrossing.pstex_t}
\caption{\small Left: Right before $r$ crosses $pq$, the circumdisc $B=B[p,q,r]$ contains all points in $P\cap \L_{pq}^+$. Right: Right after $r$ crosses $pq$, $B$ contains all points in $P\cap \L_{pq}^-$.}
\label{Fig:BeforeCrossing}
\end{center}
\end{figure}

\begin{lemma}\label{Lemma:OnceCollin}
Let $(pq,r,I=[t_0,t_1])$ be a single Delaunay crossing. Then, with the above conventions, for any $s\in P\setminus\{p,q,r\}$ the points $p,q,r,s$ define a red-blue co-circularity with respect to $pq$, which occurs during $I$ when the point $s$ either enters the cap $B[p,q,r]\cap \L_{pq}^+$, or leaves the opposite cap $B[p,q,r]\cap \L_{pq}^-$.
\end{lemma}
\begin{proof}
The proof is an adaptation of similar arguments made earlier.
By definition, $r$ crosses $pq$ at some (unique) time $t_0<t^*<t_1$ from $\L_{pq}^-$ to $\L_{pq}^+$.
The disc $B[p,q,r]$ is $P$-empty at $t_0$ and at $t_1$ and moves continuously throughout $[t_0,t^*)$ and $(t^*,t_1]$. Just before $t^*$, $B[p,q,r]$ is the entire $\L_{pq}^+$, so every point $s\in P\cap \L_{pq}^+$ at time $t^*$ must have entered $B[p,q,r]$ during $[t_0,t^*)$, forming a co-circularity with $p,q,r$ at the time it entered the disc.\footnote{If $t^*=t_0$ then
there are no red points when $r$ hits $pq$, so we consider only the second interval. The case of $t^*=t_2$ is treated symmetrically. As noted in Section \ref{Sec:Prelim}, in such cases the crossed edge $pq$ either leaves or joins the convex hull of $P$ at the time of the collinearity.} See Figure \ref{Fig:BeforeCrossing} (left). (As mentioned in Section \ref{Sec:Prelim}, this co-circularity of $p,q,r,s$ is red-blue with respect to $pq$, that is, the point $s$ enters $B[p,q,r]$ through $\partial B[p,q,r]\cap \L_{pq}^+$.) A symmetric argument (in which we reverse the direction of the time axis) shows that the same holds for all the points $s\in P$ that lie in $\L_{pq}^-$ at time $t^*$; see Figure \ref{Fig:BeforeCrossing} (right). 
\end{proof}


Our local charging schemes ``bottom out" when a carefully chosen triple of points defines two Delaunay crossings (again, possibly in a triangulation of some smaller-size sample).
Lemma \ref{Lemma:TwiceCollin} takes care of this easy case.

\begin{lemma}\label{Lemma:TwiceCollin}
The number of triples of points $p,q,r\in P$ for which there exist two time intervals $I_1,I_2$ such that either (i) both $(pq,r,I_1)$ and $(qp,r,I_2)$ are Delaunay crossings, (ii) both $(pq,r,I_1)$ and $(rq,p,I_2)$ are Delaunay crossings, or (iii) both $(pq,r,I_1)$ and $(pr,q,I_2)$ are Delaunay crossings, is at most $O(n^2)$. 
\end{lemma}

Notice that, if some triple of points $p,q,r$ in $P$ performs two distinct Delaunay crossings, both of these crossings must necessarily be single Delaunay crossings (otherwise this triple would be collinear at least three times). Hence, the statement of the lemma holds in full generality.
It is easy to check that Lemma \ref{Lemma:TwiceCollin} covers all possible scenarios (up to a permutation of $p,q,r$ and/or reversal of the time axis) where some triple $p,q,r$ is involved two single Delaunay crossings (again, because no three points of $P$ can be collinear more than twice). 

\begin{proof}
We claim that every pair $p,q\in P$ participates in at most one triple of each type.
Indeed, fix $p,q\in P$ and assume that there exist two points $r,s$ such that the triples $p,q,r$ and $p,q,s$ are involved in two (single) Delaunay crossings of the same prescribed order type (i), (ii), or (iii).
By Lemma \ref{Lemma:OnceCollin}, we encounter at least one co-circularity of $p,q,r,s$ during each of the two Delaunay crossings induced by $p,q,r$ and the two induced by $p,q,s$. If we show that these four co-circularities are distinct, we reach a contradiction to the fact that any four points can be co-circular at most three times.

If the aformentioned triples $p,q,r$ and $p,q,s$ satisfy the first condition, the resulting four crossings of $pq$ happen during pairwise disjoint intervals of time. Hence, the four co-circularities are clearly distinct.

We now proceed to establish the distinctness in the second and the third cases.
Assume next that both $(p,q,r)$ and $(p,q,s)$ fall into Case (ii); Case (iii) is handled in a fully symmetric manner.
By assumption, we have four points $p,q,r,s$ and four time intervals $I_1,I_2,I_3,I_4$, such that $(pq,r,I_1),(rq,p,I_2),(pq,s,I_3)$, and $(sq,p,I_4)$ are all Delaunay crossings. $I_1$ and $I_3$ are clearly disjoint, and Lemma \ref{Lemma:OnceCollin} yields two co-circularities of $p,q,r,s$, one occuring during $I_1$ and one during $I_3$, both red-blue with respect to $pq$.
Similarly, Lemma \ref{Lemma:OnceCollin} yields a co-circularity of $p,q,r,s$ during $I_2$ which is red-blue with respect to $qr$, and a co-circularity of the same quadruple during $I_4$, which is red-blue with respect to $qs$. Clearly, these two co-circularities are different, and are also different from the former two co-circularities, since the vertex opposite to $q$ is different in each of these co-circularities. This completes the proof of the lemma.
\end{proof}


The following lemma defines a natural order on $(p,r)$-crossings of a given orientation (clockwise or counterclockwise). 
\begin{lemma}\label{Lemma: OrderOrdinaryCrossings}
Let $(pq,r,I)$ and $(pa,r,J)$ be clockwise $(p,r)$-crossings, and suppose that $r$ hits $pq$ (during $I$) before it hits $pa$ (during $J$). Then $I$ begins (resp., ends) before the beginning (resp., end) of $J$. Clearly, the converse statements hold too. Similar statements hold for pairs of counterclockwise $(p,r)$-crossings.
\end{lemma}
\begin{proof}
In the configuration considered in the main statement of the lemma, $r$ crosses $pq$ from $\L_{pq}^-$ to $\L_{pq}^+$, and it crosses $pa$ from $\L_{pa}^-$ to $\L_{pa}^+$.
We only prove the part of the lemma concerning the ending times of the crossings, because the proof about the starting times is fully symmetric (by reversing the direction of the time axis). The statement clearly holds if $I$ and $J$ are disjoint; the interesting situation is when they partially overlap.
Note that $r$ enters $\L_{pq}^+$ only once during the Delaunay crossing of $pq$ by $r$, namely, right after $r$ hits $pq$. Indeed, by assumption, $r$ cannot exit $\L_{pq}^+$ by crossing $pq$ again during $I$, and it cannot cross $\L_{pq}\setminus pq$ because at that time $pq$, which is Delaunay in $\DT(P\setminus\{r\})$, would be Delaunay also in the presence of $r$, contrary to the definition of a Delaunay crossing.
Hence, we may assume that $r$ still lies in $\L_{pq}^+$ when it hits $pa$ during the Delaunay crossing of that edge. Indeed, otherwise the crossing of $pq$ would by then be over, so the claim would hold trivially, as noted above. In particular, $\vec{pa}$ lies clockwise to $\vec{pq}$ at that time.

\begin{figure}[htbp]
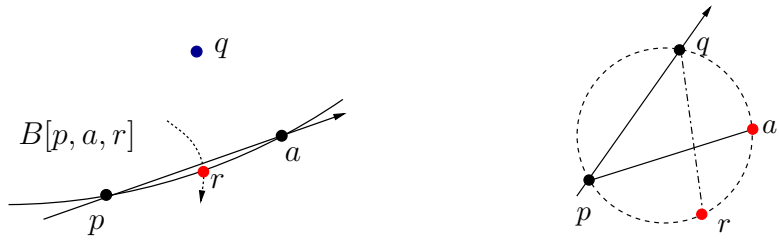

\begin{center}
\input{orderord1.pstex_t}\hspace{3cm}\input{OrderCrossings.pstex_t}
\caption{\small Proof of Lemma \ref{Lemma: OrderOrdinaryCrossings}. Left: if $r$ remains in $\L_{pq}^+$ after $I$ and before it crosses $pa$, then $q$ lies in $B[p,a,r]\cap\L_{pa}^+$ before that last collinearity. Right: The second co-circularity of $p,q,r,a$ which occurs when $q$ leaves $B[p,a,r]\cap \L_{pq}^+$. This is a red-red co-circularity with respect to $pq$, so the crossing of $pq$ is already over.}
\label{Fig:OrderOrdinaryCrossings}
\end{center}
\end{figure} 

It suffices to prove that the co-circularity of $p,q,r,a$, which (by Lemma \ref{Lemma:OnceCollin}) occurs during the Delaunay crossing of $pa$ by $r$, takes place when the crossing of $pq$ by $r$ is already finished (and, in particular, after the co-circularity of $p,q,r,a$ that occurs during the crossing of $pq$). 

Before the Delaunayhood of $pa$ is restored, we have a co-circularity $p,q,r,a$ in which $q$ leaves $B[p,a,r]\cap \L_{pa}^-$. (This is argued in the proof of Lemma \ref{Lemma:OnceCollin}: Right after the crossing, the point $q$ lies in $B[p,a,r]\cap \L_{pa}^-$, as in Figure \ref{Fig:OrderOrdinaryCrossings} (left), and has to leave that disc before it becomes empty; it cannot cross $pa$ during $J$, when this edge undergoes the Delaunay crossing by $r$). Notice that this is a red-blue co-circularity with respect to $pa$, and a red-red co-circularity with respect to $pq$; see Figure \ref{Fig:OrderOrdinaryCrossings} (right). Since no red-red or blue-blue co-circularities occur during a Delaunay crossing of an edge, the crossing of $pq$ is already over.
\end{proof}

\paragraph{Consecutive crossings.} By Lemma \ref{Lemma: OrderOrdinaryCrossings}, for any pair of points $p,r$, all the clockwise $(p,r)$-crossings can be linearly ordered by the starting times of their intervals, or by the ending times of their intervals, or by the times when $r$ hits the corresponding $p$-edge, and all three orders are indentical. We say that clockwise $(p,r)$-crossings $(pq,r,I),(pa,r,J)$ are {\it consecutive} if they are consecutive in this order.
More generally, we say that these crossings are {\it $k$-consecutive} if at most $k$ other clockwise $(p,r)$-crossings separate them in this order.

Similar notions of consecutiveness and $k$-consecutiveness apply to pairs of counterclockwise $(p,r)$-crossings
$(qp,r,I),(ap,r,J)$.

\subsection{The roadmap}\label{Subsec:Roadmap}
In Section \ref{Sec:ReduceToCrossings} we have established a pair of recurrences (\ref{Eq:FirstRecurrence}) and (\ref{Eq:MiddleRecurrence}), whose combination allows to express the maximum number $N(n)$ of Delaunay co-circularities in terms of the maximum number of Delaunay crossings $C(m)$ in smaller-size subsets, plus the maximum number of Delaunay co-circularities in smaller-size sets, plus a nearly quadratic additive term. 
Furthermore, we have seen that there can be at most quadratically many double Delaunay crossings, and quadratically many of pairs of single Delaunay crossings of the kinds considered in Lemma \ref{Lemma:TwiceCollin}.

It therefore suffices to obtain a suitable recurrence, or a system of such recurrences, that express the maximum possible number $C(n)$ of (single) Delaunay crossings only in terms of the maximum number of Delaunay co-circularities in smaller-size sets, plus a nearly quadratic additive term. (In order for the solution of such a recurrence to be near-quadratic, the respective coefficient of each recursive term of the form $N(n/k)$ must be roughly equal to $k^2$.
See \cite{Envelopes3D}, \cite[Section 7.3.2]{SA95}, and also \cite[Section 4.5]{Overlay4D} for further details on solving such systems of recurrences.)

In the predecessor paper \cite{Natan}, we used the following fairly direct charging strategy.
For each single Delaunay crossing $(pq,r,I)$ in $P$ we first checked whether it (or its immediate neighbor) is near-extremal in the order implied by Lemma \ref{Lemma: OrderOrdinaryCrossings}. Notice that $(pq,r,I)$ appears (and thus can be extremal) in two restricted families of crossings: that of the clockwise $(p,r)$-crossings, and that of the counterclockwise $(q,r)$-crossings.
If this were the case, we could charge $(pq,r,I)$ to one of the edges $pr$ and $qr$, for an overall quadratic bound.
Otherwise, we applied Theorem \ref{Thm:RedBlue} in the arrangements $\A_{pr}$ and $\A_{rq}$, and tried to charge $(pq,r,I)$, within at least one of these two arrangements, either to a shallow collinearity, or to sufficiently many shallow co-circularities.
Finally, if none of the previous chargings succeeded, we charged $(pq,r,I)$ to some triple (not necessarily $p,q,r$) which performed two Delaunay crossings in some sub-sample of $P$, so our analysis bottomed out via (the weaker analogue in \cite{Natan} of) Lemma \ref{Lemma:TwiceCollin}.

Unfortunately, the above direct approach no longer works in the present setting, where any four points can be co-circular up to three times.
Informally, its main weakness stems from the fact that Delaunay crossings involve triples of points, whereas our primary topological restriction refers to quadruples of points of $P$.
Thus, Delaunay crossings are not ``rich" enough to capture the underlying combinatorial structure of the problem.

We therefore consider several additional types of topological configurations that involve {\it quadruples} of moving points, obtained by combining two Delaunay crossings with two common points, such as $(pq,r,I)$ and $(pa,r,J)$. 
Recall that, for each Delaunay crossing $(pq,r,I)$, its edge $pq$ is almost Delaunay in $I=[t_0,t_1]$ (and fully Delaunay at the endpoints $t_0,t_1$), and the other two edges $pr$ and $rq$ are fully Delaunay in $I$ (by Lemma \ref{Lemma:MustCross}). 
The quadruples that we will shortly introduce more formally, inherit all these properties of their Delaunay crossings, but will have a rich structure, due to additional interactions between their edges and subtriples. These quadruples can be viewed as an extension of Delaunay crossings, in the sense that their edges are forced to be either Delaunay, or almost Delaunay, during various intervals whose endpoints are defined ``locally", in terms of the points and the edges of the configuration at hand. Furthermore, initially, by construction, the points of each quadruple perform at least two Delaunay crossings. The major goal of the analysis is to obtain configurations with progressively many Delaunay crossings

We next review the three types of topological configurations that arise in the course of our analysis, and highlight the intimate relations between these types of configurations, and Delaunay crossings.

\begin{figure}[htbp]
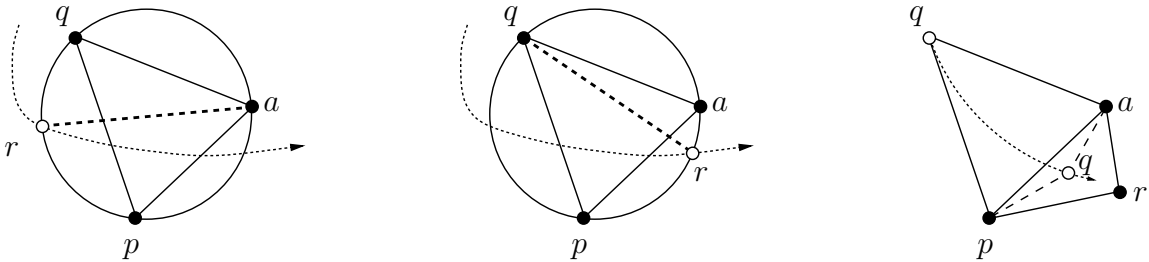

\begin{center}
\input{FirstCocircOrd.pstex_t}\hspace{2cm}\input{SecondCocircOrd.pstex_t}\hspace{2cm}\input{SpecialCrossingRoadmap.pstex_t}
\caption{\small A (clockwise) regular quadruple $\Q=(p,q,a,r)$, which is composed of clockwise $(p,r)$-crossings $(pq,r,I)$ and $(pa,r,J)$. Left and center: A possible motion of $r$, with the two co-circularities of $p,q,a,r$ that occur during $I\setminus J$ and $J\setminus I$, respectively. Right: The special crossing of $pa$ by $q$ which we enforce at the end of the analysis of regular quadruples.}
\label{Fig:TwoCocircsRoadmap}
\vspace{-0.4cm}
\end{center}
\end{figure} 

\paragraph{Regular quadruples.}
Four distinct points $p,q,a,r\in P$ form a clockwise {\it regular quadruple} (or, simply, a {\it quadruple}) $\Q=(p,q,a,r)$ in $\DT(P)$ if there exist clockwise $(p,r)$-crossings $(pq,r,I),(pa,r,J)$ that appear in this order in the sequence of clockwise $(p,r)$-crossings; refer to Figure \ref{Fig:TwoCocircsRoadmap}. We say that the quadruple is {\it consecutive} if $(pq,r,I)$ and $(pa,r,J)$ are consecutive.


Clearly, every clockwise $(p,r)$-crossing $(pq,r,I)$ forms the first part of exactly one (clockwise) consecutive quadruple, unless it is the last such $(p,r)$-crossing (with respect to the order given by Lemma \ref{Lemma: OrderOrdinaryCrossings}). The overall number of these last crossings is clearly bounded by $O(n^2)$. 
Hence, the maximum number $C(n)$ of single Delaunay crossings is asymptotically dominated by the maximum possible number $\Psi(n)$ of consecutive regular quadruples.

Let $\Q=(p,q,a,r)$ be a consecutive regular quadruple as above.
By Lemma \ref{Lemma:Crossing}, edge $pr$ of $\Q$ is Delaunay during the respective intervals $I$ and $J$ of its two $(p,r)$-crossings, whereas each of the edges $rq$ and $ra$ is (provably) Delaunay in only one of these two intervals.
In addition, the edges $pq$ and $pa$ are almost Delaunay during their respective Delaunay crossings by $r$.

Regular quadruples are studied extensively in Section \ref{Sec:CountQuad}, where we gradually extend the corresponding (almost-)Delaunayhood intervals of the respective edges $pr, rq, ra, pa$ and $pq$ of each quadruple $\Q$ until most of them cover $\conn{I}{J}={\sf conv}(I\cup J)$, including the possible gap between $I$ and $J$.
This is achieved by applying Theorem \ref{Thm:RedBlue} in the respective red-blue arrangements of these edges.
Each such application of Theorem \ref{Thm:RedBlue} is done over a carefully chosen interval, which guarantees that any shallow collinearity or co-circularity, that we encounter in the first two cases of the theorem, is charged by only few quadruples.

In Section \ref{Subseq:Quad}, we show (via Lemmas \ref{Lemma:Crossing} and \ref{Lemma:OnceCollin}) that the points of each regular quadruple $\Q=(p,q,a,r)$ are co-circular exactly once in each of the intervals $I\setminus J$ and $J\setminus I$; see Figure \ref{Fig:TwoCocircsRoadmap} (left and center). Specifically, the former co-circularity is red-blue with respect to the edges $pq$ and $ra$, and the latter co-circularity is red-blue with respect to $pa$ and $rq$. 
Notice that at least one of these co-circularities, let it be the one in $I\setminus J$, is extremal.

Arguing similarly to Section \ref{Sec:ReduceToCrossings}, we use the above co-circularities of $p,q,a,r$ (together with the additional constraints on the Delaunayhood of $rq,ra$ and $pa$) to enforce a pair of additional Delaunay crossings which occur in smaller-size point sets (which are random samples of $P$, needed for the application of the Clarkson-Shor argument \cite{CS}) and involve various sub-triples of $p,q,a,r$.
Thr analysis in Section \ref{Sec:CountQuad} is fairly involved, due to the fact that neither of the above two co-circularities of $\Q$ has to be Delaunay, or even shallow.
If some sub-triple of $\Q$ performs two Delaunay crossings, we immediately bottom out via Lemma \ref{Lemma:TwiceCollin}. 

Unfortunately, there may still exist quadruples $\Q$ whose four resulting Delaunay crossings (including the two original $(p,r)$-crossings $(pq,r,I)$ and $(pa,r,J)$) involve four distinct sub-triples $p,q,a,r$, so Lemma \ref{Lemma:TwiceCollin} cannot yet be applied. 
As our analysis shows, in this only remaining scenario, the edge $pa$ of $\Q$ undergoes a Delaunay crossing $(pa,q,\I)$ by $q$; see Figure \ref{Fig:TwoCocircsRoadmap} (right). We refer to this latter crossing as a {\it special crossing} of $pa$ by $q$, and pass the analysis of such crossings, each accompanied by a regular quadruple that induces it, to Section \ref{Sec:Special}.

\paragraph{Special quadruples.}
In Section \ref{Sec:Special} we analyze the number of special (counterclockwise) crossings by first arranging them into {\it special quadruples}. Informally, each special quadruple $\SQ=(a,p,w,q)$ is composed of two special $(a,q)$-crossings $(pa,q,\I)$ and $(wa,q,\J)$ which are consecutive in the order implied by Lemma \ref{Lemma: OrderOrdinaryCrossings}. See Figure \ref{Fig:OuterPointsRoadmap}.

\begin{figure}[htbp]
\begin{center}
\input{OuterPoints.pstex_t}
\caption{\small A (counterclockwise) special quadruple $\SQ=(a,p,w,q)$, is composed of two special crossings $(pa,q,\I)$ and $(wa,q,\J)$, which respectively correspond to some (clockwise) regular quadruples $(p,q,a,r)$ and $(w,q,a,u)$.}
\label{Fig:OuterPointsRoadmap}
\vspace{-0.4cm}
\end{center}
\end{figure} 

The treatment of (counterlockwise) special quadruples is fairly symmetric to that of (clockwise) regular quadruples, in the manner in which we extend the Delaunayhood or almost-Delaunayhood of their edges, and enforce additional (almost-)Delaunay crossings on some of their sub-triples. 
However, here we have a richer topological structure, because the two special crossings $(pa,q,\I)$ and $(wa,q,\J)$ of each special quadruple $\SQ$ are accompanied by two respective regular quadruples $\Q_1=(p,q,a,r)$ and $\Q_2=(w,q,a,u)$ that induce them.

At the final stage of the analysis (and only there), we use the above correspondence with the regular quadruples in order to charge
the surviving special quadruples $\SQ$ to especially convenient topological configurations, referred to as {\it terminal quadruples}.

\begin{figure}[htbp]
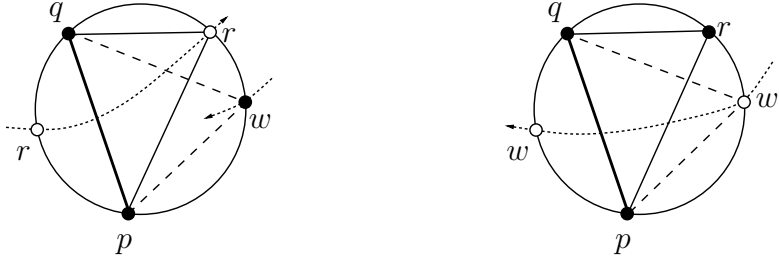

\begin{center}
\input{TerminalR.pstex_t}\hspace{3cm}\input{TerminalW.pstex_t}
\caption{\small A terminal quadruple $\TQ=(p,q,r,w)$. The points $r$ and $w$ cross $pq$ in opposite directions.
The points of $\TQ$ are co-circular three times. The extremal two co-circularities are red-blue with respect to $pq$, and the middle one is monochromatic with respect to $pq$. The left figure depicts the first and second co-circularities, and the right figure depicts the second and third co-circularities.}
\label{Fig:TerminalRoadmap} 
\vspace{-0.4cm}
\end{center}
\end{figure}

\paragraph{Terminal quadruples.} 
Each terminal quadruple $\TQ=(p,q,r,w)$ is formed by an edge $pq$, and by a pair of points $r$ and $w$ that cross $pq$ in {\it opposite} directions;\footnote{The letters $p,q,r,w$ designate the way in which a terminal quadruple is extracted from the 6-point configuration of the surviving special quadruple $\SQ=(a,p,w,q)$ and its respective pair of regular quadruples $\Q_1=(p,q,a,r)$ and $\Q_2=(w,q,a,u)$.} see Figure \ref{Fig:TerminalRoadmap}. 
In addition, $\TQ$ must satisfy several ``local" restrictions on the Delaunayhood of its various edges, and on the co-circularities and collinearities among $p,q,r,w$.
The analysis of these configurations is delegated to Section \ref{Sec:Terminal}, where we directly bound their number in terms of simpler quantities, introduced in Section \ref{Sec:Prelim}, and thereby complete the proof of Theorem \ref{Thm:OverallBound}.
(We again emphasize that the recurrences that bound the number of terminal quadruples must have only ``quadratic" terms.)

Informally, the analysis of terminal quadruples manages to bottom out (in contrast to the one of regular quadruples) because each terminal quadruple comes with {\it three} ``well-behaved" co-circularities. Specifically, the two extremal co-circularities are red-blue with respect to the crossed edge $pq$ (and thus also with respect to $rw$), and the middle one is mononochromatic with respect to $pq$; see Figure \ref{Fig:TerminalRoadmap}. These patterns allow us to use these co-circularities to enforce {\it three} additional Delaunay crossings among $p,q,r,w$ (in addition to the crossings of $pq$ by $r$ and $w$). As a result, some sub-triple among $p,q,r,w$ is involved in two Delaunay crossings, so Lemma \ref{Lemma:TwiceCollin} can always be invoked.

\section{Regular Quadruples}\label{Sec:CountQuad}

\subsection{Notation and topology}\label{Subseq:Quad}
\paragraph{Definition.}
Four distinct points $p,q,a,r\in P$ form a {\it clockwise quadruple} $\Q=(p,q,a,r)$ in $\DT(P)$ if there exist clockwise $(p,r)$-crossings $(pq,r,I),(pa,r,J)$ that appear in this order in the sequence of clockwise $(p,r)$-crossings. We say that the quadruple is {\it consecutive} if $(pq,r,I)$ and $(pa,q,J)$ are consecutive.
The definitions of a {\it counterclockwise quadruple} and of a consecutive counterclockwise quadruple are similar.

Each quadruple $\Q$ is equipped with the intervals $I_\Q=I=[t_0,t_1]$ and $J_\Q=J=[t_2,t_3]$ during which the corresponding edges $pq$ and $pa$ are absent from $\DT(P)$.

Recall that, by Theorem \ref{Thm:SpecialCrossings}, any set of $n$ moving points admits at most $O(n^2)$ double Delaunay crossings.
Clearly, every clockwise (resp., counterclockwise) single $(p,r)$-crossing forms the first part of exactly one clockwise (resp., counterclockwise) consecutive quadruple, unless it is the last such $(p,r)$-crossing (with respect to the order given by Lemma \ref{Lemma: OrderOrdinaryCrossings}). The overall number of these last crossings is clearly bounded by $O(n^2)$. 
Therefore, using $\Psi(n)$ to denote in maximum possible number of consecutive clockwise quadruples in a set of $n$ moving points, we have the following obvious bound on the maximum number $C(n)$ of all Delaunay crossings:
$$
C(n)\leq \Psi(n)+O(n^2).
$$



\paragraph{The topology of quadruples.} 
According to Lemma \ref{Lemma:OnceCollin}, the points of a clockwise quadruple $\Q$ are involved in at least one co-circularity during $I_\Q$, and in at least one co-circularity during $J_\Q$. Specifically, the former co-circularity is red-blue with respect to $pq$ (and monochromatic with respect to $pa$), so it occurs before the beginning of $J_\Q$, during $I_\Q\setminus J_\Q$.
Similarly, the latter co-circularity is red-blue with respect to $pa$ (and monochromatic with respect to $pq$), so occurs after the end of $I_\Q$, during $J_\Q\setminus I_\Q$.

Notice that the points $p,q,r,a$ are involved in exactly one co-circularity during each of the intervals $I,J$.
Indeed, recall that the point $a$ lies outside the disc $B[p,q,r]$ right before $I_\Q$ begins and right after $I_\Q$ ends. Moreover, $B[p,q,r]$ switches instantly from $\L_{pq}^+$ to $\L_{pq}^-$ only once during $I_\Q$, so $a$ hits the boundary $B[p,q,r]$ an odd number of times during $I_\Q$. A symmetric behaviour takes place during $J_\Q$, so the points $p,q,a,r$ are involved in exactly one co-circularity in each interval.

\begin{lemma}\label{Lemma:ReturnsStillCross}
Let $\Q=(p,q,a,r)$ be a clockwise quadruple with the associated Delaunay crossings $(pq,r,I_\Q=[t_0,t_1])$ and $(pa,r,J_\Q=[t_2,t_3])$ (occuring in this order). Assume also that the point $r$ hits $pq$ again after $I_\Q$ and before $r$ hits $pa$ (and enters $\L^+_{pa}$) during $J_\Q$. Then (with the conventions assumed above) the edge $rq$ is hit during $(t_1,t_3)$ by the point $a$, which crosses $\L_{rq}$ from $\L_{rq}^+$ to $\L_{rq}^-$.
\end{lemma}
Since the roles of $q$ and $a$ in $\Q$ are interchangable (by reversing the direction of the time axis), we also have a symmetric variant of the lemma, which applies if $r$ hits the edge $pa$ before $J_\Q$ but after it hits $pq$ during $I_\Q$. Symmetric versions of the lemma and this subsequent also hold if $\Q$ is a counterclockwise quadruple.
\begin{proof}
Let $\zeta_{1}$ denote the time in $J_\Q\setminus I_\Q$ when the points $p,q,a,r$ are co-circular, and recall that this co-circularity is red-blue with respect to $pa$.
Since any three points can be collinear at most twice, both points $r,a$ lie in $\L_{pq}^-$ when $r$ hits $pa$ during $J_\Q$ (this is because $r$ must lie in $\L_{pq}^-$ at that time, so $a$ also has to lie there when $r$ hits $pa$). Hence, $q$ lies then in $\L_{pa}^+$. Right before this event, $q$ lies in the cap $B[p,q,r]\cap\L_{pa}^+$. Arguing as in the proof of Lemma \ref{Lemma:OnceCollin}, the point $q$ enters the above cap at time $\zeta_1$; see Figure \ref{Fig:NotViolated} (left). In addition, the point $a$ leaves the cap $B[p,q,r]\cap\L_{rq}^-$ at the very same time $\zeta_1$.

\begin{figure}[htbp]
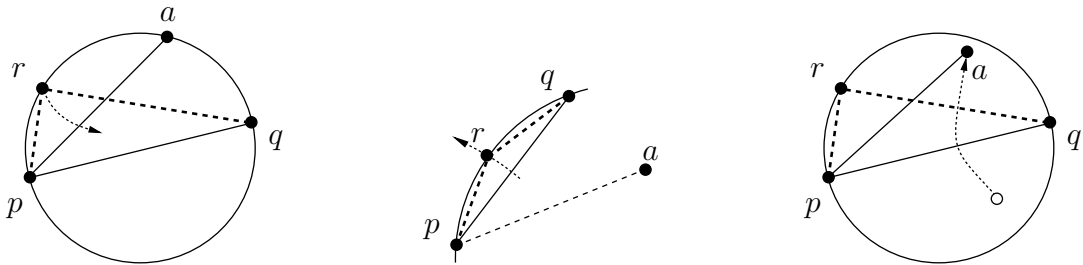

\begin{center}
\input{NotViolated.pstex_t}\hspace{2cm}\input{Beforereturn.pstex_t}\hspace{2cm}\input{EnterTheCap.pstex_t}
\caption{\small Illustrating the proof of Lemma \ref{Lemma:ReturnsStillCross}. Left: If $r$ hits $pq$ again before crossing $pa$, then $q$ enters $B[p,a,r]$ during the second co-circularity of $p,q,a,r$ (and $a$ leaves the cap $B[p,q,r]\cap \L_{rq}^-$). Center: The case where $a$ lies in the cap $B[p,q,r]\cap \L_{pq}^+$ right after $r$ returns to $\L_{pq}^-$. Right: The point $a$ can enter the cap $B[p,q,r]\cap \L_{rq}^-$ (without leaving $B[p,q,r]$) only through $rq$.}
\label{Fig:NotViolated}
\vspace{-0.3cm}
\end{center}
\end{figure} 

In particular, the preceding discussion implies that the second collinearity of $p,q,r$ occurs at some time $\tilde{t}$ before $\zeta_1$. Since $r$ can cross $\L_{pq}$ only twice, the motion of $B[p,q,r]$ remains continuous after time $\tilde{t}$ (when $B[p,q,r]$ instantly flips from $\L_{pq}^-$ to $\L_{pq}^+$). We distinguish between the following two cases.

(i) Assume first that $a$ lies in $\L_{pq}^+$ at time $\tilde{t}$, so
it lies in the cap $B[p,q,r]\cap \L_{pq}^+$ right afterwards; see Figure \ref{Fig:NotViolated} (center). 
The lemma clearly holds if the point $a$ remains in $B[p,q,r]$ during the interval $(\tilde{t},\zeta_1)$. Indeed, in this case $a$ lies in $\L_{rq}^+=\L_{pq}^+$ at time $\tilde{t}$, so it can enter the cap $B[p,q,r]\cap\L_{rq}^+$ (without leaving $B[p,q,r]$) only through the edge $rq$. See Figure \ref{Fig:NotViolated} (right). 
Furthermore, $a$ cannot leave $B[p,q,r]$ during $(\tilde{t},\zeta_1)$, because it would have to re-enter $B[p,q,r]$ before $\zeta_1$ (recall that it leaves $B[p,q,r]$ right after $\zeta_1$). But then the points of $\Q$ would have been involved in at least {\it four} distinct co-circularities, one occuring during $I_\Q$ and before time $\tilde{t}$, the two co-circularities just considered, both occuring during $(\tilde{t},\zeta_1)$, and one at $\zeta_1$ itself. This contradiction establishes the lemma in case (i).

\smallskip
(ii) Now suppose that $a$ lies in $\L_{pq}^-$ at time $\tilde{t}$. In this case, as in the proof of Lemma \ref{Lemma:OnceCollin}, $a$ lies in $B[p,q,r]$ right before $\tilde{t}$. Since $a$ lies outside $B[p,q,r]$ right after the end of $I_\Q$ (and since the motion of $B[p,q,r]$ is continuous between the two collinearities of $p,q,r$), the point $a$ has to cross the boundary of $B[p,q,r]$ after $I_\Q$ and before $\tilde{t}$. In addition, the point $a$ must now enter $B[p,q,r]$ during $(\tilde{t},\zeta_1)$, because it lies outside $B[p,q,r]$ right after $\tilde{t}$. Once again, we obtain four distinct co-circularities of $p,q,a,r$, a contradiction that shows that case (ii) is impossible, and thus completes the proof.
\end{proof}

\paragraph{Overview.}

In this section we analyze the maximum number of consecutive clockwise quadruples.
The underlying intuition behind our (admittedly, faily involved) analysis is the following. We analyze quadruples of four points $p,q,a,r$. The purpose of the analysis is to charge these quadruples to special restricted configurations that are easier to analyze. Theorem \ref{Thm:RedBlue} allows us to charge some quadruples to shallow co-circularities or collinearities, which forms the basis for various recurrences that the analysis will be deriving. In addition, Theorem \ref{Thm:SpecialCrossings} and Lemma \ref{Lemma:TwiceCollin} yield a quadratic bound 
for the number of quadruples that can be charged to a double Delaunay crossing of some triple of their points, or to two Delaunay crossings of the same triple.

Our strategy is therefore to filter away quadruples that can be charged by either of these tools, untill all quadruples are exhausted. To do so, we keep enforcing our quadruples to be involved with progressively more Delaunay crossings. Each quadruple is associated with four triples, and our goal is to force at least one triple of points to perform two Delaunay crossings, in which case Theorem \ref{Thm:SpecialCrossings} and Lemma \ref{Lemma:TwiceCollin} will yield the desired quadratic bounds.

Right from the start, a quadruple $\Q=(p,q,a,r)$ already has, by definition, two Delaunay crossings: of $pq$ by $r$, and of $pa$ by $r$. To enforce additional crossings, we need a careful (and involved) analysis of the ``topological" changes of the four moving points of $\Q$, where each event is either a collinearity of three of the points (in which case the order type of $p,q,a,r$ changes), or a co-circularity of the four points of $\Q$
(in which case the Delaunayhood of a pair of its edges ``flips").

The analysis of consecutive clockwise quadruples proceeds through six stages, numbered $0,1,\ldots,5$.
 At the $i$-th stage we consider a certain family $\F_i$ of clockwise quadruples, which are defined with respect to an underlying set $P$ of $n$ points moving as above in $\reals^2$.  (Initially, $\F_0$ consists of all consecutive quadruples in the original point set $P$. In subsequent stages, $P$ is a smaller sample from the original point set, but we continue, for simplicity, to denote it as $P$.) We assume that each quadruple $\Q$ in $\F_i$ satisfies certain topological conditions, which are formulated in terms of the four points of $\Q$, other points of $P$ (and, possibly, also nearby quadruples in $\F_i$). Our goal is 
to bound the maximum possible cardinality $\Psi_i(n)$ of $\F_i$. 
This is achieved by developing a system of recurrences, each expressing $\Psi_i$ in terms of $\Psi_{i+1}$, except for $\Psi_5$, which is analyzed in Section \ref{Sec:Special}. 
The overall solution of this system yields the desired near-quadratic bound.

\ignore{
To proceed, we apply the charging machinery of Sections \ref{Sec:Prelim} and \ref{Sec:CrossOnce} in order to dispose of a certain subset of $\F_i$. Each such argument involves either one, or several constants, say $k\ll \ell$.
If the charging does not work for a quadruple $\Q$, then we plug Theorem \ref{Thm:RedBlue} to obtain a small subset $A=A_\Q$ of at most $O(\ell)$ points so that $\Q$ satisfies a new condition within the reduced triangulation $\DT(P\setminus A)$. 
A routine sampling argument in the spirit of Clarkson and Shor \cite{CS} then leads to the following recurrence:
$$
\Psi_i(n)=O\left(\ell^4 \Psi_{i+1}(n/\ell)+k\ell^2N(n/\ell)+k^2N(n/k)+k\ell n^2\beta(n)\right),
$$
where $\Psi_{i+1}(m)$ denotes the maximum possible cardinality of the even more restricted family of quadruples (which is defined over the sample of $\lceil n/\ell\rceil$).
}

\subsection{Stage 0: Charging events in $\A_{pr}$}\label{Subsec:Stage0}
Let $\Q=(p,q,a,r)$ be a consecutive clockwise quadruple, whose two Delaunay crossings occur during the intervals $I=I_\Q=[t_0,t_1]$ and $J=J_\Q=[t_2,t_3]$. By Lemma \ref{Lemma:Crossing}, the edge $pr$ is Delaunay during each of the intervals $I,J$, but it may leave $\DT(P)$ during the possible gap between $I$ and $J$. 
\paragraph{Charging events in $\A_{pr}$.} We fix a constant $k>12$ and apply Theorem \ref{Thm:RedBlue} in $\A_{pr}$ over the interval $(t_1,t_3)$ (which covers the aforementioned gap between $I$ and $J$, if it exists).

First, assume that at least one of the Conditions (i), (ii) of Theorem \ref{Thm:RedBlue} holds.
In this case, we charge $\Q$ either to a $k$-shallow collinearity, or to $\Omega(k^2)$ $k$-shallow co-circularities, that occur in $\A_{pr}$ during $(t_1,t_3)$.
We claim that any $k$-shallow collinearity or co-circularity in $\A_{pr}$ is charged in this manner by at most $O(1)$ quadruples.
Indeed, consider the moment $t^*$ when the charged event occurs, and notice that it involves $p$ and $r$ (together with one or two additional points of $P$). After guessing $p$ and $r$ (in $O(1)$ ways), $\Q$ is the unique quadruple $(p,q,a,r)$ for which the interval $[t_1,t_3]$, delimited by the ending times of the two corresponding Delaunay crossing intervals, contains $t^*$ (by definition of consecutive quadruples, the intervals $[t_1,t_3]$ are pairwise openly disjoint, for $p$ and $r$ fixed).

Using the standard bounds on the number of $k$-shallow collinearities and co-circularities (established in Sections \ref{Sec:Prelim} and \ref{Sec:ReduceToCrossings}), in combination with the fact that each co-circularity pays only $\Theta(1/k^2)$ units when it is charged, we get that the number of such quadruples $\Q$ for which the red-blue arrangement of $pr$ satisfies one of the Conditions (i), (ii) of Theorem \ref{Thm:RedBlue}, is $O\left(k^2N(n/k)+kn^2\beta(n)\right)$. 

Assume then that the red-blue arrangement of $pr$ (during $(t_1,t_3)$) satisfies Condition (iii) of Theorem \ref{Thm:RedBlue}. That is, one can restore the Delaunayhood of $pr$ during $(t_1,t_3)$ by removing a set $A$ of at most $3k$ points of $P$ (possibly including $q$ and/or $a$).\footnote{Note that, if the gap between $I$ and $J$ does not exist, then $A=\emptyset$.}  
We now consider a random subset $R$ of $\Theta(n/k)$ points of $P$.
By the standard probabilistic argument
of Clarkson and Shor \cite{CS}, the following two events occur simultaneously with probability at least $\Theta(1/k^4)$: (1)
$p,q,a,r\in R$, and (2) none of the points of $A\setminus \{a,q\}$ belong to $R$.

\begin{figure}[htbp]
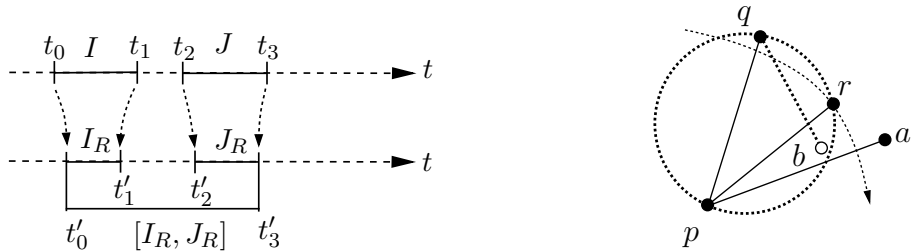

\begin{center}
\input{Intervals.pstex_t}\hspace{3cm}\input{DelQuad.pstex_t}
\caption{\small Left: The edge $pr$ of $\Q_R$ belongs to $\DT(R\setminus \{q,a\})$ throughout $\conn{I_R}{J_R}$, including the gap between $I_R$ and $J_R$. Right: Any violating pair of $pr$ in $R$, such as the pair $q,b$, must involve either $q$ or $a$.}
\label{Fig:DelQuad}
\end{center}
\vspace{-0.5cm}
\end{figure}

Condition (1) guarantees that the smaller set $R$ induces Delaunay crossings $(pq,r,I_R=[t'_0,t'_1])$ and $(pa,r,J_R=[t'_2,t'_3])$, such that $I_R\subseteq I$ and $J_R\subseteq J$. (The latter property follows because the intervals of non-Delaunayhood of $pq$ can only shrink as we pass to the triangulation $\DT(R)$ of the reduced set $R$.) 
In particular, both of these crossings are single Delaunay crossings. 
Clearly, $(pq,r,I_R)$ is followed by $(pa,r,J_R)$ in the order implied by Lemma \ref{Lemma: OrderOrdinaryCrossings}.
In other words, the four points $p,q,a,r$ define within $\DT(R)$ a clockwise quadruple $\Q_R$.
Recall that $pr$ is Delaunay during each of the intervals $I,J$.
Condition (2) guarantees that $pr$ belongs to $\DT(R\setminus \{q,a\})$ throughout the interval $[t_1,t_3]$ which covers the possible gap between $I$ and $J$. In particular, this edge belongs to $\DT(R\setminus\{q,a\})$ throughout the extended interval $\conn{I_R}{J_R}=[t'_0,t'_3]$ which consists of $I_R$, $J_R$, and the possible gap between them. See Figure \ref{Fig:DelQuad} (left). (As a matter of fact, the Delaunayhood of $pr$ in $R\setminus \{q,a\}$ extends (at least) to the bigger interval $[t_0,t_3]$.)

To recap, we can charge $\Q$ to its more refined counterpart $\Q_R$, formed by the pair of crossings $(pq,r,I_R)$ and $(pa,r,J_R)$, which shows up in the smaller triangulation $\DT(R)$, with probability at least $\Theta(1/k^4)$. 

Let $\F_R$ denote the family of all such ``hereditary" quadruples $\Q_R=(p,q,a,r)$, each of them corresponding to some consecutive clockwise quadruple $\Q=(p,q,a,r)$ in $P$, as defined above.
Notice that the quadruples of $\F_R$ are not necessarily consecutive in $R$, as the set $R$ may induce additional Delaunay crossings that do not show up in $\DT(P)$. Below we introduce a weaker notion of consecutiveness, which holds for the quadruples of $\F_R$. In the definitions below, $P$ stands for a generic set, which in general is a proper subsample of the original input.

\paragraph{Definition.} 
We say that a quadruple $\Q=(p,q,a,r)$ is {\it Delaunay} if the edge $pr$ belongs to $\DT(P\setminus \{q,a\})$ throughout the interval $\conn{I_\Q}{J_\Q}={\sf conv}(I_{\Q}\cup J_{\Q})$.

\paragraph{Definition.} Let $\F$ be a family of clockwise quadruples. We say that $\F$ is {\it nonoverlapping} if for any two quadruples $\Q_1=(p,q_1,a_1,r)$ and $\Q_2=(p,q_2,a_2,r)$, that share their first and last points, the clockwise $(p,r)$-crossings corresponding to $\Q_1$ and $\Q_2$ are distinct, except for the possibility $a_1=q_2$ or $a_2=q_1$, and occur in non-interleaving order.
That is, in the order implied by Lemma \ref{Lemma: OrderOrdinaryCrossings}, the two crossings $(pq_1,r,I_1)$ and $(pa_1,r,J_1)$ of $\Q_1$ appear either both before or both after the two crossings $(pq_2,r,I_2)$ and $(pa_2,r,J_2)$ of $\Q_2$
(again, with the possible coincidence of the second of one quadruple and the first crossing crossing of the other).

We say that a Delaunay crossing $(pq,r,I)$ is {\it in $\F$} if it is either the first or the second crossing for at least one quadruple $\Q$ in $\F$. (In total, it may show up in at most two quadruples.)




Notice that, as argued above, the ``sampled" subfamily $\F_R$ includes only Delaunay quadruples. Moreover, $\F_R$ is nonoverlapping, as the Delaunay crossings in $\F_R$ (which are defined in terms of $R$) inherit the order, implied by Lemma \ref{Lemma: OrderOrdinaryCrossings}, of their ancestors in $P$ (that is, in $\F$).

In the rest of this section, 
the underlying family $\F$ is typically fixed at each stage of our analysis, and is assumed to be nonoverlapping, and to consist only of Delaunay quadruples.
In particular, by the ``nonoverlapping" property, any ordered triple $(p,q,r)$ in $P$ will define the first (resp., second) crossing $(pq,r,I_\Q)$ (resp., $(pq,r,J_\Q)$) for at most one quadruple in $\F$.
In other words, the following condition holds:

\begin{proposition}\label{Prop:TwoTriples}
Let $\F$ be a nonoverlapping family of clockwise quadruples. Then every quadruple $\Q=(p,q,a,r)$ in $\F$ is {\it uniquely} determined by each of the ordered triples 
$(p,q,r)$ and $(p,a,r)$ of its points, which specify, respectively, the first crossing $(pq,r,I)$ and the second crossing $(pa,r,J)$ associated with $\Q$.
\end{proposition}

Let $\Psi(n)$ be the maximum number of consecutive quadruples that can be defined by a set of $n$ points moving as above in $\reals^2$. Let $\Psi_0(n)$ be the maximum cardinality of a nonoverlapping family $\F$ of Delaunay quadruples, which is defined with respect to a set of $n$ such moving points. Then the quantities $\Psi(n)$ and $\Psi_0(n)$ are related by the recurrence

\begin{equation}\label{Eq:DelQuad}
\Psi(n)=O\left(k^4\Psi_0(n/k)+k^2N(n/k)+kn^2\beta(n)\right),
\end{equation}
where $k\leq n$ is an arbitrary parameter.

\subsection{Stage 1}\label{Subsec:Stage1Reg}

To bound the above quantity $\Psi_0(n)$, we fix the underlying point set $P$ and the nonoverlapping family $\F$ of Delaunay quadruples. In addition, we fix a pair of constants $k\ll \ell$. 

Let $\Q=(p,q,a,r)$ be a Delaunay quadruple in $\F$ whose two Delaunay crossings occur during the intervals $I=I_\Q=[t_0,t_1]$ and $J=J_\Q=[t_2,t_3]$.
Recall that (by Lemma \ref{Lemma:OnceCollin}) the points of $\Q$ are involved in two co-circularities, one during $I\setminus J$ and one during $J\setminus I$. (The former co-circularity is red-blue with respect to $pq$, and the latter one is red-blue with respect to $pa$.) Denote by $\zeta_0\in I\setminus J$ and $\zeta_1\in J\setminus I$ the times when these co-circularities occur.
Clearly, at least one of these co-circularities of $p,q,a,r$ has to be extremal.
Without loss of generality, suppose that the co-circularity at time $\zeta_0$ is the first co-circularity of the points of $\Q$. 

Our analysis (at this stage) proceeds by distinguishing between several possible scenarios, and treating each of them separately. In all but the last case, we will obtain a bound in terms of quantities that were already introduced. In the last case (case (e)), the bound will also depend on the cardinality of a more specialized subfamily of quadruples, which is defined over an appropriate subsample of $P$. Such families are called {\it $1$-refined}, and their analysis is passed on to the subsequent stages. 

\bigskip
\noindent{\bf Case (a).} The edge $pr$ is hit during $[t_0,t_3]$ by at least one of the points $q,a$. In fact,  Lemma \ref{Lemma:Crossing} implies that this additional collinearity must occur during the gap $(t_1,t_2)$ (after $I$ and before $J$), so $I$ and $J$ are disjoint in this case. See Figure \ref{Fig:CaseA} (left).

Assume, for instance, that $pr$ is hit by $q$. 
Since $\Q$ is a Delaunay quadruple, the edge $pr$ belongs to $\DT(P)$ at each of the times $t_0,t_3$, and it belongs to the pruned triangulation $\DT(P\setminus \{a,q\})$ throughout $[t_0,t_3]$. 
It thus follows that the edge $pr$ undergoes a Delaunay crossing by $q$ within
the triangulation $\DT(P\setminus \{a\})$. That is, the triple $p,q,r$ defines two Delaunay crossings (of distinct order types) within this smaller triangulation. 
A routine combination of Lemma \ref{Lemma:TwiceCollin} with the probabilistic argument of Clarkson and Shor \cite{CS} (in which we sample, say, half of the points) yields an upper bound of $O(n^2)$ on the overall number of such triples $p,q,r$ in $P$ (independently of the fourth point $a$).
Since each Delaunay quadruple $(p,q,a,r)$ in $\F$ is uniquely determined by the respective ordered triple $(p,q,r)$ (as its first crossing), the same upper bound also holds for the overall number of such Delaunay quadruples in $\F$.

A similar counting argument applies if $pr$ is hit by $a$ during $[t_0,t_3]$. Namely, we argue that the edge $pr$ undergoes a Delaunay crossing by $a$ within the triangulation $\DT(P\setminus \{q\})$, so the triple $p,a,r$ defines two Delaunay crossings within that reduced triangulation, and the quadratic bound follows from Lemma \ref{Lemma:TwiceCollin}, as above. Hence we may assume, from now on, that $pr$ is not hit by $q$ or $a$ during $[t_0,t_3]$.

\begin{figure}[htbp]
\begin{center}
\input{SecondCrossing.pstex_t}\hspace{2cm}\input{NonBalanced.pstex_t}\hspace{2cm}\input{NonBalanced1.pstex_t}
\caption{\small Left: Case (a). The edge $pr$ is hit by $q$ during $(t_1,t_2)$. Center: Case (b). At least $k$ counterclockwise $(q,r)$-crossins $(uq,r,I_u)$ end during $(t_1,t_3]$. Right: Case (b) -- the symmetric scenario. At least $k$ counterclockwise $(a,r)$-crossings $(ua,r,I_u)$ begin during $[t_0,t_2)$.}
\label{Fig:CaseA}
\end{center}
\vspace{-0.5cm}
\end{figure} 

\medskip
\noindent{\bf Case (b).} At least $k$ counterclockwise $(q,r)$-crossings $(uq,r,I_u)$ end during $(t_1,t_3]$ (see Figure \ref{Fig:CaseA} (center)), or at least $k$ counterclockwise $(a,r)$-crossings $(ua,r,I_u)$ start during $[t_0,t_2)$ (see Figure \ref{Fig:CaseA} (right)).
To dispose of such quadruples $\Q$, we introduce an auxiliary counting scheme that we will use at several stages of our analysis. We first need a few definitions.

\paragraph{Chargeability.} We say that an edge $pq$ is {\it almost Delaunay} during an interval $\I=[t_0,t_1]$ if there is a set $A$ of at most $c_0$ points such that $pq$ belong to $\DT(P\setminus A)$ throughout $\I$.
Here $c_0$ is some absolute constant\footnote{This condition is similar to Condition (iii) in Theorem \ref{Thm:RedBlue}, except that here $c_0$ is a small {\it absolute} constant, whereas the parameter $k$ in the theorem can be, and is indeed set to, a suitable large value that grows as $\eps\downarrow 0$.} smaller than $8$.

\begin{figure}[htbp]
\begin{center}
\input{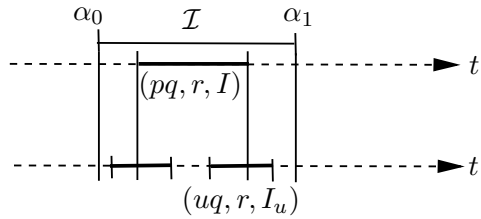}
\caption{\small The crossing $(pq,r,I)$ is $(p,r,k)$-chargeable with reference interval $\I=[\alpha_0,\alpha_1]$. At least $k$ counterclockwise $(q,r)$-crossings $(uq,r,I_u)$ occur within $\I$. By Lemma \ref{Lemma: OrderOrdinaryCrossings}, each of their respective intervals $I_u$ is contained in exactly one of the intervals $[t_0,\alpha_1]$, $[\alpha_0,t_1]$.}
\label{Fig:Chargeability}
\end{center}
\vspace{-0.5cm}
\end{figure}

We say that a Delaunay crossing $(pq,r,I)=[t_0,t_1]$ is $(p,r,k)${\it -chargeable} if there exists an interval $\I=[\alpha_0,\alpha_1]$ containing $I$ such that the following two conditions hold:
(1) the edge $pr$ is Delaunay at times $\alpha_0$ and $\alpha_1$, and almost Delaunay during the the rest of $\I$, and (2) at least $k$ counterclockwise $(q,r)$-crossings $(uq,r,I_u)$ occur within $\I$ (i.e., we have $I_u\subseteq \I$ for each of these points $u$). See Figure \ref{Fig:Chargeability}.

Similarly, we say that a Delaunay crossing $(pq,r,I)$ is $(q,r,k)${\it -chargeable} if the edge $qr$ is almost Delaunay throughout the extended interval $\I$ (and Delaunay at the endpoints of $\I$), and at least $k$ clockwise $(p,r)$-crossings $(pu,r,I_u)$ occur within $\I$.

Several remarks are in order. If $(pq,r,I)$ is a $(p,r,k)$-chargeable crossing then it need not be the {\it only} clockwise $(p,r)$-crossing to occur within the corresponding interval $\I=[\alpha_0,\alpha_1]$. Moreover, the other such $(p,r)$-crossings $(pz,r,I_z)$, that occur (if at all) within $\I$, are not necessarily $(p,r,k)$-chargeable (because this notion also depends on the other endpoint $z$ of the edge $pz$ being crossed by $r$).
Note also that, according to (a counterclockwise variant of) Lemma \ref{Lemma: OrderOrdinaryCrossings}, each of the clockwise $(q,r)$-crossings $(uq,r,I_u)$ that contribute to the $(p,r,k)$-chargeability of $(pq,r,I)$ must satisfy either $I_u\subseteq [\alpha_0,t_1]$ or $I_u\subseteq [t_0,\alpha_1]$, because the intervals $I$ and $I_u$ are either disjoint or partially overlapping (but not nested). 

Informally, the $(p,r,k)$-chargeability allows us to distribute the ``weight" of $(pq,r,I)$ over the $\Omega(k)$ arrangements $\A_{ru}$, which correspond to the above counterclockwise $(q,r)$-crossings $(uq,r,I_u)$ (each of these latter crossings is also a clockwise $(u,r)$-crossing, and is denoted this way). 
In Section \ref{Sec:Balanced} we use this idea to establish the following theorem:

\begin{theorem}\label{Thm:Balanced}
Let $k>12$ be a sufficiently large constant. Then any set $P$ of $n$ points, moving as above in $\reals^2$, induces at most
$O\left(k^2N(n/k)+kn^2\beta(n)\right)$
Delaunay crossings $(pq,r,I)$ that are either $(p,r,k)$-chargeable or $(q,r,k)$-chargeable.
\end{theorem}

We next return to the setup of the first subcase of Case (b). Since $\Q$ is a Delaunay quadruple, the edge $pr$ is almost Delaunay during the interval $[t_0,t_3]$ (it suffices to remove $q$ to $a$ to ensure Delaunayhood). According to Lemma \ref{Lemma: OrderOrdinaryCrossings}, each of the $(q,r)$-crossings $(uq,r,I_u)$ occurs entirely within $I\cup [t_1,t_3]=[t_0,t_3]$, that is, $I_u\subseteq [t_0,t_3]$. Indeed, by definition, each such $I_u$ ends before $t_3$ and after $t_1$, the end of $I$, so it has to start after $t_0$, where $I$ starts. Thus, $(pq,r,I)$ is $(p,r,k)$-chargeable (with $\I=[t_0,t_3]$). Hence, by Theorem \ref{Thm:Balanced}, the overall number of the corresponding quadruples $\Q$ is at most 
$$
O(k^2N(n/k)+kn^2\beta(n)).
$$

A symmetric argument applies if at least $k$ counterclockwise $(a,r)$-crossings $(ua,r,I_u)$ begin in $[t_0,t_2]$.
Indeed, arguing as in the preceding paragraph, each of these Delaunay crossings has to occur entirely within $[t_0,t_3]=[t_0,t_2]\cup J$, so $(pa,r,J)$ is $(p,r,k)$-chargeable.

Hence, we may assume, from now on, that at most $k$ counterclockwise $(q,r)$-crossings end in $(t_1,t_3]$, and that at most $k$ counterclockwise $(a,r)$-crossings begin in $[t_0,t_2)$.

\medskip
\noindent{\bf Case (c).} Either $rq$ is never Delaunay during $[t_3,\infty)$, or $ra$ is never Delaunay during $(-\infty, t_0]$. In the former case, by Lemma \ref{Lemma:Crossing}, no counterclockwise $(q,r)$-crossings can end in $[t_3,\infty)$, because $rq$ has to be Delaunay throughout the interval of such a crossing.
Since case (b) is ruled out, $(pq,r,I)$ is among the last $k+1$ counterclockwise $(q,r)$-crossings (with respect to the order implied by Lemma \ref{Lemma: OrderOrdinaryCrossings}). 
Clearly, this can happen for at most $O(k n^2)$ crossings $(pq,r,I)$ (and their respective quadruples $\Q$).
A fully symmetric argument applies if $ra$ never shows up in $\DT(P)$ during $(-\infty,t_0]$, in which case $(pa,r,J)$ is among the first $k+1$ counterclockwise $(a,r)$-crossings.

\begin{figure}[htbp]
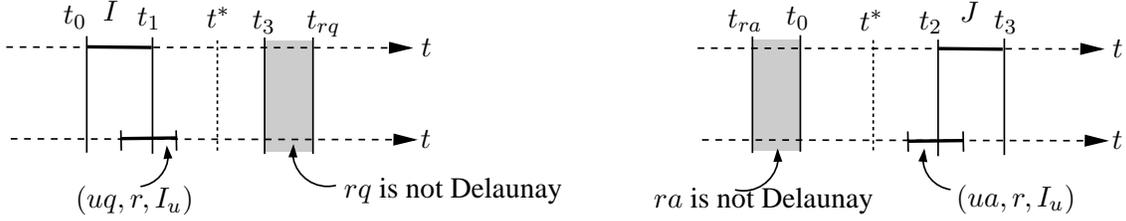

\begin{center}
\input{ChargeInRq.pstex_t}\hspace{3cm}\input{ChargeInRa.pstex_t}
\caption{\small Charging events in $\A_{rq}$ and $\A_{ra}$. Left: $t_{rq}$ is the first time in $[t_3,\infty)$ when $rq$ belongs to $\DT(P)$.
Since case (b) is ruled out, $(pq,r,I)$ is among the last $k+1$ counterclockwise $(q,r)$-crossings to end before any event in $(t_1,t_{rq})$. Right: $t_{ra}$ is the last time in $(-\infty,t_2]$ when $ra$ belongs to $\DT(P)$. After outruling case (b), $(pa,r,J)$ is among the first $k+1$ counterclockwise $(u,r)$-crossings to begin after any event in $(t_{ra},t_2)$.}
\label{Fig:ChargeInRqRa}
\end{center}
\vspace{-0.5cm}
\end{figure}

\paragraph{Preparing for cases (d) and (e).}
In the remainder of our analysis we may therefore assume that neither of the situations considered in cases (a)--(c) arises.
Let $t_{rq}$ denote the first time in $[t_3,\infty)$ when $rq$ belongs to $\DT(P)$.
Namely, we have $t_{rq}=t_3$ if $rq$ is Delaunay also at time $t_3$, and otherwise $t_{rq}$ is the first time after $t_3$ when $rq$ enters $\DT(P)$ (recall that $rq$ is Delaunay at time $t_1$);
refer to the schematic Figure \ref{Fig:ChargeInRqRa} (left).
Similarly, we let $t_{ra}$ denote the last time in $(-\infty,t_0]$ when $ra$ belongs to $\DT(P)$; see Figure \ref{Fig:ChargeInRqRa} (right).

Before proceeding to the cases (d) and (e), we first apply Theorem \ref{Thm:RedBlue} in $\A_{rq}$ over the interval $(t_1,t_{rq})$, and then apply it in $\A_{ra}$ over $(t_{ra},t_2)$, both times with the second constant parameter $\ell$.

Consider the first application of Theorem \ref{Thm:RedBlue}. If at least one of its Conditions (i), (ii) holds, we charge the quadruple $\Q$, via its first crossing\footnote{Recall that, according to Proposition \ref{Prop:TwoTriples}, $\Q$ is uniquely determined by the choice of $(p,q,r)$, which specify its first crossing.} $(pq,r,I)$, either to $\Omega(\ell^2)$
$\ell$-shallow co-circularities, or to an $\ell$-shallow collinearity in $\A_{rq}$. 
We claim that each of these $\ell$-shallow co-circularities or collinearities that occurs at some moment $t^*\in (t_1,t_{rq})$, is charged at most $O(k)$ times in this manner. Indeed, such an event must involve the points $q$ and $r$ of $\Q$ (together with one or two additional points).  
To guess the point $p$, we use the fact that
at most $k$ counterclockwise $(q,r)$-crossings end after $I$ and before $t_3$. Moreover, assuming $t_{rq}>t_3$ and recalling Lemma \ref{Lemma:Crossing}, no $(q,r)$-crossings can take place (let alone end) during $(t_3,t_{rq}]$ (when the edge $rq$ is not Delaunay).
Thus, $pq$ is among the $k+1$ edges whose counterclockwise $(q,r)$-crossings (by $r$) are the latest to end before $t^*$. 
Therefore, the overall number of quadruples $\Q$ in $\F$ for which such a charging applies is at most 
$$
O\left(k\ell^2N(n/\ell)+k\ell n^2\beta(n)\right).
$$

Finally, if Condition (iii) of Theorem \ref{Thm:RedBlue} holds, then the Delaunayhood of $rq$ can be restored, throughout the interval $I\cup [t_1,t_{rq}]=[t_0,t_{rq}]$ (recall that $rq$ is Delaunay during $I$), by removing a set $A$ of at most $3\ell$ points of $P$ (possibly including $p$ and/or $a$).

The second application of Theorem \ref{Thm:RedBlue} in $\A_{ra}$ over $(t_{ra},t_2)$ is fully symmetric. If at least one of Conditions (i), (ii) is satisfied, we dispose of $\Q$ by charging it, via its second crossing $(pa,r,J)$, either to $\Omega(\ell^2)$ $\ell$-shallow co-circularities, or to an $\ell$-shallow collinearity that occur in $\A_{ra}$ during that interval. Arguing as above, $(pa,r,J)$ is among the first $k+1$ counterclockwise $(a,r)$-crossings to begin after each charged event, which also involves $a$ and $r$. Hence, every collinearity or co-circularity is charged at most $O(k)$ times, so, as above, this charging takes place for at most $O\left(k\ell^2N(n/\ell)+k\ell n^2\beta(n)\right)$ quadruples $\Q$.
For each of the remaining quadruples we have a set $B$ of at most $3\ell$ points (possibly including $p$ and/or $q$) whose removal restores the Delaunayhood of $ra$ throughout $[t_{ra},t_2]\cup J=[t_{ra},t_3]$.

\begin{figure}[htbp]
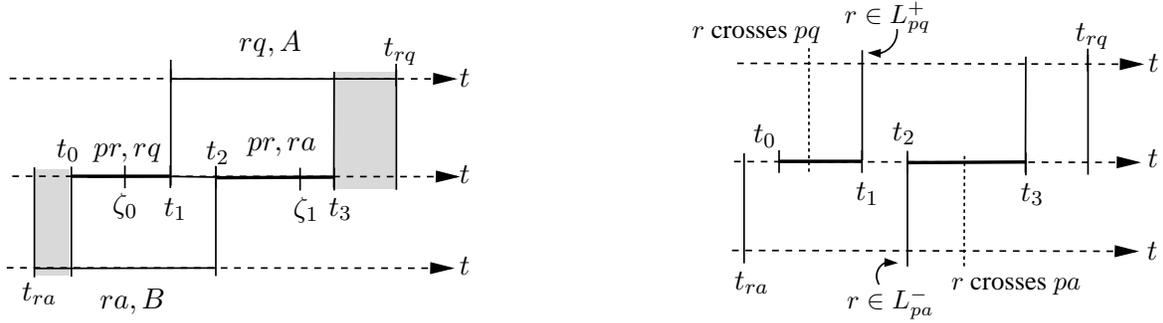

\begin{center}
\input{SummaryBeforeD.pstex_t}\hspace{3cm}\input{SummaryBeforeE.pstex_t}
\caption{\small Left: The situation when entering case (d). If we remove $A\cup B$ but retain $p,q,a,r$, then: (i) During $[t_0,t_1]$, the edges $pr$ and $rq$ are Delaunay. (ii) During $[t_2,t_3]$, the edges $pr$ and $ra$ are Delaunay. (iii) During $[t_0,t_3]$, the edge $pr$ is almost Delaunay. (iv) During $[t_0,t_{rq}]$, the edge $rq$ is almost Delaunay (and will be Delaunay if we remove $p$ and $a$). (v) During $[t_{ra},t_3]$, the edge $ra$ is almost Delaunay (and will be Delaunay if we remove $p$ and $q$). Right: The situation when entering case (e). The point $r$ can leave $\L_{pq}^+$ during $(t_1,t_{rq}]$ only through the edge $pq$. Similarly, $r$ can enter $\L_{pa}^-$ during $[t_{ra},t_2)$ only through the edge $pa$ (and otherwise remains in $\L_{pa}^-$).}
\label{Fig:SummaryBeforeD}
\vspace{-0.4cm}
\end{center}
\end{figure}

To recap, in each of remaining cases (d) and (e) we may assume the existence of the two sets $A$ and $B$ that satisfy the above properties. See Figure \ref{Fig:SummaryBeforeD} (left) for a summary of what we assume now.

\medskip
\noindent{\bf Case (d).} The point $p$ hits the edge $rq$ during $(t_1,t_{rq})$, or it hits the edge $ra$ during the symmetric interval $(t_{ra},t_2)$. Without loss of generality, we focus on the former scenario, and handle the latter one in a fully symmetric manner.

As is easy to check, the edge $rq$ undergoes a Delaunay crossing by $p$ in $\DT((P\setminus A)\cup\{p\})$, with an appropriate interval that contains the time of the actual crossing. Therefore, Lemma \ref{Lemma:TwiceCollin}, in combination with the Clarkson-Shor argument \cite{CS}, provides an upper bound of $O(\ell n^2)$ on the number of such triples $p,q,r$ (and of the corresponding quadruples $\Q$, each of which is uniquely determined by the choice of $(pq,r,I)$ as its first crossing). 

\begin{figure}[htbp]
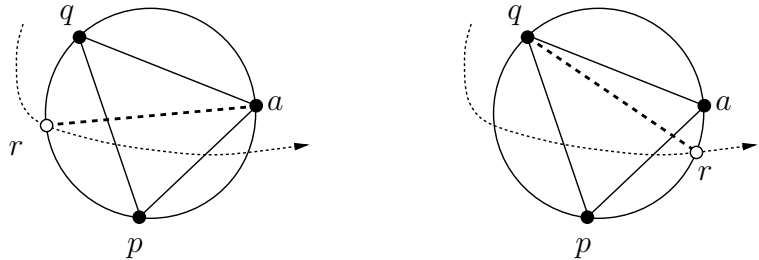

\begin{center}
\input{FirstCocircOrd.pstex_t}\hspace{2cm}\input{SecondCocircOrd.pstex_t}
\caption{\small The co-circularities at times $\zeta_0\in I\setminus J$ (left) and $\zeta_1\in J\setminus I$ (right).
In the depicted scenario, no additional collinearity of $p,q,r$ or $p,a,r$ occurs between the times when $r$ enters $\L_{pq}^+$ and $\L_{pa}^+$.}
\label{Fig:TwoCocircs}
\vspace{-0.4cm}
\end{center}
\end{figure}

\medskip
\noindent{\bf Case (e).} None of the preceding cases holds; this is the most involved case in Stage 1. See Figure \ref{Fig:SummaryBeforeD} (left and right) for a schematic summary of the following properties that we assume now.
Recall that the points of $\Q$ are involved in co-circularities at times $\zeta_0\in I\setminus J$ and $\zeta_1\in J\setminus I$ (see Figure \ref{Fig:TwoCocircs}), and that at least one of these co-circularities has to be extremal.
Without loss of generality, suppose, as already assumed earlier, that the co-circularity at time $\zeta_0$ is the {\it first} co-circularity of the points of $\Q$. 
In addition, we continue to assume that there exists a set $A$ of cardinality at most $3\ell$, such that $rq$ belongs to $\DT(P\setminus A)$ throughout the interval $[t_0,t_{rq})$. Similarly, we assume the existence of a set $B$ of at most $3\ell$ points such that $ra$ belongs to $\DT(P\setminus B)$ throughout the interval $(t_{ra},t_3]$.
Finally, since neither of the preceding cases (a), (d) holds, $r$ can re-enter the halfplane $\L_{pq}^-$ during $(t_1,t_{rq}]$ (after leaving it during $I=[t_0,t_1]$) only by crossing $pq$ again; otherwise it remains in $\L_{pq}^+$ throughout $(t_1,t_{rq}]$.
Similarly, $r$ can enter $\L_{pa}^-$ during $[t_{ra},t_2)$ (before leaving it during $J=[t_2,t_3]$) only through $pa$; otherwise it remains in $\L_{pa}^-$ throughout $[t_{ra},t_2)$.

\begin{figure}[htbp]
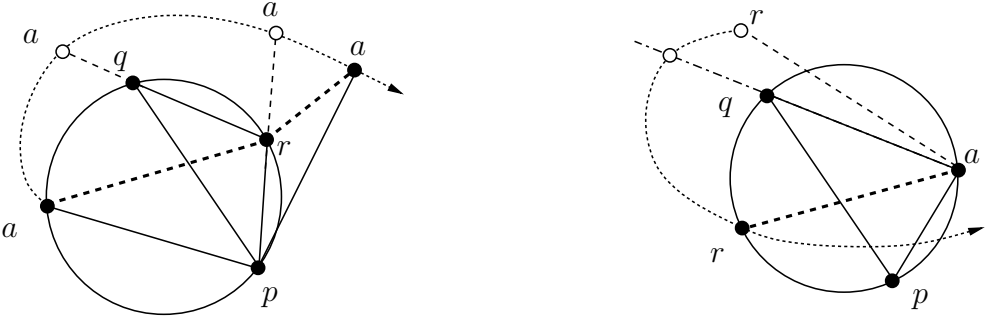

\begin{center}
\input{FirstCocircReenter.pstex_t}\hspace{3cm}\input{FirstCocirc.pstex_t}
\caption{\small  Case (e): proving that $ra$ is hit by $q$. Left: $a$ lies in $\L_{pq}^-$ when $r$ enters $\L_{pq}^+$, so $r$ has to enter $\L_{pa}^-$ (through $pa$) afterwards and before $J$. The corresponding trajectory of $a$ during $(\zeta_0,t_2)$ is depicted. Right: $a$ lies in $\L_{pq}^+$ when $r$ enters $\L_{pq}^+$, so the Delaunayhood of $ra$ is violated, right before $\zeta_0$, by $p$ and $q$.}
\label{Fig:FirstCocirc}
\vspace{-0.4cm}
\end{center}
\end{figure} 


We next argue\footnote{Here the symmetry between $q$ and $a$ breaks down, because the co-circularity at $\zeta_0$ is extremal, but the one at $\zeta_1$ is not.} that the edge $ra$ must be hit during $[t_{ra},t_2)$ by the point $q$. We distinguish between two possible scenarios (see Figure \ref{Fig:FirstCocirc}).

\smallskip
(i) If $a$ lies in $\L_{pq}^-=\L_{pr}^-$ when $r$ enters $\L_{pq}^+$ (during $I$), then $r$ has to enter $\L_{pa}^-$ before $J$. As noted above, $r$ can enter $\L_{pa}^-$ only through $pa$, as depicted in Figure \ref{Fig:FirstCocirc} (left).
Therefore, according to a suitable variant of Lemma \ref{Lemma:ReturnsStillCross}, in which the time is reversed and the points $a$ and $q$ are interchanged, the point $q$ enters the halfplane $\L_{ra}^-$ during $[t_0,t_2]$, through $ra$, as claimed.

\smallskip
(ii) Now suppose that $a$ lies in $\L_{pq}^+$ when $r$ enters this halfplane, so the first co-circularity (at time $\zeta_0$) occurs while $r$ still lies in $\L_{pq}^-$. Hence, the Delaunayhood of $ra$ is violated, right before time $\zeta_0$, by the points $q\in \L_{ra}^-$ and $p\in \L_{ra}^+$; see Figure \ref{Fig:FirstCocirc} (right). Since $ra$ is Delaunay at time $t_{ra}$ and throughout $J=[t_2,t_3]$, and since the points $p,q,a,r$ are never co-circular before $\zeta_0$,
Lemma \ref{Lemma:MustCross} implies that at least one of the points $p,q$ has to hit $ra$ during the interval $[t_{ra},\zeta_0)$, which is clearly contained in $[t_{ra},t_2)$. (Specifically, we apply Lemma \ref{Lemma:MustCross} so that the edge $pq$ in the lemma is $ra$, the points $a,b$ in the lemma are $q,p$, respectively, and the direction of the time axis is reversed.)
Moreover, since case (d) does not occur, $p$ cannot hit $ra$ during the above interval. Hence, the other point, $q$, has to cross $ra$ during $[t_{ra},\zeta_0)$, from $\L_{ra}^+$ to $\L_{ra}^-$. 

If $q$ hits $ra$ twice during $[t_{ra},t_2)$, then the triple $q,a,r$ defines either a double Delaunay crossing, or two single crossings, which occur in the smaller triangulation $\DT((P\setminus B)\cup \{q\})$. Therefore, we can use Theorem \ref{Thm:SpecialCrossings}, or Lemma \ref{Lemma:TwiceCollin}, in combination with the Clarkson-Shor technique, to show that the overall number of such triples in $P$ is at most $O(\ell n^2)$. Moreover, knowing $q,a,r$ allows us to guess $p$ in at most $O(k)$ possible ways, as $(pa,r,J)$ is one of the first $k+1$ counterclockwise $(a,r)$-crossings to begin after the above collinearity (or collinearities) of $q,a,r$ (this follows since we assume that case (b) does not arise). Hence, this scenario happens for at most $O(k\ell n^2)$ quadruples $\Q\in \F$.

Assume then that $ra$ is hit by $q$ exactly once during $(t_{ra},t_2)$. 
In this only remaining case, the edge $ra$ or, more precisely, its reversely oriented copy $ar$ undergoes (within $[t_{ra},t_2)$) exactly one (single) Delaunay crossing by $q$ in the smaller triangulation $\DT((P\setminus B)\cup \{q\})$. 
To handle these latter quadruples $\Q$, we apply a similar analysis to the edge $rq$ (keeping in mind that the co-circularity at time $\zeta_1$ is {\it not} necessarily extremal).

\begin{figure}[htbp]
\begin{center}
\input{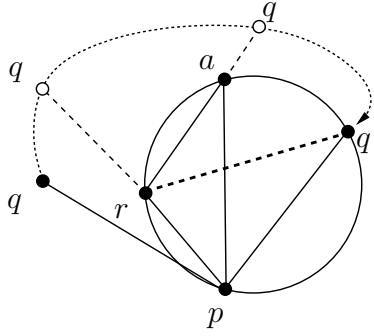}
\caption{\small  Case (e): The proposed trajectory of $q$ if $r$ re-enters $\L_{pq}^-$ before crossing $pa$. According to Lemma \ref{Lemma:ReturnsStillCross}, the point $a$ must hit the edge $rq$ during $(t_1,t_3)\subseteq (t_1,t_{rq}]$.}
\label{Fig:CannotReenter}
\vspace{-0.4cm}
\end{center}
\end{figure} 

If $rq$ is hit by $a$ during $(t_1,t_{rq}]$, then the points $q,a,r$ define two single\footnote{Since any three points can be collinear at most twice, $a$ can hit $rq$ at most once.} Delaunay crossings in the triangulation $\DT([P\setminus (A\cup B)]\cup\{q,a\})$. A routine combination of Lemma \ref{Lemma:TwiceCollin} with the probabilistic arugment of Clarkson and Shor shows that the overall number of such triples $q,a,r$ is at most $O(\ell n^2)$. Moreover, $(pq,r,I)$ is among the $k+1$ last counterclockwise $(q,r)$-crossings to end before the second collinearity of $q,a,r$. Thus, one can guess $\Q$, based on $q,a,r$, in at most $O(k)$ possible ways. In conclusion, the above scenario happens for at most $O(k\ell n^2)$ Delaunay quadruples of $\F$.

To recap, the previous chargings account for
$$
O\left(k\ell^2N(n/\ell)+k^2N(n/k)+k\ell n^2\beta(n)\right)
$$
Delaunay quadruples $\Q$ in $\F$. 
Hence, recalling that case (d) has been ruled out, we may assume, from now on, that none of the points $p,a$ hits $rq$ during the interval $[t_1,t_{rq}]$ (which contains $[t_1,t_3]$). In particular, this implies that $q$ lies in $\L_{pa}^-=\L_{pr}^-$ at the moment when $r$ enters $\L_{pa}^+$ during $J$ (i.e., $r$ lies then in $\L_{pq}^+$). Indeed, otherwise $r$ would have to first leave $\L_{pq}^+$ after $I$, necessarily through the edge $pq$ (because cases (a) and (d) do not occur), which is now impossible according to Lemma \ref{Lemma:ReturnsStillCross}. See Figure \ref{Fig:CannotReenter}. 

\begin{figure}[htbp]
\begin{center}
\input{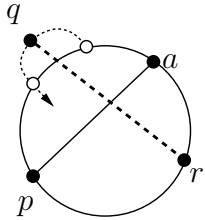}
\caption{\small Case (e). The last two co-circularities of $p,q,a,r$ that occur at times $\zeta_1\in J\setminus I$ and $\zeta_2\in (\zeta_1,t_{rq}]\setminus J$. The edges $pa$ and $rq$ intersect throughout $(\zeta_1,\zeta_2)$; that is, the order type of $p,q,a,r$ does not change there.}
\label{Fig:LastCocircs}
\vspace{-0.4cm}
\end{center}
\end{figure}

Since $q$ lies in $\L_{pa}^-$ when $r$ crosses $pa$ (during $J$) from $\L_{pa}^-$ to $\L_{pa}^+$, the Delaunayhood of $rq$ is violated right after time $\zeta_1$ by the points $p\in \L_{rq}^-$ and $a\in \L_{rq}^+$, as depicted in Figure \ref{Fig:TwoCocircs} (right). (In other words, $\zeta_1$ must occur {\it after} $r$ enters $\L_{pa}^+$, when $q$ leaves the cap $B[p,a,r]\cap \L_{pa}^-$.)
Since neither of $p,a$ can cross $rq$ during the interval $(\zeta_1,t_{rq}]$ (which is clearly contained in $(t_1,t_{rq}]$),
Lemma \ref{Lemma:MustCross} implies that the points $p,q,a,r$ are involved during this interval in a third co-circularity, at some time $\zeta_2>\zeta_1$, and the Delaunayhood of $rq$ is violated by $p$ and $a$ throughout the  interval $(\zeta_1,\zeta_2)$; see Figure \ref{Fig:LastCocircs}. As a matter of fact, the discussion preceding Lemma \ref{Lemma:ReturnsStillCross} also implies that $\zeta_2$ occurs after $J$.

Recall that each of the remaining quadruples $\Q$ is accompanied by a pair of subsets $A,B\subset P$, whose properties are detailed above.
To facilitate the subsequent stages of our analysis, we augment the above ``obstruction sets" $A$ and $B$ as follows. We add to $A$ every point $u$ for which there exists a counterclockwise $(q,r)$-crossing $(uq,r,I_u)$ that ends in $(t_1,t_{rq})$. 
(In fact, Lemma \ref{Lemma:Crossing} implies that none of these $(q,r)$-crossings end after $t_3$.) This is done to ensure that in the sampled configurations that we reach no such crossings take place. Similarly, we add to $B$ every point $u$ for which there exists a counterclockwise $(a,r)$-crossing $(ua,r,I_u)$ that begins in $(t_{ra},t_2)$. (Again, Lemma \ref{Lemma:Crossing} implies that none of these $(a,r)$-crossings begin before $t_0$.)
Since we assume that case (b) does not hold, the above augmentation increases the cardinality of each of the sets $A,B$ by at most $k\leq \ell$.

\smallskip
\noindent{\bf Remark.} 
We may assume that $a$ is not among the (at most $k$) points lately added to $A$, and that $q$ is not among the (at most $k$) points lately added to $B$.
Indeed, if 
the edge $qa$ (or its reversely oriented copy $aq$) undergoes a Delaunay crossing by $r$ then the triple $q,a,r$ defines two Delaunay crossings within $\DT((P\setminus A)\cup\{a\})$. By Lemma \ref{Lemma:TwiceCollin}, the overall number of such triples is at most $O(\ell n^2)$. Furthermore, each of these triples is shared by at most $O(k)$ quadruples that fall into case (e), so the above scenario occurs for at most $O(k\ell n^2)$ quadruples of $\F$.

\medskip
\noindent{\bf Probabilistic refinement.} 
To proceed, we consider a subset $R$ of $\lceil n/\ell \rceil$ points chosen at random from $P$. We fix a Delaunay quadruple $\Q$ as above (i.e., $\Q$ was not disposed of by the chargings of the previous cases, or by the previous chargings of case (e)), and notice that the following two events occur simultaneously, with probability at least $\Omega(1/\ell^4)$: (1) $R$ includes the four points of $\Q$, and (2) none of the points of $(A\cup B)\setminus \{p,q,a\}$ (for the augmented sets $A,B$) belong to $R$. 

Consider the triangulation $\DT(R)$ which is induced by a ``successful" sample $R$ (satisfying (1) and (2)). Notice that the four points of $\Q$ still define a Delaunay quadruple, now with respect to $R$. We continue to denote this new quadruple by $\Q$. (Note, however, that the suitably re-defined intervals $I=I_\Q$ and $J=J_\Q$ may shrink.)

Let $\F_R$ denote the family of all such ``hereditary" Delaunay quadruples $\Q$ in $R$ (such that the sample $R$ is successful for their ancestors in $\F$). Clearly, $\F_R$ is nonoverlapping.

\begin{figure}[htbp]
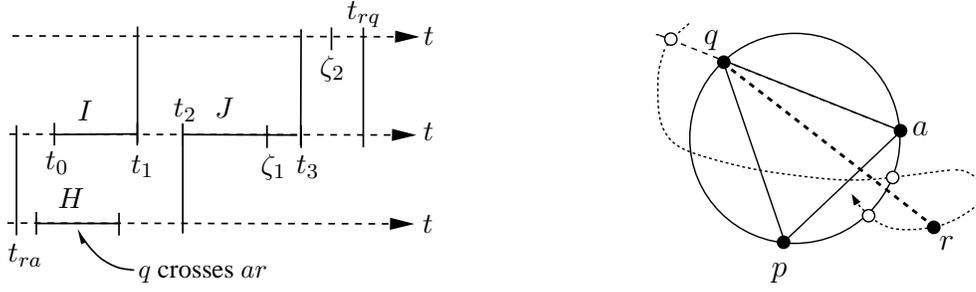

\begin{center}
\input{IntervalsExtend.pstex_t}\hspace{3cm}\input{ThirdCocircOrd.pstex_t}
\caption{\small Left: The edge $ar$ is crossed by $q$ during $[t_{ra},t_2)$. The interval $(t_3,t_{rq}]$ contains the third co-circularity $\zeta_2$.
The edges $ar$ and $rq$ are almost Delaunay during, respectively, $[t_{ra},t_2)\cup J=[t_{ra},t_3]$ and $I\cup (t_1,t_{rq}]=[t_0,t_{rq}]$.  Right: A schematic description of the trajectory of $r$.}
\label{Fig:IntervalsExtend}
\vspace{-0.4cm}
\end{center}
\end{figure}

Fix a quadruple $\Q=(p,q,a,r)$ in $\F_R$, whose two Delaunay crossings occur (within $\DT(R)$) during the intervals $I=[t_0,t_1]$, and $J=[t_2,t_3]$, and whose first two co-circularities occur at times $\zeta_0\in I\setminus J$ and $\zeta_1\in J\setminus I$.  
As before, let $t_{ra}$ denote the last time in $(-\infty,t_0]$ when $ra$ belongs to $\DT(R)$, and let $t_{rq}$ denote the first time in $[t_3,\infty)$ when $rq$ belongs the same triangulation $\DT(R)$. (Notice that, as we replace $P$ by $R$, $t_{ra}$ either remains unchanged or moves ahead, towards (the new) $t_0$. Symmetrically, $t_{rq}$ stays the same or moves back, towards (the new) $t_3$. Hence, the extended intervals $[t_{ra},t_3]$ and $[t_0,t_{rq}]$ can only shrink as we pass from $\DT(P)$ to $\DT(R)$.)
The preceding analysis implies that the following conditions hold for $\Q$:

\begin{itemize}
\item[(R1)] No counterclockwise $(a,r)$-crossings in $\F_R$ begin during $[t_{ra},t_2)$.
Moreover, the edge $ra$ belongs to $\DT(R\setminus \{p,q\})$ throughout the interval $[t_{ra},t_3]$. See Figure \ref{Fig:IntervalsExtend} (left).

\item[(R2)] No counterclockwise $(q,r)$-crossings in $\F_R$ end during $(t_1,t_{rq}]$. 
Moreover, the edge $rq$ belongs to $\DT(R\setminus \{p,a\})$ throughout the interval $[t_0,t_{rq}]$.

\item[(R3)] The set $R\setminus \{p\}$ induces a Delaunay crossing $(ar,q,H)$, whose respective interval $H_\Q=H$ is contained in $[t_{ra},t_2]$. In addition, we encounter a third co-circularity of $p,q,a,r$ at some time $\zeta_2\in [t_3,t_{rq}]$, so that the Delaunayhood of $rq$ is violated by $p\in \L_{rq}^-$ and $a\in \L_{rq}^+$ throughout $(\zeta_1,\zeta_2)$. See Figures \ref{Fig:LastCocircs} and \ref{Fig:IntervalsExtend} (right). Finally, none of the points $a,p$ crosses $rq$ during $(\zeta_2,t_{rq}]$.
\end{itemize}

\smallskip
We say that a nonoverlapping family $\F$ of Delaunay quadruples in a set $P$ is {\it $1$-refined} if its quadruples satisfy the following modified three conditions, restated with respect to $\F$ and its underlying set $P$.

\begin{itemize}
\item[\bf (Q1)] No counterclockwise $(a,r)$-crossings in $\F$ begin during $[t_{ra},t_2)$.
Moreover, the edge $ra$ belongs to $\DT(P\setminus \{p,q\})$ throughout the interval $[t_{ra},t_3]$.

\item[\bf (Q2)] No counterclockwise $(q,r)$-crossings in $\F$ end during $(t_1,t_{rq}]$. 
Moreover, the edge $rq$ belongs to $\DT(P\setminus \{p,a\})$ throughout the interval $[t_0,t_{rq}]$.

\item[\bf (Q3)] The set $P\setminus \{p\}$ induces a Delaunay crossing $(ar,q,H)$, whose respective interval $H$ is contained in $[t_{ra},t_2]$. In addition, we encounter a third co-circularity of $p,q,a,r$ at some time $\zeta_2\in [t_3,t_{rq}]$, so that the Delaunayhood of $rq$ is violated by $p\in \L_{rq}^-$ and $a\in \L_{rq}^+$ throughout $(\zeta_1,\zeta_2)$. Finally, none of the points point $a,p$ crosses $rq$ during $(\zeta_2,t_{rq}]$.
\end{itemize}

Let $\Psi_1(m)$ denote the maximum possible cardinality of a 1-refined family of Delaunay quadruples, that is defined over a set of $m$ points moving in $\reals^2$ as above.
The preceding discussion implies that the maximum cardinality $\Psi_0(n)$ of any nonoverlapping family $\F$ of Delaunay quadruples in a set of $n$ moving points satisfies the recurrence:
 
$$
\Psi_0(n)=O\left(\ell^4\Psi_1(n/\ell)+k\ell^2N(n/\ell)+k^2N(n/k)+k\ell n^2\beta(n)\right),
$$
for any pair of parameters $k\ll \ell<n$.


\begin{proposition} Let $\F$ be a 1-refined family of Delaunay quadruples. Then each quadruple $\Q=(p,q,a,r)$ in $\F$ is uniquely determined by the ordered triple $q,a,r$. (That is, there is no other quadruple in $\F$ that shares its last three points with $\Q$.)
\end{proposition}\label{Prop:UniqueArq}

\begin{proof} 
By Conditions (Q1) and (Q3), $(pa,r,J_\Q)$ is the first counterclockwise $(a,r)$-crossing (in $\F$) to begin after $q$ hits $ar$ during the corresponding interval $H=H_\Q$. 
\end{proof}


\paragraph{The subsequent chargings --- Overview.} 
To bound the above quantity $\Psi_1(n)$, we fix an underlying set $P$ of $n$ moving points and a $1$-refined family $\F$ of nonoverlapping Delaunay quadruples. In addition, we fix a quadruple $\Q=(p,q,a,r)$ in $\F$, whose Delaunay crossings occur during the intervals $I=I_\Q=[t_0,t_1]$ and $J=J_\Q=[t_2,t_3]$ (in this order). 
Recall that the points $p,q,a,r$ are involved in three co-circularities, at times $\zeta_0\in I\setminus J$, $\zeta_1\in J\setminus I$, and $\zeta_2>t_3$, and that the Delaunayhood of edge $rq$ is violated during $(\zeta_1,\zeta_2)$ by the points $p$ and $a$. Furthermore, since the co-circularities at times $\zeta_1$ and $\zeta_2$ have the same order type, the Delaunayhood of $pa$ is violated right after time $\zeta_2$ by the points $q$ and $r$.

Informally, the remainder of this section (except for Stage 5) is devoted to showing that the co-circularity at time $\zeta_2$ yields a Delaunay crossing of $pa$ by $q$.
Similarly to the crossing of $ar$ by $q$ in Condition (Q3), this crossing occurs in an appropriately reduced triangulation, and only if $\Q$ is not previously disposed of by one of the standard chargings (using Theorems \ref{Thm:RedBlue} and \ref{Thm:Balanced}).

The above implication is relatively easy to establish if $pa$ undergoes only few Delaunay crossings after $(pa,r,J)$  and before time $\zeta_2$, when it is violated by $q$ and $r$.
Indeed, following the general strategy demonstrated in Section \ref{Sec:ReduceToCrossings} (and at Stage 1), we consider three possible scenarios.

If $pa$ never re-enters $\DT(P)$ after time $\zeta_2$, then $(pa,r,J)$ (and, thereby, $\Q$) can be charged to the edge $pa$, because it is then among the few last Delaunay crossings of this edge. Otherwise, 
we consider the first time $t_{pa}$ after $\zeta_2$ when $pa$ enters $\DT(P)$ and apply Theorem \ref{Thm:RedBlue} in $\A_{pa}$ over the interval $[t_3,t_{pa}]$. Notice that, according to Lemma \ref{Lemma:MustCross}, $pa$ is crossed during this interval (or, more precisely, during its proper subinterval $(\zeta_2,t_{ra}]$) by at least one of $r$ and $q$. (This follows because no further co-circularities of $p,q,a,r$ can occur after $\zeta_2$.)

If at least one of the Conditions (i), (ii) of Theorem \ref{Thm:RedBlue} holds, we dispose of $\Q$ by charging it within $\A_{pa}$ (and, again, via its second crossing $(pa,r,J)$) either to sufficiently many shallow co-circularities, or to a shallow collinearity. As in the previous similar cases, the charging of each event in $\A_{pa}$ is almost unique, as $(pa,r,J)$ is among the few last Delaunay crossings of $pa$ to end before it. 

Finally, if Condition (iii) of Theorem \ref{Thm:RedBlue} holds, then we end up with a ``small" subset $A$ of $P$ (including at least one of $r,q$) whose removal restores the Delaunayhood of $pa$ throughout $[t_3,t_{pa}]$. Hence, $pa$ undergoes, within a suitably sampled triangulation $\DT(R)$, a Delaunay crossing by one of the points $q,r$. If $pa$ is crossed by $r$ during $[t_3,t_{pa}]$, then we can again dispose of such quadruples $\Q$ using Lemma \ref{Lemma:TwiceCollin}. Otherwise, we say that the edge $pa$ undergoes  within $\DT(R)$ a {\it special crossing} by the point $q$. By our assumption, each special crossing is charged by only a small number of triples $p,a,r$ (and quadruples $\Q$).
In Section \ref{Sec:Special} we derive a recurrence for the maximum possible number of these special crossings, which, combined with the recurrences derived in this section, and in the preceding ones, yield the asserted near-quadratic bound on the  number of Delaunay co-circularities. 

Unfortunately, the above argument does not work if the edge $pa$ of $\Q$ undergoes ``too many" Delaunay crossings during $(t_3,\zeta_2)$. In this case, we cannot easily trace the events that occur in $\A_{pa}$, back to $(pa,r,J)$ (and to $\Q$); that is, there are too many ways to guess $r$. 
At Stage 4 we use Theorems \ref{Thm:RedBlue} and \ref{Thm:Balanced} to dispose of such quadruples. 
To facilitate the fairly involved analysis of that stage, we first extend the almost-Delaunayhood of $ra$ and $rq$ from, respectively, $[t_{ra},t_3]$ and $[t_0,t_{rq}]$, to their superinterval $[t_{ra},t_{rq}]$, which covers $\zeta_0,\zeta_1,\zeta_2$ together with the aforementioned crossing of $ar$ by $q$.
These extensions are performed at the auxiliary Stages 2 and 3, and they also involve the sampling argument of Clarkson and Shor. (Hence, the instants $t_{ra}$ and $t_{rq}$ are each time redefined with respect to the underlying, progressively reduced subset of $P$.)



\subsection{Stage 2: Charging events in $\A_{pr}$ (again)} \label{Subsec:Stage2Reg} 

Before extending the almost-Delaunayhood of $ra$ and $rq$, as promised in the previous paragraph, we first tackle the edge $pr$, and extend its almost-Delaunayhood. Handling $ra$ and $rq$ will be done in the next Stage 3.

Let $\Q=(p,q,a,r)$ be a quadruple in the $1$-refined family $\F$.
Recall that the edge $pr$ is almost Delaunay during $\conn{I}{J}=[t_0,t_3]$ (and that it is in fact Delaunay if $q$ and $a$ are removed). 
We extend the almost-Delaunayhood of $pr$ to a (potentially) larger interval $[\zeta_{pr}^-,\zeta^+_{pr}]$, which covers $[t_{ra},t_{rq}]$. To do so, we fix a (new) pair of constants $k\ll \ell$.

\begin{figure}[htbp]
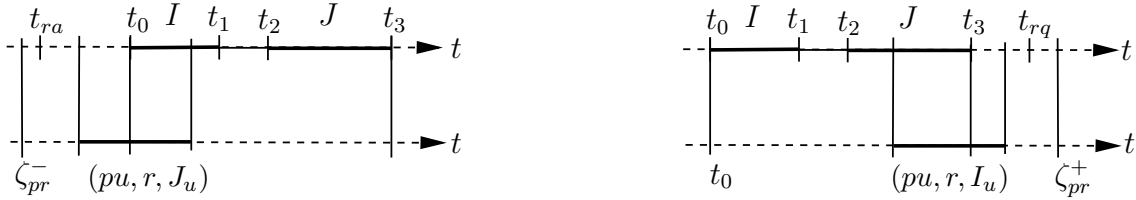

\begin{center}
\input{IntervalsExtend2.pstex_t}\hspace{3cm}\input{IntervalsExtend1.pstex_t}
\caption{\small Left: Extending the almost-Delaunayhood of $pr$ from $[t_0,t_3]$ to $(\zeta_{pr}^-,t_0)$ (left) and $(t_3,\zeta_{pr}^+]$ (right).}
\label{Fig:IntervalsExtendPr}
\vspace{-0.4cm}
\end{center}
\end{figure}

\medskip
\noindent{\bf Stage 2a.} First, we consider the interval $[t_{ra},t_3]$, where, by assumption, the edge $ra$ is almost Delaunay. Refer to Figure \ref{Fig:IntervalsExtendPr} (left).

If at least $k$ clockwise $(p,r)$-crossings $(pu,r,J_u)$ begin in $(t_{ra},t_2)$, then the Delaunay crossing $(pa,r,J)$ is $(a,r,k)$-chargeable with $\I=[t_{ra},t_3]$. Indeed, according to Lemma \ref{Lemma: OrderOrdinaryCrossings}, each of the corresponding intervals $J_u$ has to be contained in $[t_{ra},t_3]=[t_{ra},t_2]\cup J$ (since $J_u$ starts before $t_2$, the starting time of $(pa,q,J)$, it has to end before $t_3$). Hence, and according to Theorem \ref{Thm:Balanced}, the overall number of such crossings $(pa,r,J)$ is at most $O\left(k^2N(n/k)+kn^2\beta(n)\right)$. Clearly, this also bounds the overall number of such quadruples $\Q$. Therefore, we can assume, from now on, that at most $k$ clockwise $(p,r)$-crossings $(pu,r,I_u)$ begin during $(t_{ra},t_2)$.

If the edge $pr$ is never Delaunay during $(-\infty,t_{ra}]$, then $(pq,r,I)$ and $(pa,r,J)$ are among the first $k+1$ clockwise $(p,r)$-crossings, so there are only $O(kn^2)$ such crossings (and quadruples $\Q$). Otherwise, let $\zeta_{pr}^-$ denote the last time in $(-\infty,t_{ra}]$ when $pr$ belongs to $\DT(P)$. 

We now apply Theorem \ref{Thm:RedBlue} in $\A_{pr}$ over the interval $(\zeta_{pr}^-,t_2)$, with the threshold $\ell$. Note that $pr$ is Delaunay at times $\zeta_{pr}^-$ and $t_2$ (in addition to its being Delaunay throughout $I\subseteq [\zeta_{pr}^-,t_2)$). If at least one of the Conditions (i), (ii) of that theorem is satisfied, we charge $\Q$ (via $(pa,r,J)$) either to $\Omega(\ell^2)$ $\ell$-shallow co-circularities, or to an $\ell$-shallow collinearity. As in the previous such chargings, the crucial observation is that $(pa,r,J)$ is among the first $k+1$ clockwise $(p,r)$-crossings to begin after each charged event in $\A_{pr}$. Hence, 
any $\ell$-shallow co-circularity or collinearity is charged, as above, by at most $O(k)$ quadruples $\Q$.
Clearly, the above charging succeeds for at most $O\left(k\ell^2N(n/\ell)+k\ell n^2\beta(n)\right)$ quadruples $\Q$ in $\F$.

Finally, if Condition (iii) of Theorem \ref{Thm:RedBlue} holds, we end up with a set $A$ of at most $3\ell$ points so that $pr$ belongs to $\DT(P\setminus A)$ throughout the interval $[\zeta_{pr}^-,t_3]$. (Note that $A$ can include one, or both of the points $q,a$.) For each $(p,r)$-crossing $(pu,r,J_u)$ that begins in $(t_{ra},t_0)$ we add the respective point $u$ to the ``obstruction set" $A$, whose cardinality then increases by at most $k\ll \ell$. (Informally, as earlier, this allows us to assume that, in the refined configuration, no such $(p,r)$-crossings occur.)

\medskip
\noindent{\bf Stage 2b.} We next consider the interval $[t_0,t_{rq}]$ where, by assumption, edge $rq$ is almost Delaunay. Refer to Figure \ref{Fig:IntervalsExtendPr} (right). The argument is fully symmetric to the one in Stage (2a), but we repeat it for the sake of completeness.

If at least $k$ clockwise $(p,r)$-crossings $(pu,r,I_u)$ end in $(t_1,t_{rq})$, then the crossing $(pq,r,I)$ is clearly $(q,r,k)$-chargeable with $\I=[t_0,t_{rq}]$ , as each of the corresponding intervals $I_u$ begins after $t_0$ (by Lemma \ref{Lemma: OrderOrdinaryCrossings}). As before, this scenario happens for at most $O\left(k^2N(n/k)+kn^2\beta(n)\right)$ quadruples $\Q$ in $\F$. Hence, we may assume, from now on, that the above scenario does not happen for $\Q$. 

If $pr$ is never Delaunay during $[t_{rq},\infty)$, then the crossings $(pq,r,I)$ and $(pa,r,J)$ are among the last $k+1$ clockwise $(p,r)$-crossings; as above, the number of these situations is $O(kn^2)$. Otherwise, let $\zeta_{pr}^+$ denote the first time after $t_{rq}$ when $pr$ is Delaunay. 

We now apply Theorem \ref{Thm:RedBlue} in $\A_{pr}$ over the interval $(t_1,\zeta_{pr}^+)$, with the threshold $\ell$ (noting that $pr$ is Delaunay at times $t_0$ and $\zeta_{pr}^+$). If at least one of the Conditions (i), (ii) holds, we dispose of $\Q$ by charging it either to $\Omega(\ell^2)$ $\ell$-shallow co-circularities, or to an $\ell$-shallow collinearity. As before, each event in $\A_{pr}$ is charged at most $O(k)$ times, as $(pq,r,I)$ and $(pa,r,J)$ are among the last $k+1$ clockwise $(p,r)$-crossings to end before this event. Hence, the overall number of such quadruples is at most $O\left(k\ell^2N(n/\ell)+k\ell n^2\beta(n)\right)$.

Finally, if Condition (iii) of Theorem \ref{Thm:RedBlue} holds, we end up with a set $B$ of at most $3\ell$ points (possibly including and/or $a$) so that $pr$ belongs to $\DT(P\setminus B)$ throughout $[t_0,\zeta_{pr}^+]$.  For each $(p,r)$-crossing $(pu,r,J_u)$ that ends in $(t_1,t_{rq})$ we add the respective point $u$ to $B$, whose cardinality then increases by at most $k\ll \ell$.

\medskip
To recap, we may assume the existence of sets $A,B$, each of size at most $3\ell+k\leq 4\ell$, for which the edge $pr$ belongs to $\DT(P\setminus (A\cup B))$ throughout the interval $\I_{pr}=[\zeta_{pr}^-,\zeta_{pr}^+]$, which covers $[t_{ra},t_{rq}]$.
In addition, $pr$ belongs to $\DT(P)$ at times $\zeta_{pr}^-$ and $\zeta_{pr}^+$.

\medskip
\noindent{\bf Probabilistic refinement.}
Consider a subset $R$ of $\lceil n/\ell \rceil$ points, chosen at random from $P$. Fix a quadruple $\Q$ in $\F$, and note that, with probability at least $\Omega(1/\ell^4)$, 
(1) $R$ contains the four points $p,q,a,r$ of $\Q$, and (2) none of the points of $A\cup B\setminus \{q,a\}$ belong to $R$.

Assuming that the sample $R$ is successful for the chosen $\Q$, the four points $p,q,a,r$ define a Delaunay quadruple, now with respect to $R$.
We continue to denote this new quadruple by $\Q$. 
As is easy to check, the family $\F_R$ of all such ``hereditary" quadruples $\Q$ (such that the sample $R$ is successful for their ancestors in $\F$) is $1$-refined with respect to the new point set $R$.
Moreover, each quadruple in $\F_R$ satisfies the following new condition:

\medskip
\noindent{\bf (Q4)} The edge $pr$ belongs to $\DT(R\setminus \{q,a\})$ throughout an interval $\I_{pr}=[\zeta_{pr}^-,\zeta_{pr}^+]$ which covers\footnote{As in the previous step, the times $t_{ra}$ and $t_{rq}$ must be appropriately redefined with respect to the set $R$ at hand, and the interval $[t_{ra},t_{rq}]$ may shrink. The same applies to the times $\zeta_{pr}^-$ and $\zeta_{pr}^+$.} $[t_{ra},t_{rq}]$, and it belongs to $\DT(R)$ at times $\zeta_{pr}^-$ and $\zeta_{pr}^+$. Moreover, no clockwise $(p,r)$-crossings (in $\F_R$) begin in $(t_{ra},t_0)$ or end in $(t_3,t_{rq})$.

\medskip

\paragraph{Definition.} Let $\F$ be a $1$-refined family of Delaunay quadruples. We say that $\F$ is {\it $2$-refined} if its quadruples also satisfy the above condition (Q4) with respect to the underlying point set $P$ (instead of $R$).

Without loss of generality, we can put $\zeta_{pr}^-$ to be the last time in $(-\infty,t_{ra}]$ when $pr$ belongs to $\DT(R)$. Similarly, we can put $\zeta_{pr}^+$ to be the first time in $[t_{rq},\infty)$ when the edge $pr$ belongs to $\DT(R)$.

\smallskip

\medskip
Let $\Psi_2(n)$ denote the maximum cardinality of a $2$-refined family $\F$, which is defined over a set $P$ of $n$ moving points.
The preceding discussion implies the following relation between the quantities $\Psi_1(n)$ and $\Psi_2(n)$:

\begin{equation}
\Psi_1(n)=O\left(\ell^4 \Psi_2(n/\ell)+k\ell^2 N(n/\ell)+k^2N(n/k)+k\ell n^2\beta(n)\right).
\end{equation}

\subsection{Stage 3}\label{Subsec:Stage3Reg}
To bound the above quantity $\Psi_2(n)$, we fix a $2$-refined family $\F$ which is defined over a set $P$ of $n$ points moving as above in $\reals^2$, and a Delaunay quadruple $\Q$ in $\F$.

By assumption, the edges $rq$ and $ra$ of $\Q$ are almost Delaunay during the respective intervals $[t_0,t_{rq}]$ and $[t_{ra},t_3]$.
The goal of this stage is to extend the almost-Delaunayhood of these two edges to the interval $[t_{ra},t_{rq}]$.  
For the purpose of our analysis, we fix new constants $k$ and $\ell$ such that $k\ll \ell$.

\begin{figure}[htbp]
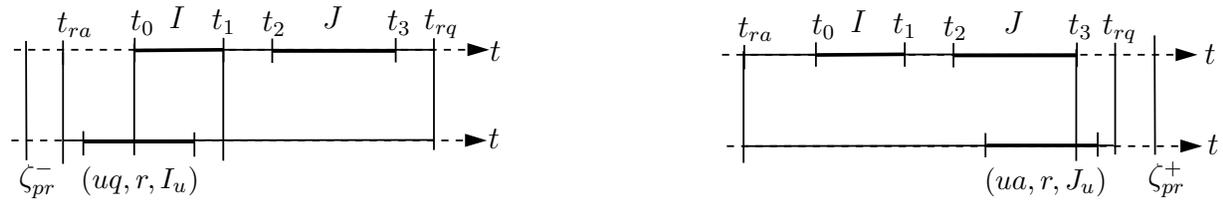

\begin{center}
\input{IntervalsExtend3.pstex_t}\hspace{3cm}\input{IntervalsExtend4.pstex_t}
\caption{\small Left: Extending the almost-Delaunayhood of $rq$ from $[t_0,t_{rq}]$ to $[t_{ra},t_{rq}]$. Right: Extending the almost-Delaunayhood of $ra$ from $[t_{ra},t_3]$ to $[t_{ra},t_{rq}]$.}
\label{Fig:IntervalsExtend3}
\vspace{-0.4cm}
\end{center}
\end{figure}  

\medskip
\noindent{\bf Charging events in $\A_{rq}$.} 
Refer to Figure \ref{Fig:IntervalsExtend3} (left).
If at least $k$ Delaunay counterclockwise $(q,r)$-crossings $(uq,r,I_u)$ begin in $(t_{ra},t_0)$, then the crossing $(pq,r,I)$ is again $(p,r,k)$-chargeable. Indeed, according to Lemma \ref{Lemma: OrderOrdinaryCrossings}, each of these crossings occurs within the larger interval $[\zeta_{pr}^-,t_0]\cup I=[\zeta_{pr}^-,t_1]$, where, by property (Q4), the edge $pr$ is assumed to be almost Delaunay. Moreover, $pr$ belongs to $\DT(R)$ at times $\zeta_{pr}^-$ and $t_1$. Therefore, Theorem \ref{Thm:Balanced} provides an upper bound of $O\left(k^2N(n/k)+kn^2\beta(n)\right)$ on the overall number of such crossings $(pq,r,I)$ (and, hence, of their corresponding quadruples $\Q$, as implied by Proposition \ref{Prop:TwoTriples}). Thus, we can assume, from now on, that the above scenario does not happen for $\Q$.
(Notice that the above application of Theorem \ref{Thm:Balanced} has been prepared by the previous Stage 2, which has extended the almost-Delaunayhood of $pr$ from $[t_0,t_3]$ to $[\zeta_{pr}^-,\zeta_{pr}^+]$.)

\smallskip
We now apply Theorem \ref{Thm:RedBlue} in $\A_{rq}$ over the interval $(t_{ra},t_0)$, with the threshold $\ell$ (noting that $rq$ is Delaunay at time $t_0$, and recalling that Theorem \ref{Thm:RedBlue} also holds if $rq$ is Delaunay at only one endpoint of the interval under consideration). If one of the Conditions (i), (ii) holds, we dispose of $\Q$ by charging it (via $(pq,r,I)$) either to $\Omega(\ell^2)$ $\ell$-shallow co-circularities or to an $\ell$-shallow collinearity. As in the previous such chargings, each event in $\A_{rq}$ is charged at most $O(k)$ times, as $(pq,r,I)$ is among the $k+1$ first counterclockwise $(q,r)$-crossings to begin after it. Hence, this charging is applicable for at most $O\left(k\ell^2N(n/\ell)+k\ell n^2\beta(n)\right)$ quadruples $\Q$ in $\F$.

Finally, if Condition (iii) of Theorem \ref{Thm:RedBlue} holds, we end up with a set $A$ of at most $3\ell$ points such that the edge $rq$ belongs to $\DT(P\setminus A)$ throughout the interval $[t_{ra},t_1]$. 

\medskip
\noindent{\bf Charging events in $\A_{ra}$.} We now apply a symmetric analysis to the edge $ra$, spelling it out for the sake of completeness. Refer to Figure \ref{Fig:IntervalsExtend3} (right).

If at least $k$ counterclockwise $(a,r)$-crossings $(ua,r,J_u)$ end during $(t_3,t_{rq})$ then the crossing $(pa,r,J)$ is $(p,r,k)$-chargeable, as each of the respective intervals $I_u$ is the contained in $(t_2,\zeta_{pr}^+]$
(when the edge $pr$ is almost Delaunay). By Theorem \ref{Thm:Balanced} (and since $pr$ is Delaunay at times $t_2$ and $\zeta_{pr}^+$), the overall number of such crossings $(pa,r,J)$ (and of their corresponding quadruples $\Q$) is at most $O\left(k^2N(n/k)+kn^2\beta(n)\right)$.

Otherwise, we apply Theorem \ref{Thm:RedBlue} in $\A_{ra}$ over the interval $(t_3,t_{rq})$ (noting that $ra$ is Delaunay at time $t_3$). If one the Conditions (i), (ii) of that theorem holds, we dispose of $\Q$ by charging it (now via $(pa,r,J)$) either to $\Omega(\ell^2)$ $\ell$-shallow co-circularities, or to an $\ell$-shallow collinearity. Once again, each event in $\A_{ra}$ is charged at most $O(k)$ times, as $(pa,r,J)$ is among the $k+1$ last counterclockwise $(a,r)$-crossings to end before it.

Finally, if Condition (iii) of Theorem \ref{Thm:RedBlue} holds, we end up with a set $B$ of at most $3\ell$ points such that the edge $ra$ belongs to $\DT(P\setminus B)$ throughout the interval $[t_1,t_{rq}]$. 

\medskip
To recap, we may assume, in what follows, that there exist sets $A,B$ as above, each of cardinality at most $3\ell$.

\medskip
\noindent{\bf Probabilistic refinement.} We consider a subset $R$ of $\lceil n/\ell\rceil$ points chosen at random from $P$. We fix a quadruple $\Q$, not disposed of by the previous chargings, and notice that the following two events occur simultaneously, with probability at least $\Omega(1/\ell^4)$:
(1) $R$ contains the four points $p,q,a,r$ of $\Q$, and (2) none of the points of $A\cup B\setminus \{q,a,r\}$ belong to $R$.

Let $\F_R$ denote the family of all hereditary quadruples $\Q$ (such that $R$ is successful for their ancestors in $\F$).
As is easy to check, $\F_R$ is $2$-refined (in $R$). Moreover, the following new conditions hold for every quadruple $\Q$ in $\F$:

\begin{itemize}
\item[\bf (Q5)] The edge $ra$ belongs to $\DT(R\setminus \{p,q\})$ throughout the interval $[t_{ra},t_{rq}]$.

\item[\bf (Q6)] The edge $rq$ belongs to $\DT(R\setminus \{p,a\})$ throughout the interval $[t_{ra},t_{rq}]$.
\end{itemize}

We say that a family $\F$ of Delaunay quadruples is {\it $3$-refined} if (1) it is $2$-refined, and (2) its quadruples satisfy Conditions (Q5) and (Q6) with respect to the underlying point set.
Let $\Psi_3(n)$ denote the maximum cardinality of a $3$-refined family of Delaunay quadruples that is defined over a set of $n$ moving points (that we keep denoting as $P$, replacing $R$ in these conditions). The preceding discussion implies the following relation between the quantities $\Psi_2(n)$ and $\Psi_3(n)$:

\begin{equation}
\Psi_2(n)=O\left(\ell^4 \Psi_3(n/\ell)+k\ell^2N(n/\ell)+k^2N(n/k)+k\ell n^2\beta(n)\right).
\end{equation}

\subsection{Stage 4}\label{Subsec:Stage4Reg}
To bound the above quantity $\Psi_3(n)$, we fix a $3$-refined family $\F$ which is defined over an underlying set $P$ of $n$ moving points. (That is, $\F$ satisfies all the six condtions (Q1)--(Q6).) 
Proposition \ref{Prop:UniqueArq} implies that every quadruple $\Q=(p,q,a,r)$ in $\F$ is uniquely determined by the ordered triple $(q,a,r)$.

For the purpose of our analysis, we also fix three new constants $k,\ell,h$ such that $12<k\ll \ell\ll h$.

\begin{figure}[htbp]
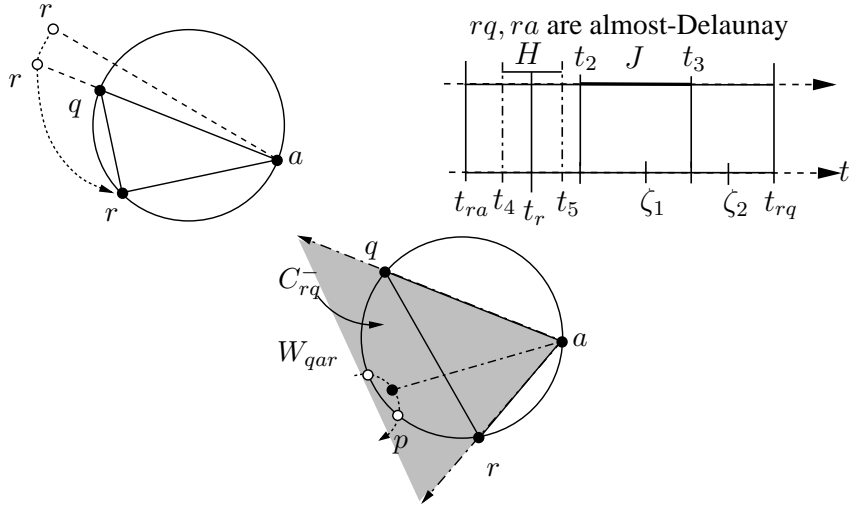

\begin{center}
\input{AlmostCrossing.pstex_t}\hspace{2cm}\input{Timeline.pstex_t}\hspace{2cm}\input{TheWedge.pstex_t}
\caption{\small The topological setup during the interval $[t_{ra},t_{rq}]$. 
Left: The edge $ar$ is hit at some time $t_r\in [t_{ra},t_2]$ by $q$. Center: we have $t_{ra}\leq t_4\leq t_r\leq t_5<t_2<\zeta_1<t_3<\zeta_2$. Right: The motion of $B[q,a,r]$ is continuous throughout $(t_r,\zeta_2]$ (the hollow circles represent the co-circularities at times $\zeta_1$ and $\zeta_2$).}
\label{Fig:TopSetup}
\vspace{-0.4cm}
\end{center}
\end{figure}

\medskip
\noindent{\bf Topological setup.} We fix a quadruple $\Q=(p,q,a,r)$ in $\F$, whose two Delaunay crossings take place during the intervals $I=[t_0,t_1]$ and $J=[t_2,t_3]$ (in this order). Refer to Figure \ref{Fig:TopSetup}.

Since $\F$ is $1$-refined, there exists a time $t_{ra}\leq t_0$ which is the last time before\footnote{If $ra$ is Delaunay at time $t_0$ then we put $t_0=t_{ra}$.} $t_0$ when the edge $ra$ belongs to $\DT(P)$, and a symmetric first time $t_{rq}\geq t_3$ when $rq$ belongs to $\DT(P)$.
Moreover, by Conditions (Q5) and (Q6), the edge $ra$ belongs to $\DT(P\setminus \{p,q\})$, and the edge $rq$ belongs to $\DT(P\setminus \{p,a\})$, throughout the interval $[t_{ra},t_{rq}]$. 

Let us summarize what we know so far about $\Q$.
By Condition (Q3), the points $p,q,a,r$ of $\Q$ are co-circular at times $\zeta_0\in I\setminus J$, $\zeta_1\in J\setminus I$, and $\zeta_2\in (t_3,t_{rq}]$. 
Moreover, the Delaunayhood of $pa$ is violated, throughout $(\zeta_1,\zeta_2)$, by the points $q\in \L_{pa}^-$ and $r\in \L_{pa}^+$.
In particular, $p$ lies throughout that interval within the wedge $W_{qar}=\L_{qa}^+\cap \L_{ra}^-$ and inside the cap $C^-_{rq}=B[q,a,r]\cap \L_{rq}^-$; see Figure \ref{Fig:TopSetup} (right). We emphasize that the order type of the quadruple $(p,q,a,r)$ remains unchanged during $(\zeta_1,\zeta_2)$, and is exactly as depicted in this figure.

In addition, by the same Condition (Q3), the smaller set $P\setminus \{p\}$ induces a (single) Delaunay crossing $(ar,q,H_\Q)$, whose interval $H=H_\Q=[t_4,t_5]$ is contained in $[t_{ra},t_2)$; see Figure \ref{Fig:TopSetup} (left and center). 
In particular, $q$ hits $ar$ at some moment\footnote{Recall from Section \ref{Subsec:Stage1Reg} that $q$ can cross $ar$ either before or after $\zeta_0$, depending on the location of $a$ when $r$ crosses $pq$. 
Our analysis only relies on the fact that $t_r<\zeta_1<\zeta_2$, which follows because $\zeta_r<t_2$ and $\zeta_1\geq t_2$.} $t_{r}\in H$, and crosses $\L_{ar}$ from $\L^-_{ar}$ to $\L^+_{ar}$. Since $q$ lies in $\L_{ar}^+$ at times $\zeta_1>t_2$ and $\zeta_2$, no further collinearities of $q,a,r$ can occur during 
$(t_{r},\zeta_2]$. (Otherwise, the point $q$ would have to re-enter $\L_{ar}^+$, after previously crossing $\L_{ar}$ back to $\L_{ar}^-$, and then the triple $q,a,r$ would be collinear three times, contrary to our assumptions.)  To recap,
the disc $B[q,a,r]$ moves continuously throughout the interval $(t_{r},\zeta_2]$, which is obviously contained in $[t_{ra},t_{rq}]$. 

\begin{figure}[htbp]
\begin{center}
\input{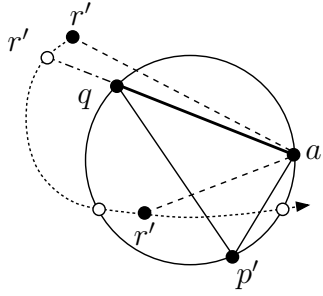}
\caption{\small A quadruple $\Q'=(p',q,a,r')$ in $\F_{qa}$.
The edge $ar'$ undergoes an $(a,q)$-crossing $(ar',q,H_{\Q'})$ within the triangulation $\DT(P\setminus \{p'\})$. 
}
\label{Fig:CandidateQuadsExample}
\vspace{-0.4cm}
\end{center}
\end{figure} 

Let $\F_{qa}$ denote the subfamily of all quadruples $\Q'=(p',q,a,r')$ in $\F$, whose middle points $q$ and $a$ are fixed and equal to those of $\Q$. (In particular, $\F_{qa}$ contains $\Q$.)
For each $\Q'=(p',q,a,r')$ in $\F_{qa}$, the appropriately pruned set $P\setminus \{p'\}$ induces the $(a,q)$-crossing $(ar',q,H_{\Q'})$; see Figure \ref{Fig:CandidateQuadsExample}. 
In what follows, we keep $\Q$ and $\F_{qa}$ fixed and distinguish between several cases.

\begin{figure}[htbp]
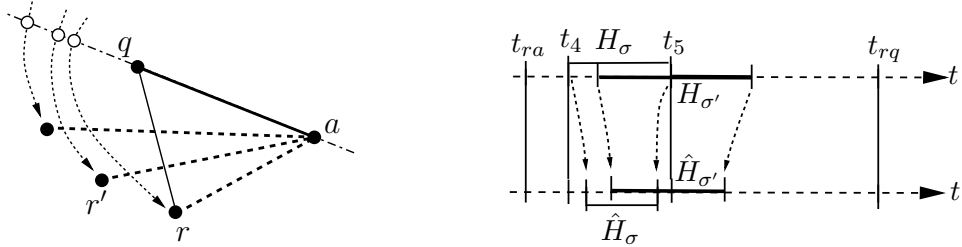

\begin{center}
\input{ManyQa.pstex_t}\hspace{2cm}\input{TimelineQa.pstex_t}
\caption{\small
Case (a). Left: At least $k$ of the crossings $(ar',q,H_{\Q'})$ end during $(t_5,t_{rq})$. Right: A successful sample $\hat{P}$ yields Delaunay crossings $(ar,q,\hat{H}_{\Q})$ and $(ar',q,\hat{H}_{\Q'})$, which occur within $[t_4,t_{rq}]$.}
\label{Fig:CandidateQuadsQa}
\vspace{-0.4cm}
\end{center}
\end{figure}

\medskip
\noindent{\bf Case (a).} 
The family $\F_{qa}$ contains at least $k$ quadruples $\Q'$ whose respective crossings $(ar',q,H_{\Q'})$ end during $(t_5,t_{rq})$. Refer to Figure \ref{Fig:CandidateQuadsQa}.
Recall that, according to Proposition \ref{Prop:UniqueArq}, the point $p'$ is uniquely determined by the choice $r'$.

Informally, we would like to dispose of $\Q$ using Theorem \ref{Thm:Balanced}, by showing that the counterclockwise $(r,q)$-crossing $(ar,q,H)$ is $(r,q,\Theta(k))$-chargeable (for the interval $\I=[t_4,t_{rq}]$). Unfortunately, the $(a,q)$-crossings $(ar',q,H_{\Q'})$ to be charged are defined with respect to (potentially) distinct sets $P\setminus \{p'\}$, and thus do not fit the definition of chargeability.

To free sufficiently many crossings $(ar',q,H_{\Q'})$ from their violating points $p'$, we  
pass from $P$ to a sample $\hat{P}$ of $\lceil n/2\rceil$ points chosen at random from $P$.
Notice though that $\F_{qa}$ can potentially include quadruples $\Q'=(p',q,a,r')$ with $p'=r$, which cannot be freed without destroying $rq$ and $(ar,q,H)$. 

Fortunately, by Proposition \ref{Prop:UniqueArq}, for any quadruple $\Q=(p,q,a,r)$ in $\F_{qa}$ there is at most one other quadruple $\Q=(p',q,a,r')$, also in $\F_{qa}$, with $r'=p$.
The pigeonhole principle then implies that at least half of the quadruples $\Q=(p,q,a,r)$ in $\F_{qa}$ satisfy the following converse condition:

\smallskip
\noindent{\bf (PH)} {\it There is at most one quadruple $\Q'=(p',q,a,r')$ in $\F_{qa}$ with $p'=r$.}
\smallskip

In more detail, consider the (possibly partial) map $\lambda: \F_{qa}\rightarrow \F_{qa}$, so that $\lambda$ maps each quadruple $\Q=(p,q,a,r)\in \F_{qa}$ to the unique quadruple $\lambda(\Q)=(\omega,q,a,p)\in \F_{qa}$  if it exists, and otherwise $\lambda$ is undefined at $\Q$. Put $\mu_\Q=|\{\Q'\mid \lambda(\Q')=\Q\}|$, for each $\Q\in \F_{qa}$. Then $\sum_{\Q\in \F_{qa}}\mu_{\Q}\leq M=|\F_{qa}|$, so the number of quadruples $\Q$ with $\mu_\Q\geq 2$ is at most $M/2$. All the remaining quadruples satisfy (PH).  

Since $q$ and $a$ are arbitrary points of $P$, (PH) holds for at least half of all quadruples in $\F$; hence we may assume that it holds for the quadruple $\Q$ under consideration.

Let $\Q'$ be a quadruple in $\F_{qa}\setminus \{\Q\}$ whose crossing $(ar',q,H'=H_{\Q'})$ ends in $(t_5,t_{rq})$.
We further assume that $p'\neq r$ and $r'\neq p$.
Then we have the following relaxed version of Lemma \ref{Lemma: OrderOrdinaryCrossings}, which can be established by observing that its original proof holds also in the new setup. 
(An alternative proof of Lemma \ref{Lemma:OrderRelaxedCrossings} can be obtained through examining the two co-circularities that are performed by $a,q,r,r'$, according to Lemmas \ref{Lemma:Crossing} and \ref{Lemma:OnceCollin}, during the intervals $H\setminus H'$ and $H'\setminus H$, and then
applying Lemma \ref{Lemma: OrderOrdinaryCrossings} for the reduced set $P\setminus \{p,p'\}$.)

\begin{lemma}\label{Lemma:OrderRelaxedCrossings}
Let $P$ be a set of points moving as above in $\reals^2$, and
let $(ar,q,H)$ and $(ar',q,H')$ be a pair of clockwise $(a,q)$-crossings that occur in the respective reduced triangulations $\DT(P\setminus \{p\})$ and $\DT(P\setminus \{p'\})$, for $p,p'\in P$.\footnote{We do not require that $p$ and $p'$ be distinct.} Furthermore, assume that $r\neq p'$ and $r'\neq p$. Then the statement of Lemma \ref{Lemma: OrderOrdinaryCrossings} holds for $(ar,q,H)$ and $(ar',q,H')$. That is, $q$ hits $ar$ (during $H$) before it hits $ar'$ (during $H'$) if and only if $H$ begins (resp., ends) before the beginning (resp., end) of $H'$. 
\end{lemma}

Clearly, the above restriction on $p'$ and $r'$ is now satisfied by at least $k-2\geq k/2$ of the quadruples $\Q'=(p',q,a,r')$ that are assumed to exist in the current case (a). Since their intervals $H'$ end in $(t_5,t_{rq})$, Lemma \ref{Lemma:OrderRelaxedCrossings} implies that, for each of them, $H'$ starts after $t_4$, and the point $q$ hits $ar'$ (during $H'$) after time $t_{r}$. 

We now return to the sample $\hat{P}$ and observe that the following two events occur simultaneously, with at least some fixed constant probability: 

\smallskip
(1) The sample $\hat{P}$ includes the three points $q,a,r$, but not $p$. Hence, $\hat{P}$ induces a single Delaunay crossing $(ar,q,\hat{H}=\hat{H}_{\Q})$ of $ar$ by $q$. 

\smallskip
(2) The sample $\hat{P}$ includes the point $r'$, but not $p'$, for at least $k/16$ of the above quadruples $\Q'=(p',q,a,r')$. For each of these $k/16$ quadruples, the sample $\hat{P}$ yields a Delaunay $(a,q)$-crossing $(ar',q,\hat{H}_{\Q'})$ with $\hat{H}_{\Q'}\subseteq H_{\Q'}$.

(To see (2), note that this property holds for any single quadruple with probability at least $1/4$, so the expected number of successful quadruples is at least $k/8$. By a variant of Markov's bound, the probability of having at least $k/16$ successful quadruples is at least $14/15$.)


\smallskip
Suppose that the sample $\hat{P}$ is indeed successful for $\Q$. 
Recall that, for each quadruple $\Q'$ in (2), $q$ hits the respective edge $ar'$ (during $H_{\Q'}$) after it hits $ar$ (during $H_{\Q}$).

We now pass to the sampled triangulation $\DT(\hat{P})$.
Lemma \ref{Lemma: OrderOrdinaryCrossings} implies, in combination with the containment $\hat{H}_{\Q'}\subseteq H_{\Q'}$, that all the Delaunay crossings $(ar',q,\hat{H}_{\Q'})$ in (2) end after $\hat{H}$ and before $t_{rq}$; see Figure \ref{Fig:CandidateQuadsQa} (right). Therefore, all of them must occur within the interval 
$H_\Q\cup [t_5,t_{rq}]\subseteq[t_4,t_{rq}]$, where the edge $rq$ is assumed to be almost Delaunay.\footnote{Notice that the times $t_{rq},t_4$ and $t_5$ are defined with respect to the original point set $P$.} In addition, the edge $rq$ belongs to $\DT(\hat{P})$ at {\it both} times $t_4$ and $t_{rq}$, because $\hat{P}$ does not include $p$. Since $\Q'$ and $(ar',q,\hat{H}_{\Q'})$ can be chosen in at least $k/16$ distinct ways, the crossing $(ar,q,\hat{H})$ is $(r,q,k/16)$-chargeable (with respect to $\hat{P}$). 

By Theorem \ref{Thm:Balanced}, the overall number of such triples $(q,a,r)$ in $\hat{P}$ is  
$O\left(k^2N(n/k)+kn^2\beta(n)\right)$.
Clearly, the same bound must hold for the overall number of quadruples $\Q$ that fall into case (a).

\medskip
\noindent{\bf Preparing for cases (b), (c): Charging events in $\A_{qa}$.} We can assume, from now on, that the family $\F_{qa}$ contains at most $k$ quadruples $\Q'$ whose ``almost Delaunay" crossings $(ar',q,H_{\Q'})$ end during $(t_5,t_{rq})$.

Before proceeding to the subsequent cases, we apply Theorem \ref{Thm:RedBlue} in $\A_{qa}$ over the interval $(t_5,t_{rq})$, now with the second constant $\ell$.
Notice that the edge $qa$ belongs to $\DT(P\setminus \{p\})$ at time $t_5$, so we omit $p$ and apply the theorem  with respect to that smaller triangulation.

If at least one of the Conditions (i), (ii) of Theorem \ref{Thm:RedBlue} is satisfied, we charge $\Q$ either to an $(\ell+1)$-shallow collinearity, or to $\Omega(\ell^2)$ $(\ell+1)$-shallow co-circularities. (Each of these events is $\ell$-shallow with respect to $P\setminus \{p\}$, and its depth can go up by $1$ when $p$ is added back.)
It remains to check that each $(\ell+1)$-shallow event, which occurs in $\A_{qa}$ at some time $t^*\in (t_5,t_{rq})$, is charged by at most $O(k)$ quadruples $\Q$.
Indeed, $q$ and $a$ are among the three or four points involved in the event.
We guess $q$ and $a$ (in $O(1)$ possible ways) and consider all ``almost Delaunay" crossings of the form $(ar',q,H_{\Q'})$, each of them associated with some (unique) ``candidate" quadruple $\Q'=(p',q,a,r')$ in $\F_{qa}$.
Since case (a) is ruled out (and since $t^*$ belongs to $(t_5,t_{rq})$), $(ar,q,H=H_{\Q})$ is among the $k$ last such ``almost Delaunay" crossings to end before time $t^*$. Since $p$ is uniquely determined by the choice of $q,a$ and $r$, we can guess $\Q$ in $O(k)$ possible ways. Hence, this scenario happens for at most
$O\left(k\ell^2N(n/\ell)+k\ell n^2\beta(n)\right)$ 
quadruples.

Now assume that Condition (iii) of Theorem \ref{Thm:RedBlue} holds. Then $P$ contains a subset $A$ of at most $3\ell$ points such that the edge $qa$ belongs to $\DT(P\setminus (A\cup \{p\}))$ throughout the interval $H\cup [t_5,t_{rq}]=[t_4,t_{rq}]$. 
In particular, the following property must hold:

{\it At most $3\ell$ points $s\in P\setminus \{p\}$ hit $qa$ during the interval $(t_r,\zeta_2)$ $(\subseteq (t_{rq},t_{rq}))$.}

\begin{figure}[htbp]
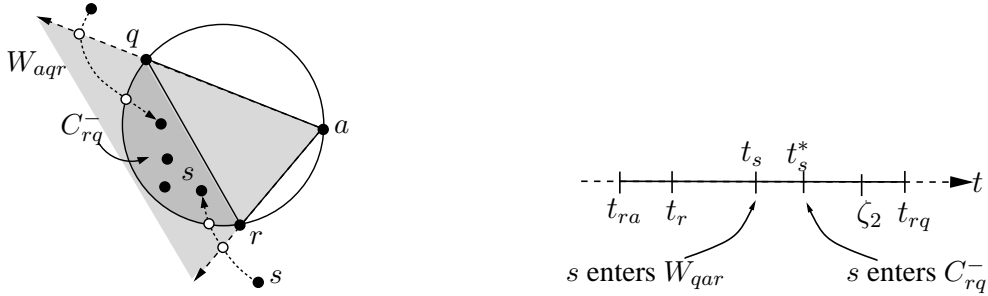

\begin{center}
\input{EnterTheWedge.pstex_t}\hspace{3cm}\input{TimelineS.pstex_t}
\caption{\small Case (b). At least $\ell$ points $s\neq p$ enter the cap $C^-_{rq}$ during $(t_r,\zeta_2)$ ($p$ is not shown). Each of the first $\ell$ of these points causes an $(\ell+1)$-shallow co-circularity with $q,a,r$.
Each of them must first enter the wedge $W_{qar}$, which is empty at time $t_r$, through one of the rays $\vec{aq},\vec{ar}$ (outside the edges $aq$ and $ar$), because none of them can cross $rq$.}
\label{Fig:EnterTheWedge}
\vspace{-0.4cm}
\end{center}
\end{figure}

\medskip
\noindent{\bf Case (b).} There exist at least $\ell$ points, distinct from $p$, 
that enter the cap $C^-_{rq}=B[q,a,r]\cap \L_{rq}^-$ during $(t_r,\zeta_2)$. We refer to Figure \ref{Fig:EnterTheWedge} and
let $s$ be any of these points. By Condition (Q6), $s$ cannot hit $rq$ during the interval $(t_r,\zeta_2)$ (which is covered by $[t_{ra},t_{rq}]$). Note also that $C_{rq}^-$ is contained in the wedge $W_{qar}=\L_{qa}^+\cap\L_{ra}^-$.
Therefore, and
since the wedge $W_{qar}$ is empty immediately after time $t_r$ (when $q,a$ and $r$ are collinear), the above point $s$ has to enter $W_{qar}$, through one its rays $\vec{ar},\vec{aq}$, during $(t_r,\zeta_2)$ and before it enters $C^-_{rq}$. 

Furthermore, Condition (Q6) implies that $s$ can enter the cap $C^-_{rq}$ only through the boundary of $B[q,a,r]$, which results in a co-circularity of $q,a,r,s$. (Recall also that $s$ enters each halfplane $\L_{qa}^+$ and $\L_{ra}^-$ at most once, so it crosses the ray $\vec{ar}$ or $\vec{aq}$ {\it outside} the respective edge $ar$ or $aq$ when entering $W_{qar}$ as above. Indeed, otherwise $s$ would be able to access $C^-_{rq}$, after crossing one of these two edges, only through the interior of $rq$.)

Assume that $s$ is
among the first $\ell$ points to enter $C^-_{rq}$ during $(t_r,\zeta_2)$. Let $t^*_s$ denote the time of the corresponding co-circularity of $q,a,r,s$, which occurs when $s$ enters $C^-_{rq}$.
Since $\Q$ satisfies Condition (Q6) (and $t_s^*$ belongs to $(t_{ra},t_{rq})$), the opposite cap $C^+_{rq}=B[q,a,r]\cap \L_{rq}^+$ contains no points of $P\setminus \{p\}$ at time $t^*_s$. (Otherwise, the Delaunayhood of $rq$ would then be violated by $s$ and another point of $P\setminus \{p\}$, contrary to (Q6).)
Therefore, and since the motion of $B[q,a,r]$ is continuous during $(t_r,\zeta_2)$, the co-circularity at time $t^*_s$ has to be $(\ell-1)$-shallow in $P\setminus \{p\}$, and thus $\ell$-shallow in $P$.

Note also that the crossing $(ar,q,H)$ has to end before $t_s^*$ (that is, $t_5<t_s^*$). Indeed, the Delaunayhood of $qr$ is violated, right after time $t_s^*$, by $s$ and $a$, which is forbidden by Lemma \ref{Lemma:Crossing} during $H$.

We distinguish between two possible subcases. In each of them we dispose of $\Q$ by charging it, within one of the arrangements $\A_{ra},\A_{qa}$, either to $\Omega(\ell^2)$ $(2\ell)$-shallow co-circularities, or to a $(2\ell)$-shallow collinearity.

\smallskip
\noindent{\bf Case (b1).} At least half of the above points $s$ cross the line $\L_{ra}$, from $\L_{ra}^+$ to $\L_{ra}^-$, during $(t_r,t^*_s)$. Since $s$ lies in $\L_{ra}^-$ at time $t_s^*$, $s$ enters $\L_{ra}^-$ exactly once during $(t_r,t_s^*)$, and it does not return to $\L_{ra}^+$ before $t_s^*$; see the motion of the marked point $s$ in Figure \ref{Fig:EnterTheWedge} (left). 
Moreover, by Condition (Q5), each of these crossings occurs outside $ra$ (i.e., within one of the outer rays of $\L_{ra}$).

To dispose of $\Q$, we again fix one of the aforementioned points $s$ and argue as in Section \ref{Sec:ReduceToCrossings}. If the halfplane $\L_{ra}^-$ contains at most $2\ell$ points of $P$ when $s$ enters it, then we encounter a $(2\ell)$-shallow collinearity of $a,r,s$.
Otherwise, the disc $B[a,r,s]$ contains at least $2\ell$ points right after the crossing, so
the three points $r,a,s$ are involved in at least $\ell$ $(2\ell)$-shallow co-circularities before time $t_s^*$ (when the open disc $B[a,r,s]$, equal to $B[a,r,q]$ at that time, contains $\ell$ or fewer points of $P$).
After repeating the above argument for each of the (at least) $\ell/2$ possible choices of $s$, we encounter in $\A_{ra}$ (during $(t_r,\zeta_2)$) either $\Omega(\ell^2)$ $(2\ell)$-shallow co-circularities, or a $(2\ell)$-shallow collinearity.
In both cases, we charge $\Q$ to these events.

We claim that each $(2\ell)$-shallow event, which occurs in $\A_{ra}$ at some time $t^*\in (t_r,\zeta_2)$, is charged by at most $O(1)$ quadruples $\Q$. Indeed, $r$ and $a$ are among the three or four points involved in every charged event. Moreover, according to Condition (Q5) and the argument in case (e) of Stage 1, $q$ is among the last two points to hit the edge $ra$ before time $t^*$. Hence, knowing $t^*$ allows us to guess the three points $q,a,r$ (which uniquely determine $\Q$) in at most $O(1)$ ways. In conclusion, the above scenario happens for at most $O\left(\ell^2 N(n/\ell)+\ell n^2\beta(n)\right)$ quadruples $\Q$ in $\F$.

\smallskip
\noindent{\bf Case (b2).} 
At least half of the above points $s$ remain in $\L_{ra}^-$ throughout the respective intervals $(t_r,t_s^*)$. 
Each of these points must enter $W_{qar}$ (during $(t_r,t_s^*)$) through the ray emanating from $q$ in direction $\vec{aq}$, thereby crossing $\L_{qa}$ from $\L_{qa}^-$ to $\L_{qa}^+$. (Recall that such a collinearity of $q,a,s$ can occur only once during $(t_r,t_s^*)$.)

Once again, we fix one of the above points $s$ and let $t_s$ denote the time in $(t_r,\zeta_2)$ when $s$ enters $W_{qar}$ through the ray emanating from $q$ in direction $\vec{aq}$. 
Arguing as in the previous case, we conclude that the three points $q,a,s$ are involved (during $(t_s,t^*_s)\subset (t_r,\zeta_2)$) either in a $(2\ell)$-shallow collinearity, or in $\Omega(\ell^2)$ $(2\ell)$-shallow co-circularities. Below we prove that each of the $(2\ell)$-shallow events, that occur in $\A_{qa}$ during $(t_r,\zeta_2)$, can be traced back to $\Q$ in at most $O(k)$ ways.\footnote{Note the difference between the two subcases: Here we only know $q,a$, and then guessing $r$ is not immediate.} Hence, it is charged at most $O(k)$ times. We then repeat the same argument for each of the remaining $\ell/2-1$ choices of $s$, and use (as in case (b1)) the standard bounds on the number of $(2\ell)$-shallow events of each type. As a result, we obtain an upper bound of 
$O\left(k\ell^2N(n/\ell)+k\ell n^2\beta(n)\right)$
of the number of such quadruples $\Q$.

To conclude, the overall number of quadruples $\Q$ that fall into Case (b) is at most

$$
O\left(k\ell^2N(n/\ell)+k\ell n^2\beta(n)\right).
$$

To complete the analysis of Case (b), we show that each $(2\ell)$-shallow event that occurs in $\A_{aq}$ during $(t_r,\zeta_2)$ is charged as above by at most $O(k)$ quadruples $\Q$ that fall into case (b2).
Let $t^*$ be the time of such an event.
First, we guess the points $q,a$, in $O(1)$ possible ways, from among the three or four points involved in the event.
Recall that, in the charging scheme of case (b2), each $(2\ell)$-shallow co-circularity or collinearity that we charge in $\A_{qa}$ is ``obtained" via some point $s$, which is also involved in this event and enters $\L^+_{qa}$ at some prior time $t_s$.
We, therefore, guess $s$ among the remaining one or two points that participate in the event under consideration.
To guess the remaining points $r$ and $p$ of $\Q$, we examine all ``candidate" quadruples $\Q'\in \F_{qa}$ whose two ``middle" points are shared with $\Q$. Recall that each of these quadruples $\Q'=(p',q,a,r')$ is accompanied by an ``almost Delaunay" crossing $(ar',q,H_{\Q'})$, where $r'$ enters $\L_{qa}^+$ at some time $t_{r'}\in H_{\Q'}$. Also recall that $\Q'$ is uniquely determined by the choice of $r'$ (as long as $q$ and $a$ remain fixed).

It suffices to consider only quadruples $\Q'=(p',q,a,r')$, in $\F_{qa}$, with the following properties:  (1) $s\neq p',r'$, (2) $t_{r'}<t_s$, and (3) $s$ lies in $\L_{ar'}^+$ during the second portion of $H_{\Q'}$ (after $t_{r'}$). This is because each of these conditions holds for $\Q'=\Q$ (and for $s$) in the charging scheme of case (b2). For example, (3) follows because we assume that case (b1) does not occur (and since $t_5<t_s^*$).
The corresponding points $r'$, which determine the above quadruples $\Q'$, are called {\it candidates} (for $r$). 


\begin{figure}[htbp]
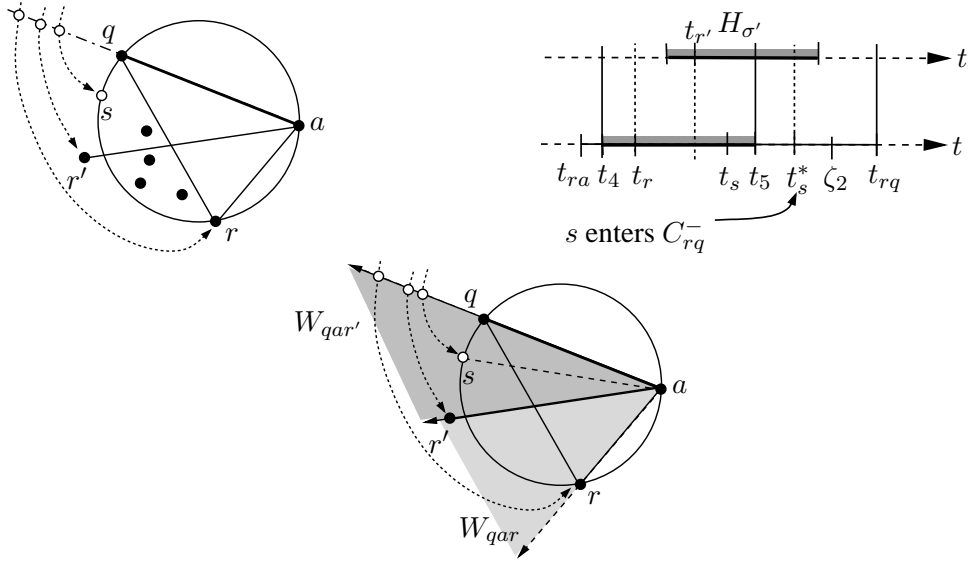

\begin{center}
\input{Candidates.pstex_t}\hspace{3cm}\input{TimelineS1.pstex_t}\hspace{3cm}\input{TwoWedges.pstex_t}
\caption{\small Top: Proposition \ref{Prop:kUnique}: $r$ is among the last $k+3$ candidates $r'$ to enter $\L_{qa}^+$ before time $t_s$; the various critical events occur in the depicted order. Bottom: Proof of Proposition \ref{Prop:kUnique}. The candidate $r'$ remains in $W_{qar}$ throughout $(t_{r'},t_s^*)$. If $H_{\Q'}$ ends after $t_s^*$, then the point $s$ remains in $W_{qar'} (\subset W_{qar})$ throughout $(t_s,t_s^*)$.}
\label{Fig:PropositionCandidates}
\vspace{-0.5cm}
\end{center}
\end{figure}

\begin{proposition}\label{Prop:kUnique}
With the above assumptions, the point $r$ is among the last $k+3$ candidates $r'$ to enter the halfplane $\L^+_{qa}$  before $t_s$ (each candidate at the respective time $t_{r'}$).
\end{proposition}
\begin{proof}
Refer to Figure \ref{Fig:PropositionCandidates}.
Assume to the contrary that the proposition does not hold (for $\Q$ and $s\neq p,q,a,r$ as above).
Hence, we have at least $k$ candidates $r'$ such that $t_r<t_{r'}<t_s$ and $r'\not\in \{p,r\}$, and such that the  points $p'$ of their respective quadruples $\Q'=(p',q,a,r')$ are distinct from $r$. (We continue to assume that $\Q$ satisfies property (PH), introduced in case (a), so the last two restrictions on $p'$ and $r'$ exclude from our consideration at most three candidates $r'$, with their quadruples $\Q'$.)

To establish the proposition, we fix a candidate $r'$ and its corresponding quadruple $\Q'=(p',q,a,r')$, as above, and argue that the respective interval $H_{\Q'}$ ends during $(t_5,t_{rq})$. Repeating the same argument for the remaining $k-1$ possible choices of $r'$ will imply that the quadruple $\Q$ falls into case (a) and thereby reach a contradiction.

Indeed, since $t_r<t_{r'}$, Lemma \ref{Lemma:OrderRelaxedCrossings} shows that
the interval $H_{\Q'}$ ends after $H=[t_4,t_5]$. (As in case (a), the lemma relies on the assumption that $p\neq r'$ and $r\neq p'$.)
It remains to check that $H_{\Q'}$ ends before $t_{rq}$.

If $H_{\Q'}$ ends before $t_s^*$, then we are done (as $t_s^*<t_{rq}$). Hence, we may also assume that both times $t_{r'}$ and $t_s^*$ belong to the interval $H_{\Q'}$ (as depicted in Figure \ref{Fig:PropositionCandidates} (top-right)).
This, and the choice of $r'$ as a candidate for $r$, implies that $r'$ remains in the halfplanes $\L_{qa}^+, \L_{sa}^+$ throughout the interval $(t_{r'},t_s^*)\subseteq H_{\Q'}$. Indeed, $r'$ cannot re-enter $\L_{qa}^-$ during the second portion of $H_{\Q'}$, after entering $\L_{qa}^+$ at time $t_{r'}\in H_{\Q'}$. (This is because $q,a,r'$ perform only one collinearity during the crossing $(ar',q,H_{\Q'})$.) Similarly, since $\Q'$ satisfies property (3), the point $s$ remains in $\L_{ar'}^+$ throughout $(t_{r'},t_s^*)$ (so $r'$ remains in $\L_{sa}^+$). 
We thus conclude that $s$ lies inside $W_{qar'}=\L_{qa}^+\cap \L_{r'a}^-$ throughout the interval $(t_s,t_s^*)$; see Figure \ref{Fig:CandidatesProp} (bottom).

Also notice that, with the above assumptions, $r'$ must lie, throughout the longer interval $(t_{r'},t^*_s)\subseteq H_{\Q'}$, inside the wedge $W_{qar}=\L_{qa}^+\cap \L_{ra}^-$. Indeed, $r'$ enters $W_{qar}$ at time $t_{r'}\in (t_r,t^*_s)(\subseteq (t_r,\zeta_2)\cap H_{\Q'})$ and cannot again cross the ray $\vec{aq}$ during $H_{\Q'}$. 
Moreover, if $r'$ leaves $W_{qar}$ (during $(t_{r'},t^*_s)$) through the other ray $\vec{ar}$, then the edge $ar'$ is hit by $r$, or the edge $ar$ is hit by $r'$. Clearly, the former crossing is forbidden by Lemma \ref{Lemma:Crossing} during the interval $H_{\Q'}$ (where $ar'$ experiences a Delaunay crossing by $q$), and the latter one is ruled out by Condition (Q5). (As a matter of fact, in the second case $r'$ must also cross $rq$, thereby entering $\triangle qar$, before it reaches $ra$. This collinearity is also impossible by Condition (Q6).)


\begin{figure}[htbp]
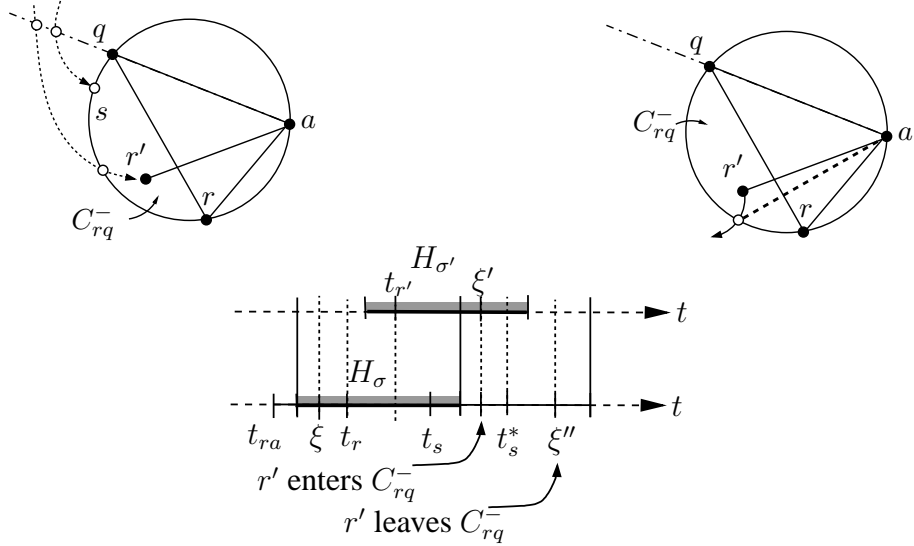

\begin{center}
\input{Candidates2.pstex_t}\hspace{4cm}\input{Candidates3.pstex_t}\hspace{4cm}\input{TimeLineStillEnds.pstex_t}
\caption{\small Proof of Proposition \ref{Prop:kUnique}: The scenario where $r'$ lies within $B[q,a,r]$ at time $t_s^*$. Left: $r'$ enters $C_{rq}^-$ during $(t_r,t^*_s)$ through the arc $C[q,a,r]\cap \L_{rq}^-$, at some time $\xi'$ (left). Right: $r'$ must leave $C_{rq}^-$ before $t_{rq}$ (and after $H_{\Q'}$). Below: The points $q,a,r,r'$ are co-circular at times $\xi\in H_{\Q}\setminus H_{\Q'},\xi'\in H_{\Q'}\setminus H_{\Q}$ and $\xi''\in (\xi',t_{rq}]$.
The interval $H_{\Q'}$ ends before $\xi''$ (and, thus, before $t_{rq}$).}
\label{Fig:CandidatesProp}
\vspace{-0.4cm}
\end{center}
\end{figure} 

To recap, we can assume that $H_{\Q'}$ ends {\it after} $t_s^*$, and that the edges $aq,as,ar'$ and $ar$ appear, at time $t_s^*$, in counterclockwise order around $a$.
To show that $H_{\Q'}$ ends before $t_{rq}$, we distinguish between two possible cases. 

\smallskip
\noindent(1) If $r'$ lies outside $B[q,a,s]=B[q,a,r]$ at time $t_s^*$, then the Delaunayhood of the edge $ar'$ is violated, at that very moment, by the points $s\in \L^+_{ar'}$ and $r\in \L^-_{ar'}$ (as depicted in Figure \ref{Fig:PropositionCandidates} (left)). Since $p'\not\in\{s,r\}$, the crossing $(ar',q,H_{\Q'})$ (occurring in $\DT(P\setminus \{p'\})$)
has to end before $t_s^*$, which is contrary to our assumptions.

\smallskip
\noindent(2) Now suppose that $r'$ lies at time $t_s^*$ within $B[q,a,r]$, as depicted in Figure \ref{Fig:CandidatesProp} (left).
Since $r'$ remains in $W_{qar}$ throughout $(t_{r'},t^*_s]$ (and since $r'$ lies outside $B[q,a,r]$ at time $t_{r'}$, when it enters $W_{qar}$), it can enter $B[q,a,r]$ (or, more precisely, its cap $C_{rq}^-$) during $(t_{r'},t_s^*)$ only through the circular arc $C[q,a,r]\cap \L_{rq}^-$. When that happens,
we encounter a co-circularity of $q,a,r,r'$ at some time $\xi'\in(t_{r'},t^*_s]\subseteq H_{\Q'}$, right after which the Delaunayhood of $rq$ is violated by $r'\in \L_{rq}^-$ and $a\in \L^+_{rq}$. 
Since $p\neq r'$ and $r\neq p'$, this co-circularity occurs after $H=H_{\Q}$. 

Applying Lemma \ref{Lemma:OnceCollin} to $(ar,q,H)$ shows that another co-circularity of $q,a,r,r'$ (red-blue with respect to $ar$ and thus monochromatic with respect to $ar'$) must occur at some time $\xi<\xi'$ during the symmetric interval $H_{\Q}\setminus H_{\Q'}$.
As is easy to check\footnote{Note, for instance, that $(a,r,r',q)$ is a counterclockwise quadruple in $\DT(P\setminus \{p,p'\})$, so the argument preceding Lemma \ref{Lemma:ReturnsStillCross} applies to it.}, $\xi$ and $\xi'$ are the only co-circularities of $q,a,r,r'$ to occur during $H_{\Q}$ and $H_{\Q'}$.

To complete our analysis, we apply Lemma \ref{Lemma:MustCross} for the edge $rq$, with the reference interval $(\xi',t_{rq}]$. By Conditions (Q3) and (Q6), neither of $a,r'$ can cross $rq$ during the larger interval $[t_r,t_{rq}]$. Therefore,
we encounter a third co-circularity of $q,a,r,r'$ at some time $\xi''$ in $(\xi',t_{rq}]$, which occurs when $r'$ leaves the cap $C_{rq}^-$. See Figure \ref{Fig:CandidatesProp} (right). Since $\xi$ and $\xi'$ are the only co-circularities to occur during $H_\Q\cup H_{\Q'}$, the third co-circularity $\xi''$ must occur {\it after} $H_{\Q'}$. (See Figure \ref{Fig:CandidatesProp} (bottom).)
Hence, $H_{\Q'}$ has to end before $t_{rq}$ also in this last case.
\end{proof}

\begin{figure}[htbp]
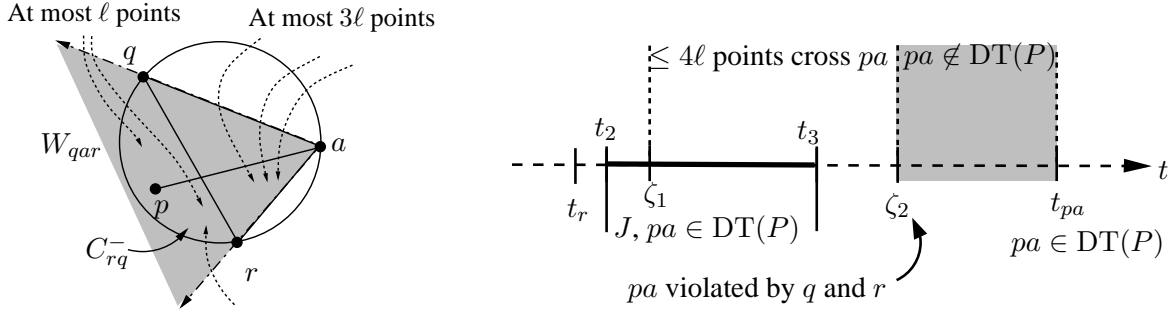

\begin{center}
\input{TheWedgeFinal.pstex_t}\hspace{2cm}\input{PaDiagram.pstex_t}
\caption{\small Left: Case (c). At most $\ell$ points of $P\setminus \{p\}$ enter $C^-_{rq}$, and at most $3\ell$ points of $P\setminus \{p\}$ cross $qa$, during $(t_r^,\zeta_2)$. Hence, at most $4\ell$ points cross $pa$ during $(\zeta_1,\zeta_2)$. Right: a schematic summary of our setup in case (c).}
\label{Fig:CrossPa}
\vspace{-0.4cm}
\end{center}
\end{figure}

\medskip
\noindent{\bf Case (c).} 
Assume that none of the previous cases or preliminary chargings applies to $\Q$.
In particular, since the charging within $\A_{qa}$ following case (a) does not apply, at most $3\ell$ points of $P\setminus \{p\}$ cross $qa$ during $(t_r,\zeta_2)$. 
Furthermore, since case (b) does not occur, at most $\ell$ points of $P\setminus \{p\}$ enter the cap $C^-_{rq}=B[q,a,r]\cap \L^-_{rq}$, during the interval $(t_r,\zeta_2)$. See Figure \ref{Fig:CrossPa} (left).

We again emphasize that, by condition (Q5), no point in $P\setminus \{p,q\}$ can hit the edge $ra$ during the interval $[t_{ra},t_{rq}]$ (which contains $[t_r,\zeta_2]$). Similarly, condition (Q6) implies that no point in $P\setminus \{p,a\}$ can hit the edge $rq$ during that interval.

We claim that at most $4\ell$ points of $P\setminus\{p,a\}$ can hit the edge $pa$ during the interval $(t_3,\zeta_2)$ ($\subseteq (\zeta_1,\zeta_2)$).
Indeed, fix any of these points $s$. 
Recall the edge $pa$ is contained during the interval $(\zeta_1,\zeta_2)$ in the region $B[q,a,r]\cap W_{qar}$; see Figures \ref{Fig:TopSetup} (right) and \ref{Fig:CrossPa} (left). Hence, $s$ has to lie in $B[q,a,r]\cap W_{qar}$ when it hits $pa$, as well.
Since $W_{qar}$ contains no points of $P$ at time $t_r$, the point $s$ has to enter this wedge during $(t_r,\zeta_2)$ through one of the rays $\vec{ar},\vec{ap}$. 
If $s$ crosses $pa$ within $\L_{rq}^-$ then, in particular, it has to enter the cap $C^-_{rq}$ during $(t_r,\zeta_2)$.
Otherwise, if $s$ hits $pa$ within $\L_{rq}^+$, then it must have previously entered the triangle $\triangle qar$ through the edge $qa$. (By Conditions (Q5) and (Q6), $s$ cannot crosses either of the edges $ra,rq$ during $(t_{r},\zeta_2)$.) 
We thus conclude that the overall number of points in $P$ that cross $pa$ during $(t_3,\zeta_2)$ cannot exceed $\ell+3\ell=4\ell$. 

\smallskip
\noindent{\bf Charging events in $\A_{pa}$.} The above analysis implies, in particular, that the edge $pa$ undergoes at most $4\ell$ Delaunay crossings within $(t_3,\zeta_2)$. If the edge $pa$ never re-enters $\DT(P)$ after time $\zeta_2$, then $(pa,r,J)$ is among the last $4\ell+1$ Delaunay crossings of $pa$. Clearly, this scenario happens for at most $O(\ell n^2)$ quadruples $\Q$. 

Otherwise, let $t_{pa}$ be the first time after $\zeta_2$ when $pa$ re-enters $\DT(P)$. Refer to the schematic Figure \ref{Fig:CrossPa} (right). Since the co-circularity at time $\zeta_2$ is the {\it last} co-circularity of the points of $\Q$, Lemma \ref{Lemma:MustCross} implies that the edge $pa$ is hit during $(\zeta_2,t_{pa}]\subseteq (t_3,t_{pa}]$ by at least one of the remaining two points $q$ and $r$. 

We apply Theorem \ref{Thm:RedBlue} in $\A_{pa}$ over the interval $(t_3,t_{pa})$, with the third constant parameter $h$ (noting that $pa$ is Delaunay at both endpoints of that interval). If one of the Conditions (i), (ii) holds, we charge $\Q$ (via $(pa,r,J)$) either to an $h$-shallow collinearity, or to $\Omega(h^2)$ $h$-shallow co-circularities (where each charged event occurs during $(t_3,t_{pa})$ and involves $p$ and $a$, together with one or two additional points of $P$).
Any such $h$-shallow event is charged by at most $O(\ell)$ quadruples. Indeed, the two points $p,a$ can be guessed in at most $O(1)$ possible ways out of the three or four points involved in it, and $(pa,r,J)$ is among the last $4\ell+1$ Delaunay crossings of $pa$ to end before the respective time of the event.
Therefore, the above charging accounts for at most $O\left(\ell h^2N(n/h)+\ell h n^2\beta(n)\right)$ quadruples $\Q$.

\begin{figure}[htbp]
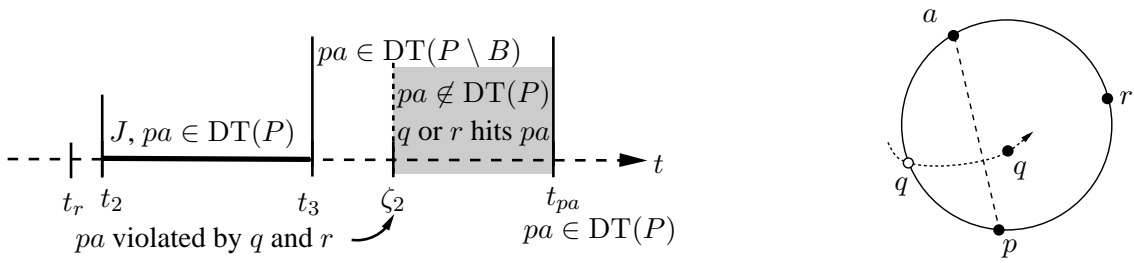

\begin{center}
\input{PaDiagram1.pstex_t}\hspace{3cm}\input{SpecialCrossing1.pstex_t}
\caption{\small Left: If Condition (iii) of Theorem \ref{Thm:RedBlue} holds, then we have a subset $B$ of at most $3h$ points whose removal restores the Delaunayhood of $pa$ throughout $[t_2,t_{pa}]=J\cup [t_3,t_{pa}]$. Right: If $q$ hits $pa$ during $[t_3,t_{pa}]$, then $(P\setminus B)\cup \{q\}$ induces a Delaunay crossing of $pa$ by $q$.}
\label{Fig:CrossPa1}
\vspace{-0.4cm}
\end{center}
\end{figure}

Assume then that Condition (iii) of Theorem \ref{Thm:RedBlue} holds. That is, $P$ contains a subset $B$ of at most $3h$ points (possibly including one, or both of the points $q,r$) such that the edge $pa$ belongs to $\DT(P\setminus B)$ throughout the interval $J\cup [t_3,t_{pa}]=[t_2,t_{pa}]$. See Figure \ref{Fig:CrossPa1} (left).

If $pa$ is crossed by $r$ during $[t_3,t_{pa}]$, then the smaller set $(P\setminus B) \cup \{r\}$ yields two Delaunay crossings of $pa$ by the same point $r$. The routine combination of Lemma \ref{Lemma:TwiceCollin} with the probabilistic argument of Clarkson and Shor implies that the overall number of such triples $p,a,r$ in $P$ is at most $O(h n^2)$. Clearly, this also bounds the overall number of such quadruples $\Q$.

Assume then that $pa$ is hit by $q$, as depicted in Figure \ref{Fig:CrossPa1} (right). If this happens twice during $(t_3,t_{pa})$ then the smaller set $(P\setminus B)\cup \{q\}$ induces either two single Delaunay crossings or one double Delaunay crossing, of the edge $pa$ by $q$. In each of these cases, we can show, as usual, that the overall number of such triples $p,q,a$ in $P$ is at most $O(h n^2)$ by combining Lemma \ref{Lemma:TwiceCollin} or Theorem \ref{Thm:SpecialCrossings} with the probabilistic argument of Clarkson and Shor.
Furthermore, $(pa,r,J)$ is among the last $4\ell+1$ Delaunay crossings that the edge $pa$ undergoes before being hit by $q$. Hence, this scenario occurs for at most $O(\ell h n^2)$ Delaunay quadruples $\Q$ in $\F$.

To recap, we may assume that $q$ hits the edge $pa$ only once during $(t_3,t_{pa})$, so this edge undergoes a single Delaunay crossing by $q$ within $(P\setminus B)\cup \{q\}$. 

\paragraph{Probabilistic refinement.}
Consider a random sample $R$ of $\lceil n/h\rceil$ points chosen at random from $P$. 
Notice that the following two conditions hold simultaneously, with probability at least $\Omega(1/h^4)$:
(1) the four points of $\Q$ belong to $R$, and (2) $R$ includes none of the points of $B\setminus \{q,r\}$.

If the sample $R$ is indeed successful, the four points $p,q,a,r$ define a Delaunay quadruple with respect to $R$.
Let $\F_R$ be the resulting family of such hereditary Delaunay quadruples in $R$. Clearly, $\F_R$ is $3$-refined
(with respect to the underlying set $R$).
In addition, each quadruple $\Q$ in $\F_R$ satisfies the following new condition:

\smallskip
\noindent{\bf (Q7)} The edge $pa$ belongs to the triangulation $\DT(R\setminus \{q,r\})$ throughout the interval $(t_2,t_{pa})$, where $t_{pa}$ denotes the first time after $\zeta_2$ when the edge $pa$ re-enters $\DT(R)$.
Moreover, $pa$ is hit in $(t_3,t_{pa}]$ by $q$, but not by $r$, and this occurs only once during $(t_3,t_{pa}]$. 
In particular, the point set $R\setminus \{r\}$ induces a single Delaunay crossing $(pa,q,\I_r)$, whose interval $\I_r$ is contained in $(t_3,t_{pa}]$.

\medskip
We say that a family $\F$ of quadruples is {\it $4$-refined} if (1) it is $3$-refined, and (2) its quadruples satisfy the above condition (Q7) with respect to the underlying point set $P$ (i.e., with $R$ replaced by $P$).
For each quadruple $\Q$ in such a $4$-refined family $\F$, we refer to the corresponding crossing $(pa,q,\I_r)$ (which figures in condition (Q7)) as the {\it special crossing} of $pa$ by $q$ in $\F$.

As in the previous conditions, when regarding $R$ as an underlying point set, some of the critical times (e.g., $t_{pa}$) may shift.
As is easy to check Condition (Q7), we have the following analogue of Propositions \ref{Prop:TwoTriples} and \ref{Prop:UniqueArq}, showing that the notion of a special crossing is well defined:

\begin{proposition}\label{Prop:UniquePaq}
Let $\F$ be a $4$-refined family of Delaunay quadruples. Then every quadruple $\Q=(p,q,a,r)$ in $\F$ is uniquely determined by its triple $(p,a,q)$. 
Hence, there is one-to-one correspondence between Delaunay quadruples of $\F$ and their special crossings, so it remains to bound the number of the latter.
\end{proposition}
\begin{proof}
We are to show that the fourth point, $r$, of $\Q$, is uniquely determined by the first three points $p,a,r$.
Indeed, by condition (Q7), $r$ the {\it last} point of $P$ to cross $pa$, from $\L_{pq}^-$ to $\L_{pq}^+$, before $q$ performs this same type of crossing.
\end{proof}


\medskip
Let $\Psi_4(m)$ denote the maximum cardinality of a $4$-refined family $\F$ of Delaunay quadruples that is defined with respect to a set of $m$ moving points.
The preceding discussion implies the following recurrence:
\begin{equation}
\Psi_3(n)=O\left(h^4\Psi_4(n/h)+\ell h^2N(n/h)+k\ell^2N(n/\ell)+k^2N(n/k)+\ell h n\beta(n)\right),
\end{equation}
for any triple of parameters $12\ll k\ll \ell\ll h$.

\medskip
By the above Proposition \ref{Prop:UniquePaq}, there is one-to-one correspondence between Delaunay quadruples $\Q=(p,q,a,r)$ of a $4$-refined family $\F$, and their respective triples $(p,q,a)$, which yield the corresponding special crossings, so it suffices to bound the number of the latter configurations. This is indeed done in Section \ref{Sec:Special}, whose analysis is formulated mainly in the terms of {\it special} crossings. However, before we proceed in that direction, one last refinement is in order.

\subsection{Stage 5: Extending the almost-Delaunayhood of $pq$}
Let $\F$ be a $4$-refined family of Delaunay quadruples, which is defined over a set $P$ of $n$ moving points.
Let $\Q=(p,q,a,r)$ be a Delaunay quadruple in $\F$, which satisfies all the seven conditions (Q1)--(Q7) that were enforced in the course of the preceding four stages. 

Note that the edge $pq$ belongs to $\DT(P\setminus \{r\})$ throughout the interval $I$ of its Delaunay crossing by $r$. Furthermore, by condition (Q7), the edge $pa$ undergoes in $P\setminus \{r\}$ a Delaunay crossing $(pa,q,\I_r=[\t_0,\t_1])$. Hence, Lemma \ref{Lemma:MustCross} implies that $pq$ belongs to $\DT(P\setminus \{r\})$ also during $\I_r$. We next extend the almost-Delaunayhood of $pq$ from $I$ and $\I_r$ to the rest of $\conn{I}{\I_r}={\sf conv}(I\cup \I_r)$.

\begin{figure}[htbp]
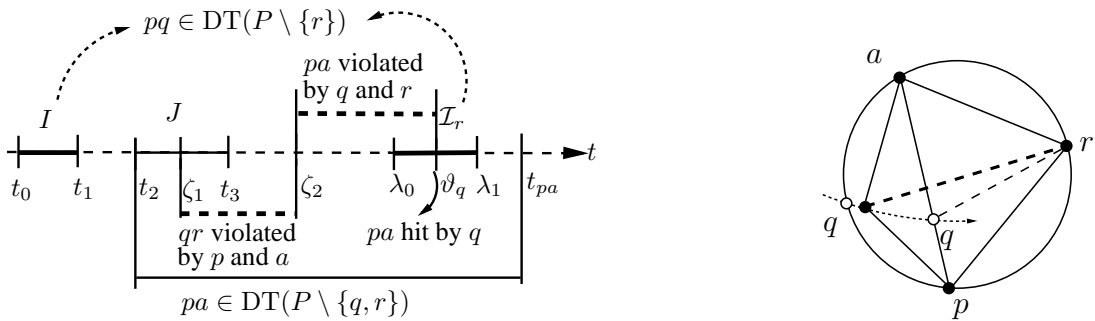

\begin{center}
\input{PqSetupDiagram.pstex_t}\hspace{3cm}\input{SpecialCrossing.pstex_t}
\caption{\small Left: The setup at the beginning of Stage 5. Note that the edge $pq$ belongs to $\DT(P\setminus \{r\})$ throughout each of the intervals $I$ and $\I_r$. The Delaunayhood of $rq$ is violated by $p$ and $a$ between the last two co-circularities $\zeta_1,\zeta_2$.
The edge $pa$ is hit by $q$ at some time $\sp_q\in (\zeta_2,t_{pa})$, and its Delaunayhood is violated by $q$ and $r$ throughout the interval $(\zeta_2,\sp_q)$. Right: A possible motion of $q$ during $(\zeta_2,\sp_q)$.}
\label{Fig:PqSetupDiagram}
\vspace{-0.4cm}
\end{center}
\end{figure}

\paragraph{Setup.} Refer to Figure \ref{Fig:PqSetupDiagram}. By condition (Q3), the Delaunayhood of $rq$ is violated by $p\in \L_{rq}^-$ and $a\in \L_{rq}^+$ between the last two co-circularities $\zeta_1\in J$ and $\zeta_2>t_3$ of $p,q,a,r$ (both of them red-blue with respect to $pa$ and $rq$).
Right after time $\zeta_2$ (when $rq$ is freed from the above violation by $p$ and $a$), the Delaunayhood of $pa$ is violated by $q\in \L_{pa}^-$ and $r\in \L_{pa}^+$.
By condition (Q7), $pa$ re-enters $\DT(P)$ at some time $t_{pa}>\zeta_2$ (which is the first such time after $\zeta_2$), and belongs to $\DT(P\setminus \{r,q\})$ throughout $(t_3,\t_{pa}]$. Finally, $pa$ is hit at some time in $(t_3,t_{pa}]$ by $q$ but not by $r$. Hence, applying Lemma \ref{Lemma:MustCross} from time $\zeta_2$, we conclude that $q$ crosses $pa$ from $\L_{pa}^-$ to $\L_{pa}^+$ at some moment $\sp_q\in (\zeta_2,t_{pa}]$, with the property that the Delaunayhood of $pa$ is violated by $q\in \L_{pq}^-$ and $r\in \L_{pq}^+$ throughout $(\zeta_2,\sp_q)$.
In particular, the aforementioned special crossing $(pa,q,\I_r)$ in $P\setminus \{r\}$ occurs entirely during $(t_3,t_{pa}]$, and its interval $\I_r$ contains the above time $\sp_q$ when $q$ enters $\L_{pa}^+$. 
(However, $\I_r$ need not necessarily contain $\zeta_2$.)

The preceding discussion implies that the intervals $I=[t_0,t_1]$ and $\I_r=[\t_0,\t_1]$ (where $pq$ is known to be almost Delaunay) are indeed disjoint.
We also emphasize that the edges $pa$ and $rq$ intersect throughout $(\zeta_1,\sp_q)=(\zeta_1,\zeta_2)\cup (\zeta_2,\sp_q)$. 

To enforce the almost-Delaunayhood of $pq$ in the resulting gap $(t_1,\t_0)$, we fix a pair of constants $12<k\ll \ell$ and proceed in two steps. 

\medskip
\noindent{\bf Charging events in $\A_{rq}$.}
As a preparation,
we first extend the almost-Delaunayhood of $rq$. 
Recall that, by condition (Q6), $rq$ belongs to $\DT(P\setminus \{p,a\})$ throughout the interval $(t_{ra},t_{rq})$. Here $t_{rq}$ denotes the first time after $t_3$ when $rq$ is Delaunay, and $t_{ra}$ denotes the last time before (or at) $t_0$ when $ra$ is Delaunay.
Note that $(t_{ra},t_{rq})$ contains the respective times $\zeta_0,\zeta_1$ and $\zeta_2$ of the three co-circularities co-circularities of $p,q,a,r$.
Recall also that $\zeta_2$ occurs after the ending time $t_3$ of $J$. Hence, the inequality $t_{rq}>t_3$ is strict, so $rq$ is not Delaunay right before time $t_{rq}$.

We next extend the almost-Delaunayhood of $rq$ to a potentially larger interval $(t_{ra},\sp_q)$ (where, as above, $\sp_{q}$ denotes the time in $\I_r$ when $q$ enters the halfplane $\L_{pa}^+$ through $pa$). 
We can assume, with no loss of generality, that $t_{rq}<\sp_q$. (Otherwise, we are done.)
Therefore, and since $\zeta_2<t_{rq}$, the Delaunayhood of $pa$ is violated by $q\in \L_{pa}^-$ and $r\in \L_{pa}^+$ throughout the interval $(t_{rq},\sp_q)\subset (\zeta_2,\sp_q)$.

We apply Theorem \ref{Thm:RedBlue} in $\A_{rq}$
over the interval $(t_{rq},\sp_q)$, and with the first constant $k$. (This is possible because $rq$ is Delaunay at time $t_{rq}$.)
In the first two cases of Theorem \ref{Thm:RedBlue}, we charge $\Q$ (via $(pq,r,I)$) either to a $k$-shallow collinearity, or to $\Omega(k^2)$ $k$-shallow co-circularities. 
Below we argue that any event in $\A_{rq}$ is charged as above by at most $O(1)$ quadruples $\Q$. 

\begin{figure}[htbp]
\begin{center}
\input{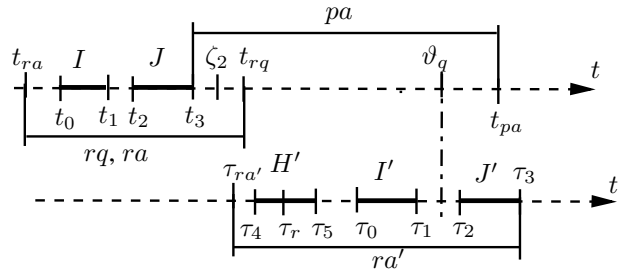}
\caption{\small Proposition \ref{Prop:ExtendRq}: The subfamily $\Gamma_{qr}$ contains at most $3$ quadruples $\Q'=(p',q,a',r)$ whose respective crossings $(p'q,r,I')$ end in $(t_1,\sp_q)$. To establish the proposition, we fix such a quadruple $\Q'$, with $p'\neq a$ and $a'\neq p$, and argue that the second crossing $(p'a',r,J')$ of $\Q'$ ends after $\sp_q$.}
\label{Fig:RqSetupExtend}
\vspace{-0.4cm}
\end{center}
\end{figure} 

Note that the respective points $q$ and $r$ of $\Q$ can be chosen in $O(1)$ possible ways from among the three or four points involved in the event.
Now consider the subfamily $\Gamma_{qr}$ of all quadruples $\Q'=(p',q,a',r)\in \F$ whose second and fourth points  are equal to $q$ and $r$, respectively. (In particular, $\Gamma_{qr}$ includes the quadruple $\Q=(p,q,a,r)$ under consideration.) Notice that each $\Q'\in \Gamma_{qr}$ is composed of two clockwise $(p',r)$-crossings $(p'q,r,I'=[\tau_0,\tau_1])$, $(p'a',r,J'=[\tau_2,\tau_3])$, and comes with a counterclockwise $(r,q)$-crossing $(a'r,q,H'=[\tau_4,\tau_5])$ (which occurs in the smaller set $P\setminus \{p'\}$, and before $J'$ begins). Note also that the first crossing $(p'q,r,I')$ of $\Q'$ is also a counterclockwise $(q,r)$-crossing.

Proposition \ref{Prop:ExtendRq} below implies that the first crossing $(pq,r,I)$ of $\Q$ is among the last four such $(q,r)$-crossings $(p'q,r,I')$ to end before any event that occurs in $\A_{rq}$ during $(t_{rq},\sp_q)$. (See Figure \ref{Fig:RqSetupExtend} for a schematic illustration.)

\begin{proposition}\label{Prop:ExtendRq}
With the above notation, the family $\Gamma_{qr}$ contains at most $3$ quadruples $\Q'=(p',q,a',r)$ whose respective first crossings $(p'q,r,I'=[\tau_0,\tau_1])$ end in $(t_1,\sp_q)$.
\end{proposition}

Hence, any $k$-shallow co-circularity or $k$-shallow collinearity is charged as above by at most $O(1)$ quadruples of $\Gamma_{qr}$, so the above charging accounts for at most $O\left(k^2 N(n/k)+kn^2\beta(n)\right)$ quadruples  $\Q\in\F$.

We can assume, then, that Condition (iii) of Theorem \ref{Thm:RedBlue} holds, so there is a set $A_{rq}$ of at most $3k$ point whose removal restores the Delaunayhood $rq$ throughout $(t_{rq},\sp_q)$.

\paragraph{Proof of Proposition \ref{Prop:ExtendRq}.}
Propositions \ref{Prop:TwoTriples}, \ref{Prop:UniqueArq} and \ref{Prop:UniquePaq} imply that (i) there exist at most $2$ quadruples $\Q'=(p',q,a',r)\in \Gamma_{qr}$ with $p''=a$ or $a''=p$, and (ii) for any other choice of $\Q'\in \Gamma_{qr}\setminus \{\Q\}$, all the six points $p,q,a,r,p',a'$ are distinct. 

Consider all the quadruples quadruples $\Q'\in \Gamma_{qr}$ that fall into the second category, and whose first crossings $(p'q,r,I')$ end in 
$(t_{rq},\sp_q)$. Let $\Q'$ be the unique quadruple of this kind whose respective first crossing $(p'q,r,I'=[\tau_0,\tau_1])$ ends {\it first}. (That is, there is no other quadruple $\Q''=(p'',q,a'',r)\in \Gamma_{qr}$ that satisfies $\{p'',a''\}\cap \{p,a\}=\emptyset$, and whose first crossing $(p''q,r,I'')$ ends in $(t_{rq},\tau_1)$.) Refer to Figure \ref{Fig:RqSetupExtend}.



Let $\tau_{ra'}$ denote the last time before (or at) the beginning $\tau_0$ of $I'$ when the edge $ra'$ is Delaunay.
Since $\Q'$ is $4$-refined, the respective intervals $I'=[\tau_0,\tau_1],J'=[\tau_2,\tau_3]$, and $H'$, of $\Q'$, are all contained in $[\tau_{ra'},\tau_3]$. 
Condition (Q6) on $\Q'$ implies that $rq$ belongs to 
$\DT(P\setminus \{p',a'\})$ throughout $[\tau_{ra'},\tau_3]$.
Therefore, and since both $I'$ and $J'$ end after $t_{rq}$, we get that $\zeta_2<\tau_{ra'}$.
(Otherwise, we would get $\tau_{ra'}<\zeta_2<t_{rq}<\tau_1<\tau_3$, so the above interval $[\tau_{ra'},\tau_3]$ would contain the time $\zeta_2$, right before which the Delaunayhood of $rq$ is violated by $p$ and $a$).

By the choice of $\Q'$, any quadruple $\Q''=(p'',q,r,a'')\in \Gamma_{qr}$ whose respective $(q,r)$-crossing $(p''q,r,I'')$ ends in $(t_{rq},\tau_1)$, must satisfy $p''=a$ or $a''=p$.
Furthermore, Condition (Q2) (on $\Q'$) implies that there exist no quadruples $\Q''\in \Gamma_{qr}$ whose respective $(q,r)$-crossings $(p''q,r,I'')$ end in $(\tau_1,\tau_3)$.
It, therefore, suffices to show that $\tau_3>\sp_q$ (that is, that the second crossing $(p'a',r,J')$ of $\Q'$ ends after $q$ enters $\L_{pa}^+$).

\begin{figure}[htbp]
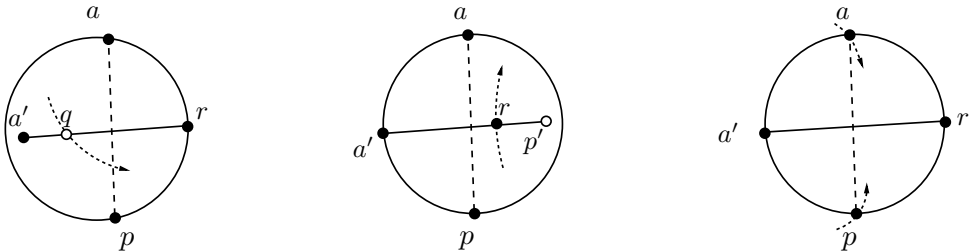

\begin{center}
\input{ChargeRq.pstex_t}\hspace{2cm}\input{ChargeRq1.pstex_t}\hspace{2cm}\input{ChargeRq2.pstex_t}
\caption{\small Proof of Proposition \ref{Prop:ExtendRq}. We assume, for a contradiction, that $\tau_3<\sp_q$, so both crossings $(a'r,q,H')$ and $(p'a',r,J')$ occur within $(\zeta_2,\sp_q)$.
Left: At the time $\tau_q\in H'$ when $q$ hits $a'r$, the Delaunayhood of $pa$ is violated by $a$ and $r'$.
Center: If $ar'$ and $pa$ still intersect at the time in $J'$ when $r$ hits $p'a'$, then the Delaunayhood of $pa$ is violated by $p'$ and $a'$ at some moment during $(\zeta_2,\sp_q)\subset (t_3,t_{pa})$. Right: The last scenario, where $pa$ recovers from its previous violation by $a'$ and $r$ through a co-circularity.}
\label{Fig:RqProp}
\vspace{-0.4cm}
\end{center}
\end{figure}

Indeed, assume for a contradiction that $\tau_3<\sp_q$. Then, recalling that $\tau_{ra'}>\zeta_2$, we conclude that $[\tau_{ra'},\tau_3]$ is contained in the interval $(\zeta_2,\sp_q)$, where the Delaunayhood of $pa$ is violated by $q$ and $r$.

By condition (Q1) on $\Q'$, its edge $ra'$ belongs to $\DT(P\setminus \{p',q\})$ throughout $[\tau_{ra'},\tau_3]$. Hence, at the time $\tau_r\in H'\subset [\tau_{ra'},\tau_3]$ when $q$ enters $\L_{a'r}^+$, the edge $pa$ is intersected by $a'r=a'q\cup qr$, so the Delaunayhood of $pa$ is violated then by $r$ and $a'$. See Figure \ref{Fig:RqProp} (left). (Otherwise, the Delaunayhood of $ra'$ would be violated by $p$ and $a$, which is impossible during $[\tau_{ra'},\tau_3]$.)

If $ra'$ still intersects $pa$ at the time in $J'\subset (\zeta_2,\sp_q)$ when $r$ hits $p'a'$ during the second crossing of $\Q'$, then the same argument shows that Delaunayhood of $pa$ is violated then by $p'$ and $a'$, contrary to condition (Q7) on $\Q$. (See Figure \ref{Fig:RqProp} (center).)
Otherwise, there is a time in $(\tau_{r'},\tau_3)$ when the edge $pa$ recovers from its previous violation by $r$ and $a'$.
Notice that, by condition (Q7), none of $r,a'$ can hit $pa$ during the above interval (which is contained in $(\zeta_2,\sp_q)\subset(t_3,t_{pa})$).
Applying Lemma \ref{Lemma:MustCross} for $\{p,a,r,a'\}$,  we get that the four points $p,a,r,a'$ are involved during $(\tau_{ra'},\tau_3)$ in a red-blue co-circularity with respect to $pa$ and $ra'$ (as depicted in Figure \ref{Fig:RqProp} (right)), contrary to the almost-Delaunayhood of $ra'$ in $(\tau_{ra'},
\tau_3)$. This final contradiction completes the proof of Proposition \ref{Prop:ExtendRq}. $\Box$

\smallskip
We thus can assume, in what follows, that there is a subset $A_{rq}$ of at most $3k$ points whose removal restores the Delaunayhood of $rq$ throughout $(t_{rq},\sp_q)$. 

\paragraph{Charging events in $\A_{pq}$.}
We apply Theorem \ref{Thm:RedBlue} in $\A_{pq}$ over the interval $(t_1,\sp_q)$, which covers the gap $(t_1,\t_0)$ between $I$ and $\I_r$. 

In cases (i) and (ii) of Theorem \ref{Thm:RedBlue}, we charge $\Q$ within $\A_{pq}$ either to an $\ell$-shallow collinearity or to $\Omega(\ell^2)$ $\ell$-shallow co-circularities. We claim that any such event, which occurs in $\A_{pq}$ during $(t_1,\sp_q)$, is charged in this manner by at most $O(k)$ quadruples $\Q=(p,q,a,r)$.

Indeed, the points $p$ and $q$ of $\Q$ can be guessed in $O(1)$ possible ways among the three or four points involved in the event. 
Let $\FQ_{pq}$ denote the sub-family of all quadruples $\Q'=(p,q,a',r')\in \F$ whose first two points are equal to $p$ and $q$, respectively.
Note that $\FQ_{pq}$ includes the quadruple $\Q$ under consideration, and that, for each $\Q'\in \FQ_{pq}$, its first crossing is of the form $(pq,r',I')$.
Proposition \ref{Prop:ChargePq} (below) implies that the first crossing $(pq,r,I)$ of $\Q$ is among the last $6k+3$ such crossings to end before any $\ell$-shallow event that we charge in $\A_{pq}$. Hence, the above charging applies to at most $O\left(k \ell^2 N(n/\ell)+k \ell n^2\beta(n)\right)$ quadruples. 

\begin{figure}[htbp]
\begin{center}
\input{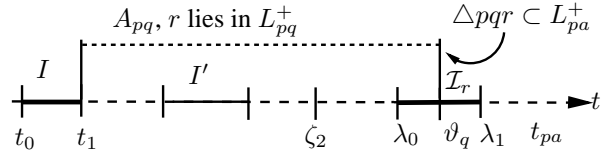}
\caption{\small Extending the almost-Delaunayhood of $pq$ to $(t_1,\t_0)$. We apply Theorem \ref{Thm:RedBlue} over the larger interval $(t_1,\sp_q)$. By Proposition \ref{Prop:ChargePq}, the family $\FQ_{pq}$ contains at most $6k+2$ quadruples $\Q'=(p,q,a',r')$ whose respective first crossings $(pq,r',I')$ end in $(t_1,\sp_q)$.}
\label{Fig:ApplyPq}
\vspace{-0.4cm}
\end{center}
\end{figure} 

Finally, if Condition (iii) of Theorem \ref{Thm:RedBlue} is satisfied, we end up with a subset $A_{pq}$ of at most $3\ell$ points whose removal restores the Delaunayhood of $pq$ throughout $(t_1,\sp_q)$. In this case, we can ``free" $\Q$ from the points of $A_{pq}\setminus \{a,r\}$ (thereby extending the almost-Delaunayhood of $pq$ to $(t_1,\t_0)\subset (t_1,\sp_q)$) through the standard probabilistic argument.

\begin{proposition}\label{Prop:ChargePq}
The family $\FQ_{pq}$ contains at most $6k+2$ quadruples $\Q'=(p,q,a',r')$ whose respective crossings $(pq,r',I')$ end in $(t_1,\sp_q)$.
\end{proposition}
\begin{proof}
Since $p$ and $q$ are fixed, Propositions \ref{Prop:TwoTriples} and \ref{Prop:UniquePaq} imply that any $\Q'\in \FQ_{pq}$ is uniquely determined by each of its respective points $a',r'$.
Hence, we have at most two $\Q'=(p,q,a',r')\in \FQ_{pq}$ that satisfy $a'=r$ or $r'=a$, and, for any other quadruple in
$\FQ_{pq}$, none of its respective points $a'$ and $r'$ is equal to $a$ or $r$. 

Fix $\Q'=(p,q,a',r')\in \FQ_{pq}$ whose respective first crossing $(pq,r',I')$ ends in $(t_1,\sp_q)$, and with the property that $\{a,r\}\cap \{a',r'\}\neq \emptyset$. To establish the proposition, it suffices to show that, for any such $\Q'$, at least one of its points $a',r'$ belongs to the set $A_{rq}$ (obtained at the end of the preceding step) of cardinality is at most $3k$. Indeed, we have at most $3k$ quadruples $\Q'$ with $a'\in A_{rq}$, and at most $3k$ quadruples $\Q'$ with $r'\in A_{rq}$, and each $\Q'\in \FQ_{pq}$ is uniquely determined by each of its respective points $a'$ and $r'$.

We, therefore, proceed to establishing the latter property.
Notice that the intervals $I$ and $I'$ are disjoint, so we have $I'\subset (t_1,\sp_q)$.
Note also that $r$ lies in $\L_{pq}^+$ right after time $t_1$, and also at the later time $\sp_q$ when $q$ hits $pa$
(thereby freeing $pa$ from its previous violation by $q\in \L_{pa}^-$ and $r\in \L_{pa}^+$).
Hence, $r$ has to remain in $\L_{pq}^+$ throughout $(t_1,\sp_q)$ (or, else, it would cross $\L_{pq}$ three times during $I\cup (t_1,\sp_q)$); see Figure \ref{Fig:ApplyPq}. We thus conclude that, at the moment in $I'$ when $r'$ hits $pq$, $r'$ enters the triangle $\triangle pqr$ (whose order type remains fixed throughout $(t_1,\sp_q)$).

\begin{claim}\label{Claim:MustLeavePQR}
Let $t'$ be the time in $I'$ when the above point $r'\in P\setminus P_{\Q}$ enters $\triangle pqr$ through the interior of $pq$. Then $r'$ must leave $\triangle pqr$ during $(t',\sp_q)$.
\end{claim}

\begin{proof} 
Assume for a contradiction that $r'$ remains in $\triangle pqr$ throughout $(t',\sp_q)$.
Recall that $pa$ is intersected by $rq$ throughout $(t_3,\sp_q)\subset (\zeta_1,\zeta_2)\cup (\zeta_2,\sp_q)$, with $q\in \L_{pa}^-$ and $r\in \L_{pa}^+$. Observe that there is a time in $[t_3,\sp_q)$ when $r'$ lies within $\triangle pqr\cap \L_{pa}^-$.
Indeed, this property clearly holds if $r'$ enters $\L_{pa}^+$ in the interval $(t_3,\sp_q)$, where $pq$ is contained in $\L_{pa}^-$; see Figure \ref{Fig:MustLeavePqr} (center). Assume then that
$r'$ enters $\triangle pqr$ before $t_3$ (i.e., $t_3\in (t',\sp_q)$). However, in this case $r'$ has to lie at time $t_3$ within $\triangle pqr\cap \L_{pa}^-$, as depicted in Figure \ref{Fig:MustLeavePqr} (left). 
(Otherwise, $r'$ would lie at that moment in the cap $B[p,a,r]\cap \L_{pa}^+\supset \triangle pqr\cap \L_{pa}^+$, which is known to be $P$-empty throughout the second portion of $J=[t_2,t_3]$.)

To see a contradiction, notice that $\triangle pqr$ lies at time $\sp_q$ entirely within the closure of $\L_{pa}^+$; see Figure \ref{Fig:MustLeavePqr} (right). Therefore, $r'$ too has to enter $\L_{pa}^+$ during $(t_3,\sp_q)$. However, $r'$ cannot cross $\L_{pa}$ during $(t_3,\sp_q)$
through one of its rays outside $pa$ and while remaining inside the triangle $\triangle pqr$ (because the segments $pa$ and $rq$ intersect there), and condition (Q7) on $\Q$ implies that $r'$ cannot hit $pa$ during $(t_3,\sp_q)\subset [t_3,t_{pa}]$. This contradiction completes the proof of Claim \ref{Claim:MustLeavePQR}.
\end{proof}

\smallskip
Consider the first time in $(t',\sp_q)$ when $r'$ leaves $\triangle pqr$, through one of the edges $pr,pq,rq$. (Here, as before, $t'$ denotes the time when $r'$ hits $pq$ during the first Delaunay crossing $(pq,r',I')$ of $\Q'=(p,q,a',r')$.)
Recall that $r'$ cannot cross $pr$ during $(t_1,t_3)$, because $\Q$ is a Delaunay quadruple (that is, $pr$ belongs to $\DT(P\setminus \{q,a\})$ throughout $\conn{I}{J}=[t_0,t_3]$). 
Furthermore, $r'$ cannot cross $pr$ in $(t_3,\sp_q)$ either: otherwise $r'$ would first have to enter $\L_{pa}^+$ through the relative interior of $pa$, contrary to condition (Q7) on $\Q$. We, thereby, conclude that $r'$ can leave $\triangle pqr$ during $(t',\sp_q)\subset (t_1,\sp_q)$ only through one of the remaining edges $qr$ and $pq$.

\begin{figure}[htbp]
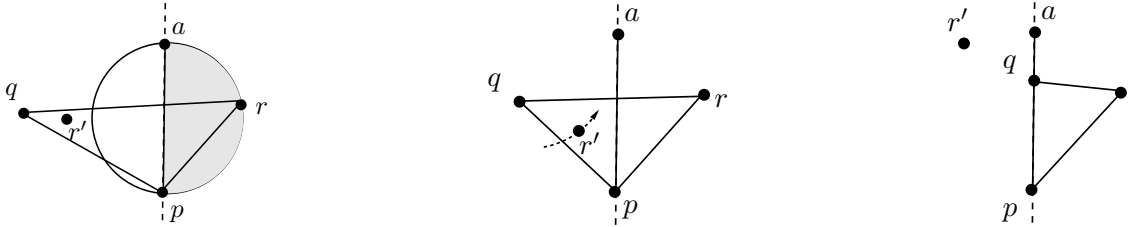

\begin{center}
\input{MustLeavePqr.pstex_t}\hspace{3cm}\input{MustLeavePqr1.pstex_t}\hspace{3cm}\input{MustLeave2.pstex_t}
\caption{\small Proof of Claim \ref{Claim:MustLeavePQR}. Left: If $r'$ enters $\triangle pqr$ during $[t_3,\sp_q)$, this can happen only within $\L_{pa}^-$. Center: If $t'<t_3$ then $r'$ lies in $\triangle pqr\cap \L_{pa}^-$ at time $t_3$ (because the rest of $\triangle pqr$ lies inside the $P$-empty cap $B[p,a,r]\cap \L_{pa}^+$). 
Right: In both cases, $r'$ must exit $\triangle pqr$ before time $\sp_q$ (at which $\triangle pqr$ passes entirely to $\L_{pa}^+$).}
\label{Fig:MustLeavePqr}
\vspace{-0.4cm}
\end{center}
\end{figure}

If $r'$ exits $\triangle pqr$ during $(t',\sp_q)\subset (t_1,\sp_q)$ through the relative interior $rq$, then, by condition (Q2), this can occur only in the smaller interval $(t_{rq},\sp_q)$ (and only if $\sp_q>t_{rq}$). Hence, in this case $q$ belongs to $A_{rq}$, and we are done.

Assume, then, that $r'$ leaves $\triangle pqr$ through the edge $pq$, as depicted in Figure \ref{Fig:PqProp}.
Consider the second Delaunay crossing $(pa',r',J')$ of $\Q'=(p,q,a',r')$. Recall that $I'$ begins after $t_1$ and before the beginning of $J'$, so $(pa',r',J')$ occurs too after the end of $I$. 
Since $\Q'\in \FQ_{pq}$ is $4$-refined, the point $r'$ remains $\L_{pq}^+$ after $t'$ and until the end of $J'$ (or, else, $r'$ would cross $\L_{pq}$ three times). Therefore, $J'$ ends before $r'$ exits $\triangle pqr$ through $pq$ (and, in particular, before $\sp_q$).
To conclude, the second crossing $(pa',r',J')$ of $\Q'$ occurs entirely within $(t_1,\sp_q)$.
To complete our analysis, we distinguish between the following two sub-cases:

If $a'$ lies in $\L_{rq}^+$ at the time in $J'$ when $r'$ hits $pa'$, then $rq$ is intersected at that moment by the Delaunay edge $r'a'$; see Figure \ref{Fig:PqProp} (left). Hence, Delaunayhood of $rq$ is violated at some moment in $J'\subset (t_1,\sp_q)$ by $r'$ and $a'$. Furthermore, condition (Q2) on $\Q$ implies that the above violation is possible only during $(t_{rq},\sp_q)$, so
at least one of $a',r'$ must belong to $A_{rq}$.

\begin{figure}[htbp]
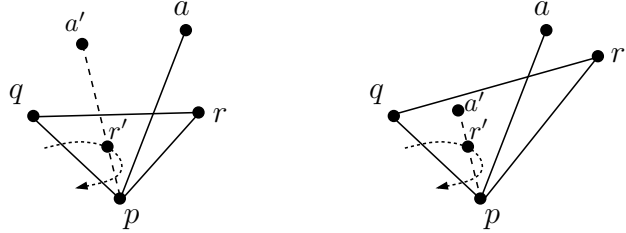

\begin{center}
\input{AgainViolated.pstex_t}\hspace{2cm}\input{AgainViolated1.pstex_t}
\caption{\small Proof of Proposition \ref{Prop:ChargePq}. The second $(p,r')$-crossing $(pa',r',J')$ of $\Q'$ ends before $r'$ hits $pq$ again. The two possible scenarios are depicted.}
\label{Fig:PqProp}
\vspace{-0.4cm}
\end{center}
\end{figure} 

Assume, then, that $a'$ lies in $\L_{rq}^-$ when $r'$ hits $pa'$ during $J'$. Hence, both points $r',a'$ lie at that time inside the triangle $\triangle pqr$; see Figure \ref{Fig:PqProp} (right).
Arguing as before, we conclude that $a'$ leaves $\triangle pqr$ before $\sp_q$ through one of the edges $rq$ and $pq$. However, condition (Q7) on $\Q'$ implies that $a'$ cannot leave $\triangle pqr$ through the edge $pq$: otherwise $q$ would enter the halfplane $\L_{pa}^+$ twice (once during the respective special crossing of $\Q'$, and another time through one of the outer rays of $\L_{pa}\setminus pa$).
Therefore, in this case $a'$ can leave $\triangle pqr$ before $\sp_q$ only through the relative interior of $rq$. Arguing as before, we conclude that $a'$ again belongs to $A_{rq}$. 
\end{proof}

To recap, the previous chargings within $\A_{pq}$ and $\A_{rq}$ altogether account for at most $O(k\ell^2 N(n/\ell)+k^2 N(n/k)+k\ell n^2\beta(n))$ quadruples in our $4$-refined family $\F$. Each surviving quadruple $\Q=(p,q,a,r)$ in $\F$ comes with a subset $A_{pq}$ of at most $3\ell$ points so that $pq$ is Delaunay in $P\setminus A_{pq}$ throughout the gap $(t_1,\t_0)\subset (t_1,\sp_q)$ between the respective intervals $I$ and $\I_r$ of $\Q$.

\paragraph{Probabilistic refinement.}
We apply the probabilistic argument of Clarkson and Shor \cite{CS} one more time.

We say that a family $\F$ of Delaunay quadruples is {\it 5-refined} or, simply, {\it refined} if it is $4$-refined with respect to the underlying point set $P$, and each quadruple $\Q$ in $\F$ satisfies the following new condition:

\medskip
\noindent{\bf (Q8)} The edge $pq$ belongs to $\DT(P\setminus \{a,r\})$ throughout the respective interval $\conn{I}{I_r}=[t_0,\t_1]$. (Here, as above, $I=[t_0,t_1]$ is the interval of the first $(p,r)$-crossings of $\Q$, and $\I_r=[\t_0,\t_1]$ is the interval of the special crossing of $pa$ by $q$.)
\smallskip

That is, we require that the family $\F$ is nonoverlapping, and that its quadruples are Delaunay and satisfy all the $8$ conditions (Q1) -- (Q8).

\medskip
Let $\Psi_5(n)$ denote the maximum cardinality of a refined family of Delaunay quadruples, that can be defined over an underlying set of $n$ moving points. 

The routine sampling argument of Clarkson and Shor \cite{CS} leads to the following recurrence:

$$
\Psi_4(n)\leq O\left(\ell^4\Psi_5(n/\ell)+k\ell^2N(n/\ell)+k^2N(n/k)+kn^2\beta(n)\right).
$$

As argued in the previous section, there is one-to-one correspondence between (1) quadruples $\Q=(p,q,a,r)$ in a refined family $\F$, (2) their respective triples $(p,q,a)$, and (3) the special crossings $(pa,q,\I_r)$ performed by these triples.

As reviewed in the beginning of this section, the analysis of $\Psi_5(m)$ is delegated to Section \ref{Sec:Special}, which primarily deals with the third type of configurations.
 


\section{Special Crossings and Special Quadruples}\label{Sec:Special}



In the preceding section we have established a sequence of recurrences implying that the maximum number $\Psi(n)$ of consecutive quadruples (and, hence, the maximum number $N(n)$ of Delaunay co-circularities) in a set $P$ of $n$ moving points is (asymptotically) dominated by the maximum possible cardinality $\Psi_5(m)$ of a {\it refined} family $\F$ of Delaunay quadruples that is defined over of a certain $m$-size subsample $R\subset P$.

To bound the above quantity $\Psi_5(n)$, for any $n>0$, we fix a set $P$, and a refined family $\F$ of (clockwise) Delaunay quadruples that is defined over $P$. That is, $\F$ is nonoverlapping, and each of its quadruples $\Q=(p,q,a,r)$ satisfies the eight conditions (Q1) -- (Q8) (stated in terms of $p,q,a,r$, $\F$ and $P$). 

In particular, every triple of points of $\Q=(p,q,a,r)\in \F$ yield a Delaunay crossing, which sometimes occurs within a {\it reduced} triangulation obtained by omitting from $P$ the remaining fourth point of $\Q$.
Indeed, recall that $\Q$, as any clockwise quadruple, is formed by a pair of clockwise $(p,r)$-crossings $(pq,r,I)$ and $(pa,r,J)$. The two additional crossings $(ar,q,H)$ and $(pa,q,\I_r)$ have been enforced at Stages 1 and 4 of Section \ref{Sec:CountQuad}, as parts of the respective conditions (Q3) and (Q7), and they occur within the respective appropriately {\it reduced} triangulations $\DT(P\setminus \{p\})$ and $\DT(P\setminus \{r\})$.

Recall also that, according to Propositions \ref{Prop:TwoTriples}, 
\ref{Prop:UniqueArq}, and \ref{Prop:UniquePaq}, each quadruple $\Q$ in $\F$ is {\it uniquely} determined by {\it each} of the four ordered triples $(p,q,r),(p,a,r)$, $(a,r,q)$, and $(p,a,q)$, which realize its four Delaunay crossings. (That is, in each triple the third point performs a clockwise Delaunay crossing of the edge connecting the first two points.)

To bound the cardinality of $\F$, we focus, for each quadruple $\Q=(p,q,a,r)$ in $\F$, on the last type of crossing $(pa,q,\I_r)$, realized by its first three points $p,q,a$, and referred to as the special crossing of $pa$ by $q$. 
We emphasize that $(pa,q,\I_r)$ is also a regular Delaunay crossing which occurs in the smaller triangulation $\DT(P\setminus \{r\})$. For convenience of notation, we refer to $r$ as the {\it outer point} of $(pa,q,\I_r)$.

We further label each special crossing $(pa,q,\I_r)$ as a {\it clockwise (special) $(p,q)$-crossing}, and as a {\it counterclockwise (special) $(a,q)$-crossing}. 
Notice that Lemma \ref{Lemma: OrderOrdinaryCrossings} need not hold for {\it special} $(p,q)$-crossings of the same type (that is, either clockwise or counterclockwise), because these are defined with respect to reduced point sets, each omitting the respective outer point $r$.
As a matter of fact, the respective outer points of any two such $(p,q)$-crossings are always {distinct}, because, as noted above, their ancestor quadruples in $\F$ are uniquely determined by the respective triples $(p,q,r)$. Hence, any two $(p,q)$ crossings (of the same type) are always defined with respect to {\it distinct} point sets.
Instead, we use Lemma \ref{Lemma:OrderRelaxedCrossings}, which imposes certain restrictions on the almost-Delaunay crossings that can be compared by it.
For example, two counterclockwise special $(a,q)$-crossings $(pa,q,\I_r)$ and $(wa,q,\I_u)$, with respective outer points $r$ and $u$, become incompatible if and only if $r=w$ or $p=u$.

We first perform a preliminary pruning step that will ensure, in particular, that Lemma \ref{Lemma:OrderRelaxedCrossings} indeed applies to any pair of surviving counterclockwise special $(a,q)$-crossings. This will be done by considering all possible pairs of {\it distinct} such $(a,q)$-crossings $(pa,q,\I_r)$ and $(wa,q,\I_u)$, and by omitting from $\F$ their corresponding quadruples $\Q=(p,q,a,r)$ and $\Q'=(w,q,a,u)$ if 
they share one or more additional points, apart from $q$ and $a$.
A similar pruning step will ensure that any two clockwise special $(p,q)$-crossings $(pa,q,\I_r)$ and $(pw,q,\I_{u})$ share only the pair $(p,q)$.

The crucial observation is that the overall number of quadruples that we omit from $\F$, at both steps, does not exceed $O(n^2)$.
Indeed, assume, for instance, that a pair $(pa,q,\I_r)$ and $(wa,q,\I_u)$ of counterclockwise special $(a,q)$-crossings share an additional, third point (again, apart from $q$ and $a$).
Recalling that each quadruple $\Q$ in $\F$ is uniquely determined by any ordered sub-triple of its points, we conclude that $p\neq w$ and $r\neq u$. 
That is, we have 
$r=w$ or $p=u$. Assume, with no loss of generality, that $r=w$. Recall that each ordered sub-triple in $\Q$ or in $\Q'$ performs a Delaunay crossing (perphaps within a suitably reduced triangulation). 
We therefore get from $\Q$ the crossing $(ar,q,H)$, within $P\setminus \{p\}$, and we get from $\Q'$ the crossing 
$(wa,q,\I_{u})=(ra,q,\I_{u})$, within $P\setminus \{u\}$.
We thus obtain two {\it distinct}\footnote{Indeed, recall that, in our notation, $q$ crosses $ar$ (during $H$) from $\L_{ar}^-$ to $\L_{ar}^+$, and it crosses the reversely oriented copy $ra$ of $ar$ (during $\I_{u}$) from $\L_{ra}^+=\L_{ar}^-$ to $\L_{ra}^-$.} Delaunay crossings which are performed by the {\it same} triple $(a,r=w,q)$ and within the same reduced triangulation $\DT(P\setminus\{p,u\})$. Hence, a routine combination of Lemma \ref{Lemma:TwiceCollin} with the probabilistic argument of Clarkson and Shor implies that the underlying point set $P$ contains at most $O(n^2)$  such triples $(a,q,r)$. Clearly, this also bounds the overall number of such quadruples $\Q=(p,q,a,r)$ and $\Q'=(w=r,q,a,u)$ that we omit. 
A symmetric analysis is peformed for pairs $(pa,q,\I_r)$ and $(pw,q,\I_u)$ of clockwise $(p,q)$-crossings that have a third point in common, and their respective quadruples $\Q=(p,q,a,r)$ and $\Q'=(p,q,w,u)$.


To conclude, we can assume, from now on, that any pair which consists of any two counterclockwise special $(a,q)$-crossings, or of any two special clockwise $(p,q)$-crossings, involves six distinct points (including the two outer points) and, therefore, satisfies the conditions of Lemma \ref{Lemma:OrderRelaxedCrossings}.
Therefore, all the remaining counterclockwise special $(a,q)$-crossings, with $a,q$-fixed, can be linearly ordered by the starting times of their intervals, or by the ending times of their intervals, or by the times when $q$ hits the corresponding $a$-edge, and all three orders are identical. Furthermore, Lemma \ref{Lemma:OrderRelaxedCrossings} imposes a similar order on the remaining clockwise special $(p,q)$-crossings, with $p,q$ fixed.


\paragraph{Special quadruples.}
We say that two counterclockwise special $(a,q)$-crossings are {\it consecutive} if they are consecutive with respect to the natural order induced by Lemma \ref{Lemma:OrderRelaxedCrossings}. That is, no other counterclockwise special $(a,q)$-crossings appear in this order between them.

\begin{figure}[htbp]
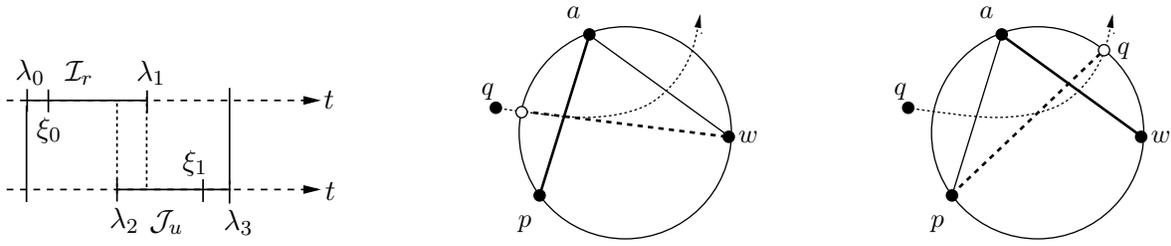

\begin{center}
\input{SpecialCrossingIntervals.pstex_t}\hspace{2cm}\input{Xi0.pstex_t}\hspace{2cm}\input{Xi1.pstex_t}
\caption{\small The special quadruple $\SQ=(a,p,w,q)$. The respective intervals $\I_r$ and $\J_u$ of the two special crossings associated with $\SQ$ are either disjoint, or partially overlapping (left). The points of $\SQ$ are co-circular at times $\xi_0\in \I_r\setminus \J_u$ (center) and $\xi_1\in \J_u\setminus \I_r$ (right).
}
\label{Fig:SpecialQuad}
\vspace{-0.4cm}
\end{center}
\end{figure}

Four points $a,p,w,q$ form a {\it special quadruple} $\SQ=(a,p,w,q)$ if we encounter two (distinct) counterclockwise special $(a,q)$-crossings $(pa,q,\I_r)$ and $(wa,q,\J_u)$, with the respective outer points $r$ and $u$, that occur in this order (that is, $q$ crosses $pa$ before $wa$); these crossings need not be consecutive. Refer to Figure \ref{Fig:SpecialQuad}.
We then use $P_{\SQ}$ to denote the set which consists of the four points $a,p,w,q$ of $\SQ$, and of the two outer points $r$ and $u$. 

\smallskip
\noindent{\bf Remark.} Our notation requires some understanding from the reader: Whenever we talk about a special quadruple $\SQ=(a,p,w,q)$, we also need to specify the two outer points $r$ and $u$. We generally do so, but do not consider them as an integral part of the quadruple, because, until Stage 4, they do not play any role in the topological changes that the quadruple undergoes. However, the outer points will ``return to life" in Stage 4, and then their presence will lead to so called {\it terminal quadruples} which we will use to finish up the analysis. See also the overview below.

Fix a special quadruple $\SQ=(a,p,w,q)$, as above. 
Lemma \ref{Lemma:OnceCollin} implies\footnote{Since the crossings of $\SQ$ are defined with respect to reduced points sets $P\setminus \{r\}$ and $P\setminus \{u\}$, this implication critically relies on the assumption that $p\neq u$ and $w\neq r$.} that the four points $a,p,w,q$ are involved in at least one co-circularity during $\I_r$, and in at least one co-circularity during $\J_u$. 
Specifically, the former co-circularity is red-blue with respect to the edges $pa$ and $qw$, so it must occur before the beginning of $\J_u$, during $\I_r\setminus \J_u$. (See Figure \ref{Fig:SpecialQuad} (center).) Similarly, the latter co-circularity is red-blue with respect to the edges $wa$ and $pq$, so it must occur after the end of $\I_r$, during $\J_u\setminus \I_r$. (See Figure \ref{Fig:SpecialQuad} (right).)
Furthermore, the same argument as in Section \ref{Subseq:Quad} shows that the points of $\SQ$ are involved in {\it exactly one} co-circularity during each of the intervals $\I_r$ and $\J_u$, and we denote the respective times of these co-circularities as $\xi_0\in \I_r\setminus \J_u$ and $\xi_1\in \J_u\setminus \I_r$.

It is also instructive to note that the triangulation $\DT(P\setminus \{r,u\})$ contains an ordinary  counterclockwise quadruple $(a,p,w,q)$, with the associated Delaunay crossings $(pa,q,\I)$ and $(wa,q,\J)$, such that $\I\subseteq\I_r$ and $\J\subseteq \J_u$. This immediately implies that the statement of Lemma \ref{Lemma:ReturnsStillCross} (or, more precisely, of its counterclockwise variant) must hold also for the counterclockwise {\it special} quadruples.

\begin{figure}[htbp]
\begin{center}
\input{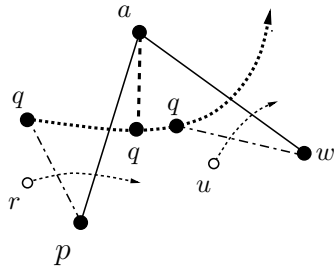}
\caption{\small A consecutive counterclockwise special quadruple $\SQ=(a,p,w,q)$, composed of two special crossings $(pa,q,\I_r)$ and $(wa,q,\J_u)$, with respective outer points $r$ and $u$. The special crossings of $\SQ$ correspond to regular Delaunay quadruples $(p,q,a,r)$ and $(w,q,a,u)$ in $\F$.}
\label{Fig:OuterPoints}
\vspace{-0.4cm}
\end{center}
\end{figure}

\smallskip
\noindent{\bf Consecutive special quadruples.} 
We say that the special quadruple $\SQ=(a,p,w,q)$, as above, is {\it consecutive} if its counterclockwise $(a,q)$-crossings $(pa,q,\I_r)$ and $(wa,q,\J_u)$ are consecutive in the previously established order (implied by Lemma \ref{Lemma:OrderRelaxedCrossings}). 
In this case, $\SQ=(a,p,w,q)$ is uniquely determined by each of its crossings $(pa,q,\I_r),(wa,q,\J_u)$.
This, combined with Propositions \ref{Prop:TwoTriples}, \ref{Prop:UniqueArq} and \ref{Prop:UniquePaq}, implies that $\SQ$ is uniquely determined by every (ordered) triple of points that are chosen from the {\it same} quadruple  $(p,q,a,r)$ or $(w,q,a,u)$; see Figure \ref{Fig:OuterPoints}. That is, the following statement holds (with the above assumptions):
\begin{proposition}\label{Prop:TriplesSpecialQuad}
Let $\SQ=(a,p,w,q)$ be a consecutive special quadruple, and let $(pa,q,\I_r)$ and $(wa,q,\J_u)$ be the special crossings associated with $\SQ$, with respective outer points $r$ and $u$. Then $\SQ$ is uniquely determined by each of the following eight triples:
$(p,q,a), (p,q,r),$ $(p,a,r),$ $(a,r,q),$ $(w,q,a),$ $(w,q,u),$ $(w,a,u)$, and $(a,u,q)$. 
\end{proposition}

Let $\Phi(n)$ denote the maximum number of consecutive special quadruples that can be induced by a set of $n$ moving points and a refined family $\F$ of Delaunay quadruples.  The preceding discussion implies the following relation between the maximum possible numbers of special crossings (identified with their respective {\it ordinary} quadruples in $\F$) and consecutive {\it special} quadruples:
$$
\Psi_5(n)=\Phi(n)+O(n^2).
$$

\paragraph{Overview.} The analysis of special consecutive (counterclockwise) quadruples proceeds through five stages, numbered $0,1,\ldots,4$.

At the $i$-th stage we consider a certain subclass of consecutive (counterclockwise) special quadruples, defined with respect to a refined family $\F$, which is constructed over the underlying set $P$ of $n$ moving points. 
We assume that each quadruple $\SQ=(a,p,w,q)$ under consideration satisfies certain topological conditions, which are formulated in terms of the extended set $P_\SQ$ (including the outer points $r$ and $u$ of the two special crossings associated with $\SQ$), $\F$, and $P$. At each new stage we enforce one, or several new conditions, so our special quadruples become progressively constrained.

The first four stages $i=0,\ldots, 3$ are almost identical to the corresponding stages described in Section \ref{Sec:CountQuad}. 
Informally, we put the outer points $r$ and $u$ aside and then gradually enforce upon our quadruples $\SQ$ the counterclockwise variants of the six conditions (Q1)--(Q6), which arise in the similar stages of Section \ref{Sec:CountQuad}. 
As noted above, this requires some caution, as the corresponding special $(a,q)$-crossings $(pa,q,\I_r)$ and $(wa,q,\J_u)$ are defined in terms of the (distinct) reduced point sets $P\setminus \{r\}$ and $P\setminus \{u\}$.

At each of these four stages, we first invoke Theorems \ref{Thm:RedBlue}, \ref{Thm:SpecialCrossings} and \ref{Thm:Balanced}, and Lemma \ref{Lemma:TwiceCollin}, in order to dispose of all special quadruples that fail to satisfy the newly enforced conditions, even after removal of a small-size subset of $P$. 
The surviving quadruples are passed on to the next stage, after an appropriate probabilistic refinement. 


At the last Stage 4 we follow the same strategy and first apply a sequence of preparatory chargings, similar to those described in Section \ref{Subsec:Stage4Reg}.
To handle the remaining quadruples $\SQ$ (that are not disposed of by these chargings) we re-introduce the corresponding outer points $r,u$ of their special crossings $(pa,q,\I_r)$ and $(wa,q,\J_u)$ to our analysis. This allows us to charge
such quadruples $\SQ$ to especially convenient topological configurations, referred to as {\it terminal quadruples}.

Informally, each terminal quadruple is formed by an edge, say $e=pq$, and by a pair of points that cross $e$ in {\it opposite} directions (i.e., one of them crosses $e$ from $\L^-_{pq}$ to $\L_{pq}^+$, and the other crosses $e$ from $\L_{pq}^+$ to $\L^-_{pq}$). 
The analysis of these configurations is delegated to Section \ref{Sec:Terminal}, where we directly bound their number in terms of simpler quantities, introduced in Section \ref{Sec:Prelim} (and thereby complete the proof of Theorem \ref{Thm:OverallBound}).
To do so, we show that, for ``most" terminal quadruples (if their number is at least superquadratic), some three of their four points perform two Delaunay crossings, again allowing us to use Lemma \ref{Lemma:TwiceCollin}, to obtain a quadratic bound on their number.

 The emergence of terminal quadruples can be attributed to the following interplay between special quadruples and their outer points. Fix a consecutive (counterclockwise) special quadruple $\SQ=(a,p,w,q)$, as above. Recall that the four points of $\SQ$ are co-circular at some times $\xi_0\in \I_r\setminus \J_u$ and $\xi_1\in \J_u\setminus \J_u$. Assume, with no loss of generality, that the co-circularity at time $\xi_0$ is the first co-circularity of $a,p,w,q$, and has index $1$. (A similar assumption was made for ordinary quadruples in Section \ref{Sec:CountQuad}.)  
At Stage 1 we enforce upon such special quadruples $\SQ$ a suitable counterclockwise variant of condition (Q3), according to which the edge $qw$ undergoes a Delaunay crossing by $p$ (where it crosses $wq$ from $\L_{wq}^+$ to $\L_{wq}^-$). Recall, however, that the underlying family $\F$ includes the ordinary quadruple $(w,q,a,u)$, so the reversely-oriented copy $wq$ of $qw$ undergoes a Delaunay crossing by the outer point $u$ (which then crosses $\L_{wq}$ from $\L_{wq}^-$ to $\L_{wq}^+$). This makes $(w,q,u,p)$ an obvious candidate for a terminal quadruple that $\SQ$ can charge. A symmetric behaviour occurs if the co-circularity at time $\xi_1$ has index $3$.

\subsection{Stage 0: Charging events in $\A_{qa}$}\label{Subsec:Stage0Sp}
Fix a consecutive special quadruple $\SQ=(a,p,w,q)$, whose two special $(a,q)$-crossings $(pa,q,\I_r)$ and $(wa,q,\J_u)$, with respective outer points $r$ and $u$, correspond to quadruples $(p,q,a,r)$ and $(w,q,a,u)$ in the underlying refined family $\F$. See Figure \ref{Fig:OuterPoints}. Recall that, according to Proposition \ref{Prop:TriplesSpecialQuad}, $\SQ$ is uniquely determined by each of the ordered triples $(p,a,q)$, $(w,a,q)$, which perform its two special $(a,q)$-crossings $(pa,q,\I_r=[\t_0,\t_1]$ and $(wa,q,\J_u=[\t_2,\t_3])$.
Our goal is to extend the almost-Delaunayhood of $qa$ to the possible gap $[\t_1,\t_2]$ between $\I_r$ and $\J_u$. 
To do so, we fix a suitable constant $k$ and apply Theorem \ref{Thm:RedBlue} in $\A_{qa}$ over the interval $(\t_1,\t_3)$, which covers the aforementioned gap (if it exists). Notice that the edge $qa$ is not necessarily Delaunay (in $\DT(P)$) at times $\t_1$ and $\t_3$, so we apply this theorem with respect to the smaller set $P\setminus \{r\}$ (where, by Lemma \ref{Lemma:Crossing}, $\DT(P\setminus \{r\})$ clearly contains $qa$ at time $\t_1$.

If at least one of the Conditions (i) or (ii) of Theorem \ref{Thm:RedBlue} holds, we can charge $\SQ$ either to a $k$-shallow collinearity, or to $\Omega(\k^2)$ $k$-shallow co-circularities, which are encountered in the reduced red-blue arrangement $\A_{qa}^{(r)}$, defined over $P\setminus \{r\}$, during $(\t_1,\t_3)$. (Each of these events is $(k+1)$-shallow in in $\A_{qa}$ when defined over the entire set $P$.)
It suffices to check that each $(k+1)$-shallow collinearity or co-circularity, that occurs in the larger arrangement $\A_{qa}$ at some time $t^*\in (\t_1,\t_3)$, is charged by at most $O(1)$ special quadruples $\SQ$. Indeed,
the points $q$ and $a$ of $\SQ$ can be guessed in at most $O(1)$ ways among the three or four points involved in the shallow event.
Furthermore, no counterclockwise special $(a,q)$-crossings $(p'a,q,\I_{r'})$ end in $(\t_1,\t_3)$, so
$(pa,q,\I_r)$ is the last such $(a,q)$-crossing to end before time $t^*$. This gives us the third point $p$, and Proposition \ref{Prop:TriplesSpecialQuad} then completes the proof of the claim.
To conclude, the Clarkson-Shor probabilistic argument implies that the above scenario happens for at most 
$$
O\left(\k^2N(n/\k)+\k n^2\beta(n)\right)
$$ 
special quadruples $\SQ$.

Now suppose that Condition (iii) of Theorem \ref{Thm:RedBlue} is satisfied. Then there is a subset $A$ of at most $3\k$ points (not including $r$) such that the edge $qa$ belongs to $\DT(P\setminus (A\cup \{r\}))$ throughout the interval $\I_r\cup [\t_1,\t_3]=[\t_0,\t_3]$.

To proceed, we consider a random subset $R$ of $\lceil n/\k\rceil$ points of $P$.
Let $\F_R$ denote the induced family of surviving (regular) Delaunay quadruples.
Namely, a (regular) quadruple $\Q$ in $\F$ yields a counterpart in $\F_R$ if and only if $R$ includes the four points of $\Q$.
As is easy to check, $\F_R$ is also refined with respect to its underlying set $R$. 
Furthermore, it can be viewed as a subset of $\F$, because each of its quadruples has a (unique) ancestor in $\F$.
Therefore, $\F_R$ yields no new Delaunay crossings, whose counterparts did not arise already in the context of $\F$.

Note that the following two events occur simultaneously with probability at least $\Omega(1/\k^6)$: (1) $R$ includes the six points of $P_{\SQ}$, and (2) none of the points of $A\setminus P_{\SQ}$ belongs to $R$. 

Assume that the sample $R$ is indeed successful (for the chosen special quadruple $\SQ$). Then the family $\F_R$ still contains the quadruples $(p,q,a,r)$ and $(w,q,a,u)$. Hence, $\F_R$ still yields the special crossings of $pa$ and $wa$ by $q$ (with the same outer points $r$ and $u$). We continue to denote these crossings by $(pa,q,\I_{r})$ and $(wa,q,\J_u)$ but observe that the corresponding intervals $\I_r=[\t_0,\t_1]$ and $\J_u=[\t_2,\t_3]$ may shrink in the process. (See Section \ref{Subsec:Stage0} for more details.) Therefore, $R$ and $\F_R$ also yield the (counterclockwise) special quadruple $(a,p,w,q)$, which we continue to denote by $\SQ$. Furthermore,
$\SQ$ is again consecutive with respect to $R$ and $\F_R$ (because the underlying family $\F_R$ induces no new special crossings, which did not arise in the context of $\F$). Moreover, since $R'$ contains none of the points $r,u$, the edge $qa$ now belongs to $\DT(R\setminus \{r,u,p,w\})$ throughout the extended interval $\conn{\I_r}{\J_u}=[\t_0,\t_3]$; see Figure \ref{Fig:DelQuadSp}.

\begin{figure}[htbp]
\begin{center}
\input{IntervalsSp.pstex_t}
\caption{\small After replacing the underlying set $P$ by its subsample $R$, the edge $qa$ belongs to $\DT(R\setminus \{r,u,p,w\})$ throughout $\conn{\I_r}{\J_u}={\sf conv}(\I_r\cup \J_u)$, including the gap between $\I_r$ and $\J_u$.
(The intervals $\I_r$ and $\J_u$ may shrink in the process.)}
\label{Fig:DelQuadSp}
\end{center}
\vspace{-0.5cm}
\end{figure}

\paragraph{Definition.} Let $P$ be a (finite) set of moving points, and let $\F$ be a refined family constructed over $P$.
We say that a consecutive special quadruple $\SQ=(a,p,w,q)$, formed by counterclockwise special $(a,q)$-crossings $(pa,q,\I_r=[\t_0,\t_1])$ and $(wa,q,\J_u=[\t_2,\t_3])$ (both of them in $\F$) is {\it Delaunay} (again, with respect to $P$ and $\F$), if its edge $aq$ belongs to $\DT(P\setminus \{p,w,r,u\})$ throughout the extended interval $\conn{\I_r}{\J_u}=[\t_0,\t_3]$.


\smallskip
Let $\Phi_0(m)$ denote the maximum number of consecutive Delaunay special quadruples that can be induced by a refined family $\F$ defined over $n$ moving points. The preceding discussion implies the following recurrence.
\begin{equation}
\Phi(n)\leq O\left(\k^6\Phi_0(n/\k)+\k^2N(n/\k)+\k n^2\beta(n)\right),
\end{equation}
for any constant parameter $k\geq 12$.

\subsection{Stage 1}\label{Subsec:Stage1Special}
To bound the above quantity $\Phi_0(n)$, we fix an underlying set $P$ of $n$ moving points, a refined family $\F$, and a consecutive Delaunay special quadruple $\SQ=(a,p,w,q)$, obtained from the corresponding special crossings $(pa,q,\I_r=[\t_0,\t_1])$ and $(wa,q,\J_u=[\t_2,\t_3])$; $r$ and $u$ are the corresponding outer points. See Figure \ref{Fig:OuterPoints}. By definition, the edge $qa$ belongs to $\DT(P\setminus \{p,w,r,u\})$ throughout the interval $[\t_0,\t_3]$.

As in Section \ref{Sec:CountQuad}, we fix constants $12<k\ll \ell$ and distinguish between five possible scenarios, where the roles of the edges $pq$ and $wq$ are mostly symmetric.
In all but the last case, we will be able to bound the number of (the relevant) Delaunay special quadruples in terms of quantities that were already introduced in Section \ref{Sec:Prelim}. In the last case (case (e)), our bound will also depend on the number of special quadruples of a more restricted type, which are defined over an appropriate subsample of $R$ of $P$. Such quadruples will be called {\it $1$-restricted}, and their analysis will be passed on to the subsequent stages. 

\medskip
\noindent {\bf Case (a).} The edge $qa$ is hit during $[\t_0,\t_3]$ by at least one of the points $p,w$.
Clearly, this collinearity can happen only during the gap between $\I_r$ and $\J_u$ (if it exists). 

If $qa$ is hit by $p$ then the triple $p,a,q$ defines two distinct (single) Delaunay crossings within the smaller triangulation $\DT(P\setminus \{w,r,u\})$. (Here we exploit the fact that the crossed edge $qa$ is almost Delaunay throughout $[\t_0,\t_3]$.) According to Lemma \ref{Lemma:TwiceCollin}, combined with the Clarkson-Shor argument, where we use a sample of size $n/2$, the overall number of such triples $(p,q,a)$ (and, hence, of such special quadruples $\SQ=(a,p,w,q)$, each of them uniquely determined by its corresponding triple $(p,q,a)$) is at most $O(n^2)$.

If $qa$ is hit by $w$ then we similarly argue that the triple $(w,a,q)$ defines two distinct Delaunay crossings within $\DT(P\setminus \{p,r,u\})$, so the number of such special quadruples $\SQ$ (each of them uniquely determined by the corresponding triple $(w,q,a)$) is at most $O(n^2)$ too.

\medskip
\noindent {\bf Case (b).} At least $k$ clockwise special $(p,q)$-crossings $(pa',q,\I_{r'})$ end in $(\t_1,\t_3]$, or at least $k$ clockwise special $(w,q)$-crossings $(wa',q,\J_{u'})$ begin in $[\t_0,\t_2)$; each of these crossings comes with its respective outer point $r'$ or $u'$.

\begin{figure}[htbp]
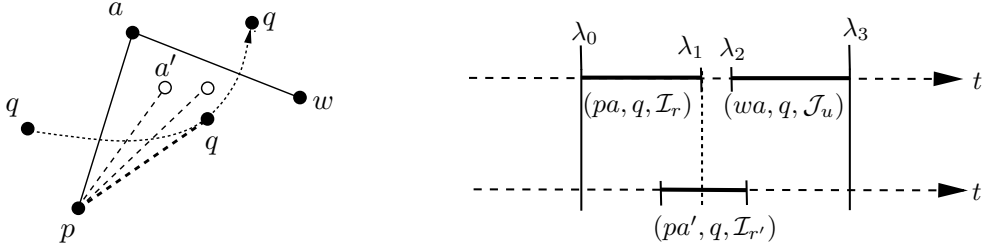

\begin{center}
\input{UnbalancedSpecial.pstex_t}\hspace{2cm}\input{ChargeabilitySpecial.pstex_t}
\caption{\small Case (b): at least $k$ clockwise special $(p,q)$-crossings $(pa',q,\I_{r'})$ end in $(\t_1,\t_3]$.}
\label{Fig:UnbalancedSpecial}
\end{center}
\vspace{-0.5cm}
\end{figure} 

Without loss of genarality, we consider only the former scenario, and handle the latter one in a fully symmetric manner. 
Recall that a special $(p,q)$-crossing $(pa',q,\I_{r'})$ is uniquely determined by each of the triples 
$(p,a',q)$ and $(p,q,r')$. Hence, at most one of these special crossings has $a'$ equal to $u$. 
Moreover, the preliminary pruning (applied to clockwise special $(p,q)$-crossings) guarantees that none of them can have $r'=a$ or $a'=r$.

We apply Theorem \ref{Thm:Balanced}, in combination with the standard argument of Clarkson and Shor, in order to dispose of such special quadruples $\SQ$.
To do so, we consider a random subset $R$ of $\lceil n/4\rceil$ points of $P$ and notice that the following two conditions hold simultaneously with probability $\Omega(1)$: (1) $R$ includes the points $p,q$ and $a$, but none of $r,u$, and (2) for at least a constant fraction of the above special $(p,q)$-crossings $(pa',q,\I_{r'})$, the set $R$ includes the point $a'$ but not $r'$.

Specifically, (1) holds with some constant probability close to $(1/4)^3(3/4)^2$.
Concerning (2), assume without loss of generality that the number of relevant crossings $(pq',q,\I_{r'})$ is exactly $k$ (so at least $k-1$ of them satisfy $a'\neq u$). Then, conditioning on the success of (1), the expected number of these crossings that satisfy the property in (2) is very close to $(k-1)(3/16)$, or larger. Hence, Markov's inequality implies that, with an appropriate choice of parameters, the probability of (2), conditioned on the success of (1), is also some fixed constant. Hence, the probability that both (1) and (2) hold is also $\Omega(1)$, as claimed.

If the sample $R$ is successful (for the given $\SQ$),
then it clearly yields an (ordinary) Delaunay crossing $(pa,q,\I)$, whose respective interval $\I$ is contained in $[\t_0,\t_1]$ (as $R\subseteq P\setminus \{r,u\}$). 
It remains to check that this crossing is $(a,q,\Theta(k))$-chargeable, with respect to the interval $[\t_0,\t_3]$. 

To see the latter property, note that each of the above special $(p,q)$-crossings $(pa',q,\I_{r'})$, for which the sample $R$ includes $a'$ but not $r'$, yields the Delaunay crossing $(pa',q,\I')$ in $R$, with $\I'\subseteq \I_{r'}$. 
Therefore, Lemma \ref{Lemma: OrderOrdinaryCrossings} implies that $(pa',q,\I')$ occurs within $[\t_0,\t_1]\cup \I_{r'}\subseteq [\t_0,\t_3]$.
Moreover, $aq$ belongs to $\DT(R)$ at times $\t_0$ and $\t_3$ (in addition to its almost-Delaunayhood in $\DT(R)$, with only two points $p,w$ removed, during $[\t_0,\t_3]$).

Theorem \ref{Thm:Balanced} implies, then, that the overall number of such triples $(p,q,a)$ in $R$ is only 
$$
O\left(k^2N(n/k)+kn^2\beta(n)\right).
$$
Clearly, this also bounds the overall number of Delaunay special quadruples $\SQ$ falling into case (b).

\smallskip
To conclude, we can assume, from now on, that case (b) does not occur. That is, fewer than $k$ clockwise special $(p,q)$-crossings end in $(\t_1,\t_3]$, and fewer than $k$ clockwise special $(w,q)$-crossings begin in the symmetric interval $[\t_0,\t_2)$.

\medskip
\noindent{\bf Case (c).} 
No clockwise special $(p,q)$-crossings $(pa',q,\I_{r'})$, with $r'\not\in \{w,u\}$, end during $[\t_3,\infty)$, or no clockwise special $(w,q)$-crossings $(wa',q,\J_{u'})$, with $u'\not\in \{p,r\}$, begin during $(-\infty,\t_0]$. 

Without loss of generality, we consider only the first subcase and handle the other one in a fully symmetric manner.
Note that the preliminary pruning (combined with the fact that $(pa,q,\I_{r})$ is uniquely determined by the triple $(p,q,r)$) guarantees that no clockwise special $(p,q)$-crossing $(pa',q,\I_{r'})$ can have $r'$ in $\{r,a\}$. 


Since case (b) does not occur, $(pa,q,\I_r)$ is among the $k+3$ last clockwise special $(p,q)$-crossings (in the standard order provided by Lemma \ref{Lemma:OrderRelaxedCrossings}). Indeed, at most $k$ such special crossings $(pa',q,\I_{r'})$ end in $(\t_1,\t_3]$, and at most two of them can end after $\t_3$, namely those whose outer point is either $u$ or $w$ (recalling that this outer point, together with $p,q$, uniquely determines the crossing).
Therefore, we can charge $(pa,q,\I_r)$ and $\SQ$ to the edge $pq$, so this situation happens
for at most $O(kn^2)$ special quadruples $\SQ$.

\begin{figure}[htbp]
\begin{center}
\input{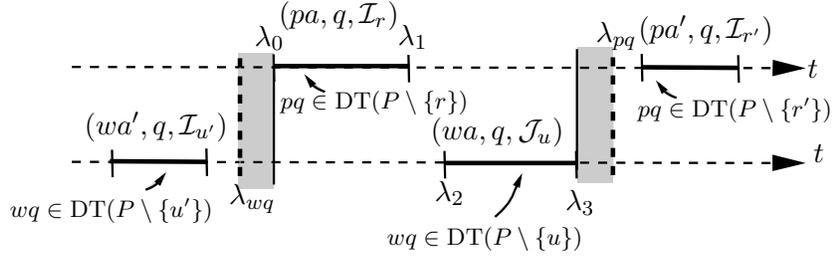}
\caption{\small Assuming (c) does not hold, we put $\t_{pq}$ to be the first time in $[\t_3,\infty)$ when $pq$ belongs to some reduced triangulation $\DT(P\setminus \{r'\})$, for $r'\not \in P_{\SQ}$.
Similarly, we put $\t_{wq}$ to be the last time in $(-\infty,\t_0]$ when $wq$ belongs to some reduced triangulation 
$\DT(P\setminus \{u'\})$, for $u'\not\in P_{\SQ}$.}
\label{Fig:SpecialAfterC}
\end{center}
\vspace{-0.8cm}
\end{figure}

\paragraph{Preparing for cases (d) and (e).}
For the remainder of this stage, we assume that none of the cases (a), (b) or (c) occurs. In particular, there is a special $(p,q)$-crossing $(pa',q,\I_{r'})$, whose outer point $r'$ satisfies $r'\not\in \{w,u\}$, that ends after $\t_3$. (Refer to Figure \ref{Fig:SpecialAfterC}.) 
Therefore, and according to Lemma \ref{Lemma:Crossing}, $pq$ belongs to $\DT(P\setminus \{r'\})$ either at time $\t_3$ or at some later time. Moreover, $r'$ does not belong to $P_{\SQ}$ because, after the preliminary pruning, there remain no clockwise special $(p,q)$-crossings $(pa',q,\I_{r'})$ with $r'\in \{a,r\}$.
Let $\t_{pq}$ be the first time in $[\t_3,\infty)$ when $pq$ belongs to some triangulation $\DT(P\setminus \{r'\})$, for some $r'\not\in P_{\SQ}$. More precisely, we put $\t_{pq}=\t_3$ if $pq$ belongs to such a triangulation at time $\t_3$, and otherwise we set $\t_{pq}$ to be the first time after $\t_3$ when $pq$ enters $\DT(P\setminus \{r'\})$ (for some $r'\not\in P_{\SQ}$). 

A symmetric argument (adapted for clockwise special $(w,q)$-crossings) shows that there is a special $(w,q)$-crossing $(wa',q,\J_{u'})$, with an outer point $u'\not\in\{a,r\}$, that begins before $\t_0$ (so $wq\in \DT(P\setminus \{u'\})$ at some time before or at $\t_0$).
We define $\t_{wq}$ to be the last time in $(-\infty,\t_0]$ when the edge $wq$ belongs to some triangulation $\DT(P\setminus \{u'\})$, for some $u'\not\in P_{\SQ}$.
In what follows, we use $r'$ and $u'$ to denote 
a fixed\footnote{Notice that we do not claim that the choice $r'$ and $u'$ is unique.} pair of points, outside $P_{\SQ}$, whose removal restores the Delaunayhood of $pq$ and $wq$ at respective times 
$\t_{pq}$ and $\t_{wq}$, and for which $\t_{pq}$ is smallest and $\t_{wq}$ is largest.

Before proceeding to the cases (d) and (e), we first apply Theorem \ref{Thm:RedBlue} in $\A_{pq}$ over the interval $(\t_1,\t_{pq})$, and then apply it in $\A_{wq}$ over the symmetric interval $(\t_{wq},\t_2)$, both times with the second constant $\ell\gg k$.


Consider the first application of Theorem \ref{Thm:RedBlue}. It is performed with respect to the {\it reduced} triangulation $\DT(P\setminus \{r,r'\})$, which contains $pq$ at time $\t_{pq}$. 
If (at least) one of the first two conditions of Theorem \ref{Thm:RedBlue} holds, we charge $\SQ$, via $(pa,q,\I_r)$, either to $\Omega(\ell^2)$
$(\ell+2)$-shallow co-circularities, or to an $(\ell+2)$-shallow collinearity. (Each of these events is $\ell$-shallow with respect to $P\setminus\{r,r'\}$.) As before, the crucial observation is that each co-circularity or collinearity, which occurs at some time $t^*\in (\t_1,\t_{pq})$, is charged in the above manner by at most $O(k)$ special quadruples $\SQ$. Indeed, the points $p$ and $q$ of $\SQ$ can be chosen in $O(1)$ ways among the three or four points involved in the event.
Furthermore, recall that $\SQ$ is uniquely determined by the triple $(a,p,q)$, so it suffices to guess $a$ (for the chosen $p,q$ and $t^*$).

Since case (b) has been ruled out, at most $k$ clockwise special $(p,q)$-crossings $(pa',q,\I_{r'})$ end in $(\t_1,\t_3)$. Moreover, assuming $\t_{pq}>\t_3$, no such crossing can end in $(\t_3,\t_{pq}]$ unless its respective outer point $r'$ belongs to $\{w,u\}$ (which happens for at most two special $(p,q)$-crossings). 
Therefore, $(pa,q,\I_r)$ is among the last $k+3$ clockwise special $(p,q)$-crossings to end before time $t^*$. 

To conclude, the above charging accounts for at most $O\left(k\ell^2N(n/\ell)+k\ell n^2\beta(n)\right)$ special quadruples $\SQ$.

\begin{figure}[htbp]
\begin{center}
\input{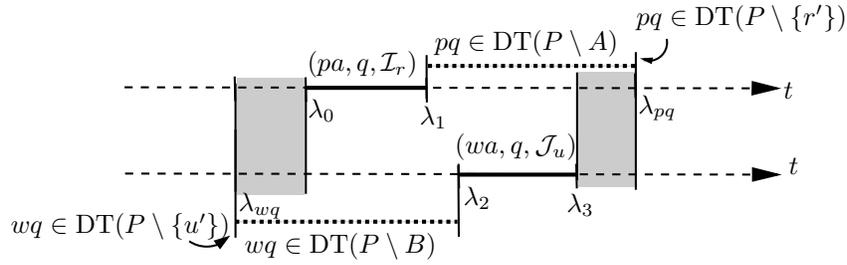}
\caption{\small Extending the almost-Delaunayhood of $pq$ and $wq$, in preparation for cases (d) and (e), respectively, from $\I_r=[\t_0,\t_1]$ to $[\t_0,\t_{pq}]$, and from $\J_u=[\t_2,\t_3]$ to $[\t_{wq},\t_3]$.}
\label{Fig:ExtendBeforeDSp}
\end{center}
\vspace{-0.5cm}
\end{figure}

Finally, if Condition (iii) of Theorem \ref{Thm:RedBlue} holds, then the Delaunayhood of $pq$ can be restored throughout the interval $[\t_1,\t_{pq}]$ by removing a subset $A$ of at most $3\ell+2$ points of $P$ (including $r$ and $r'$); see Figure \ref{Fig:ExtendBeforeDSp}.

The second application of Theorem \ref{Thm:RedBlue} in $\A_{wq}$ is fully symmetric, and it is done with respect to the set $P\setminus \{u,u'\}$ in the interval $(\t_{wq},\t_2)$. If at least one of the Conditions (i), (ii) is satisfied, we dispose of $\SQ$ by charging it (via $(wa,q,\J_u)$) to $(\ell+2)$-shallow collinearities and co-circularities that occur in $\A_{wq}$ during $(\t_{wq},\t_2)$. (Since case (b) has been ruled out, $(wa,q,\J_u)$ is among the first $k+3$ special counterclockwise $(w,q)$-crossings to begin after each charged event. Hence, every collinearity or co-circularity is charged at most $O(k)$ times.) As before, this accounts for at most $O\left(k\ell^2N(n/\ell)+k\ell n^2\beta(n)\right)$ special quadruples $\SQ$.

For each of the remaining special quadruples we have a set $B$ of at most $3\ell+2$ points (including $u$ and $u'$) whose removal restores the Delaunayhood of $wq$ throughout $[\t_{wq},\t_2]$; see Figure \ref{Fig:ExtendBeforeDSp} again.

To recap, in each of the remaining cases (d) and (e), we may assume the existence of the first time $\t_{pq}\geq \t_3$ when $pq$ belongs to some reduced triangulation $\DT(P\setminus \{r'\})$, and of the symmetric last time $\t_{wq}\leq \t_0$ when $wq$ belongs to a similarly reduced triangulation $\DT(P\setminus \{u'\})$, where $u'$ and $r'$ are fixed points outside $P_{\SQ}$. 
In addition, there exist sets $A$ (including $r$ and $r'$) and $B$ (including $u$ and $u'$), both of cardinality at most $3\ell+2$, whose removal restores the Delaunayhood of $pq$ and $wq$ throughout the respective intervals $[\t_1,\t_{pq}]$ and $[\t_{wq},\t_2]$ (and, therefore, extends the almost-Delaunayhood of these edges to the respective larger intervals $[\t_0,\t_{pq}]=I_r\cup [\t_1,\t_{pq}]$ and $[\t_{wq},\t_3]=[\t_{wq},\t_2]\cup \J_u$).

\medskip
\noindent{\bf Case (d).} The point $a$ hits the edge $pq$ during $[\t_1,\t_{pq}]$, or it hits the edge $wq$ during the symmetric interval $[\t_{wq},\t_2]$. 

In the former scenario, the triple $(a,p,q)$ defines two Delaunay crossings within $\DT((P\setminus A)\cup\{p\})$, and, in the latter, the symmetric triple $(a,w,q)$ defines two Delaunay crossings within $\DT((P\setminus B)\cup\{w\})$.
In both cases, we can use Lemma \ref{Lemma:TwiceCollin}, in combination with the sampling argument of Clarkson and Shor, to show that the overall number of the relevant triples in $P$ is at most $O(\ell n^2)$. As in case (a), this also bounds the overall number of such special quadruples $\SQ=(a,p,w,q)$.

\medskip
\noindent{\bf Case (e).} None of the previous cases (a)--(d) occurs, and none of the preliminary charging arguments apply to $\SQ$. 

In particular, since cases (a) and (d) have been ruled out, either the point $q$ either remains in $\L_{pa}^+$ after the end $\t_1$ of $\I_r$ and until crossing $wa$ (during $\J_u$), or else it re-enters $\L_{pa}^-$ during that period, through the relative interior of $pa$. 
Similarly, $q$ must remain in $\L_{wa}^-$ after crossing $pa$ (during $\I_r$) and until the beginning $\t_2$ of $\J_u$, unless it crosses $wa$ (from $\L_{wa}^+$ to $\L_{wa}^-$) during that period.

In addition, we assume the existence of the sets $A$ and $B$, as above, whose removal restores the Delaunayhood of $pq$ and $wq$ throughout the respective intervals $[\t_{wq},\t_3]$ and $[\t_0,\t_{pq}]$.


Recall that, according to Lemma \ref{Lemma:OnceCollin}, the four points of $\SQ$ are co-circular at times $\xi_0\in \I_r\setminus \J_u$ and $\xi_1\in \J_u\setminus \I_r$ (see, e.g., Figure \ref{Fig:SpecialQuad}). 
Clearly, at least one of these co-circularities is extremal. 
We therefore distinguish between two subcases (whose treatment remains fully symmetric untill the beginning of Stage 4).

\medskip
\noindent\textbf{Case (e1): The co-circularity at time $\xi_1$ has index $3$.}
In this case, we say that $\SQ$ is a {\it right special quadruple}. We claim that in this case the edge $pq$ is hit during $(\t_1,\t_{pq}]$ by the point $w$, which crosses it from $\L_{pq}^+$ to $\L_{pq}^-$. To show this, we distinguish between two sub-scenarios.

\begin{figure}[htbp]
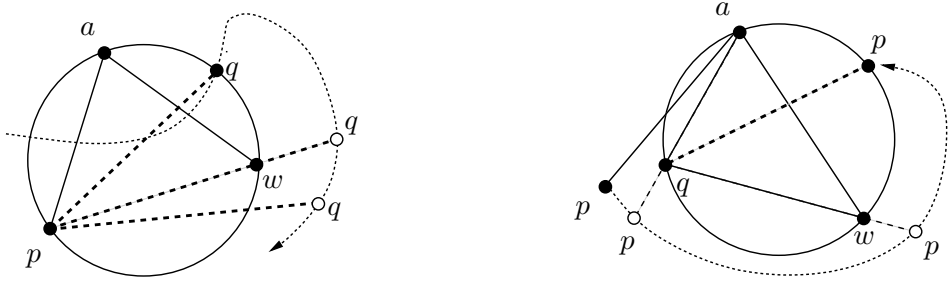

\begin{center}
\input{RightCrossSp.pstex_t}\hspace{3cm}\input{RightCrossSpReturn.pstex_t}
\caption{\small Case (e1): $\xi_1$ is the last co-circularity of $a,p,w,q$. Arguing that the edge $qp$ is crossed by $w$ during $(\t_1,\t_{pq}]$. Left: A possible motion of $q$ if $p\in \L_{wa}^-$ when $q$ crosses $wa$ (during $\J_u$). Right: A possible motion of $p$ (after $\I_r$) if $q$ re-enters $\L_{pa}^-$ through $pa$.}
\label{Fig:CocircsSpRight}
\vspace{-0.4cm}
\end{center}
\end{figure}

\smallskip
\noindent (i) If $p$ lies in $\L_{wa}^-$ when $q$ enters the opposite halfplane $\L_{wa}^+$ (during $\J_u$), then the Delaunayhood of $pq$ is violated, right after time $\xi_1$, by $w\in \L_{pq}^+$ and $a\in \L_{pq}^-$. See Figure \ref{Fig:CocircsSpRight} (left). Hence, $pq$ is hit by at least one of these two points during $(\xi_1,\t_{pq}]\subseteq (\t_2,\t_{pq}]$, as prescribed in cases (i) and (ii) of Lemma \ref{Lemma:MustCross} (case (iii) thereof cannot arise since $\xi_1$ has index $3$). Since case (d) has been ruled out, $a$ cannot hit $pq$ during $(\t_0,\t_{pq}]$. Hence, $pq$ must be hit by $w$, which then crosses it from $\L_{pq}^+$ to $\L_{pq}^-$ (this crossing direction is also prescribed by the lemma).

\smallskip
\noindent(ii) If $p$ lies in $\L_{wa}^+$ when $q$ enters this halfplane, then $q$ must re-enter $\L_{pa}^-$ after $\I_r$ and before it reaches $\L_{wa}^+$. Hence, the co-circularity at time $\xi_1$ is as depicted in Figure \ref{Fig:CocircsSpRight} (right); that is, it occurs with $p\in \L_{wa}^+$ and $q\in \L_{wa}^-$. Since none of the preceding cases (a), (d) holds, $q$ can re-enter $\L_{pa}^-$ during this interval only through the edge $pa$. Therefore, the counterclockwise variant of Lemma \ref{Lemma:ReturnsStillCross} (adapted for special quadruples, as described in the introduction to this section) implies that in this case too $w$ crosses $pq$ from $\L_{pq}^+$ to $\L_{pq}^-$, during $(\t_1,\t_3]\subseteq (\t_1,\t_{pq}]$; see Figure \ref{Fig:CocircsSpRight} (right).
(As a matter of fact, this collinearity must occur during $(\t_1,\xi_1)$.)

\medskip
To conclude, in both sub-scenarios the edge $qp$ undergoes a Delaunay crossing by $w$ within the smaller triangulation $\DT((P\setminus A)\cup\{w\})$, and the respective interval $\H=[\t_4,\t_5]$ of that crossing is contained 
in $[\t_1,\t_{pq}]$. 
(We again emphasize that $A$ includes both points $r,r'\neq w$, so the edge $pq$ belongs to $\DT((P\setminus A)\cup \{w\})$ throughout $\I_r=[\t_0,\t_1]$ and at time $\t_{pq}$.) 


If $w$ hits $pq$ {\it twice} during $(\t_1,\t_{pq}]$, then $pq$ undergoes within $\DT((P\setminus A)\cup\{w\})$ either two (single) Delaunay crossings, or a double Delaunay crossing, by the same point $w$.
We thus charge $\SQ$ to the respective triple $(p,q,w)$ and use Theorem \ref{Thm:SpecialCrossings} or Lemma \ref{Lemma:TwiceCollin}, in combination with the probabilistic argument of Clarkson and Shor, to show that the overall number of such triples $(p,q,w)$ is at most $O(\ell n^2)$.
Since case (b) does not occur, $(pa,q,\I_r=[\t_0,\t_1])$ is among the last $k+3$ special clockwise $(p,q)$-crossings to end before the above crossings of $pq$ by $w$. (Namely, at most $k$ such $(p,q)$-crossings end during $(\t_1,\t_3]$, and at most two of them can end in $(\t_3,\t_{pq}]$, if $\t_3\neq\t_{pq}$; see the analysis preceding case (e) for more details.)
In particular, any triple $(p,q,w)$ is shared by at most $k+3$ charging quadruples $\SQ$. Hence, the above additional collinearities of $p,q,w$ are encountered for at most $O(k\ell n^2)$ special quadruples.

A similar argument applies if the edge $wq$ is hit by $p$ during\footnote{Since $p\neq u$, the point $p$ cannot cross $wq$ during $\J_u=[\t_2,\t_3]$, as $wq$ belongs to $\DT(P\setminus \{u\})$ during that interval.} $[\t_{wq},\t_2)$. In this case, the triple $(p,q,w)$ performs two distinct single Delaunay crossings within the triangulation $\DT((P\setminus (A\cup B))\cup\{w,p\})$ (namely, the crossing of $qp$ by $w$, and the crossing of $wq$ by $p$). The same bound $O(k\ell n^2)$ holds in this case too.


We thus assume, from now on, that
$w$ hits $pq$ exactly once during $(\t_1,\t_{pq}]$, and that $p$ does not cross $wq$ during the symmetric interval $[\t_{wq},\t_2)$.
In particular, this implies that $q$ lies in $\L_{wa}^-$ when it enters $\L_{pa}^+$ during $\I_r$. Indeed, otherwise $q$ would have to cross $\L_{wa}$ (thereby leaving $\L_{wa}^+$) between the times when it enters the halfplanes $\L_{pa}^+$ and $\L_{wa}^+$ (both times during the respective special crossings). Since neither of the cases (a), (d) holds, $q$ can cross $\L_{wa}$, for the first time, only within $wa$.
However, in this latter case the counterclockwise variant Lemma \ref{Lemma:ReturnsStillCross} would imply that $p$ hits $wq$ during $[\t_{wq},\t_2)$ (which has been ruled in the previous paragraph).

\begin{figure}[htbp]
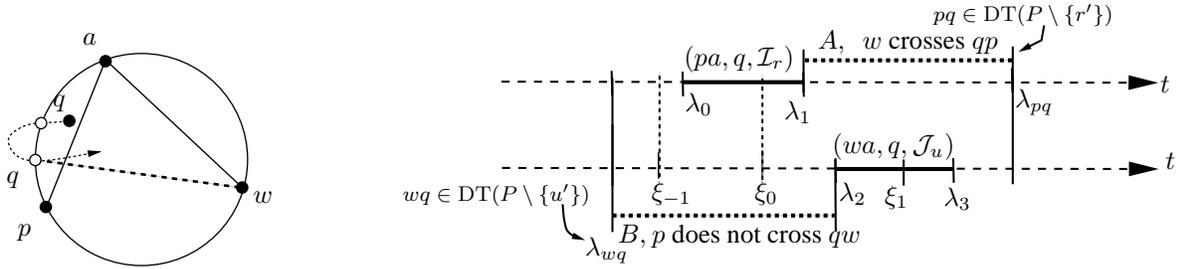

\begin{center}
\input{FirstCocircsRight.pstex_t}\hspace{1.9cm}\input{TimelineAfterEsp.pstex_t}
\caption{\small Case (e1). Left: A possible motion of $q$ before and during $\I_r$. The Delaunayhood of $wq$ is violated, right before $\xi_0$, by $p$ and $a$. The points of $\SQ$ are involved, at some time $\xi_{-1}\in [\t_{wq},\xi_0)$ in another co-circularity (of index $1$). The order type of $\SQ$ remains fixed throughout $[\xi_{-1},\xi_0]$. Right: A schematic summary of what we eventually assume at the end of case (e1).} 
\label{Fig:FirstCocircsRight}
\vspace{-0.4cm}
\end{center}
\end{figure}

We may therefore assume that $w$ lies in $\L_{pa}^+=\L_{qa}^+$ when $q$ crosses $pa$ (during $\I_r$). See Figure \ref{Fig:FirstCocircsRight} (left). Arguing as in the previous similar situations, we conclude that the Delaunayhood of $wq$ is violated, right before $\xi_0$, by $p\in \L_{wq}^-$ and $a\in \L_{wq}^+$. 
(That is, $w$ enters the cap $B[p,q,a]\cap\L_{pa}^+$ at time $\xi_0$.)
By Lemma \ref{Lemma:MustCross} (applied with respect to $\DT(P\setminus \{u,u'\})$), and since none of the points $a,p$ is allowed to cross $wq$ during $[\t_{wq},\t_2]$, the four points $p,q,a,w$ must be co-circular at some time $\xi_{-1}\in [\t_{wq},\xi_0)$, right before which the Delaunayhood of $pa$ is violated by $q\in \L_{pa}^-$ and $w\in \L_{pa}^+$. (We must have $\t_{wq}\leq \xi_{-1}<\xi_0<\t_2<\xi_1$.) Moreover, $wq$ is intersected by $pa$ throughout $[\xi_{-1},\xi_0]$. (In other words, the order type of $a,p,w,q$ remains fixed there.)

A schematic summary of what we assume in case (e1) (by the end of its analysis) is given in Figure \ref{Fig:FirstCocircsRight} (right).

\medskip
\noindent\textbf{Case (e2): The co-circularity at time $\xi_0$ has index $1$.} In this case, we say that $\SQ$ is a {\it left special quadruple}. We apply a fully symmetric topological analysis (in which we switch the roles of $pq$ and $wq$, and reverse the direction of the time axis). 

\begin{figure}[htbp]
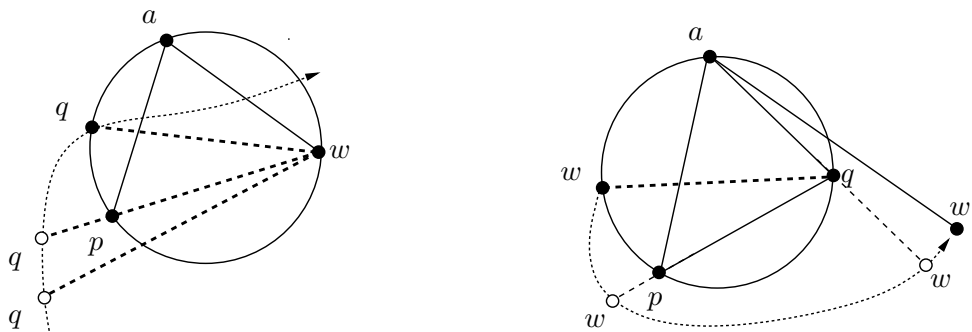

\begin{center}
\input{LeftCrossSp.pstex_t}\hspace{3cm}\input{LeftCrossReturn.pstex_t}
\caption{\small Case (e2): $\xi_0$ is the {\it first} co-circularity of $a,p,w,q$. Arguing that $p$ hits $qw$ in $[\t_{wq},\t_2)$. Left: A possible motion of $q$ if $w$ lies in $\L_{pa}^+=\L_{qa}^+$ when $q$ hits $pa$ (during $\I_r$). Right: A possible motion of $w$ if $q$ hits $wa$ also before $\J_u$ (and after its hits $pa$ in $\I_r$).}
\label{Fig:CocircsSpLeft}
\vspace{-0.4cm}
\end{center}
\end{figure}

Briefly, we use one of the Lemmas \ref{Lemma:MustCross} or \ref{Lemma:ReturnsStillCross} to show that $p$ crosses $wq$, from $\L_{wq}^+$ to $\L_{wq}^-$, during the interval $[\t_{wq},\t_2]$. As in case (e1), we distinguish between two possible scenarios, now depending on the location of $w$ when $q$ crosses $pa$ (during $\I_r$).

(i) If $w$ lies in $\L_{pa}^+$ when $q$ crosses $pa$ during $\I_r$ then the Delaunayhood of $wq$ is violated, right before $\xi_0$, by $p\in \L_{wq}^-$ and $a\in \L_{wq}^+$. Hence, the promised crossing follows from the {\it time-reversed} variant of Lemma \ref{Lemma:MustCross} (and because case (d) has been ruled out); see Figure \ref{Fig:CocircsSpLeft} (left).

We again emphasize that, in this subscenario of case (e2), the crossing of $wq$ by $p$ occurs after $\t_{wq}$ and {\it before} $\xi_0$. (Note that Figure \ref{Fig:CocircsSpLeft} (left) depicts a possible trajectory of $q$ in the standard time direction. In the time-reversed application of Lemma \ref{Lemma:MustCross}, the point $q$ moves {\it backwards}, so $p$ crosses $wq$ from $\L_{wq}^-$ to $\L_{wq}^+$. In the standard time direction, the crossing is from $\L_{wq}^+$ to $\L_{wq}^-$, as asserted.)

\smallskip 
(ii) If $w$ lies in $\L_{pa}^-=\L_{pq}^-$ (i.e., $q$ and $p$ lie in $\L_{wa}^+$) when $q$ crosses $pa$, then $q$ will have to enter $\L_{wa}^-$ before the beginning of $\J_u$ (and only through the interior of $wa$, as cases (a) and (d) have been ruled out). Therefore, the asserted crossing of $wq$ by $p$ now follows from a suitable (counterclockwise and time-reversed) variant of Lemma \ref{Lemma:ReturnsStillCross}; see Figure \ref{Fig:CocircsSpLeft} (right).

(Again, Figure \ref{Fig:CocircsSpLeft} (right) depicts a possible trajectory of $w$ in the standard time direction. In the time-reversed application of Lemma \ref{Lemma:ReturnsStillCross}, the roles of $p$ and $w$ in the statement of the lemma are switched, and $p$ crosses $wq$ in the opposite direction, from $\L_{wq}^-$ to $\L_{wq}^+$.)

\smallskip
If $p$ hits $wq$ twice during $[\t_{wq},\t_2)$, or if $w$ hits $pq$ during $(\t_1,\t_{pq}]$, then we can dispose of $\SQ$ using Theorem \ref{Thm:SpecialCrossings} or Lemma \ref{Lemma:TwiceCollin}. Namely, we then argue that the triple $(p,w,q)$ is involved within $\DT((P\setminus (B\cup A))\cup\{q,w\})$ either in two distinct single Delaunay crossings, or in one double Delaunay crossing. Hence, the overall number of such triples in $P$ is at most $O(\ell n^2)$. Furthermore, any triple $(p,w,q)$ is shared by at most $k+3$ special quadruples $\SQ$ (namely, such special quadruples $\SQ=(a,p,w,q)$ whose second crossings $(wa,q,\J_u)$ are among the first $k+3$ clockwise special $(w,q)$-crossings to begin after $p$ crosses $qw$ from $\L_{qw}^-$ to $\L_{qw}^+$); see case (e1) for a fully symmetric argument.

\begin{figure}[htbp]
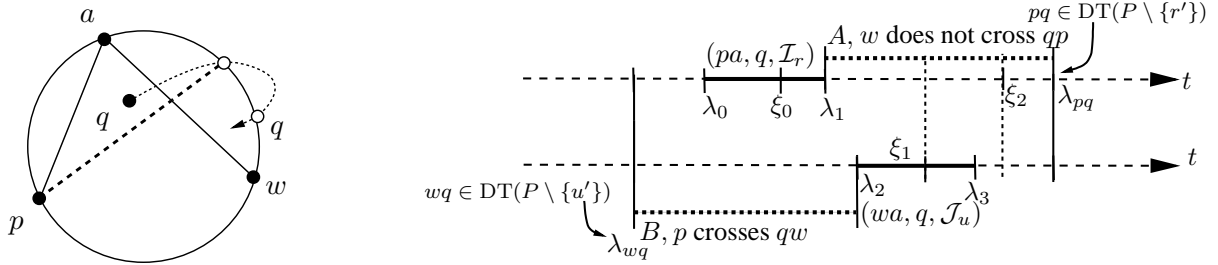

\begin{center}
\input{LastCocircLeft.pstex_t}\hspace{1.9cm}\input{TimelineAfterEleft.pstex_t}
\caption{\small Case (e2). Left: A possible motion of $q$ during $\J_u$, and afterwards. The Delaunayhood of $pq$ is violated, right after $\xi_1$, by $a$ and $w$. The points of $\SQ$ are involved, at some time $\xi_{2}\in (\xi_1,\t_{pq}]$ in another co-circularity (of index $3$). The order type of $\SQ$ remains fixed throughout $[\xi_{1},\xi_2]$. Right: A schematic summary of what we eventually assume at the end of case (e2).} 
\label{Fig:FirstCocircsLeft1}
\vspace{-0.4cm}
\end{center}
\end{figure} 

To conclude, we may assume that $p$ hits $wq$ only once during $[\t_{wq},\t_2)$ (crossing it from $\L_{wq}^+$ to $\L_{wq}^-$), and that  $w$ does not cross $pq$ during $[\t_1,\t_{pq}]$. 
Lemma \ref{Lemma:MustCross} then implies that the points of $\SQ$ are co-circular at some time $\xi_2\in (\t_1,\t_{pq}]$, and that $pq$ is intersected by $aw$ throughout $[\xi_1,\xi_2]$; see Figure \ref{Fig:FirstCocircsLeft1} (left). A schematic summary of what we assume by the end of case (e2) is given in Figure \ref{Fig:FirstCocircsLeft1} (right).

\medskip
\paragraph{Probabilistic refinement.} For each clockwise special $(p,q)$-crossing $(pa',q,\I_{r'})$ that ends during $(\t_1,\t_{pq})$ we add the corresponding point $a'$ to the obstruction set $A$ of $pq$. Similarly, for each clockwise special $(w,q)$-crossing $(wa',q,\J_{u'})$ that begins during $(\t_{wq},\t_2)$ we add the point $a'$ to the obstruction set $B$ of $wq$. 
As in Section \ref{Sec:CountQuad}, this is done in order to dispose of the corresponding special $(p,q)$- and $(w,q)$-crossings. Since we add at most $k+2$ elements to each set, and since $k\ll \ell$, each of the sets $A,B$ still contains at most $4\ell$ points of $P$.

Consider a subset $R$ of $\lceil n/\ell\rceil$ points chosen at random from $P$. Let $\F_R$ denote the refined family induced by $\F$ over $R$. Notice that the following two conditions hold simultaneously with probability at least $\Omega(1/\ell^6)$:
(1) The 6 points of $P_{\SQ}$ belong to $R$, and (2) $R$ includes none of the points of $(A\cup B)\setminus P_{\SQ}$.

Assume that the above sample $R$ is indeed successful for the chosen $\SQ=(a,p,w,q)$. Then the points of $P_{\SQ}$ still yield a Delaunay consecutive special quadruple (of the same topological type, which can be either right or left) with respect to $R$ and $\F_R$. We continue to denote this new quadruple as $\SQ$ but note
that the respective intervals $\I_r$ and $\J_u$ of the special crossings $(pa,q,\I_r)$ and $(wa,q,\J_u)$ may shrink as we pass from $\DT(P)$ to $\DT(R)$. We next review the additional properties gained by $\SQ$ in $\DT(R)$.

\smallskip
First, recall that the old time $\t_{pq}$ (defined after case (c) in terms of $P$) was accompanied by a point $r'\not\in P_{\SQ}$, whose removal restored the Delaunayhood of $pq$ at that time. Since $r'$ is among the omitted points of $A$, we can redefine $\t_{pq}$ as the first time in $[\t_3,\infty)$ when $pq$ belongs to $\DT(R)$.  Similarly, we redefine $\t_{wq}$ as the last time in $(-\infty,\t_0]$ when $wq$ belongs to $\DT(R)$. (In both cases, we refer to the new values of $\t_0$ and $\t_3$.) By what has just been noted, the new value of $\t_{pq}$ (resp., of $\t_{wq}$) decreases (resp., increases) from its old value.

Second, the following three conditions hold with respect to $R$ and $\F_R$, and with the new values of $\t_0,\t_1,\t_2,\t_3,\t_{pq}$ and $\t_{wq}$ (see Figure \ref{Fig:ThreeConditionsSp} for a schematic summary):

\medskip
\noindent{\bf (S1)} The edge $pq$ belongs to $\DT(R\setminus \{a,r,w,u\})$ throughout the interval $[\t_0,\t_{pq}]$. Furthermore, no clockwise special $(p,q)$-crossings $(pa',q,\I_{r'})$ end during $(\t_1,\t_{pq})$ (except perhaps for the special crossings of $pu$ and $pw$ by $q$).

\medskip
\noindent{\bf (S2)} The edge $wq$ belongs to $\DT(R\setminus \{p,a,r,u\})$ throughout the interval $[\t_{wq},\t_3]$. Furthermore, no clockwise special $(w,q)$-crossings $(wa',q,\I_{u'})$ begin during $(\t_{wq},\t_2)$ (except perhaps for the special crossings of $wr$ and $wp$ by $q$).

\medskip
\noindent{\bf (S3a)} If $\SQ$ is a {\it right} quadruple, then the set $P\setminus \{a,r,u\}$ induces a Delaunay crossing $(qp,w,\H)$ which occurs within $(\t_1,\t_{pq}]$. Furthermore, $w$ hits $pq$ only once during $(\t_1,\t_{pq}]$, so this is a single Delaunay crossing.  Moreover, the points of $\SQ$ are co-circular at some time $\xi_{-1}\in [\t_{wq},\xi_0)$, and the edge $qw$ is violated by $a\in \L_{qw}^+$ and $p\in \L_{qw}^-$ throughout the interval $(\xi_{-1},\xi_0)$. Finally, $p$ does not cross $qw$ in $[\t_{wq},\t_2]$.

\smallskip
\noindent{\bf (S3b)} If $\SQ$ is a {\it left} quadruple, then the set $P\setminus \{a,r,u\}$ induces a Delaunay crossing $(qw,p,\H)$, which occurs within $[\t_{wq},\t_2)$. Furthermore, $p$ hits $wq$ only once during $[\t_{wq},\t_2)$, so this is a single Delaunay crossing. Moreover, the points of $\SQ$ are co-circular at some time $\xi_{2}\in (\xi_1,\t_{pq}]$, and the edge $pq$ is violated by $a\in \L_{pq}^-$ and $w\in \L_{pq}^+$ throughout the interval $(\xi_{1},\xi_2)$. Finally, $w$ does not cross $qp$ in $[\t_1,\t_{pq}]$.

\begin{figure}[htbp]
\begin{center}
\input{RightRefined.pstex_t}\\
\vspace{1cm}
\input{LeftRefined.pstex_t}
\caption{\small A schematic summary of the properties of $\SQ$ within $\DT(R)$. The edge $pq$ is Delaunay at time $\t_{pq}$, and it is almost Delaunay (with the omission of only $a,r,u$) throughout $[\t_0,\t_{pq}]$. The edge $wq$ is Delaunay at time $\t_{wq}$, and it is almost Delaunay (with the same omission) throughout $[\t_{wq},\t_3]$.
Top: If $\SQ$ is a right quadruple, then $qp$ undergoes the crossing $(qp,w,\H)$ within $(\t_1,\t_{pq}]$, and we encounter an additional co-circularity of $a,p,w,q$ at some time $\xi_{-1}\in [\t_{wq},\xi_0)$. Bottom: If $\SQ$ is a left quadruple, then $qw$ undergoes the crossing $(qw,p,\H)$ within $[\t_{wq},\t_2)$, and the additional co-circularity occurs at some time $\xi_2\in (\xi_1,\t_{pq}]$ (below).} 
\label{Fig:ThreeConditionsSp}
\vspace{-0.4cm}
\end{center}
\end{figure}

\paragraph{Definition.} Assume that we are given a set $P$ of moving points, and a refined family $\F$.
Let $\SQ=(a,p,w,q)$ be a consecutive Delaunay special quadruple that is defined with respect to $\F$ and $P$.
We say that $\SQ$ is {\it $1$-restricted} if it satisfies the above three conditions (S1), (S2), and (S3a) or (S3b), where the reference sets $R$ and $\F_R$ are replaced by $P$ and $\F$, respectively. (We also implicitly require that the values $\t_{pq}$ and $\t_{wq}$, mentioned in conditions (S1) and (S2), actually exist.)

\smallskip

Let $\Phi_1(m)$ denote the maximum number of $1$-restricted special quadruples that can be defined over a set of $n$ moving points (and a refined family of regular Delaunay quadruples).
Then the following recurrence holds:
$$
\Phi_0(n)\leq O\left(\ell^6 \Phi_1(n/\ell)+k\ell^2N(n/\ell)+k^2N(n/k)+k\ell n^2\beta(n)\right).
$$

\begin{proposition}\label{Prop:UniquePqw}
With the above assumptions, any ordered triple $(p,q,w)$ can be shared by at most three 1-restricted special quadruples $\SQ=(a,p,w,q)$ of each topological type (i.e., right or left).
\end{proposition}
\begin{proof}
Let $\SQ=(a,p,w,q)$ be a $1$-restricted right special quadruple.
By Conditions (S2) and (S3a), $(pa,q,\I_r)$ is among the $3$ last special counterclockwise $(p,q)$-crossings to end before $w$ enters $\L_{pq}^+$ (during $\H$). Hence, $a$ is determined, up to three possible values, by the choice of $(p,q,w)$.
A fully symmetric argument applies if $\SQ$ is a left special quadruple.
\end{proof}


\paragraph{The subsequent stages --- Overview.} 
Fix a refined family $\F$, defined with respect to an underlying set $P$ of $n$ moving points. Let $\SQ=(a,p,w,q)$ be a $1$-refined Delaunay quadruple, consistent with $P$ and $\F$ and induced by special crossings
$(pa,q,\I_r)$ and $(wa,q,\J_u)$.
The correspondence between special crossings and their ordinary quadruples in $\F$ implies that the edges 
$pq$ and $wq$ undergo Delaunay crossings by the respective outer points $r$ and $u$; see Figure \ref{Fig:OuterPoints}.
Furthermore, if $\SQ$ is a right special quadruple, then condition (S3a) implies that $pq$ or, more precisely, its reversely oriented copy $qp$, undergoes a Delaunay crossing (in the reduced triangulation $\DT(P\setminus \{a,r,u\})$) by $w$, so the points $r$ and $w$ cross $pq$ {\it in opposite directions}.
Similarly, if $\SQ$ is a left special quadruple, then the edge $wq$ undergoes two oppositely oriented Delaunay crossings, by $u$ and $p$ (the latter occuring within $\DT(P\setminus \{a,r,u\})$, as above).

Our general strategy is to charge $\SQ$ to one of the above configurations $(p,q,r,w)$ or $(w,q,u,p)$ (depending on the right or left nature of $\SQ$), which will be referred to as {\it terminal quadruples}. Notice that each of those configurations involves one of the outer points $r$ and $u$, in addition to some three regular points of $\SQ$.
Nevertheless, several preparatory restrictions need to be enforced upon our special quadruples before actually charging them to terminal quadruples. Informally, this is done to further restrict the arising terminal quadruples and, consequently, to facilitate their eventual treatment at Stage 4 and in Section \ref{Sec:Terminal}.

At the subsequent Stages 2 and 3, we do not distinguish between left and right special quadruples $\SQ=(a,p,w,q)$. The topological restrictions enforced during these stages on special quadruples are fairly analogous to the ones enforced on ordinary quadruples during the parallel stages in Section \ref{Sec:CountQuad}.
Namely, for each $\SQ$ as above we extend the almost-Delaunayhood of its three edges $aq$, $pq$, and $wq$ from, respectively, $[\t_0,\t_3]$, $[\t_0,\t_{pq}]$, and $[\t_{wq},\t_3]$ to larger intervals, which cover $[\t_{wq},\t_{pq}]$.
The intimate correspondence between special crossings and ordinary quadruples is largely ignored throughout these technical stages, and
the outer points $r$ and $u$ do not play any meaningful role.

At the last Stage 4, 
we finally distinguish between left and right special quadruples. In both cases, we exploit the interplay between our quadruples and their respective outer points $r$ and $u$, which re-enter the analysis and finally give rise to terminal quadruples. (As  noted above, for each of the two types, only one outer point is used.)
This analysis is preceded by several preparatory charging arguments, analogous to the ones described in Section \ref{Subsec:Stage4Reg}.

\subsection{Stage 2}\label{Subsec:Stage2Special}
Let $\SQ=(a,p,w,q)$ be a $1$-restricted (Delaunay) special quadruple.
Our next goal is to extend the almost Delaunayhood of $qa$ from $[\t_0,\t_3]=\conn{\I_r}{\J_u}$ to some larger interval $[\xi_{qa}^-,\xi_{qa}^+]$, which covers $[\t_{wq},\t_{pq}]$.
As in the parallel Section \ref{Subsec:Stage2Reg}, we proceed in two steps, after fixing the constant parameters $12<k\ll \ell$. 

\begin{figure}[htbp]
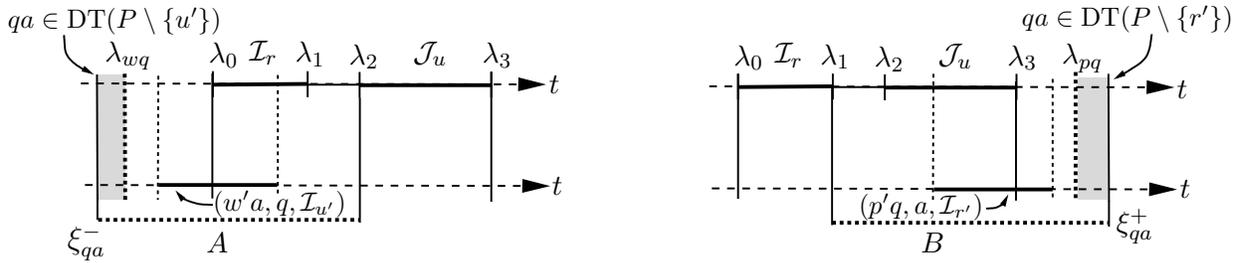

\begin{center}
\input{IntervalsExtendSpecial1.pstex_t}\hspace{2cm}\input{IntervalsExtendSpecial2.pstex_t}
\caption{\small Extending the almost-Delaunayhood of $qa$ from $[\t_0,\t_3]$ to $[\xi_{qa}^-,\t_0]$ (left) and to $[\t_3,\xi_{qa}^+]$ (right).}
\label{Fig:IntervalsExtendSpecialAq}
\vspace{-0.4cm}
\end{center}
\end{figure}

\medskip
\noindent{\bf Stage (2a).} First, we consider the interval $[\t_{wq},\t_3]$ where, by assumption, $wq$ is almost Delaunay. (It is in fact Delaunay in $P\setminus \{u\}$ throughout $\J_u=[\t_2,\t_3]$ and at time $\t_{wq}$.) Refer to Figure \ref{Fig:IntervalsExtendSpecialAq} (left).
If at least $k$ special counterclockwise $(a,q)$-crossings $(w'a,q,\J_{u'})$ (in $\F$) begin during $[\t_{wq},\t_2)$, then we can bound the overall number of such special quadruples $\SQ$ via the already routine combination of Theorem \ref{Thm:Balanced} with random sampling. 

 
Note, as a preparation, that the preliminary pruning (described in at beginning of this section) ensures that each of the above special $(a,q)$-crossings $(w'a,q,\J_{u'})$, where $u'$ is its respective outer point, satisfies $\{w',u'\}\cap P_{\SQ}=\emptyset$. Therefore, Lemma \ref{Lemma:OrderRelaxedCrossings} implies that $q$ hits each of the respective edges $w'a$ (during $\J_{u'}$) before it hits $wa$ (during $\J_u$).

To set the stage for an application of Theorem \ref{Thm:Balanced}, we consider a random subset $\hat{P}\subset P$ of $\lceil n/2 \rceil$ points, and argue that, with some fixed positive probability, $(wa,q,\J_u)$ becomes a $(w,q,\Theta(k))$-chargeable Delaunay crossing in $\hat{P}$ (with a potentially shrunk interval $\J_u$), with the reference interval $[\t_{wq},\t_3]$ (the proof of this property is identical to that given in Sections \ref{Subsec:Stage4Reg} and \ref{Subsec:Stage1Special}).
Briefly, this follows because $\hat{P}$ satisfies the following two conditions with probability $\Omega(1)$: (1) $\hat{P}$ includes $a,w,q$ but not $u$, and (2) for at least a constant fraction of the above special $(a,q)$-crossings $(w'a,q,\J_{u'})$, $\hat{P}$ includes $w'$ but not $u'$. The former condition guarantees that $\hat{P}$ yields a Delaunay crossing $(wa,q,\J)$, for some interval $\J\subseteq \J_u$, and that $qa$ belongs to $\DT(\hat{P})$ at times $\t_{wq}$ and $\t_3$. The latter condition implies that $\Omega(k)$ (ordinary) counterclockwise $(a,q)$-crossings occur within $[\t_{wq},\t_2)\cup \J\subseteq [\t_{wq},\t_3]$.

Theorem \ref{Thm:Balanced} now implies that the overall number of the above triples $(w,q,a)$ in $\hat{P}$ is at most 
$O\left(k^2N(n/k)+kn^2\beta(n)\right)$. By Proposition \ref{Prop:TriplesSpecialQuad}, this yields the same bound on the maximum number of the special quadruples $\SQ$ that fall into the present scenario.
Assume, then, that at most $k$ clockwise special $(a,q)$-crossings $(w'a,q,\J_{u'})$ begin during $[\t_{wq},\t_2)$. 

If no clockwise special $(a,q)$-crossings begin in $(-\infty,\t_{wq}]$, then 
$(wa,q,\J_u)$ is among the first $k+1$ such special $(a,q)$-crossings $(w'a,q,\J_{u'})$, so it can be charged to the pair $(a,q)$. (After the preliminary pruning, there remain no counterclockwise special $(a,q)$-crossings $(w'a,q,\J_{u'})$ with $u'\in P_{\SQ}$. Furthermore, Lemma \ref{Lemma:Crossing} implies that no other such $(a,q)$-crossings, with $u'\not\in P_{\SQ}$, can begin in $[\xi_{qa}^-,\t_{wq})$.)
Therefore, and because of Proposition \ref{Prop:TriplesSpecialQuad}, this happens for at most $O(kn^2)$
special quadruples $\SQ$. 

Assume next that some clockwise special $(a,q)$-crossing $(w'a,q,\J_{u'})$ begins in $(-\infty,\t_{wq}]$. Therefore, using Lemma \ref{Lemma:Crossing}, there is a last time $\xi_{qa}^-$ in $(-\infty,\t_{wq}]$ when the edge $qa$ belongs to some reduced triangulation $\DT(P\setminus \{u'\})$, for $u'\not\in P_{\SQ}$. 
In what follows, we use $u'$ to denote such a (fixed) point whose removal restores the Delaunayhood of $qa$ at (the last possible) time $\xi_{qa}^-$.

To proceed, we apply Theorem \ref{Thm:RedBlue} in $\A_{qa}$ over the interval $(\xi^-_{qa},\t_2)$. We do this for the above, reduced triangulation $\DT(P\setminus \{u'\})$, and with the second constant $\ell$. If at least one of the Conditions (i), (ii) of that theorem holds, we charge $\SQ$ (via $(wa,q,\J_u)$) either to an $(\ell+1)$-shallow collinearity or to $\Omega(\ell^2)$ $(\ell+1)$-shallow co-circularities. (Each of these events is $\ell$-shallow in $\DT(P\setminus \{u'\})$.) The choice of $\xi^-_{qa}$ implies that no special $(a,q)$-crossing $(w'a,q,\J_{u'})$ begins in $[\xi_{qa}^-,\t_{wq})$, and therefore, arguing as above, it guarantees that any event in $\A_{qa}$ is charged by at most $O(k)$ quadruples. Hence, this charging accounts for at most $O\left(k\ell^2 N(n/\ell)+k\ell n^2\beta(n)\right)$ quadruples $\SQ$.

Finally, if Condition (iii) of Theorem \ref{Thm:RedBlue} holds, then there is a set $A$ of at most $3\ell+1$ points (including $u'$) whose removal restores the Delaunayhood of $qa$ throughout $[\xi^-_{qa},\t_3]$.

\medskip
\noindent{\bf Stage (2b).} 
We similarly use Theorem \ref{Thm:Balanced} to extend the almost-Delaunayhood of $qa$ from $\I_r=[\t_0,\t_1]$ to the interval $(\t_1,\t_{pq}]$ where, by assumption, the edge $pq$ is almost Delaunay. (It is Delaunay in $P\setminus \{r\}$ throughout $\I_r=[\t_0,\t_1]$ and at time $\t_{pq}$.)
The argument is fully symmetric to the one in Stage (2a), but we briefly repeat it for the sake of completeness.

Refer to Figure \ref{Fig:IntervalsExtendSpecialAq} (right).
If at least $k$ special $(a,q)$-crossings $(p'a,q,\I_r)$ end in $(\t_1,\t_{pq}]$ then we again use Theorem \ref{Thm:Balanced} to show that the number of such special quadruples is at most $O\left(k^2N(n/k)+kn^2\beta(n)\right)$. In short, we argue that a random subset of $\lceil n/2\rceil$ points yields a $(p,q,\Theta(k))$-chargeable Delaunay crossing of $pa$ by $q$, with probability $\Theta(1)$.) Hence, we can assume that at most $k$ special $(a,q)$-crossings, as above, end during $(\t_1,\t_{pq}]$.

If no clockwise special $(a,q)$-crossings begin in $[\t_{pq},\infty)$, then 
$(wa,q,\J_u)$ is among the last $k+1$ such special $(a,q)$-crossings $(p'a,q,\I_{r'})$, so it can be charged to the pair $(a,q)$. Clearly,
that scenario occurs for at most $O(kn^2)$ special quadruples $\SQ$.

Otherwise, we choose the first time $\xi_{qa}^+$ in $[\t_{pq},\infty)$ when the edge $qa$ belongs to some reduced triangulation $\DT(P\setminus \{r'\})$, with $r'\not\in P_{\SQ}$. 
In what follows, we use $r'$ to denote such a (fixed) point whose removal restores the Delaunayhood of $qa$ at time $\xi_{qa}^+$.
We then apply Theorem \ref{Thm:RedBlue} in $\A_{qa}$ over the interval $(\t_1,\xi_{qa}^+)$. This is done with respect to the point set $P\setminus \{r'\}$, and with the constant $\ell$. 

If at least one of the Conditions (i), (ii) is satisfied, we dispose of $\SQ$ by charging it (via $(pa,q,\I_r)$) to $(\ell+1)$-shallow events in $\A_{qa}$, and argue, as above, that each event is charged by at most $O(k)$ quadruples. Hence, the above charging accounts for at most $O\left(k\ell^2 N(n/\ell)+k\ell n^2\beta(n)\right)$ special quadruples.

Finally, if none of the preceding scenarios occur, we end up with a subset $B$ of at most $3\ell+1$ points (including $r'$) whose removal restores the Delaunayhood of $aq$ throughout $[\t_0,\xi_{qa}^+]$.

\medskip
\noindent{\bf Probabilistic refinement.} We say that a special quadruple $\SQ=(a,p,w,q)$ is {\it $2$-restricted} if (1) it is $1$-restricted with respect to the underlying set $P$ and refined family $\F$, and (2) it satisfies the following new condition:

\medskip
\noindent{\bf (S4)} The edge $qa$ belongs to $\DT(P\setminus \{p,w,u,r\})$ throughout the interval $[\xi_{aq}^-,\xi_{aq}^+]$, where $\xi_{aq}^-$ (resp., $\xi_{aq}^+$) denotes the last time in $(-\infty,\t_{wq}]$ (resp., first time in $[\t_{pq},\infty)$) when the edge $aq$ is Delaunay (and where we assume that the times $\xi_{qa}^-,\xi_{qa}^+$ exist).

\medskip
Let $\Phi_2(m)$ denote the maximum number of $2$-restricted special quadruples that can be defined over a set of $m$ moving points (and a refined family $\F$). The preceding analysis, combined with the standard sampling argument of Clarkson and Shor, leads to the following recurrence:

\begin{equation}
\Phi_1(n)=O\left(\ell^6 \Phi_2(n/\ell)+k\ell^2 N(n/\ell)+k^2N(n/k)+k\ell n^2\beta(n)\right).
\end{equation}

\subsection{Stage 3}
To bound the above quantity $\Phi_2(n)$, we fix a set $P$ of $n$ moving points, and a refined family $\F$.
In addition, we fix a $2$-restricted special quadruple $\SQ=(a,p,w,q)$ (with outer points $r$ and $u$), which is defined with respect to $P$ and $\F$.

Recall that the edge $pq$ is Delaunay at time $\t_{pq}$, and that it is almost Delaunay during $[\t_0,\t_{pq}]$ (it is Delaunay with the omission of $a,w,r$ and $u$). Similarly, $wq$ is Delaunay at time $\t_{wq}$, and it is almost Delaunay during $[\t_{wq},\t_3]$ (it is Delaunay with the omission of $a,p,r,u$).
Our goal in this stage is (i) to extend the almost-Delaunayhood of $pq$ to a (possibly) larger interval 
$[\xi_{pq},\t_{pq}]$, for some $\xi_{pq}\leq\xi^-_{qa}\leq\t_{wq}$, and (ii) to extend the almost-Delaunayhood of $wq$ to an interval 
$[\t_{wq},\xi_{wq}]$, for some $\xi_{wq}\geq\xi_{qa}^+\geq\t_{pq}$.

Our analysis consists of two symmetric arguments, similar to the ones used in Section \ref{Subsec:Stage1Special} (cases (b) and (c)). Both arguments use Theorem \ref{Thm:Balanced} (in combination with the almost-Delaunayhood of $qa$ in $[\xi^-_{qa},\xi^+_{qa}]$) and refer to the same pair of constant parameters $12<k\ll \ell$.

\begin{figure}[htbp]
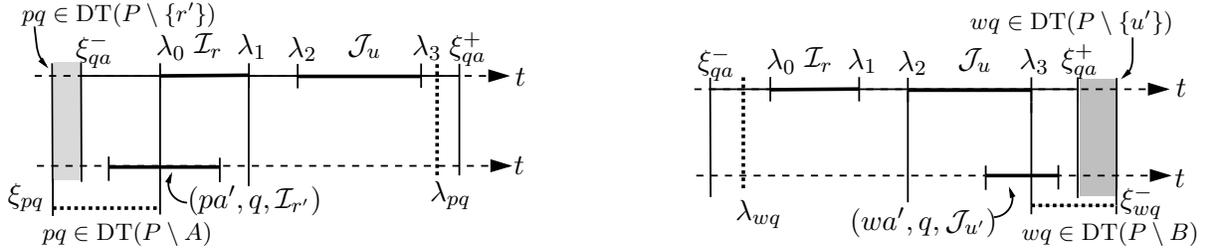

\begin{center}
\input{IntervalsExtend3Special.pstex_t}\hspace{2.3cm}\input{IntervalsExtend3SpeciallWq.pstex_t}
\caption{\small Left: Extending the almost-Delaunayhood of $pq$ from $[\t_0,\t_{pq}]$ to $[\xi_{pq},\t_{pq}]$. Right: Extending the almost-Delaunayhood of $wq$ from $[\t_{wq},\t_3]$ to $[\t_{wq},\xi_{wq}]$.
}
\label{Fig:IntervalsExtend3Special}
\vspace{-0.4cm}
\end{center}
\end{figure}

\paragraph{Extending the almost-Delaunayhood of $pq$.} 
Refer to Figure \ref{Fig:IntervalsExtend3Special} (left).
If at least $k$ special $(p,q)$-crossings $(pa',q,\I_{r'})$ begin during $[\xi^-_{qa},\t_0)$, then we can invoke Theorem \ref{Thm:Balanced} to show that the number of such special quadruples $\SQ$ is at most 
$O\left(k^2N(n/k)+kn^2\beta(n)\right)$.

Specifically, recall that the edge $qa$ is Delaunay at times $\xi_{qa}^-,\xi_{qa}^+$, and that it is almost Delaunay (with only four potentially obstructing points $p,w,u,r$) during  $[\xi_{qa}^-,\xi_{qa}^+]\supset [\xi_{qa}^-,\t_0)\cup \I_r=[\xi_{qa}^-,\t_1]$.
Hence, a random subset $\hat{P}\subset P$ of $\lceil n/2\rceil$ points would make $(pa,q,\I_r)$, with some fixed positive probability, an $(a,q,\Theta(k))$-chargeable crossing in $\hat{P}$ with $[\xi_{qa}^-,\t_1]$ as a reference interval (where $\xi_{qa}^-$ and $\t_1$ are still defined with respect to $P$, and $\I_r$ is possibly shrunk in $\hat{P}$).

Assume then that at most $k$ clockwise special $(p,q)$-crossings begin during $[\xi^-_{qa},\t_0)$.
If no such $(p,q)$-crossings $(pa',q,\I_{r'})$, with $r'\not \in P_{\SQ}$, begin before $\t_0$, then $(pa,q,\I_r)$ is among the first $k+3$ clockwise special $(p,q)$-crossings (including such crossings whose respective outer point $r'$ belongs to $P_{\SQ}$).\footnote{Recall from Section \ref{Subsec:Stage1Special} that at most two such crossings have $r'\in P_{\SQ}$.}
Clearly, the overall number of such quadruples $\SQ$ is at most $O(kn^2)$.

We may therefore assume that
the previous sub-scenario does not occur. In particular, there exists $\xi_{pq}$ which is the last time in $(-\infty,\xi_{qa}^-]$ when $pq$ belongs to some reduced triangulation $\DT(P\setminus \{r'\})$, for $r'\not\in P_{\SQ}$. In what follows, we use $r'$ to denote such a fixed point, whose removal restores the Delaunayhood of $pq$ at (the latest possible) time $\xi_{pq}$.

To proceed, we apply Theorem \ref{Thm:RedBlue} in $\A_{pq}$ over the interval $(\xi_{pq},\t_0)$. We do so with the second constant $\ell$, and with respect to the reduced point set $P\setminus \{r'\}$. 

If at least one of the Conditions (i), (ii) of Theorem \ref{Thm:RedBlue} holds, we charge $\SQ$ (via $(pa,q,\I_r)$) either to 
$\Omega(\ell^2)$ $(\ell+1)$-shallow co-circularities, or to an $(\ell+1)$-shallow collinearity. As above, the choice of $\xi_{pq}$ guarantees that $(pa,q,\I_r)$ is among the first $k+3$ special $(p,q)$-crossings to begin after any event that we charge within $\A_{pq}$, so any event is charged as above by at most $O(k)$ quadruples $\SQ$. Hence, the above charging is applicable for at most $O\left(k\ell^2 N(n/\ell)+k\ell n^2\beta(n)\right)$ special quadruples. 

Finally, if Condition (iii) of Theorem \ref{Thm:RedBlue} is satisfied, we have a set $A$ of at most $3\ell+1$ points (including $r'$, and perhaps some of $a,w,r,u$) whose removal restores the Delaunayhood
of $pq$ throughout $[\xi_{pq},\t_0]$. We further add to our conflict set $A$ every point $a'$ whose respective $(p,q)$-crossing $(pa',q,\I_{r'})$ begins in $[\xi_{pq},\t_0)$. (This is done to ensure that these $(p,q)$-crossings do not arise in the following Stage 4. Note that at most $2$ such crossings $(pa',q,\I_{r'})$ begin in $[\xi_{pq},\xi_{qa}^-)$, and each of them satisfies $r'\in P_\SQ$.)
Since there are at most $k+2$ crossings of this kind, and since $k\ll \ell$, the cardinality of the augmented set $A$ does not exceed $4\ell$.

\paragraph{Extending tha almost-Delaunayhood of $wq$.} The argument is fully symmetric to the one that was used for $pq$, but we briefly repeat it for the sake of completeness.

Refer to Figure \ref{Fig:IntervalsExtend3Special} (right).
If at least $k$ special $(w,q)$-crossings $(wa',q,\J_{u'})$ end during $(\t_3,\xi_{qa}^+]$, we consider a random subset of $\lceil n/2\rceil$ points and argue as before that 
$(wa,q,\J_u)$ becomes, with some fixed positive probability, an $(a,q,\Theta(k))$-chargeable special crossing (now with $[\t_2,\xi^+_{qa}]$ as the reference interval).
Therefore, Theorem \ref{Thm:Balanced} implies that the number of such special quadruples $\SQ$ is at most $O\left(k^2N(n/k)+kn^2\beta(n)\right)$.

Assume then that at most $k$ clockwise special $(w,q)$-crossings $(wa',q,\J_{u'})$ end during $(\t_3,\xi_{qa}^+]$.
Furthermore, we may assume that there exists $\xi_{wq}$, which is the first time in $[\xi_{qa}^+,\infty)$ when the edge $wq$ belongs to some triangulation $\DT(P\setminus \{u'\})$, for $u'\not\in \{a,r,w,u\}$. 
(Otherwise, $(wa,q,\J_u)$ would be among the last $k+3$ clockwise special $(w,q)$-crossings, which can happen for at most $O(kn^2)$ special quadruples of the kind considered here.)
In what follows, we use $u'$ to denote a fixed point whose removal restores the Delaunayhood of $wq$ at time $\xi_{wq}$. 

To proceed, we apply Theorem \ref{Thm:RedBlue} in $\A_{wq}$ over the interval $(\t_3,\xi_{wq})$, with the second parameter $\ell$ and respect to the point set $P\setminus \{u'\}$.

If at least one of the Conditions (i), (ii) of Theorem \ref{Thm:RedBlue} holds, we dispose of $\SQ$ by charging it (via $(wa,q,\J_u)$) to $(\ell+1)$-shallow events in $\A_{wq}$. The choice of $\xi_{wq}$ guarantees that each collinearity or co-circularity is charged in this manner by at most $O(k)$ quadruples $\SQ$. Hence, the above charging is applicable for at most $O\left(k\ell^2 N(n/\ell)+k\ell n^2\beta(n)\right)$ special quadruples. 

Finally, if Condition (iii) of Theorem \ref{Thm:RedBlue} is satisfied, we end up with a subset $B$ of at most $3\ell+1$ points (including $u'$ and perhaps some of $a,p,r,u$) whose removal restores the Delaunayhood
of $wq$ throughout the interval $[\t_3,\xi_{wq}]$. We add to $B$ every point $a'$ whose respective crossing $(wa',q,\J_{u'})$ ends in $(\t_3,\xi_{qa}^+]$. (As before, this is done to ensure that these $(w,q)$-crossings do not arise in the following Stage 4.) As above, the cardinality of the augmented set $B$ does not exceed $4\ell$.

\medskip
\noindent{\bf Probabilistic refinement.} We say that a special quadruple $\SQ$ is {\it $3$-restricted} if (1) it is $2$-restricted, and (2) it satisfies the following additional conditions:

\medskip
\noindent{\bf (S5)} The edge $pq$ belongs to $\DT(P\setminus \{a,w,u,r\})$ throughout the interval $[\xi_{pq},\t_{pq}]$, where $\xi_{pq}$ denotes the last time in $(-\infty,\xi_{qa}^-]$ when the edge $pq$ is Delaunay (and we assume the existence of such a time $\xi_{pq}$). In addition, at most two special $(p,q)$-crossings $(pa',q,\I_{r'})$ begin during $[\xi_{pq},\t_0)$ (namely, the possible crossings of $pw$ and $pu$ by $q$).

\medskip
\noindent{\bf (S6)} The edge $wq$ belongs to $\DT(P\setminus \{a,p,u,r\})$ throughout the interval $[\t_{wq},\xi_{wq}]$, where $\xi_{wq}$ denotes the first time in $[\xi_{qa}^+,\infty)$ when the edge $wq$ is Delaunay (and we assume the existence of such a time $\xi_{wq}$). In addition, at most two special $(w,q)$-crossings $(wa',q,\J_{u'})$ end during $(\t_3,\xi_{wq}]$ (namely, the possible crossings of $wp$ and $wr$ by $q$).

\medskip
Let $\Phi_3^R(m)$ (resp., $\Phi_3^L(m)$) denote the maximum number of $3$-restricted right (resp., left) special quadruples that can be defined over a set of $m$ moving points (and a fixed refined family $\F$).
The preceding analysis, in combination with the routine sampling argument of Clarkson and Shor, implies the following recurrence:

\begin{equation}
\Phi_2(n)=O\left(\ell^6\Phi_3^R(n/\ell)+\ell^6\Phi_3^L(n/\ell)+k\ell^2 N(n/\ell)+k^2N(n/k)+k\ell n^2\beta(n)\right)
\end{equation}

\subsection{Stage 4: The number of right quadruples}\label{Subsec:Stage4Right}
To bound the maximum possible number $\Phi_3^R(n)$ of $3$-restricted right special quadruples, we fix the underlying set $P$ of $n$ moving points, and a refined family $\F$. 

\medskip
\noindent{\bf Topological setup.} 
According to Proposition \ref{Prop:UniquePqw}, any $3$-restricted quadruple $\SQ=(a,p,w,q)$ shares its triple $(p,q,w)$ with at most two other such quadruples. (In other words, it suffices to bound the overall number of the corresponding triples $(p,q,w)$.) 
We strengthen the above property, by considering, without loss of generality, at most {\it one} $3$-restricted right quadruple for each triple $(p,q,w)$.
Therefore, in what follows every special quadruple $\SQ=(a,p,w,q)$ under consideration will be uniquely determined by its triple $(p,q,w)$.

To proceed, we fix a $3$-restricted right special quadruple $\SQ=(a,p,w,q)$, with respect to $P$ and $\F$, whose two special $(a,q)$-crossings take place during the intervals $\I_r=[\t_0,\t_1]$ and $\J_u=[\t_2,\t_3]$ (in this order), where $r$ and $u$ are the respective outer points. 
Recall that the original ``regular" family $\F$ includes the quadruples $\Q_1=(p,q,a,r)$ and $\Q_2=(w,q,a,u)$.

Refer to Figure \ref{Fig:TopSetupRight}.
Since $\SQ$ is $3$-restricted, there exist a time $\t_{wq}\leq \t_0$ which is the last time before\footnote{If $wq$ is Delaunay at time $\t_0$ then we put $\t_{wq}=\t_0$.} $\t_0$ when the edge $wq$ belongs to $\DT(P)$, and a symmetric first time $\t_{pq}\geq \t_3$ when $pq$ belongs to $\DT(P)$.
By Condition (S4), there exist the first time $\xi_{qa}^+$ in $[\t_{pq},\infty)$, and the last time $\xi_{qa}^-$ in $(-\infty,\t_{wq}]$ when the edge $qa$ is Delaunay, so that this edge is almost-Delaunay during the interval $[\xi_{qa}^-,\xi_{qa}^+]$ (with only $p,w,u,r$ as the possible obstructing points).
Moreover, by Conditions (S5) and (S6), there exist the first time $\xi_{wq}\in [\xi_{qa}^+,\infty)$, and the symmetric last time $\xi_{pq}\in (-\infty,\xi^-_{qa}]$ when the respective edges $wq$ and $pq$ are Delaunay. Moreover, $wq$ and $pq$ are almost Delaunay during, respectively, $[\t_{wq},\xi_{wq}]$ and $[\xi_{pq},\t_{pq}]$
(each with four obstructing points, as specified in these conditions).

\begin{figure}[htbp]
\begin{center}
\input{AlmostCrossingSpecial.pstex_t}\hspace{2cm}\input{TimeLineSpecial.pstex_t}\hspace{2cm}\input{TheWedgeRight.pstex_t}
\caption{\small The topologt76ical setup during the interval $(\xi_{-1},\t_q)\subseteq [\t_{wq},\t_{pq}]$. 
Left: The edge $qp$ is hit at some time $\t_q\in [\t_1,\t_{pq}]$ by $w$, so it undergoes a Delaunay crossing $(qp,w,\H=[\t_4,\t_5])$ within $\DT(P\setminus \{a,r,u\})$. Right: We have $\t_{wq}\leq \xi_{-1}<\xi_0<\t_4<\t_q<\t_5<\xi_{wq}$. Bottom: The motion of $B[p,q,w]$ is continuous throughout $[\xi_{-1},\t_q)$ (the hollow circles represent the co-circularities at times $\xi_{-1}$ and $\xi_0$).}
\label{Fig:TopSetupRight}
\vspace{-0.4cm}
\end{center}
\end{figure}

Let us summarize what we know so far about the motion of $a,p,w,q$.
By Condition (S3a), these points are co-circular at times $\xi_{-1}\in [\t_{wq},\t_0)$, $\xi_0\in \I_r\setminus \J_u$, and $\xi_1\in \J_u\setminus \I_r$. 
Moreover, the Delaunayhood of $wq$ is violated, throughout $(\xi_{-1},\xi_0)$, by the points $a\in \L_{wq}^+$ and $p\in \L_{wq}^-$ (so, in particular, neither of these points crosses $wq$ during this period).
Hence, $a$ lies throughout that interval within the wedge $W_{qpw}=\L_{pq}^+\cap \L_{pw}^-$ and inside the cap $C^-_{qw}=B[p,q,w]\cap \L_{qw}^-$.
We emphasize that the order type of the quadruple $(q,p,w,a)$ remains unchanged during $(\xi_{-1},\xi_0)$.

In addition, by the same Condition (S3a), the smaller set $P\setminus \{a,r,u\}$ yields a (single) Delaunay crossing $(qp,w,\H_{\SQ})$, whose interval $\H=\H_{\SQ}=[\t_4,\t_5]$ is contained in $(\t_{1},\t_{pq}]$. 
In particular, $w$ hits $pq$ at some moment\footnote{Recall from Section \ref{Subsec:Stage1Special} that $w$ can cross $qp$ either before or after $\xi_2$, depending on the location of $p$ when $q$ crosses $wa$. 
Our analysis only relies on the fact that $\t_q>\xi_0>\xi_{-1}$.} $\t_{q}\in \H$, when $w$ crosses $\L_{pq}$ from $\L^+_{pq}$ to $\L^-_{pq}$. Since $w$ lies in $\L_{pq}^+$ at times $\xi_{-1}$ and $\xi_{0}$, no further collinearities of $p,w,q$ can occur during 
$[\xi_{-1},\t_q)$. (Otherwise, the point $w$ would have to re-enter $\L_{pq}^+$ before $\t_q$, and then the triple $p,q,w$ would be collinear three times, contrary to our assumptions.)  To conclude,
the disc $B[p,q,w]$ moves continuously throughout the interval $[\xi_{-1},\t_q)$, which is obviously contained in $[\xi_{pq},\t_{pq}]\cap [\t_{wq},\xi_{wq}]=[\t_{wq},\t_{pq}]$.

\paragraph{Overview.} We fix three constant parameters $k,\ell,h$, such that $12<k\ll \ell \ll h$, and distinguish between four possible cases. The first two cases (a)--(b) are fairly similar to the cases (a)--(b) that we encountered 
in Section \ref{Subsec:Stage4Reg} when handling ordinary quadruples, and case (c) is very similar to the preceding case (b). 
In case (a) we bound the number of right special quadruples $\SQ$, that fall into it, using Theorem \ref{Thm:Balanced}.
In each of the subsequent cases (b) and (c), we manage to bound the number of special quadruples $\SQ$, that fall into that case, by charging 
them within the arrangements $\A_{pw}$, $\A_{pq}$ and $\A_{wq}$.
(The crucial difference between the two setups is that the extremal co-circularity among $\xi_0$ and $\xi_1$ now occurs during the {\it second} crossing $(wa,q,\J_u)$, so the topological analysis of Section \ref{Subsec:Stage4Reg} must be performed in a ``time-reversed" manner.)

In the final, most involved, case (d), we re-introduce at last the outer point $r$. (The other outer point $u$ is not used in the analysis of right special quadruples.) The correspondence between $(pa,q,\I_r)$ and its ancestor quadruple 
$\Q=(p,q,a,r)$ in $\F$ implies that the points $r$ and $w$ cross the same edge $pq$ in opposite directions.
Hence, $\SQ$ can be charged to the resulting so-called {\it terminal quadruple} $(p,q,r,w)$.
In Section \ref{Sec:Terminal} we express the number of these terminal quadruples in terms of more elementary quantities, that were introduced in Section \ref{Sec:Prelim}. This, combined with a parallel (and mostly symmetric, although considerably simplified) analysis of $3$-restricted {\it left} special quadruples, finally produces a complete system of recurrences whose solution is $O(n^{2+\eps})$, for any $\eps>0$.

\begin{figure}[htbp]
\begin{center}
\input{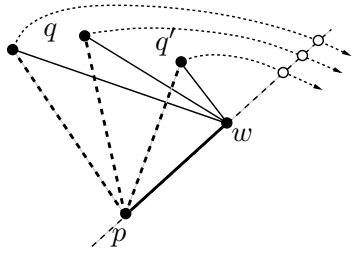}
\caption{\small Each right special quadruple $\SQ'=(a',p,w,q')\in \G_{pw}^R$ (with respective outer points $r'$ and $u'$) comes with a counterclockwise $(p,w)$-crossing $(q'p,w,\H_{\SQ'})$, which occurs within $\DT(P\setminus \{a',r',u'\})$.}
\label{Fig:PwCrossings}
\vspace{-0.4cm}
\end{center}
\end{figure}

\bigskip
In what follows, we consider the family $\G^R_{pw}$ of all $3$-restricted right special quadruples of the form $\SQ'=(a',p,w,q')$, which share their middle pair with $\SQ$. 
We may assume that each $\SQ'=(a',p,w,q')\in \G^R_{pw}$ is uniquely determined by the choice of $q'$ (as the only ``free" point in the triple $(p,q',w)$). 
Note that the set $P_{\SQ'}$ of each $\SQ'$ includes, in addition to the four points $a',p,w,q'$ of $\SQ'$, the respective outer points $r'$ and $u'$ of its special crossings $(pa',q',\I_{r'})$ and $(wa',q',\J_{u'})$.
Furthermore, each of these quadruples $\SQ'\in \G^R_{pw}$ is accompanied by a counterclockwise $(p,w)$-crossing $(q'p,w,\H_{\SQ'}=\H')$, which occurs within the smaller triangulation $\DT(P\setminus\{a',r',u'\})$. See Figure \ref{Fig:PwCrossings}.
We use $\t_{q'}$ to denote the time in $\H'$ when the respective point $q'$ of $\SQ'$ enters the halfplane $\L_{pw}^+$ (or, equivalently, when $w$ crosses $q'p$ from $\L_{pq'}^+=\L_{q'p}^-$ to $\L_{q'p}^+$).

Notice that Lemma \ref{Lemma:OrderRelaxedCrossings} readily generalizes to the above $(p,w)$-crossings. Namely, a pair of such crossings $(qp,w,\H_{\SQ})$ and $(q'p,w,\H_{\SQ'})$, which occur within the respective triangulations $\DT(P\setminus \{a,r,u\})$ and $\DT(P\setminus \{a',r',u'\})$, are {\it compatible}, provided that $q'\neq a,r,u$ and $q\neq a',r',u'$, in the sense that the orders in which the intervals $\H_{\SQ}$ and $\H_{\SQ'}$ begin or end 
are both consistent with the time stamps $\t_q$ and $\t_{q'}$.

To proceed, we distinguish between four possible cases.

\begin{figure}[htbp]
\begin{center}
\input{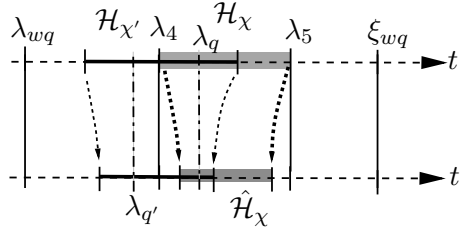}
\caption{\small Case (a): At least $k$ counterclockwise $(p,w)$-crossings $(q'p,w,\H_{\SQ'})$ either begin in $[\t_{wq},\t_4)$ or end in $(\t_5,\xi_{wq}]$ (one such crossing of the former type is depicted). Then, with some fixed and positive probability, the sample $\hat{P}$ yields a Delaunay crossing $(qp,w,\hat{\H}_{\SQ})$ that is $(q,w,\Theta(k))$-chargeable with respect to $[\t_{wq},\xi_{wq}]$.}
\label{Fig:TimelinePw}
\vspace{-0.4cm}
\end{center}
\end{figure}

\medskip
\noindent{\bf Case (a).} For at least $k$ of the above quadruples $\SQ'=(a',p,w,q')\in \G^R_{pw}$, their respective $(p,w)$-crossings $(q'p,w,\H')$ either begin in $[\t_{wq},\t_4]$, or end in $[\t_5,\xi_{wq}]$. Refer to Figure \ref{Fig:TimelinePw}.
Recall that, by condition (S6), the edge $qw$ is Delaunay at each of the times $\t_{wq}$ and $\xi_{wq}$, and that it is almost Delaunay during the entire interval $[\t_{wq},\xi_{wq}]$. 

To bound the number of such quadruples $\SQ$, we wish to argue that the crossing $(qp,w,\H)$ is $(q,w,\Theta(k))$-chargeable, for the reference interval $[\t_{wq},\xi_{wq}]$.
Unfortunately (and we have already encountered this technical issue before, e.g., in Section \ref{Subsec:Stage4Reg}), the crossing $(qp,w,\H)$ occurs within the reduced triangulation $\DT(P\setminus \{a,r,u\})$, whereas each of the above crossings $(q'p,w,\H')$ occurs within a possibly different (and also reduced) triangulation $\DT(P\setminus \{a',r',u'\})$. 

As in the previous similar situations (including the matching scenario (a) in Section \ref{Subsec:Stage4Reg}), we can free sufficiently many crossings $(q'p,w,\H')$ from their ``violators" $a',r'$ and $u'$ by passing to a smaller triangulation $\DT(\hat{P})$, which is induced by a random subset $\hat{P}\subset P$ of $\lceil n/4\rceil$ points.
Note though that $\G^R_{pw}$ can potentially include many quadruples $\SQ'$ with $q\in \{a',r',u'\}$, which cannot be freed without destroying $(qp,w,\H)$.

Fortunately, for any special quadruple $\SQ=(a,p,w,q)\in \G^R_{pw}$ (with outer points $r$ and $u$) the family $\G^R_{pw}$ includes at most three other quadruples $\SQ'=(a',p,w,q')$ whose respective points $q'$ are equal to  
one of $a,r$ or $u$. The pigeonhole principle then implies that at least {\it one quarter} of all quadruples $\SQ=(a,p,w,q)$ in $\G^R_{pw}$ satisfy the following condition:

\smallskip
\noindent {\bf (PHR1)} \noindent{\it There exist at most three quadruples $\SQ'\in \G^R_{pw}$ with $q\in \{a',r',u'\}$.}

\smallskip
\noindent(See Section \ref{Subsec:Stage4Reg} for the short proof of a similar claim, with the matching condition (PH).)

\smallskip
Since $p$ and $w$ are arbitrary points of $P$, (PHR1) holds for at least a quarter of all $3$-restricted right special quadruples under consideration; hence we may assume that it holds for the special quadruple $\SQ$ at hand.
Therefore, at least $k-6\geq k/2$ of the relevant quadruples $\SQ'=(a',p,w,q')\in \G^R_{pw}\setminus \{\SQ\}$ (with respective outer points $r'$ and $u'$, and with $(q'p,w,\H')$ starting in $[\t_{wq},\t_4]$ or ending in $[\t_5,\xi_{wq}]$) satisfy 
(i) $q\not\in \{a',r',u'\}$, and (ii) $q'\not\in \{a,r,u\}$. 

A suitable extension of Lemma \ref{Lemma:OrderRelaxedCrossings} then implies that at least $k/2$ of the above crossings $(q'p,w,\H')$ fully occur within $[\t_{wq},\xi_{wq}]$. 
Returning to the sampled triangulation $\DT(\hat{P})$, it is easy to check
that the following two conditions hold simultaneously with some fixed probability (see Stage 1 of this section for a similar argument): (1) the set $\hat{P}$ includes $p,q$ and $w$, but none of $a,r,u$, and (2) for at least $\Theta(k)$ of the above quadruples $\SQ'$ (with $\H_{\SQ'}$ starting in $[\t_{wq},\t_4)$ or ending in $(\t_5,\xi_{wq}]$), the sample $\hat{P}$ includes their respective points $q'$, but none of $a',r',u'$.

In the case of success, $\hat{P}$ yields a $(q,w,\Theta(k))$-chargeable (ordinary) Delaunay crossing of $qp$ by $w$, for the reference interval $[\t_{wq},\xi_{wq}]$. To see this, recall that $wq$ is Delaunay at both times $\t_{wq}$ and $\xi_{wq}$, and that it is almost Delaunay in $(\t_{wq},\xi_{wq})$ (it is Delaunay with the omission of $a,p,r,u$).
Then, according to condition (1), the sample $\hat{P}$ yields some single Delaunay crossing $(qp,w,\hat{\H}_{\SQ})$, whose respective interval $\hat{\H}_{\SQ}$ is contained in $\H_{\SQ}$ (as depicted in Figure \ref{Fig:TimelinePw}). Similarly, according to condition (2), $\hat{P}$ yields at least $\Theta(k)$ counterclockwise Delaunay $(p,w)$-crossings that occur within $[\t_{wq},\xi_{wq}]$.

To conclude, Theorem \ref{Thm:Balanced} implies that the overall number of such triples $(p,q,w)$ in $P$ does not exceed
$$
O\left(k^2N(n/k)+kn^2\beta(n)\right),
$$ 
which immediately also bounds the overall number of the corresponding $3$-restricted quadruples $\SQ$.

\paragraph{Preparing for cases (b) and (c): Charging events in $\A_{pw}$.}
We may assume, from now on, that there exist at most $k$ special quadruples $\SQ'\in \G_{pw}^R$ whose respective $(p,w)$-crossings $(q'p,w,\H')$ either begin in $[\t_{wq},\t_4]$, or end in $[\t_5,\xi_{wq}]$. 

Before proceeding to the following cases, we apply Theorem \ref{Thm:RedBlue} in $\A_{pw}$ in order to extend the almost-Delaunayhood of $pw$ from $\H=[\t_4,\t_5]$ to $[\t_{wq},\xi_{wq}]$. Notice that $[\t_{wq},\xi_{wq}]\setminus \H$
consists of two, possibly empty, intervals $[\t_{wq},\t_4)$ and $(\t_5,\xi_{wq}]$, and we consider each of them separately.
Note also that the edge $pw$ belongs during $\H$ to the reduced triangulation $\DT(P\setminus \{a,r,u\})$ (but not necessarily to $\DT(P)$), so Theorem \ref{Thm:RedBlue} must be applied with respect to this smaller set.

Consider, for instance, the interval $[\t_{wq},\t_4)$. We apply Theorem \ref{Thm:RedBlue} within $\A_{pw}$ over $(\t_{wq},\t_4)$, with our second parameter $\ell$, and with respect to the reduced set $P\setminus \{a,r,u\}$, noting that $pw$ belongs to $\DT(P\setminus \{a,r,u\})$ at the end of this interval.

If at least one of the Conditions (i), (ii) holds, we charge $\SQ$ within $\A_{pw}$, via $(qp,w,\H)$, either to an $(\ell+3)$-shallow collinearity, or to $\Omega(\ell^2)$ $(\ell+3)$-shallow co-circularities in $P$. (Each of these events is $\ell$-shallow with respect to the reduced set $P\setminus \{a,r,u\}$.) 
Notice that the points $p$ and $w$ are involved in each of these events, and
since case (a) has been ruled out, at most $k$ other $(p,w)$-crossings $(q'p,w,\H')$ of this kind begin after the respective time $t^*$ of any charged event and before $(qp,w,\H)$. That is, $(qp,w,\H)$ is among the first $k+1$ such $(p,w)$-crossings to begin after $t^*$. 
Hence, any $(\ell+3)$-shallow collinearity or co-circularity is charged in the above manner by at most $O(k)$ special quadruples $\SQ$.
To conclude, the above scenario occurs for at most $O\left(k\ell^2 N(n/\ell)+k\ell n^2\beta(n)\right)$ quadruples $\SQ$. 

Otherwise, if Condition (iii) holds, one can restore the Delaunayhood of $pw$ throughout $[\t_{wq},\t_4]$
by removing at most $3\ell+3$ points of $P$ (including $a,r,u$).

A fully symmetric argument can be used to extend the almost-Delaunayhood of $pw$ to the symmetric interval $(\t_5,\t_{wq}]$.
At the end, we have either disposed of $\SQ$ through conditions (i), (ii) of Theorem \ref{Thm:RedBlue} or ended up with a set $A_{pw}$ of at most $6\ell+3$ points (including $a,r,u$) whose removal restores the Delaunayhood of $pw$ throughout $[\t_{pq},\xi_{wq}]$. 
Hence, we may assume, in what follows, that the above set $A_{pw}$ exists.

\begin{figure}[htbp]
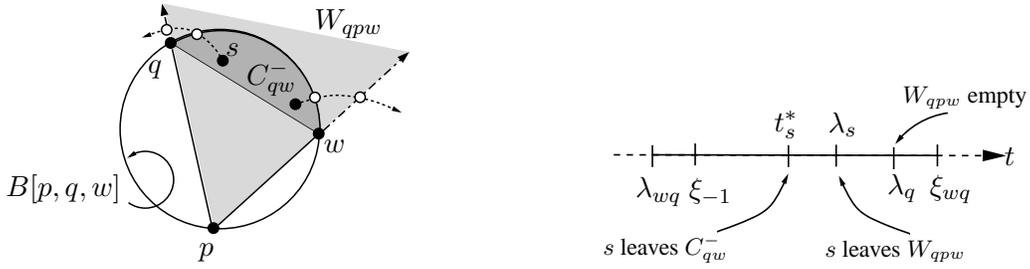

\begin{center}
\input{LeaveTheWedgeRight.pstex_t}\hspace{2.5cm}\input{TimelineLeaveRight.pstex_t}
\caption{\small Case (b). A total of at least $\ell$ points $s\neq a,r,u$ appear in the cap $C^-_{qw}$ during $(\xi_{-1},\t_q)$. 
Each of them must leave the cap $C_{qw}^-$ (through the boundary of $B[p,q,w]$) and then leave the wedge $W_{qpw}$ (through one of the rays $\vec{pq},\vec{pw}$, outside the respective edges $pq$ and $pw$) before time $\t_q$. Left: The geometric scenario. Right: A symbolic summary of the corresponding events.}
\label{Fig:LeaveTheWedgeRight}
\vspace{-0.4cm}
\end{center}
\end{figure} 

\medskip
\noindent{\bf Case (b).} There exist a total of at least $\ell$ points of $P$, distinct from $a,r,u$, such that each of them appears in the cap $C^-_{qw}=B[p,q,w]\cap \L_{qw}^-$ at some time during the interval $(\xi_{-1},\t_q)$. (Note that some of these points may belong to $A_{pw}$.) Recall that $\t_q$ denotes the time in $\H$ when $w$ enters $\L_{pq}^-$, through $pq$, and that no additional collinearities of $p,q,w$ can occur during $(\xi_{-1},\t_q)$, so the motion of $B[p,q,w]$ is fully continuous there.

Refer to Figure \ref{Fig:LeaveTheWedgeRight}.
Let $s\in P\setminus \{a,r,u\}$ be one of the points that visit $C^-_{qw}$ during $(\xi_{-1},\t_q)$. Since the above cap $C^-_{qw}$ is fully contained in the wedge $W_{qpw}=\L_{pq}^+\cap \L_{pw}^-$ during that interval, $s$ must leave $W_{qpw}$ before time $\t_q$ (when $W_{qpw}$ shrinks to the single ray $\vec{pq}=\vec{pw}$) through one of the rays $\vec{pw},\vec{pq}$. We also note that, by condition (S6) (and since $(\xi_{-1},\t_{q})\subseteq [\t_{wq},\xi_{wq}]$), $wq\in \DT(P\setminus \{a,p,r,u\})$ throughout $(\xi_{-1},\t_{q})$, so $s$, which has to leave $C^-_{qw}$ before it leaves $W_{qpw}$, can do so only through the boundary of $B[p,q,w]$. This results in a co-circularity of $p,q,w,s$, and is easily seen to imply that $s$ leaves $W_{qpw}$ by crossing one of the rays $\vec{pw}$ or $\vec{pq}$ {\it outside} the respective edge $pw$ or $pq$. 

In what follows, we assume that $s$ is among the last $\ell$ points to leave $C_{qw}^-$ during $(\xi_{-1},\t_q)$.
Let $t_{s}^*$ denote the time of the corresponding co-circularity of $p,q,w,s$, which occurs when $s$ leaves $C^-_{qw}$ through the boundary of $B[p,q,w]$. Since $\SQ$ satisfies condition (S6), the opposite cap $C^+_{qw}=B[p,q,w]\cap \L_{qw}^+$ 
contains no points of $P\setminus \{a,r,u\}$ at time $t_s^*$. (Otherwise, the Delaunayhood of $wq$ would be violated, at time $t_s^*$, by $s$ and any of these points.) Therefore, the co-circularity at time $t_s^*$ has to be $(\ell-1)$-shallow in $P\setminus \{a,r,u\}$, and thus $(\ell+2)$-shallow in $P$.

Note also that the co-circularity at time $t_s^*$ is red-blue with respect to the edge $wq$, which is violated right before it by $p$ and $s$. Lemma \ref{Lemma:Crossing}, together with the choice of $s\neq a,p,r,u$, imply 
that this co-circularity cannot occur during the crossing $(qp,w,\H_{\SQ}=[\t_4,\t_5])$ (which occurs in $P\setminus \{a,r,u\}$), so $t_s^*<\t_4$. 
(However, condition (S6) does not rule out the violation of $wq$ by $p$ and $s$ during the larger interval $[\t_{wq},\xi_{wq}]\setminus \H$, because the Delaunayhood of $wq$ is assumed to hold there only under the omission of $a,r,u$, and of $p$.)

To proceed, we distinguish between two possible subcases. In each of them we manage to dispose of $\SQ$ by charging it, within one of the arrangements $\A_{pq}, \A_{pw}$, either to $\Omega(\ell^2)$ $(2\ell)$-shallow co-circularities, or to a $(2\ell)$-shallow collinearity.

\medskip
\noindent {\bf Case (b1).} At least half of the above points $s$ cross the line $\L_{pq}$, from $\L_{pq}^+$ to $\L_{pq}^-$, during $(t_s^*,\t_q)$. (This also includes points $s$ that possibly cross $\L_{pq}$ outside the ray $\vec{pq}$, after leaving $W_{qpw}$ through the other ray $\vec{pw}$.) By Condition (S5) (and since $(t_s^*,\t_q)\subseteq (\xi_{-1},\t_q)\subseteq [\xi_{pq},\t_{pq}]$), each of these crossings occurs outside $pq$, within one of the corresponding outer rays of $\L_{pq}$.

For each $s$ we argue, exactly as in Section \ref{Subsec:Stage4Reg}, that the points $p,q,s$ are involved during $(t_s^*,\t_q)\subseteq (\xi_{-1},\t_q)$ either in a $(2\ell)$-shallow collinearity, or in $\Omega(\ell)$ $(2\ell)$-shallow co-circularities. That is, as $s$ approaches $\L_{pq}$, the disc $B[p,q,s]$ ``swallows" the entire halfplane $\L_{pq}^+$. If the disc, which contains at most $\ell+2$ points at the beginning of the process, ``swallows" at least $\ell-2$ points in this process, then each of the first $\ell-2$ resulting co-circularities are $(2\ell)$-shallow (in $P$). Otherwise, the collinearity of $q,p,s$ is $(2\ell)$-shallow.

Since $s$ can be chosen in at least $\Omega(\ell)$ different ways, the points $p$ and $q$ are involved during $(\xi_{-1},\t_q)$ either in $\Omega(\ell^2)$ $(2\ell)$-shallow co-circularities, or in a $(2\ell)$-shallow collinearity.
In both cases, we charge $\SQ$ to these events. 

Note that each $(2\ell)$-shallow event, which occurs in $\A_{pq}$ at some time $t^*\in (\xi_{-1},\t_q)$, can be traced back to $(qp,w,\H)$ (and, by Proposition \ref{Prop:UniquePqw}, also to $\SQ$) in at most $O(1)$ possible ways because $w$ is among the first four points to hit the edge $pq$ after time $t^*$, according to condition (S5). 
Hence, the above scenario happens for at most $O(\ell^2N(n/\ell)+\ell n^2\beta(n))$ special quadruples $\SQ$.

\medskip
\noindent {\bf Case (b2).} At least half of the above points $s\neq a,r,u$ remain in $\L_{pq}^+$ throughout the respective intervals $(t_s^*,\t_q)$. Each of these points must leave $W_{qpw}=\L_{pq}^+\cap \L_{pw}^-$, also during $(t_s^*,\t_q)$, through the ray emanating from $w$ in direction $\vec{pw}$, thereby crossing $\L_{pw}$ from $\L_{pw}^-$ to $\L_{pw}^+$. 
(Recall that $s$ can cross $\L_{pw}$ from $\L_{pw}^+$ to $\L_{pw}^-$ at most once, because the triple $p,w,s$ can be collinear at most twice.)

We again fix one of these points $s$, and use $\t_s$ to denote the corresponding time in $(t_s^*,\t_q)$ when $s$ leaves $W_{qpw}$ through the ray emanating from $w$ in direction $\vec{pw}$. As in the previous case, we conclude that either the collinearity of $p,w,s$ at time $\t_s$ is $(2\ell)$-shallow, or the points $p,w,s$ are involved in $\Omega(\ell)$ $(2\ell)$-shallow co-circularities during the preceding interval $(t_s^*,\t_s)$. 
As in Section \ref{Subsec:Stage4Reg}, the main challenge is to argue that each of the above $(2\ell)$-shallow events, which occur in $\A_{pw}$ during $(t_s^*,\t_s]\subseteq (\xi_{-1},\t_q)$, can be traced back to $\SQ$ in at most $O(k)$ ways.\footnote{As in Section \ref{Subsec:Stage4Reg}, the multiplicity of the chargings is the major difference between case (b1) and the present case (b2).}

To show this, let $t^*\in (\xi_{-1},\t_q)$ be the time of a $(2\ell)$-shallow collinearity or co-circularity that occurs in $\A_{pw}$. 
First, we guess the points $p$ and $w$ of $\SQ$ in $O(1)$ possible ways among the three or four points involved in the event. 
We next recall that, in the charging scheme of case (b2), each $(2\ell)$-shallow co-circularity or collinearity that we charge in $\A_{pw}$ is obtained via some point $s$, which is also involved in the event, that leaves $\L_{pw}^-$ at the respective time $\t_s$. We therefore guess $s$ among the remaining one or two points involved in the event.
To guess the remaining points $a$ and $q$ of $\SQ$, we examine all ``candidate" special quadruples $\SQ'\in \G_{pw}^R$ whose two middle points $(p,w)$ are shared with $\SQ$. 
Recall that each of these quadruples is accompanied by the $(p,w)$-crossing $(q'p,w,\H'=\H_{\SQ'})$, where $q'$ enters $\L_{pw}^+$ at the respective time $\t_{q'}\in \H'$. 
Recall also that $\SQ'$ is uniquely determined by the choice of $q'$ (as long as $p$ and $w$ remain fixed).

Clearly, with $s$ fixed, it suffices to consider only special quadruples $\SQ'=(a',p,w,q')$ in $\G_{pw}^R$ with the following properties: 
(1) $s\neq a',r',u'$, where $r'$ and $u'$ are the outer points of $\SQ'$, (2) $\t_{q'}>\t_s$, and (3) $s$ lies in $\L_{pq'}^+$ during the first portion of $\H_{\SQ'}$ (before $\t_{q'}$). This is because each of these conditions holds for $\SQ$ and $s$ in the charging scheme of case (b2). For example, (3) follows because case (b1) does not occur for $s$ (and since $t_s^*<\t_4$). 

If a special quadruple $\SQ'=(a',p,w,q')\in \G_{pw}^R$ satisfies the above three conditions (1)--(3), we say that the respective point $q'$ (which uniquely determines $\SQ'$) is a {\it candidate} (for being $q$).



Proposition \ref{Prop:kUniqueSpecial} below guarantees that each $(2\ell)$-shallow event, which occurs in $\A_{pw}$ at some time $t^*\in (\xi_{-1},\t_q)$, is charged by at most $k+7$ quadruples in $\SQ'\in \G_{pw}^R$, because the corresponding points $q'$ of these quadruples are among the first $k+7$ candidates to leave $\L_{pw}^-$ after time $\t_s$. 
Repeating the same argument for each of the $\Omega(\ell)$ possible choice of $s$ shows that at most
$O\left(k\ell^2N(n/\ell)+k\ell n^2\beta(n)\right)$
special quadruples can fall into case (b2). 

\begin{proposition}\label{Prop:kUniqueSpecial}
With the above assumptions, the point $q$ is among the first $k+6$ candidates $q'$ to leave the halfplane $\L_{pw}^-$ after $\t_s$.
\end{proposition}
\begin{proof}
The fairly technical proof of this proposition is symmetric to the one of Proposition \ref{Prop:kUnique}, so we only briefly review it.

\begin{figure}[htbp]
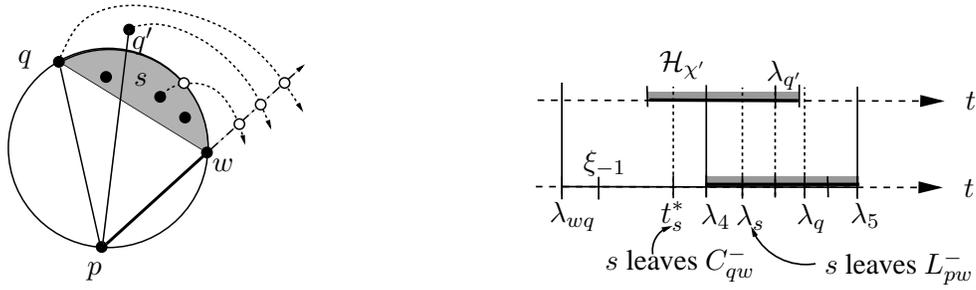

\begin{center}
\input{CandidateProp.pstex_t}\hspace{3cm}\input{TimeLineRightProp.pstex_t}
\caption{\small Proposition \ref{Prop:kUniqueSpecial}. Left: $q$ is among the first $k+7$ candidates $q'$ to leave $\L_{pw}^-$ after time $\t_s$. The figure depicts a point $q'$ lying outside $B[p,q,w]$ at the time $t_s^*$ when $s$ leaves the cap $C_{qw}^-$. Right: The various critical events occur in the depicted order. Note that $\t_s$  occurs either before $\t_4$, or in (the first part, preceding $\t_q$, of) $\H=[\t_4,\t_5]$.}
\label{Fig:PropositionCandidatesRight}
\vspace{-0.5cm}
\end{center}
\end{figure}

Assume to the contrary that the proposition does not hold (for $\SQ$ and $s\neq a,r,u$ as above).
Hence, we have at least $k$ candidates $q'$ such that $\t_s<\t_{q'}<\t_q$ and $q'\not\in \{a,r,u\}$, and such that the first points $a'$, and the outer points $r'$ and $u'$, of their quadruples $\SQ'=(a',p,w,q')$ are all distinct from $q$. (We continue to assume that $\SQ$ satisfies property (PHR1), introduced in case (a), so the last two restrictions $q'\neq \{q,a,r\}$ and $q\neq \{a',r',u'\}$ (the latter using (PHR1)) exclude from our consideration at most six candidates $q'\neq q$ together with their quadruples $\SQ'$.)

To establish the proposition, we fix a candidate $q'$ and its corresponding quadruple $\SQ'=(a',p,w,q')$ (with outer points $r'$ and $u'$), as above, and argue that the respective interval $\H_{\SQ'}$ begins during $(\t_{wq},\t_4)$. See Figure \ref{Fig:PropositionCandidatesRight} (right). Repeating the same argument for the remaining $k-1$ possible choices of $q'$ will imply that the quadruple $\SQ$ falls into case (a) and we would thereby reach a contradiction.

Indeed, since $\t_{q'}<\t_q$ (and $q'\neq a,r,u$ and $q\neq a',r',u'$), a suitable variant of Lemma \ref{Lemma:OrderRelaxedCrossings} shows that
the interval $\H_{\SQ'}$ begins before $\H_{\SQ}=[\t_4,\t_5]$. 
It thus remains to check that $\H_{\SQ'}$ begins after $\t_{wq}$.

If $\H_{\SQ'}$ begins after $t_s^*$, then we are done (as $t_s^*>\t_{wq}$). Hence, we may also assume that both times $t_s^*$ and $\t_{q'}$ belong to the interval $\H_{\SQ'}$. (More precisely, $t_s^*$ belongs to the first part of $\H_{\SQ'}$, before $\t_{q'}$; this is the situation considered in Figure \ref{Fig:PropositionCandidatesRight} (right).)
This, and the above conditions (2)--(3) (which hold for $\SQ'$ because $q'$ is a candidate point), imply that $q'$ remains in the halfplanes $\L_{pw}^-, \L_{ps}^-$ throughout the interval $(t_s^*,\t_{q'})$. 
Therefore, $s$ lies inside $W_{q'pw}=\L_{pq'}^+\cap \L_{pw}^-$ throughout the interval $(t_s^*,\t_s)$.

In addition, the standard properties of $\SQ$ and $\SQ'$ as $3$-restricted special quadruples imply that $q'$ must lie, throughout the longer interval $(t^*_s,\t_{q'})\subseteq \H_{\SQ'}\cap (\xi_{-1},\t_q)$, inside the wedge $W_{qpw}=\L_{pq}^+\cap \L_{pw}^-$. (Otherwise either the points $q',p$ and $w$ would be collinear more than once during $\H_{\SQ'}$, or the edge $q'p$ would be hit by $q$, or the edge $qp$ would be hit by $q'$. The first two cases are impossible by the definition of $(q'p,w,\H_{\SQ'})$, and the last one is ruled out by condition (S5).)

To recap, we may assume that $\H_{\SQ'}$ begins {\it before} $t_s^*$, and that the edges $pq,pq',ps$, and $pw$ appear, at time $t_s^*$, in this clockwise order around $p$.
To show that $\H_{\SQ'}$ begins after $\t_{wq}$, we distinguish between two possible cases. 

\medskip
\noindent(1) If $q'$ lies outside $B[p,w,s]=B[p,q,w]$ at time $t_s^*$ (as depicted in Figure \ref{Fig:PropositionCandidatesRight} (left)), then the Delaunayhood of $pq'$ is violated, at that very moment, by $s$ and $q$. Hence, the crossing $(q'p,w,\H_{\SQ'})$ (occurring in $\DT(P\setminus \{a',r',u'\})$) has to begin after $t_s^*$, contrary to our assumptions.

\begin{figure}[htbp]
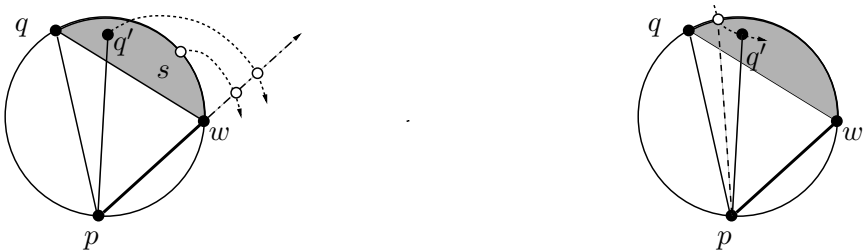

\begin{center}
\input{CandidatePropRight1.pstex_t}\hspace{3cm}\input{CandidatePropRight2.pstex_t}
\caption{\small Proof of Proposition \ref{Prop:kUniqueSpecial}: Left: The scenario where $q'$ lies within $B[p,q,w]$ at time $t_s^*$. Right: The candidate $q'$ must have entered $C_{qw}^-$, through $B[p,q,w]$, after $\t_{wq}$ (and before $t_s^*$, $\H=\H_{\SQ}$ and $\H_{\SQ'}$).}
\label{Fig:PropositionCandidatesRight1}
\vspace{-0.5cm}
\end{center}
\end{figure} 

\medskip
\noindent(2) If $q'$ lies at time $t_s^*$ within $B[p,q,w]$ (as depicted in Figure \ref{Fig:PropositionCandidatesRight1} (left)), then the interplay between the $(p,w)$-crossings $(qp,w,\H_{\SQ})$ and $(q'p,w,\H_{\SQ'})$ yields three co-circularities of the points $p,w,q,q'$. Namely, the last two co-circularities occur during $\H_{\SQ'}\setminus \H_{\SQ}$ and $\H_{\SQ}\setminus \H_{\SQ'}$. The first co-circularity occurs when $q'$ enters $C_{qw}^-$ after time $\t_{wq}$, when $wq$ is fully Delaunay, and before $t_s^*$, when the Delaunayhood of $wq$ is violated by $q'\in C_{qw}^-$ and $p\in B[p,q,w]\cap \L_{qw}^+$. (Briefly, this follows since, by conditions (S2) and (S5), none of $p,q'$ can cross $wq$ in the interval $[\t_{wq},\t_q]$; see the proof of Proposition \ref{Prop:kUniqueSpecial} for a fully symmetric argument.) As is easy to check, the points $p,w,q$ and $q'$ are co-circular only once during each of the intervals $\H_{\SQ'}$ and $\H_{\SQ}$, so their first co-circularity occurs before $\H_{\SQ'}$; see Figure \ref{Fig:PropositionCandidatesRight1} (right). Hence, to allow room for the first co-circularity to occur, $\H_{\SQ'}$ has to begin after $\t_{wq}$ also in this case.
As noted above, this completes the proof of the proposition.
\end{proof}


\medskip
\noindent{\bf Case (c).} A total of at least $\ell$ points $s\in P\setminus A_{pw}$ appear in the cap $C^+_{pw}=B[p,q,w]\cap \L_{pw}^+$ at some time during $(\xi_{-1},\t_q)$. Here $A_{pw}$ continues to denote the subset of at most $6\ell+3$ points, including $a,r$ and $u$, whose removal restores the Delaunayhood of $pw$ throughout the interval $[\t_{wq},\xi_{wq}]$. (Recall that $A_{pw}$ was obtained by applying Theorem \ref{Thm:RedBlue} in $\A_{pw}$, after ruling out case (a).)

\begin{figure}[htbp]
\begin{center}
\input{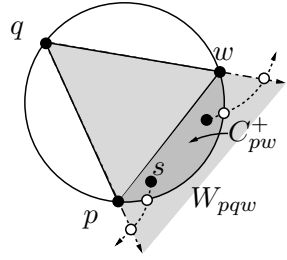}
\caption{\small Case (c). A total of at least $\ell$ points $s\in P\setminus A_{pw}$ appear in the cap $C^+_{pw}$ during $(\xi_{-1},\t_q)$. 
Each of them must leave the cap $C_{pw}^+$ (through the boundary of $B[p,q,w]$) and then exit the wedge $W_{pqw}$ (through one of the rays $\vec{qp},\vec{qw}$, outside the respective edges $pq$ and $wq$) before time $\t_q$.}
\label{Fig:LeaveTheWedgePw}
\vspace{-0.4cm}
\end{center}
\end{figure} 

Clearly, $C^+_{pw}$ is contained in the wedge $W_{pqw}=\L_{pq}^+\cap \L_{wq}^-$, which shrinks at time $\t_q$ to the ray $\vec{qp}=\vec{qw}$. Hence, each of these points $s$ has to leave $C^+_{pw}$ and $W_{pqw}$ (in this order) before time $\t_q$. Furthermore, $s$ can leave $C^+_{pw}$ only through the boundary of $B[p,q,w]$, at a co-circularity of $p,q,w,s$. (Otherwise $s$ would have to hit $pw$ and, therefore, belong to $A_{pw}$.)
In addition, $s$ can leave $W_{pqw}$ only through one of the rays $\vec{qp}$ and $\vec{qw}$ (outside the respective segments $qp,qw$). 
See Figure \ref{Fig:LeaveTheWedgePw}.

As in the previous case (b), we restrict our attention to the last $\ell$ such points $s$ of $P\setminus A_{pw}$ to leave $C_{pw}^+$ during $(\xi_{-1},\t_q)$, and use $t_s^*$ to denote the time of the respective co-circularity.
Clearly, the opposite cap $C^-_{pw}=B[p,q,w]\cap \L_{pw}^-$ contains then no points of $P\setminus A_{pw}$. Indeed, otherwise the Delaunayhood of $pw$ would be violated by $s$ and any one of these points (contrary to our assumption that $pw\in \DT(P\setminus A_{pw})$ throughout $[\t_{wq},\xi_{wq}]\supset (\xi_{-1},\t_q)$).
Hence, the resulting co-circularity of $p,q,w,s$ at time $t_s^*$ is $(7\ell+2)$-shallow in $P$, because, at the time of co-circularity, the circumdisc $B[p,q,w]=B[p,s,w]$ can contain in its interior at most the $6\ell+3$ points of $A_{pw}$ and at most $\ell-1$ points of $P\setminus A_{pw}$.

\medskip
\noindent {\bf Case (c1).} If at least half of the above points $s$ cross the line $\L_{pq}$ (from $\L_{pq}^+$ to $\L_{pq}^-$) during their respective intervals $(t_s^*,\t_q)$,
then we argue exactly as in subcase (b1). 
Namely, we fix one of these points $s$ and notice that the points $p,q,s$ are involved, during $(t_s^*,\t_{q})$, either in an $(8\ell)$-shallow common collinearity, or in $\Omega(\ell)$ $(8\ell)$-shallow co-circularities, occuring within the whole set $P$. That is, as $s$ approaches $\L_{pq}$, the disc $B[p,q,s]$ ``swallows" the entire halfplane $\L_{pq}^+$. If the disc, which contains at most $7\ell+2$ points at the beginning of the process, ``swallows" at least $\ell-2$ points in this process, then each of the first $\ell-2$ resulting co-circularities are $(8\ell)$-shallow (in $P$). Otherwise, the collinearity of $q,p,s$ is $(8\ell)$-shallow.

We thus repeat the above argument for each of the (at least) $\ell/2$ possible choices of $s$ and charge $\SQ$ within $\A_{pq}$ (via $(qp,w,\H)$) either to $\Omega(\ell^2)$ $(8\ell)$-shallow co-circularities, or to an $(8\ell)$-shallow collinearity. 
As in case (b1), each $(8\ell)$-shallow collinearity or co-circularity occurs during $(\xi_{-1},\t_q)$, and involves $p$ and $q$, so it is charged by at most $O(1)$ special quadruples $\SQ$ (because $\SQ$ is uniquely determined by $(p,q,w)$ and $w$ is among the first four points to hit $pq$ after the respective time $t^*$ of the event, because of condition (S5)). 

\medskip
\noindent {\bf Case (c2).} We may assume, then, that at least half of the above points $s$ leave $W_{pqw}$ through the ray $\vec{qw}$ (outside the segment $qw$).
For each of these points $s$, a symmetric variant of the argument in case (c1) implies that the points $q,w,s$ are involved during $(t_s^*,\t_q)$ either in an $(8\ell)$-shallow collinearity, or in $\Omega(\ell)$ $(8\ell)$-shallow co-circularities. As before, we repeat the above argument for the (at least) $\ell/2$ eligible choices of $s$ and charge $\SQ$, within $\A_{wq}$, either to $\Omega(\ell^2)$ $(8\ell)$-shallow co-circularities or to an $(8\ell)$-shallow collinearity. 

We claim that each of the resulting $(8\ell)$-shallow events, which occur in $\A_{wq}$ during $(\xi_{-1},\t_q)$, can be traced back to $\SQ$ in at most $O(1)$ possible ways.
Indeed, fix any of the above events, which occurs in $\A_{wq}$ at some time $t^*\in (\xi_{-1},\t_q)$.
We first guess $w$ and $q$ in $O(1)$ possible ways among the three or four points involved in the event. 
To guess the point $a$ (which would then uniquely determine $(wa,q,\J_u)$ and thereby also $\SQ$), we consider all special $(w,q)$-crossings $(wa',q,\J_{u'})$ (in $\F$) and recall that, according to conditions (S2) and (S6), at most $O(1)$ such crossings can begin during $[\t_{wq},\t_{2})$ or end during $(\t_3,\xi_{wq}]$. Notice also that the interval $[\t_{wq},\xi_{wq}]$, which covers $(\xi_{-1},\t_q)$, is the union of $[\t_{wq},\t_{2})$, $\J_u=[\t_2,\t_3]$, and $(\t_3,\xi_{wq}]$. 

To guess $a$ (based on $t^*,q$ and $w$), we distinguish between two possible situations.

\begin{figure}[htbp]
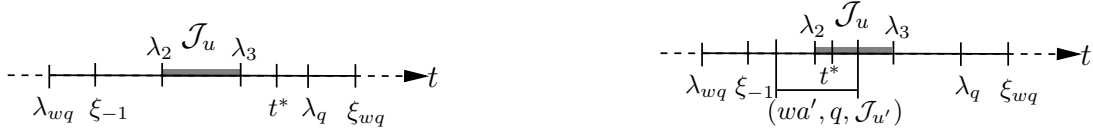

\begin{center}
\input{CaseCRight1.pstex_t}\hspace{3cm}\input{CaseCRight2.pstex_t}
\caption{\small Case (c2): Guessing $a$ based on $t^*$, $w$ and $q$. Left: If $t^*\in (\t_3,\t_q)$, then $(wa,q,\J_u=[\t_2,\t_3])$ is among the last three special clockwise $(w,q)$-crossings to 
end before $t^*$. Right: If $t^*\in (\xi_{-1},\t_3]$, then $(wa,q,\J_u)$ is among the first $O(1)$ special clockwise $(w,q)$-crossings to end after $t^*$. Any other such $(w,q)$-crossing $(wa',q,\J_{u'})$ (with $u'\neq p$), that ends in $(t^*,\t_3)\subset (\xi_{-1},\t_3)$, must begin after $\xi_{-1}$ (and, therefore, in $[\t_{wq},\t_2)$).}
\label{Fig:CaseCRight}
\vspace{-0.4cm}
\end{center}
\end{figure}

\noindent(i) If $t^*$ belongs to $(\t_3,\t_q)\subseteq (\t_3,\xi_{wq}]$ then $(wa,q,\J_u=[\t_2,\t_3])$ is among the last three special clockwise $(w,q)$-crossings to 
end before $t^*$, because $\SQ$ satisfies condition (S6). See Figure \ref{Fig:CaseCRight} (left).

\noindent(ii) If $t^*$ belongs to the interval $(\xi_{-1},\t_3]$, which is contained in $[\t_{wq},\t_2)\cup \J_u$,  
then we resort to a more subtle argument, in which we show that $(wa,q,\J_u=[\t_2,\t_3])$ is among the first $O(1)$ special clockwise $(w,q)$-crossings to 
end after $t^*$. See Figure \ref{Fig:CaseCRight} (right).

Our goal is to bound the number of special clockwise $(w,q)$-crossings that end in $(t^*,\t_3)$.
Note that the preliminary pruning (peformed before the definition of special quadruples) guarantees that each of these crossings $(wa',q,\J_{u'})$ satisfies $a'\neq u$ and $u'\neq a$, and therefore begins before $\J_u=[\t_2,\t_3]$ (by Lemma \ref{Lemma:OrderRelaxedCrossings}). 
Furthermore, note that we have $u'=p$ for at most one of these crossings $(wa',q,\J_{u'})$, because each of them is uniquely determined by the respective triple $w,q,u'$.  
We claim that each of the remaining $(w,q)$-crossings $(wa',q,\J_{u'})$ under consideration (satisfying also $u'\neq p$) must begin in $(\xi_{-1},\t_2)\subseteq [\t_{wq},\t_2)$. This, together with condition (S2), implies that their number is $O(1)$ too.

To see this final claim, note that if a $(w,q)$-crossing $(wa',q,\J_{u'})$, as above, begins before $\xi_{-1}$, then its respective interval $\J_{u'}$ contains the time $\xi_{-1}$ (because it ends after $t^*>\xi_{-1}$), right after which the Delaunayhood of $wq$ is violated by $p$ and $a$. This, however, is impossible because, by Lemma \ref{Lemma:Crossing}, $wq$ is Delaunay throughout $\J_{u'}$ in $P\setminus \{u'\}$, and $u'\neq p,a$.

\medskip
To recap, in each of the cases (c1) and (c2) we charge $\SQ$ either to $\Omega(\ell^2)$ $(8\ell)$-shallow co-circularities, or to an $(8\ell)$-shallow collinearity, which occur in one of the arrangements $\A_{pq}, A_{wq}$ during the interval $(\xi_{-1},\t_q)$. Furthermore, each $(8\ell)$-shallow event is charged by at most $O(1)$ special quadruples. Hence, at most 
$O\left(\ell^2N(n/\ell)+\ell n^2\beta(n)\right)$ 
special quadruples $\SQ$ fall into case (c).


\bigskip
\noindent{\bf Case (d).} Assume that none of the preceding cases occurs. 
In particular, there is a subset $A_{pw}$ of at most $6\ell+3$ points (including $a$, $r$ and $u$) whose removal restores the Delaunayhood of $pw$ throughout the interval $[\t_{wq},\xi_{wq}]$. Furthermore, a total of fewer than $\ell$ points of $P\setminus \{a,r,u\}$ ever appear in the cap $C^-_{qw}$ during $(\xi_{-1},\t_q)$, and a total of fewer than $\ell$ points of $P\setminus A_{pw}$ points ever appear in the cap $C^+_{pw}$ during that interval.\footnote{Note the built-in asymmetry between $qw$ and $pw$ in the analysis: The former is almost Delaunay in the interval $[\t_{wq},\xi_{wq}]$ (and Delaunay at both endpoints $\t_{wq},\xi_{wq}]$), whereas the latter becomes Delaunay there only after the removal of $A_{pw}$ (which includes $a,p,r,u$).}

In this last remaining scenario, we finally consider the interplay between the special quadruple $\SQ$ under consideration and the ordinary Delaunay quadruple $\Q=(p,q,a,r)$ in $\F$, which corresponds to the {\it first} special $(a,q)$-crossing $(pa,q,\I_r)$ of $\SQ$. At the end of this section, we shall charge $\SQ$ to the terminal quadruple $\varrho=(p,q,r,w)$, which is composed of the edge $pq$, and of the two points $r$ and $w$ that cross $pq$ in opposite directions. (The outer point $u$ of the second special $(a,q)$-crossing $(wa,q,\J_u)$ is not used for right quadruples; it will be used in the mostly symmetric analysis of left quadruples, given in Section \ref{Subsec:Stage4Left}.)

Before charging $\SQ$ to the above terminal quadruple $\varrho$, we enforce a Delaunay crossing of one of the edges $pr,qr$ by the point $w$. In addition, we shall have to enforce two more crossings performed by the points of $\varrho$ in order to ensure that at least two of the resulting {\it five} crossings are performed by the same sub-triple of $\varrho$ (so as to allow us to apply our cornerstone Lemma \ref{Lemma:TwiceCollin} and thereby obtain a quadratic bound on the number of such quadruples).

To facilitate the forthcoming analysis, we first establish several auxiliary claims.

\begin{figure}[htbp]
\begin{center}
\input{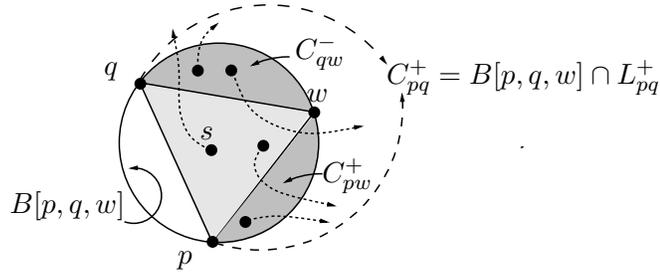}
\caption{\small Lemma \ref{Prop:ShallowCup}: A total of at most $8\ell+1$ points $s$ of $P$ appear in the cap $C_{pq}^+=B[p,q,w]\cap \L_{pq}^+$ (consisting of all the shaded portions) during $(\xi_{-1},\t_q)$. 
All of them must leave $C_{pq}^+$ before $\t_q$. None of these points $s$ can leave 
$C_{pq}^+$ through $pq$, unless it is one of $a,r,u$.}
\label{Fig:PqCapShallow}
\vspace{-0.4cm}
\end{center}
\end{figure}

\begin{lemma}\label{Prop:ShallowCup}
With the above assumptions, a total of at most $8\ell+1$ points of $P$ appear in the cap $C_{pq}^+=B[p,q,w]\cap \L_{pq}^+$ during $(\xi_{-1},\t_q)$.
\end{lemma}
\begin{proof}
Refer to Figure \ref{Fig:PqCapShallow}.
Recall that the motion of $B[p,q,w]$ is continuous throughout $(\xi_{-1},\t_q)$. Notice that the above cap $C_{pq}^+=B[p,q,w]\cap \L_{pq}^+$ (which contains $w$ on its boundary) is empty right before time $\t_q$, when the edge $pq$ is crossed by $w$. 
Hence, any point $s$ that appears in this cap during $(\xi_{-1},\t_q)$ has to leave it before $\t_q$. Furthermore, condition (S5) (together with the inclusion $(\xi_{-1},\t_q)\subseteq [\xi_{pq},\t_{pq}]$) implies that $s$ cannot escape $C_{pq}^+$ through the edge $pq$, unless it is equal to one of $a,r,u$.
Therefore, any such point $s\neq a,r,u$ has to leave $C_{pq}^+$ through one of the circular arcs bounding the earlier caps $C_{qw}^-,C_{pw}^+$, so it must first appear in one of the caps $C_{qw}^-$ or $C_{pw}^+$.
Since cases (b) and (c) have been ruled out, and since $a,r,u$ belong to the set $A_{pw}$, the overall number of such points cannot exceed $(\ell-1)+(\ell-1)+(6\ell+3)=8\ell+1$.
\end{proof}

\begin{figure}[htbp]
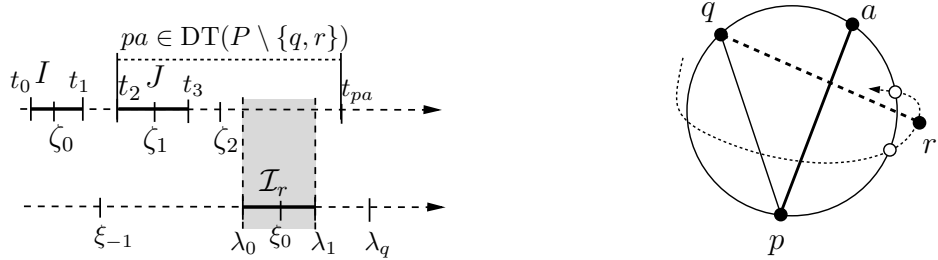

\begin{center}
\input{TwoQuadsInterplay.pstex_t}\hspace{3cm}\input{TwoCocircsRegular.pstex_t}
\caption{\small The (regular) clockwise quadruple $\Q=(p,q,a,r)$ of $(pa,q,\I_r)$ is composed of two $(p,r)$-crossings $(pq,r,I=[t_0,t_1])$, $(pa,r,J=[t_2,t_3])$. The points $p,q,a,r$ are co-circular at times $\zeta_0\in I\setminus J,\zeta_1\in J\setminus I$, and $\zeta_2>t_3$ (left). The last two co-circularities have the same order type, and the Delaunayhood of $rq$ is violated by $p$ and $a$ throughout $(\zeta_1,\zeta_2)$ (right).}
\label{Fig:TwoQuadsInterplay}
\vspace{-0.6cm}
\end{center}
\end{figure}

We next consider the ordinary quadruple $\Q=(p,q,a,r)$ in $\F$, which corresponds to the first special crossing $(pa,q,\I_r)$ of $\SQ$. Refer to Figure \ref{Fig:TwoQuadsInterplay}.
We continue to denote the two Delaunay crossings of $\Q$ by $(pq,r,I=[t_0,t_1])$ and $(pa,r,J=[t_2,t_3])$. 
Recall that the points of $\Q$ are co-circular at times $\zeta_0\in I\setminus J,\zeta_1\in J\setminus I$ and $\zeta_2>t_3$.
By condition (Q3) on $\Q$, the last two co-circularities of $\Q$ (at times $\zeta_1$ and $\zeta_2$) have the same order type, and the Delaunayhood of $rq$ is violated by $p\in \L_{rq}^-$ and $a\in \L_{rq}^+$ throughout the interval $(\zeta_1,\zeta_2)$ (see Figure \ref{Fig:TwoQuadsInterplay} (left)). Therefore, the Delaunayhood of $pa$ is violated right after time $\zeta_2$ by $r$ and $q$.

\smallskip
\noindent{\it Remark:} Note that $\Q$ and $\SQ$ have ``opposite" topological behaviour, in the sense that the additional co-circularity of $\Q$ (outside $I$ and $J$) occurs at time $\zeta_2$, after the respective second interval $J$ of $\Q$, whereas the corresponding additional co-circularity of $\SQ$ (outside $\I_r$ and $\J_u$) occurs at time $\xi_{-1}$, before the respective first interval $\I$ of $\SQ$.

By condition (Q7), the edge $pa$ re-enters $\DT(P)$ at some time $t_{pa}\geq \zeta_2>t_2$. Furthermore, $pa$ belongs to $\DT(P\setminus \{r,q\})$ throughout the interval $[t_2,t_{pa}]=J\cup[t_3,t_{pa}]$, which covers $J$ (including $\zeta_1\in J\setminus I$) and $\zeta_2$.
Moreover, we recall that (using the Delaunayhood of $pa$ at time $t_3$, and the extremality of $\zeta_2$, via Lemma \ref{Lemma:MustCross}), $q$ crosses $pa$ from $\L_{pa}^-$ to $\L_{pa}^+$ during $(t_3,t_{pa}]$. As argued in Section \ref{Subsec:Stage4Reg}, this yields the Delaunay crossing $(pa,q,\I_r=[\t_0,\t_1])$ in $P\setminus \{r\}$ as the unique special crossing of $\Q$, with $\I_r\subset (t_3,t_{pa}]$.

To conclude, the second crossing $(pa,r,J=[t_2,t_3])$ of $\Q$ and the first crossing $(pa,q,\I_r=[\t_0,\t_1])$ of $\SQ$ occur during disjoint intervals and in this order.\footnote{Note that, even though $q$ hits $pa$ after $\zeta_2$, during the above special crossing, it is not known whether the last co-circularity $\zeta_2$ of $\Q$ occurs in $\I_r$ or beforehand, in $(t_3,\t_0]$.}

Finally, by condition (Q8), the edge $pq$ belongs to $\DT(P\setminus \{r,a\})$ throughout the interval $[t_0,\t_1]=\conn{I}{\I_r}(={\sf conv}(I\cup \I_r))$. Therefore, the almost-Delaunayhood of $pq$ extends from $[\xi_{pq},\t_{pq}]$ to the potentially larger interval $[t_0,\t_{pq}]$ (assuming $\xi_{pq}>t_0$, that is, $I=[t_0,t_1]$ is not contained in $[\xi_{pq},\t_{pq}]$).

The following claim is crucial for understanding the interplay between $\Q$ and $\SQ$.

\begin{lemma}\label{Claim:Interplay}
With the above assumptions, we have $\zeta_1\in (\xi_{-1},\xi_0)$. 
\end{lemma}

\begin{figure}[htbp]
\begin{center}
\input{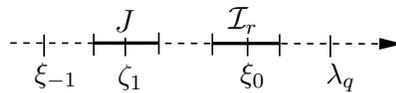}
\caption{\small Lemma \ref{Claim:Interplay} claims that $\zeta_1\in (\xi_{-1},\xi_0)$.}
\label{Fig:ClaimInterplay}
\vspace{-0.6cm}
\end{center}
\end{figure}

\begin{proof} 
The inequality $\zeta_1<\xi_0$ follows because $\zeta_1$ occurs during the second crossing $(pa,r,J)$ of $\Q$, whereas $\xi_0$ occurs during the first special crossing $(pa,q,\I_r)$ of $\SQ$ (which begins after $J$). 
See Figure \ref{Fig:ClaimInterplay} (left) and Figure \ref{Fig:ClaimInterplay}.

To establish the inequality $\zeta_1>\xi_{-1}$, let us assume for a contradiction that $\zeta_1\leq \xi_{-1}$; see Figure \ref{Fig:InterplayContradict}.
Since $\Q=(p,q,a,r)$ belongs to the refined family $\F$ (and, therefore, satisfies condition (Q3)), its point $q$ remains in $B[p,a,r]\cap\L_{pa}^-$ after $r$ enters $\L_{pa}^+$ during $J=[t_2,t_3]$ and until time $\zeta_1\in J$ (when $q$ leaves the cap $B[p,a,r]\cap \L_{pa}^-$). Also note that, with the above assumption that $\zeta_1<\xi_{-1}$, the point $q$ cannot leave $\L_{pa}^-$ during $(\zeta_1,\xi_{-1})$. Indeed, 
$q$ lies in $\L_{pa}^-$ at both endpoints of that interval, because the quadruples $\Q$ and $\SQ$ satisfy the respective conditions (Q3) and (S3a),
and it can enter the halfplane $\L_{pa}^+$ only once (which occurs during $\I_r$ and after $\xi_{-1}$).

\begin{figure}[htbp]
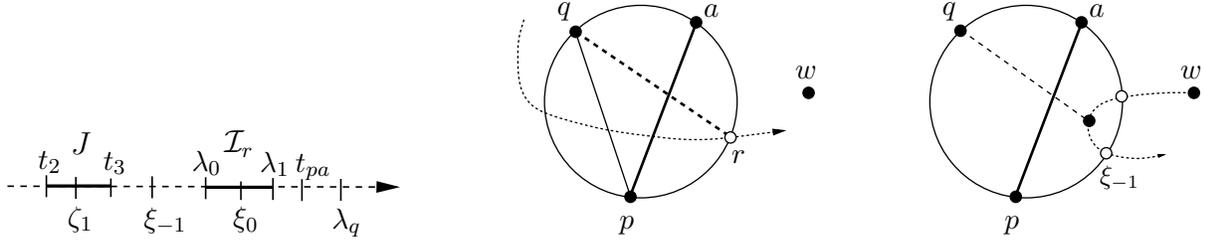

\begin{center}
\input{TimelineInterplayContradict.pstex_t}\hspace{1.5cm}\input{Zeta1Interplay.pstex_t}\hspace{1.5cm}\input{XiMinusInterplay.pstex_t}
\caption{\small Proof of Lemma \ref{Claim:Interplay}. If $\zeta_1<\xi_{-1}$ (left) then $w$ has to enter $B[p,q,a]\cap \L_{pa}^+$, which is empty at time $\zeta_1$ (center), before leaving it at time $\xi_{-1}$ (right). By Condition (Q7), $w$ can enter $B[p,q,a]\cap \L_{pa}^+$ during $(\zeta_1,\xi_{-1})\subset (t_2,t_{pa})$ only through the boundary of $B[p,q,a]$.}
\label{Fig:InterplayContradict}
\vspace{-0.6cm}
\end{center}
\end{figure}

The above reasoning implies that the motion of $B[p,q,a]$ is continuous throughout $(\zeta_1,\xi_{-1})$. Furthermore, $w$ lies outside the cap $B[p,q,a]\cap \L_{pa}^+$ at time $\zeta_1$, for otherwise the Delaunayhood of $pa$ would be violated by $q$ and $w$ (which cannot happen during the interval $J$, where $pa$ belongs to $\DT(P\setminus \{r\})$); see Figure \ref{Fig:InterplayContradict} (center).
By condition (S3a), $w$ leaves the cap $B[p,q,a]\cap \L_{pa}^+$ at time $\xi_{-1}$.
Therefore, $w$ must have previously entered that cap, in the interval $(\zeta_1,\xi_{-1})$. Note that, since $\xi_{-1}<\t_0$, the latter interval is contained in $(\zeta_1,t_{pa})$, where $t_{pa}$ denotes the first time after $\zeta_1$ and $\zeta_2$ when $pa$ again belongs to $\DT(P)$.

Since $\Q$ satisfies condition (Q7), $w$ cannot enter $B[p,q,a]\cap \L_{pa}^+$ during $(\zeta_1,\xi_{-1})\subseteq (t_2,\t_1)\subseteq (t_2,t_{pa})$ through the edge $pa$. Furthermore, $w$ cannot enter $B[p,q,a]\cap \L_{pa}^+$ during that interval through the boundary of $B[p,q,a]$, as that would cause a forbidden fourth co-circularity of $p,q,a,w$; see Figure \ref{Fig:InterplayContradict} (right). Hence, we have reached 
a contradiction, and the claim follows.
\end{proof}

By Lemma \ref{Lemma:Crossing}, none of the co-circularities $\xi_{-1},\xi_{0}$ can occur during $J$, so we have $J\subset (\xi_{-1},\xi_0)$.
This, combined with the properties (S1)--(S3a) of $\SQ$, implies that $(\xi_{pq}<)\xi_{-1}<t_2<\zeta_1<t_3<\t_0<\xi_0<\t_1<\t_q(<\t_{pq})$. See Figure \ref{Fig:InterplayPqSetup} (left).

Since $\Q$ satisfies condition (Q3), $r$ cannot return to $\L_{pq}^-$ (after leaving it during $I$) before time $\zeta_1$ (when $r$ leaves the cap $B[p,q,a]\cap \L_{pq}^+$), for otherwise the triple $p,q,r$ would be collinear at least three times.

Furthermore, if $r$ re-enters $\L_{pq}^-$ through $pq$ during the subsequent interval $(\zeta_1,\t_{pq}]$, then the edge $pq$ undergoes two Delaunay crossings by $r$ within the triangulation $\DT(P\setminus \{a,u,w\})$. 
Indeed, Lemma \ref{Claim:Interplay} implies that $(\zeta_1,\t_{pq})$ is contained in $[\xi_{-1},\t_{pq}]\subset [\xi_{pq},\t_{pq}]$, and the edge $pq$ belongs to $\DT(P\setminus \{a,r,u,w\})$ throughout the latter interval by condition (S5) (in addition to its being Delaunay at the endpoints $\xi_{pq}$ and $\t_{pq}$).
By Lemma \ref{Lemma:TwiceCollin} and Proposition \ref{Prop:TwoTriples}, this happens for at most $O(n^2)$ special quadruples $\SQ$.

\begin{figure}[htbp]
\begin{center}
\input{TimelinePqInterplay.pstex_t}\hspace{2cm}\input{TimelinePqInterplayNotReturn.pstex_t}
\caption{\small Left: The setup implied by Lemma \ref{Claim:Interplay}. We have $\xi_{pq}<\xi_{-1}<\zeta_1<\xi_0<\t_q<\t_{pq}$, and $J\subset (\xi_{-1},\xi_0)$. The point $r$ remains in $\L_{pq}^+$ throughout $(t_1,\zeta_1)$. Right: If $r$ were to hit $pq$ also in $(\zeta_1,\t_{pq})$, then $pq$ would undergo two Delaunay crossings by $r$ within $\DT(P\setminus \{a,w,u\})$. Hence, we can assume that no such collinearity occurs.}
\label{Fig:InterplayPqSetup}
\vspace{-0.6cm}
\end{center}
\end{figure}

To conclude, ignoring the favourable quadruples just considered, we may assume that the above scenario does not occur, so $r$ does not cross $pq$ in the interval $(t_1,\t_{pq}]$. (However, $r$ can still return to $\L_{pq}^-$ during $(t_1,\t_{pq}]$, or, more precisely, during $(\zeta_2,\t_{pq}]$, by crossing one of the outer rays of $\L_{pq}$, outside $pq$.) See Figure \ref{Fig:InterplayPqSetup} (right).

\medskip
\noindent{\bf The three co-circularities of $p,q,r,w$.} 
We now argue that the four points $p,q,r,w$ are involved in exactly three co-circularities, and characterize the order types of these co-circularities.
First, recall that one such co-circularity occurs at some time $\delta_0\in I$, according to Lemma \ref{Lemma:OnceCollin}. Since this co-circularity is induced by the crossing of $pq$ by $r$, it is red-blue with respect to $pq$ and to $rw$. Moreover, as will follow from the subsequent analysis, this is the first co-circularity of this quadruple; see Figure \ref{Fig:Delta0}.

\begin{figure}[htbp]
\begin{center}
\input{DeltaZero.pstex_t}
\caption{\small The co-circularity of $p,q,r,w$ occurring at some time $\delta_0\in I$. It is red-blue with respect to the edges $pq$ and $rw$.}
\label{Fig:Delta0}
\vspace{-0.6cm}
\end{center}
\end{figure}

\smallskip
To obtain the second co-circularity of $p,q,r,w$, 
we recall that (as reviewed at the beginning of this section, and depicted in Figure \ref{Fig:TopSetupRight} (bottom)) the Delaunayhood of $wq$ is violated by $p\in \L_{wq}^-$ and $a\in \L_{wq}^+$ throughout the interval $(\xi_{-1},\xi_0)$, and the order type of $p,q,w$ remains fixed (i.e, $w$ lies in $\L_{pq}^+$) throughout the larger interval $(\xi_{-1},\t_q)$. 

\begin{figure}[htbp]
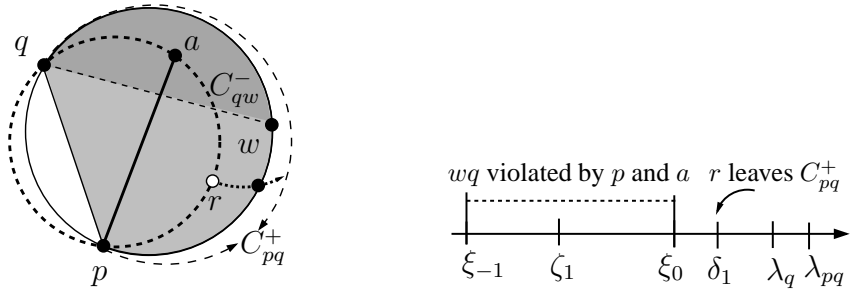

\begin{center}
\input{ZetaDeltaInterplay.pstex_t}\hspace{2cm}\input{TimeLineDelta1.pstex_t}
\caption{\small Obtaining the second co-circularity $\delta_1$ of $p,q,r,w$. The co-circularity of $p,q,a,r$ at time $\zeta_1$ is red-red with respect to $pq$, and belongs to the interval $(\xi_{-1},\xi_0)$, during which $a$ lies in $C_{qw}^-(\subset C_{pq}^+)$. 
Hence, $r$ lies at that time within the cap $C_{pq}^+=B[p,q,w]\cap \L_{pq}^+$, so $\delta_1$ necessarily occurs in $(\zeta_1,\t_q)$, when $r$ escapes the above cap $C_{pq}^+$ (without crossing $pq$). This is also a red-red co-circularity with respect to $pq$.}
\label{Fig:Delta1}
\vspace{-0.6cm}
\end{center}
\end{figure}

By Lemma \ref{Claim:Interplay}, the interval $(\xi_{-1},\xi_0)$ contains $\zeta_1$, so $a$ lies at that time in the cap $C_{qw}^-\subset C_{pq}^+$ (after it enters $C_{qw}^-$ at time $\xi_{-1}$, and before escaping it at time $\xi_0$).
Since the points $p,q,a,r$ are involved at time $\zeta_1$ in a red-red co-circularity with respect to $pq$
(as prescribed by condition (Q3) on $\Q$), both $a$ and $r$ lie at time $\zeta_1$ within the cap $C_{pq}^+=B[p,q,w]\cap \L_{pq}^+$; see Figure \ref{Fig:Delta1} (left).

Since $w$ remains in $\L_{pq}^+$ throughout the longer interval $(\xi_{-1},\t_q)$, 
the four points $p,q,r,w$ are involved, during $(\zeta_1,\t_q)$, in a co-circularity, which occurs when $r$ leaves the above cap $B[p,q,w]\cap \L_{pq}^+$. (Otherwise $r$ would have to escape $B[p,q,w]\cap \L_{pq}^+$ through the interior of $pq$ before this cap shrinks to $pq$ at time $\t_q$, which cannot happen during $(t_1,\t_q]\subseteq (t_1,\t_{pq}]$ by condition (Q8).) Clearly, this co-circularity is red-red with respect to the edge $pq$, and occurs after $I$ and between $\zeta_1\in J\setminus I$ and $\t_q$. 
We denote by $\delta_1$ the time of the {\it first} such co-circularity event in $(\zeta_1,\t_q)$, at which $r$ leaves $B[p,q,w]\cap \L_{pq}^+$. (As will soon turn out, this is the second co-circularity of $p,q,r,w$.)

\smallskip
\noindent{\it Remark.} We again emphasize that $\delta_1\in (\zeta_1,\t_q)\subset (\xi_{-1},\t_q)$, and that
$r$ remains in the cap $B[p,q,w]\cap \L_{pq}^+$ throughout the interval $[\zeta_1,\delta_1)$. (However, the order between $\delta_1$ and $\xi_0$ is not known, and is immaterial for our analysis.)


\smallskip
We next claim that the points $p,q,r,w$ are involved in a third co-circularity, red-blue with respect to $pq$, at some time $\delta_2\in (\delta_1,\t_{pq}]$. Notice that the desired co-circularity cannot be obtained by simply applying Lemma \ref{Lemma:OnceCollin} to the crossing $(qp,w,\H)$, because it is defined only with respect to the reduced point set 
$P\setminus \{a,r,u\}$. 

Instead, we consider the four-point triangulation $\DT(\{p,q,r,w\})$, and observe that the edge $qp$ undergoes there a Delaunay crossing by $w$, which takes place during some sub-interval of $(\delta_1,\t_{pq}]$ that contains $\t_q$ (the time of the actual collinearity of the three points).
Indeed, $pq$ is Delaunay in $\{p,q,r,w\}$ at times $\delta_1<\t_q$ and $\t_{pq}\geq \t_q>\delta_1$, and it is Delaunay in $\{p,q,r\}$ throughout $(\delta_1,\t_{pq}]$ (because $r$ is assumed not to cross $pq$ in the even larger interval $(t_1,\t_{pq}]$).

\begin{figure}[htbp]
\begin{center}
\input{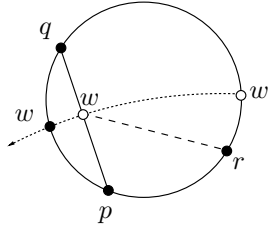}
\caption{\small The third co-circularity of $p,q,r,w$ occurs at some time $\delta_2\in (\delta_1,\t_{pq}]$,
and is red-blue with respect to the edges $pq$ and $rw$. 
This co-circularity is part of a Delaunay crossing of $qp$ by $w$, which occurs within the four-point triangulation $\DT(\{p,q,r,w\})$, during some subinterval of $(\delta_1,\t_{pq}]$ that contains $\t_q$.}
\label{Fig:Delta2}
\vspace{-0.6cm}
\end{center}
\end{figure}

Furthermore, the above crossing in $\DT(\{p,q,r,w\})$ must be single. Indeed, since $w$ lies in $C_{wq}^-\subset\L_{pq}^+$ throughout the interval $(\xi_{-1},\xi_0)$ which contains $\zeta_1$, it has to remain in $\L_{pq}^+$ throughout $[\zeta_1,\t_q]\supset [\delta_1,\t_q]$ (or, else, $w$ would cross $\L_{pq}$ three times).
Furthermore, $w$ does not cross $pq$ again in $(\t_q,\t_{pq}]$ (by condition (S3b)).
We hence apply Lemma \ref{Lemma:OnceCollin} to this single crossing, which gives us the desired third 
co-circularity (see Figure \ref{Fig:Delta2}).

\medskip
To conclude, the four points $p,q,r,w$ are involved in three co-circularities, which occur at times $\delta_0\in I=[t_0,t_1]$, $\delta_1\in (\zeta_1,\t_{q})(\subset (\xi_{-1},\t_q))$, and $\delta_2\in (\delta_1,\t_{pq})$. The two extremal co-circularities (which occur at times $\delta_0$ and $\delta_2$) are red-blue with respect to the edges $pq$ and $wr$, and thus monochromatic with respect to $pr$, $qr$, $pw$, $qw$. The middle co-circularity (at time $\delta_1$) is red-red with respect to $pq$.\footnote{This alternation in the order type is crucial of the forthcoming analysis.}

\medskip
We are now ready to establish the following important consequence of Lemma \ref{Prop:ShallowCup}.

\begin{lemma}\label{Prop:FewPr}
With the above assumptions, at most $8\ell+1$ clockwise (Delaunay) $(p,r)$-crossings $(pq',r,I')$ in $\F$, and at most $8\ell+1$ counterclockwise (Delaunay) $(q,r)$-crossings $(p'q,r,I')$ in $\F$, can end in the interval $(t_1,\delta_{1})$.
\end{lemma}

\begin{figure}[htbp]
\begin{center}
\input{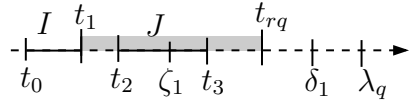}
\caption{\small Preparing for the proof of Lemma \ref{Prop:FewPr}. By conditions (Q2) and (Q4), no clockwise $(p,r)$-crossings, and no counterclockwise $(q,r)$-crossings in $\F$ end in the shaded interval between $t_1$ and $t_{rq}>t_3(>\zeta_1)$, where $t_{rq}$ is the first such time after $\zeta_1$ when $rq$ belongs to $\DT(P)$. Unfortunately, the order of $\delta_1$ and $t_{rq}$ is not known.}
\label{Fig:TimelineFewPr}
\vspace{-0.6cm}
\end{center}
\end{figure}

Recall that an (ordinary) Delaunay crossing is in $\F$ if it is either the first or the second crossings of some Delaunay quadruple in $\F$.
In Section \ref{Sec:CountQuad} we have already enforced comparable restrictions (via conditions (Q2) and (Q4)), which
imply that no clockwise $(p,r)$-crossings $(pq',r,I')$ in $\F$, and no counterclockwise $(q,r)$-crossings $(p'q,r,I')$ in $\F$, end after time $t_1$ and before time $t_{rq}>t_3(>\zeta_1)$, which is first such time after $\zeta_1$ when the edge $rq$ belongs to $\DT(P)$. See Figure \ref{Fig:TimelineFewPr}. (In addition, conditions (Q2) and (Q3) imply that $rq$ belongs to $\DT(P\setminus \{a,p\})$ throughout $(t_1,t_{rq}]$, and neither $a$ nor $p$ can hit $rq$ in that interval.)
Unfortunately, the order of $t_{rq}$ and $\delta_1$ is not known, so condition (Q4) does not immediately imply the above property.

\begin{proof}[Proof of Lemma \ref{Prop:FewPr}.]
We first consider clockwise $(p,r)$-crossings.
Let $(pq',r,I')$ be such a Delaunay crossing that ends in $(t_1,\delta_{1})$. Note that
the point $q'$ has to be distinct from $a$ (for, otherwise, $(pq',r,I')$ would co-incide with $(pa,r,J)$), and that the points $p,q,q',r$ form an (ordinary, not necessarily consecutive quadruple) clockwise quadruple.
Recall also that $r$ remains in $\L_{pq}^+$ after entering that halfplane during $I=[t_0,t_1]$ and until time $\delta_1$ (when $r$ escapes $C_{pq}^+=B[p,q,w]\cap \L_{pq}^+$).
In particular, $q$ lies in $\L_{pr}^-=\L_{pq'}^-$ when $r$ enters $\L_{pq'}^+$ (during $I'$). Hence, the points $p,q,r,q'$ are involved in a co-circularity at some time $\zeta'\in I'\setminus I$, right after which the Delaunayhood of $rq$ is violated by $p$ and $q'$. See Figure \ref{Fig:FewPrQr} (left).

\begin{figure}[htbp]
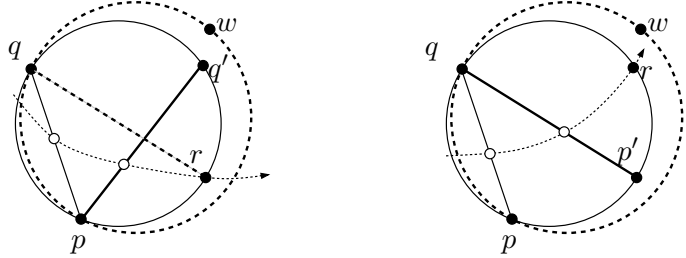

\begin{center}
\input{FewPrCrossings.pstex_t}\hspace{2cm}\input{FewQrCrossings.pstex_t}
\caption{\small Proof of Lemma \ref{Prop:FewPr}. Left: $(pq',r,I')$ is a clockwise $(p,r)$-crossing that ends 
$(t_1,\delta_1)$. The points $p,q,r,p'$ are co-circular at some time $\zeta'\in I'\setminus I$. If $\zeta'$ occurs in $(\zeta_1,\delta_1)$, then $p'$ lies in $C_{pq}^+=B[p,q,w]\cap \L_{pq}^+$ at that moment. Right: $(p'q,r,I')$ is a counterclockwise $(q,r)$-crossing that occurs within $(\zeta_1,\delta_1]$. The points $p,p',q,r$ are co-circular at some time $\zeta'\in I'\setminus I$, when both $r$ and $p'$ lie inside $C_{pq}^+$.}
\label{Fig:FewPrQr}
\vspace{-0.6cm}
\end{center}
\end{figure}

We first argue that $\zeta'$ cannot occur before $\zeta_1$. Indeed, otherwise, applying Lemma \ref{Lemma:MustCross} for the edge $rq$, from time $\zeta'$,
would imply that at least one of the following events must occur between $\zeta'$ and $t_{rq}$ (which is the first time after $\zeta_1$ when $rq$ belongs $\DT(P)$): 
(1) $q'$ hits $rq$, 
(2) $p$ hits $rq$, or
(3) the four points $p,q,q',r$ are involved in an additional co-circularity of the same order type.

However, cases (1), (2) are impossible by conditions (Q2) and (Q3) on $\Q$ (using that $\zeta_1<t_{rq}$). Moreover, the co-circularity in (3) can occur only after the end of both $I$ and $I'$ (because $p,q,q'$ and $r$ form a regular clockwise quadruple; see Section \ref{Subsec:SingleProp}), in which case $(pq',r,I')$ has to end before $t_{rq}$, contrary to condition (Q4) on $\Q$. Hence, $\zeta'$ must occur after $\zeta_1$.

We may thus assume that $\zeta'$ belongs to the interval $(\zeta_1,\delta_{1})$ which, by Lemma \ref{Claim:Interplay}, is contained in $(\xi_{-1},\t_q)$, so both $q'$ and $r'$ lie at time $\zeta'$ within the cap $C_{pq}^+=B[p,q,w]\cap \L_{pq}^+$. According to Lemma \ref{Prop:ShallowCup}, the overall number of such points $q'$ is at most $8\ell+1$.

The treatment of counterclockwise $(q,r)$-crossings (also in $\F$) is similar (but somewhat simpler). 
Indeed, let $(p'q,r,I')$ be such a crossing. Condition (Q2) implies that it cannot end in the interval $(t_1,\zeta_1]$ (because $\zeta_1$ belongs to $J\setminus I\subset (t_1,\t_{rq})$). Furthermore, Lemma \ref{Lemma:Crossing} implies that any counterclockwise $(q,r)$-crossing $(p'q,r,I')$ that ends after $\zeta_1$ has to begin also after $\zeta_1$. (Otherwise, its respective interval $I'$ would contain the time $\zeta_1$ of a red-blue co-circularity with respect to $rq$, contrary to the Delaunayhood of $rq$ during $I'$.)
We consider the co-circularity of $p,p',q,r$, which must occur at some time $\zeta'\in I'\setminus I$ and notice, as in the previous case, that both $r$ and $p'$ lie at that moment in the cap $C_{pq}^+$;  see Figure \ref{Fig:FewPrQr} (right). Therefore, the overall number of such points $p'$ does not exceed $8\ell+1$.
\end{proof}

\begin{figure}[htbp]
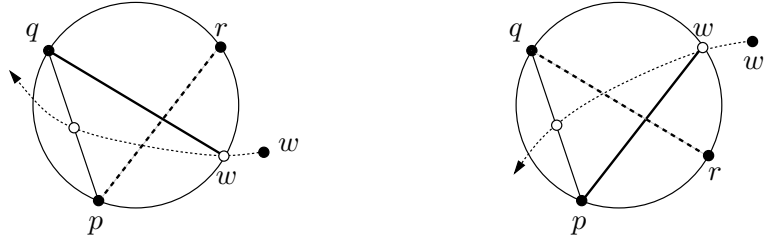

\begin{center}
\input{CaseD1.pstex_t}\hspace{3cm}\input{CaseD2.pstex_t}
\caption{\small Left: Case (d1). The co-circularity at time $\delta_1$ is red-blue with respect to the edges $pr$ and $wq$. Right afterwards, the Delaunayhood of $pr$ is violated by $q$ and $w$. Right: Case (d2). The co-circularity at time $\delta_1$ is red-blue with respect to the edges $rq$ and $pw$. Right afterwards, the Delaunayhood of $rq$ is violated by $p$ and $w$.}
\label{Fig:CasesD12}
\vspace{-0.6cm}
\end{center}
\end{figure}

\paragraph{Cases (d1) and (d2): Overview.}
To proceed, we distinguish between two possible subcases. In subcase (d1), we assume that the middle co-circularity, which occurs at time $\delta_1$, is red-blue with respect to the edges $pr$ and $wq$ (see Figure \ref{Fig:CasesD12} (left)), and then use it to enforce (via Lemma \ref{Lemma:MustCross}) the following two additional crossings: (i) a Delaunay crossing of $pr$ by at least one the points $w,q$, and (ii) a Delaunay crossing of $wq$ by at least one of the points $p,r$.
(For the second crossing, it will suffice to argue that $wq$ is hit by one of the points $p,r$ in the interval $[\t_{wq},\t_{pq}]\subseteq [\t_{wq},\xi_{wq}]$.) However, this can easily be established by applying Lemma \ref{Lemma:MustCross} to $wq$ {\it backwards} from the second co-circularity $\delta_1\in [\t_{wq},\t_{pq}]$ of $p,q,r,w$.)

Therefore, the points $p,q,r,w$ (or, more precisely, their sub-triples) will perform four distinct Delaunay crossings---the two new crossings just promised and the two ``old" ones, of $pq$ by $r$ and by $w$. If a pair of these crossings is performed by the {\it same} triple,
we will use Lemma \ref{Lemma:TwiceCollin} to bound the overall number of such special quadruples $\SQ$. Otherwise we will charge $\SQ$ to the (probabilistically refined) terminal quadruple $\varrho=(p,q,r,w)$, whose four possible sub-triples are involved in {\it four} Delaunay crossings, namely, the crossings of $pq$ by $r$ and $w$, the crossing of $pr$ by $w$, and the crossing of $wq$ by $r$.

In Section \ref{Sec:Terminal} we will use the third co-circularity $\delta_2$ to enforce, for each terminal quadruple $\varrho=(p,q,r,w)$ of the above kind, an additional, fifth crossing (namely, a crossing of $rw$ by $p$ or $q$). As a result, some sub-triple of $p,q,r,w$ will be involved in two Delaunay crossings, which will allow us to obtain a ``quadratic" recurrence for the number of such quadruple, via Lemma \ref{Lemma:TwiceCollin}.

In subcase (d2), we assume the co-circularity at time $\delta_1$ to be red-blue with respect to the edges $rq$ and $pw$ (see Figure \ref{Fig:CasesD12} (right)), and then use it to enforce a Delaunay crossing of $rq$ by at least one of $p$ and $w$. 
If $rq$ is crossed by $p$, we can dispose of $\SQ$ via Lemma \ref{Lemma:TwiceCollin}. Otherwise, we charge $\SQ$ to the (probabilistically refined) terminal quadruple $\varrho=(p,q,r,w)$ (whose points are known, so far, to perform only {\it three} crossings).

In Section \ref{Sec:Terminal} we will enforce, for each terminal quadruple $\varrho=(p,q,r,w)$ of the latter type, 
{\it two} additional crossings, namely, a crossing of $pw$ by one of $r,q$, and a crossing of $rw$ by one of $p,q$.
Hence, once again we will be able to use Lemma \ref{Lemma:TwiceCollin} to handle such terminal quadruples too.

\medskip
\noindent{\bf Case (d1).} The co-circularity at time $\delta_{1}$ is red-blue with respect to the edge $pr$ whose Delaunayhood is violated right afterwards by $q$ and $w$ (see Figure \ref{Fig:CasesD12} (left)). 

Note that the above violation of $pr$ does not hold either right before, or right after time $\t_q$.
More precisely, it does not hold for that side of $\t_q$ when $w$ and $r$ lie in the same side of $\L_{pq}$, in which case the segments $pq$ and $rw$ do not even intersect; see Figure \ref{Fig:Delta2}.

\begin{figure}[htbp]
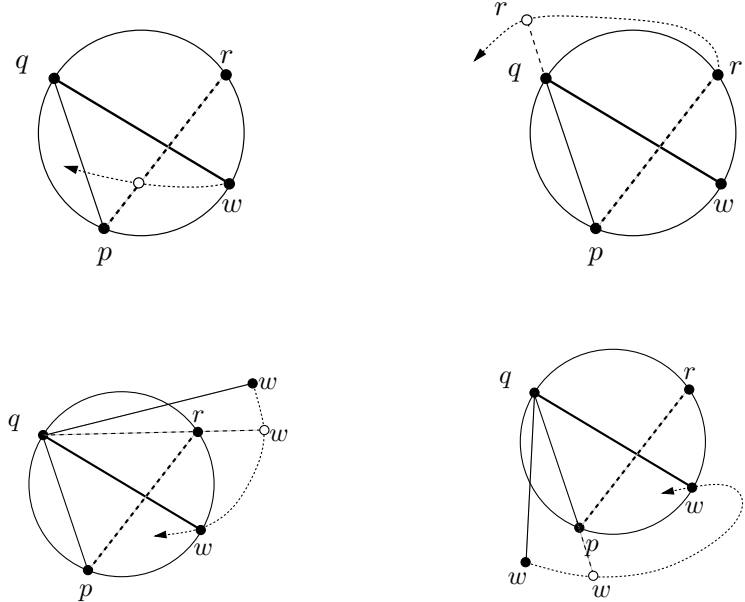

\begin{center}
\input{CaseD1HitW.pstex_t}\hspace{3cm}\input{CaseD1HitQ.pstex_t}\\
\vspace{1cm}
\input{CaseD1HitR.pstex_t}\hspace{3cm}\input{CaseD1HitP.pstex_t}
\caption{\small Lemma \ref{Prop:HitQw}. Top: Possible trajectories of $w$ (left) or $r$ (right) during $(\delta_1,\t_q)$, which realize the crossing of $pr$ by the respective point. Bottom: Possible trajectories of $w$ during $(\upsilon_{pq},\delta_1)$, which realize the crossing of $wq$ by $r$ (left) or by $p$ (right).}
\label{Fig:TrajectoriesD1}
\vspace{-0.6cm}
\end{center}
\end{figure}

Therefore, and since $\delta_1$ is the {\it only} red-red co-circularity of $p,q,r,w$ with respect to $pq$, applying Lemma \ref{Lemma:MustCross} over the interval $(\delta_1,\t_q)$, within the triangulation $\DT(\{p,q,r,w\})$, shows that $pr$ is hit during $(\delta_1,\t_q)$ by at least one of $q$ or $w$. See Figure \ref{Fig:TrajectoriesD1} (top).

A very similar argument shows that the edge $wq$ is hit by one of $p$ or $r$ after $r$ enters $\L_{pq}^+$ (during $I$) and before $\delta_1$.
Indeed, let $\upsilon_{pq}$ denote the time in $I$ when $r$ hits $pq$.
Note that the edge $wq$ is violated right before $\delta_{1}$ by $p$ and $r$, and that the above violation did not hold at time $\upsilon_{pq}$. Therefore, another application of Lemma \ref{Lemma:MustCross} in $\DT(\{p,q,r,w\})$, from time $\delta_1$ backwards, shows that the edge $wq$ is hit during $(\upsilon_{pq},\delta_1)$ by at least one of the two points $p$ or $r$. See Figure \ref{Fig:TrajectoriesD1} (bottom).

To conclude, we have established the following claim.

\begin{lemma}\label{Prop:HitQw}
With the above notation, the following two properties hold in case (d1):

(i) The edge $wq$ is hit in $(\upsilon_{pq},\delta_1)$ by at least one of the points $p,r$. Namely, either $r$ crosses $wq$ from $\L_{wq}^-$ to $\L_{wq}^+$, or $p$ crosses $wq$ in the reverse direction. Moreover,
the Delaunayhood of $wq$ is violated by $p$ and $r$ after the last such crossing and until $\delta_1$.

(ii) The edge $pr$ is hit in $(\delta_1,\t_q)$ by at least one of the points $w,q$. Namely, either $w$ crosses $pr$ from $\L_{pr}^+$ to $\L_{pr}^-$, or $q$ crosses $pr$ in the reverse direction. Moreover, the Delaunayhood of $pr$ is violated by $w$ and $q$ after $\delta_1$ and until the first such crossing.
\end{lemma}

\begin{figure}[htbp]
\begin{center}
\input{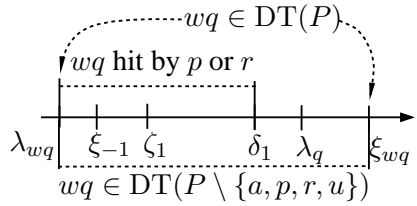}
\caption{\small Case (d1)--obtaining a crossing of $wq$ by at least one of $p,r$. The edge $wq$ is Delaunay at times $\t_{wq}$ and $\xi_{wq}$.
Since the Delaunayhood of $wq$ is violated by $p$ and $r$ right before time $\delta_1\in [\t_{wq},\xi_{wq}]$, it is hit by one of these points during $[\t_{wq},\delta_1)$.}
\label{Fig:CrossWq}
\vspace{-0.6cm}
\end{center}
\end{figure}

\noindent {\bf Case (d1) -- the crossing of $wq$ by $p$ or $r$.} We next turn the crossing in Lemma \ref{Prop:HitQw} (i) into a Delaunay crossing of $wq$ by $r$. Recall that $\delta_1$ belongs to the interval $(\t_{wq},\xi_{wq})$.
Therefore, and since $wq$ is Delaunay at time $\t_{wq}$ (and at time $\xi_{wq}$), the crossing in Lemma \ref{Prop:HitQw} (i) has to occur in the interval $[\t_{wq},\delta_1)$; see Figure \ref{Fig:CrossWq}.
Therefore, and since $wq$ is Delaunay in $\DT(P\setminus \{a,p,r,u\})$ during $[\t_{wq},\xi_{wq}]$ (by condition (S6)), $wq$ undergoes within that latter interval a Delaunay crossing by $p$ or $r$ within a suitably reduced triangulation $\DT(P\setminus \{a,r,u\})$ or $\DT(P\setminus \{a,p,u\})$. 

If $wq$ is hit by $p$ during $[\t_{wq},\delta_1]$, then the points $p,q,w$ define two Delaunay crossings within the reduced triangulation $\DT(P\setminus \{r,a,u\})$.  
A routine combination of Lemma \ref{Lemma:TwiceCollin} with the Clarkson-Shor probabilistic argument implies 
that the overall number of such triples $(p,q,w)$ in $P$ is $O(n^2)$. By Proposition \ref{Prop:UniquePqw}, this also bounds the overall number of such special quadruples $\SQ$.

We may therefore assume that $wq$ is hit during $[\t_{wq},\delta_1)$ by the point $r$, in which case the smaller set $P\setminus\{a,p,u\}$ induces a Delaunay crossing of $wq$ by $r$. 
Note that each such triple $(q,w,r)$ is shared by at most $O(1)$ special quadruples $\SQ$ as above. Indeed, by Lemma \ref{Lemma:Crossing}, $r$ cannot hit $wq$ during the crossing $(wa,q,\J_u)$ (which is defined with respect to $P\setminus \{u\}$). If $r$ hits $wq$ in $[\t_{wq},\t_2)$ then, by condition (S2), $(wa,q,\J_u)$ is among the first three clockwise special $(w,q)$-crossings to begin after that collinearity. Otherwise,
if $r$ hits $wq$ in $(\t_3,\xi_{wq}]$, then condition (S6) similarly implies that $(wa,q,\J_u)$ is 
among the last three such $(w,q)$-crossings to end before this collinearity.
We thus have established the following claim:

\begin{lemma}\label{Claim:UniqueQrw}
With the above assumptions, for any given triple $(q,w,r)$ there remain at most six $3$-restricted special quadruples $\SQ=(a',p',w',q')$, with respective outer points $r'$ and $u'$, that satisfy $(q',w',r')=(q,w,r)$.
\end{lemma}

In other words, any triple $(q,w,r)$ is shared by at most {\it six} special quadruples that have survived the previous chargings (after falling into case (i)). Hence, the special quadruple $\SQ$ under consideration is almost-uniquely determined by the choice of $(q,w,r)$.

In what follows, we therefore assume that the edge $wq$ undergoes (within a suitably reduced triangulation $\DT(P\setminus\{a,p,u\})$) a Delaunay crossing by $r$, and that $\SQ$ and $\varrho$ are almost uniquely determined by this additional crossing triple $(q,w,r)$.

\medskip
\noindent{\bf Case (d1)--the crossing of $pr$ by $q$ or $w$.} We next turn the crossing in Lemma \ref{Prop:HitQw} (ii) into a Delaunay crossing of $pr$ by $w$.
If $pr$ does not re-enter $\DT(P)$ after time $\delta_1$ then, by Lemma \ref{Prop:FewPr}, $(pq,r,I)$ is among the $O(\ell)$ last (regular) $(p,r)$-crossings (because $pr$ is Delaunay during each of these crossings). By Proposition \ref{Prop:TriplesSpecialQuad}, this can happen for at most $O(\ell n^2)$ special quadruples $\SQ$.
Therefore, we may assume that $pr$ re-enters $\DT(P)$ after $\delta_1$.

\begin{figure}[htbp]
\begin{center}
\input{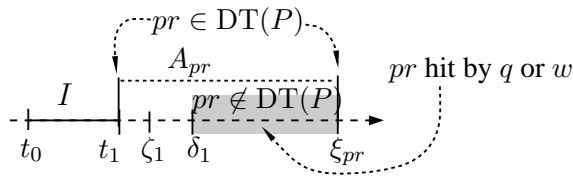}
\caption{\small Case (d1)--enforcing a crossing of $pr$ by one of the points $q,w$. The edge $pr$ is Delaunay throughout $I=[t_0,t_1]$ and at time $\xi_{pr}>\delta_1$, which is the first such time after $\delta_1$ when $pr$ re-enters $\DT(P)$.
The Delaunayhood of $pr$ is violated by $q$ and $w$ right after $\delta_1\in (t_1,\xi_{pr}]$, so it is hit by one of these points during $(\delta_1,\xi_{pr}]$.}
\label{Fig:CrossPr}
\vspace{-0.6cm}
\end{center}
\end{figure}

Let $\xi_{pr}$ denote the first time in $[\delta_1,\infty)$ when the edge $pr$ is again Delaunay (in $P$); see Figure \ref{Fig:CrossPr}. 
Clearly, the time when $pr$ is hit by one of $q,w$ (as prescribed by Lemma \ref{Prop:HitQw} (ii)) belongs to
the interval $(\delta_1,\xi_{pr}]$, which is contained in $(\zeta_1,\xi_{pr}]\subseteq (t_1,\xi_{pr}]$.
To turn this crossing into a Delaunay crossing, we apply Theorem  \ref{Thm:RedBlue} in $\A_{pr}$ over the interval $(t_1,\xi_{pr})$, with the third constant $h\gg \ell$. 

If at least one of the Conditions (i), (ii) of Theorem \ref{Thm:RedBlue} holds, we can charge $\SQ$, within $\A_{pr}$, either to an $h$-shallow collinearity or to $\Omega(h^2)$ $h$-shallow co-circularities. Lemma \ref{Prop:FewPr} ensures that each $h$-shallow event, that occurs in $\A_{pr}$ at some time $t^*\in (t_1,\xi_{pr})$, is charged in this manner by at most $O(\ell)$ special quadruples. Indeed, the corresponding points $p$ and $r$ are involved in the event, so we can guess them in $O(1)$ possible ways, and $(pq,r,I)$ is among the last $8\ell+2$ clockwise $(p,r)$-crossings to end before time $t^*$.
Therefore, the above charging accounts for at most $O\left(\ell h^2N(n/h)+\ell hn^2\beta(n)\right)$ special quadruples $\SQ$.

We may assume, then, that Condition (iii) of Theorem \ref{Thm:RedBlue} holds. That is, there is a subset $A_{pr}$ of at most $3h$ points (perhaps including some of $q,a,u$ and $w$) whose removal restores the Delaunayhood of $pr$ throughout the interval $[t_1,\xi_{pr}]$. 

If $pr$ is crossed during $(\delta_1,\xi_{pr}]$ by $q$ (from $\L_{pr}^-$ to $\L_{pr}^+$), then the triple $p,q,r$ performs two Delaunay crossings within the triangulation $\DT((P\setminus A_{pr})\cup \{q\})$. A routine combination of Lemma \ref{Lemma:TwiceCollin} with the probabilistic argument of Clarkson and Shor implies that $P$ contains at most $O(hn^2)$ triples $p,q,r$ of this kind. By Proposition \ref{Prop:TriplesSpecialQuad}, this also bounds the overall number of such special quadruples $\SQ$. 

To conclude, we are left with the case where the edge $pr$ is crossed during $(\delta_1,\xi_{pr}]$ by $w$ (from $\L_{pr}^+$ to $\L_{pr}^-$).
Hence, the reversely oriented copy $rp$ of $pr$ undergoes within the smaller triangulation $\DT((P\setminus A_{pr})\cup \{w\})$ a Delaunay crossing $(rp,w,\T=[\tau_0,\tau_1])$, where $\T\subseteq [t_1,\xi_{pr}]$ (the crossing must begin after $t_1$, since $pr$ is Delaunay during $I$, by Lemma \ref{Lemma:Crossing}).


\begin{lemma}\label{Claim:UniquePrw}
With the above assumptions, for any given triple $(p,r,w)$ there remain at most $8\ell+2$ $3$-restricted special quadruples $\SQ=(a',p',w',q')$, with respective outer points $r'$ and $u'$, that fall into case (d1) and satisfy $(p',r',w')=(p,r,w)$.
\end{lemma}

\begin{proof}
By Proposition \ref{Prop:TriplesSpecialQuad}, each $\SQ$ as above is uniquely determined by $(pq,r,I)$ which, according to Lemma \ref{Prop:FewPr}, is among the last $8\ell+2$ clockwise $(p,r)$-crossings to end before $w$ hits $pr$ (as prescribed by Lemma \ref{Prop:HitQw}).
\end{proof}

If the above Delaunay crossing of $rp$ by $w$, which occurs within the reduced triangulation $\DT((P\setminus A_{pr})\cup\{w\})$, is a double Delaunay crossing, then we can charge $\SQ$ to this crossing. A standard combination of Lemma \ref{Lemma:TwiceCollin} with the probabilistic argument of Clarkson and Shor implies that the overall number of such triples $(p,r,w)$ in $P$ is only $O(h n^2)$, so the overall number of such special quadruples $\SQ$ does not exceed $O(\ell h n^2)$. Therefore, we may assume, in what follows, that the above crossing $(rp,w,\T=[\tau_0,\tau_1])$ is a {\it single} Delaunay crossing.

To facilitate the subsequent steps of the analysis, we augment the above conflict set $A_{pr}$ as follows. For each clockwise $(p,r)$-crossing $(pq',r,I')$ (in $\F$) that ends during $(t_1,\delta_1)$ we add the respective point $q'$ to $A_{pr}$. 
Informally, this is done to get rid of these $(p,r)$-crossings $(pq',r,I')$ (see below for details).
Since there are only at most $8\ell+1$ such points $q'$ (and since $\ell \ll h$), the overall cardinality of $A_{pr}$, after the augmentation, is at most $3h+8\ell+1\leq 4h$.

To conclude, in case (d1), after disposing of $O\left(N(\ell h^2 N(n/h)+\ell hn^2\beta(n)\right)$ special quadruples, we may assume that the four points of $\varrho=(p,q,r,w)$ perform at least four Delaunay crossings, namely, $(pq,r,I)$, $(qp,w,\H)$,
the crossing of $wq$ by $r$ (which occurs in $P\setminus \{a,p,u\}$ and within $[\t_{wq},\xi_{wq}]$), and the lately enforced single Delaunay crossing $(rp,w,\T=[\tau_0,\tau_1])$ (which occurs in $(P\setminus A_{pr})\cup \{w\}$).

\medskip
\noindent{\bf Case (d1) -- converging.} 
In Section \ref{Subsec:TypeA}, we will exploit the third co-circularity of $p,q,r,w$, which occurs at time $\delta_2\in (\delta_1,\t_{pq}]$ and is red-blue with respect to $pq$ and $rw$, to enforce the crossing of $rw$ by at least one of $p$ and $q$. As a result, one of the triples $(p,r,w)$ of $(q,r,w)$ will perform two Delaunay crossings in an appropriately refined triangulation, and our analysis will bottom out into a quadratic bound via Lemma \ref{Lemma:TwiceCollin}.


To obtain the above crossing of $rw$, we will first apply Theorem \ref{Thm:RedBlue} in the red-blue arrangement of this edge, so as to extend the (almost-)Delaunayhood interval of $rw$ from $\T=[\tau_0,\tau_1]$ (where $rp$ undergoes an almost-Delaunay crossing by $w$) to a larger interval that will contain both $\delta_2$, and a time when $rw$ is hit by $p$ or $q$.
At the end of the analysis, we will manage either to charge the quadruple $(p,q,r,w)$ within $\A_{rw}$ (in cases (i) and (ii) of Theorem \ref{Thm:RedBlue}), or else to extract the desired Delaunay crossing of $rw$.

The above use of Theorem \ref{Thm:RedBlue} will be prepared by applying Theorem \ref{Thm:Balanced} for the  clockwise $(r,w)$-crossing $(rp,w,\T=[\tau_0,\tau_1])$, so as to ensure that each event in $\A_{rw}$ be charged by only few other such terminal quadruples $\varrho'=(p',q',r,w)$, via the respective $(r,w)$-crossings $(rp',w,\T')$. That is, if we encounter too many $(r,w)$-crossings $(rp',w,\T')$ that can charge such an event, the crossing $(rp,w,\T)$ will become $(p,w)$-chargeable, and can thus be accounted for by Theorem \ref{Thm:RedBlue}.

In order for the crossing $(rp,w,\T)$ to be $(p,w)$-chargeable, we need an appropriate time $\xi_{pw}$ after $\delta_2$ when the edge $pw$ is Delaunay (or, at least, almost Delaunay, with none of the obstruction points equal to $r,p,w$). In addition, the edge $pw$ must be almost Delaunay throughout the entire interval where Theorem \ref{Thm:Balanced} is applied. We next proceed to accomplish all these steps in more detail.

\begin{figure}[htbp]
\begin{center}
\input{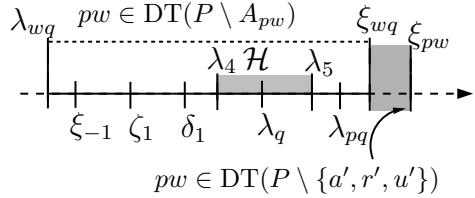}
\caption{\small 
In the preparation for cases (b) and (c), we have extended the Delaunayhood of $pw$ from $\H=[\t_4,\t_5]$ (where it belongs to $\DT(P\setminus \{a,r,u\})$) to 
the larger interval $[\t_{wq},\xi_{wq}]$. We next extend the almost-Delaunayhood of $pw$ beyond $\xi_{wq}$, until some time $\xi_{pw}$ when $pw$ belongs to some reduced triangulation $\DT(P\setminus\{a',r',u'\})$ (for $a',r',u'\not \in\{q,r\}$).}
\label{Fig:ExtendPw}
\vspace{-0.4cm}
\end{center}
\end{figure}

\medskip
\noindent{\bf Charging even more events in $\A_{pw}$.} Our first step is to extend the almost-Delaunayhood of $pw$. Refer to Figure \ref{Fig:ExtendPw}.
Recall that, in preparation for cases (b) and (c), we have already extended the almost-Delaunayhood of $pw$ from $\H=\H_{\SQ}=[\t_4,\t_5]$ (where $qp$ is crossed by $w$) to the interval $[\t_{wq},\xi_{wq}]$, which covers $\H=[\t_4,\t_5],(\xi_{-1},\t_q)$ and $\t_{pq}$. (In particular, $[\t_{wq},\xi_{wq}]$ contains $\delta_1\in (\zeta_1,\t_q)\subset (\xi_{-1},\t_q)$ and $\delta_2\in (\delta_1,\t_{pq}]$.) This has been achieved at the cost of removing a certain subset $A_{pw}$, which consists of at most $6\ell+3$ points, including $a,r,u$.
Unfortunately, the above obstruction set $A_{pw}$ contains $r$ (and perhaps also $w$), so removing $A_{pw}$ in its entirety would destroy the Delaunay crossing $(rp,w,\T)$ (instead of facilitating its $(p,w)$-chargeability in a smaller triangulation).




We next obtain a time $\xi_{pw}>\xi_{wq}$ when $pw$ belongs to some reduced triangulation $\DT(P\setminus \{a',r',u'\})$, for $a',r',u'\not\in \{q,r\}$, and extend the almost-Delaunayhood of $pw$ from $\t_5$ beyond $\xi_{wq}$, until $\xi_{pw}$.


To do so, we return to the family $\G_{pw}^R$ of $3$-restricted right special quadruples $\SQ'=(a',p,w,q')$ that share their middle points $p,w$ with $\SQ$. (In particular, $\G_{pw}^R$ includes $\SQ$.) 

Recall that each special quadruple $\SQ'\in \G_{pw}^R$ is accompanied by a $(p,w)$-crossing $(q'p,w,\H')$, which is defined with respect to the corresponding set $P\setminus \{a',r',u'\}$.
Without loss of generality, we assume that all quadruples in $\G_{pw}^R$ fall into case (d1), and that none of them have been disposed of by the previous chargings within $\A_{pr}$. (In addition, we continue to assume that the special quadruple $\SQ$ under consideration satisfies condition (PHR1).)


By Lemma \ref{Claim:UniquePrw}, any triple $(p,r',w)$ can be shared by at most $8\ell+2$ special quadruples $\SQ'=(a',p,w,q')\in \G_{pw}^R$ under consideration (each with its respective outer points $r'$ and $u'$).
Therefore, the pigeonhole principle implies that at least some fixed fraction of all $3$-restricted quadruples $\SQ=(a,p,w,q)\in \G_{pw}^R$ under consideration (again, with respective outer points $r$ and $u$) satisfy the following condition:

\smallskip
\noindent{\bf (PHR2)} {\it At most $O(\ell)$ other $3$-restricted quadruples $\SQ'=(a',p,w,q')\in \G_{pw}^R$ (each with its respective outer points $r'$ and $u'$) can satisfy $r\in \{a',r',u'\}$.}
\smallskip

(Briefly, this can be shown by considering the {\it multi-function} $\mu:\G_{pw}^R\rightarrow \G_{pw}^R$ mapping each special quadruple $\SQ=(a,p,w,q)$, with respective outer points $r$ and $u$, to at most $(8\ell+2)\times 3=O(\ell)$ other quadruples $\SQ'=(a',p,w,q')$, whose respective outer points $r'$ are chosen from $a,r,u$. Hence, average ``in-degree" of each quadruple $\SQ\in \G_{pw}^R$, which is exactly the number of quadruples $\SQ'$ so that at least one of their respective points $a',r',u'$ is equal to the first outer point $r$ of $\SQ$, is also $O(\ell)$.)

\begin{figure}[htbp]
\begin{center}
\input{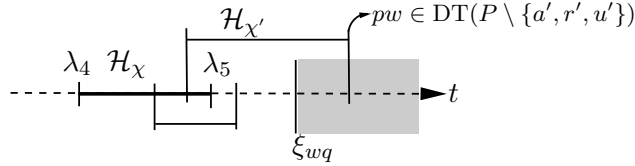}
\caption{\small 
The family $\G_{pw}^R$ contains at most $O(\ell)$ quadruples $\SQ'$ with non-empty intersection $\{a',r',u'\}\cap \{q,r\}$, and at most $k$ quadruples $\SQ'$ whose respective $(p,w)$-crossings end in $(\t_5,\xi_{wq}]$.
If $\G_{pw}^R$ contains no special quadruples $\SQ'$ that satisfy $\{a',r',u'\}\cap \{q,r\}=\emptyset$, and whose respective $(p,w)$-crossings $(q'p,w,\H_{\SQ'})$ end after $\xi_{wq}$, then $(qp,w,\H)$ is among the last $O(\ell)$ such $(p,w)$-crossings.}
\label{Fig:PwD1Last}
\vspace{-0.4cm}
\end{center}
\end{figure}

\smallskip
Therefore, we can assume, in what follows, that the above condition holds for $\SQ$ at hand. 
Combining this\footnote{As a matter of fact, our previous inability to enforce (PHR2) was the only reason why the present analysis in $\A_{pw}$ had not been applied right after handling case (a), in a more general context.} with (PHR1) shows that all but at most 
$3+O(\ell)=O(\ell)$ special quadruples $\SQ'=(a',p,w,q')\in \G_{pw}^R$, with respective outer points $r'$ and $u'$, have $\{q,r\}\cap\{a',r',u'\}=\emptyset$. 
Recall also that, since case (a) has been ruled out, $\G_{pw}^R$ contains at most $k$ quadruples $\SQ'$ whose respective $(p,w)$-crossings $(q'p,w,\H_{\SQ'})$ end in $(\t_5,\xi_{wq}]$. See Figure \ref{Fig:PwD1Last}.

Assume first that $\G_{pw}^R$ contains no special quadruples $\SQ'=(a',p,w,q')$ (with respective outer points $r'$ and $u'$) that satisfy $\{a',r',u'\}\cap \{q,r\}=\emptyset$, and whose respective $(p,w)$-crossings $(q'p,w,\H_{\SQ'})$ end after $\xi_{wq}$.
Therefore, $\G_{pw}^R$ contains at most $k+O(\ell)=O(\ell)$ such quadruples $\SQ'$ whose $(p,w)$-crossings $(q'p,w,\H_{\SQ'})$ end after the ending time $\t_5$ of $\H=\H_{\SQ}$ (including the at most $k$ such quadruples whose $(p,w)$-crossings $\SQ'$ end in $(\t_5,\xi_{wq}]$, and the at most $O(\ell)$ such quadruples $\SQ'$ with non-empty intersection $\{a',r',u'\}\cap \{q,r\}$). Hence, we can charge $\SQ$, via its respective $(p,w)$-crossing $(qp,w,\H_{\SQ}=\H)$, to the edge $pw$, so the above scenario occurs for at most $O(\ell n^2)$ special quadruples $\SQ$ under consideration.

Assume, then, that, for some $\SQ'\in\G_{pw}^R$, with $\{q,r\}\cap\{a',r',u'\}=\emptyset$, its respective $(p,w)$-crossing $(q'p,w,\H_{\SQ'})$ ends after $\xi_{wq}$.
By Lemma \ref{Lemma:Crossing}, $pw$ belongs to $\DT(P\setminus \{a',r',u'\})$ throughout $\H_{\SQ'}$.
In particular, we can choose a time $\xi_{pw}\in [\xi_{wq},\infty)$, which is the first such time when the edge $pw$ belongs to some reduced triangulation $\DT(P\setminus \{a',r',u'\}$), where $a',r',u'\in P\setminus \{r,q\}$.
In what follows, we use $a',r'$ and $u'$ to denote the above three points $a',r',u'$, whose removal restores the Delaunayhood of $pw$ at time $\xi_{pw}$.

The preceding discussion implies that at most $O(\ell)$ of the above $(p,w)$-crossings $(q'p,w,\H_{\SQ'})$ can end in $(\xi_{wq},\xi_{pw}]$ (and that, for each of those crossings, its respective obstruction set $\{a',r',u'\}$ intersects $\{q,r\}$).
Therefore, and since case (a) has been ruled out, at most $k+O(\ell)=O(\ell)$ of the above $(p,w)$-crossings can end in $(\t_5,\xi_{pw})$. 

\begin{figure}[htbp]
\begin{center}
\input{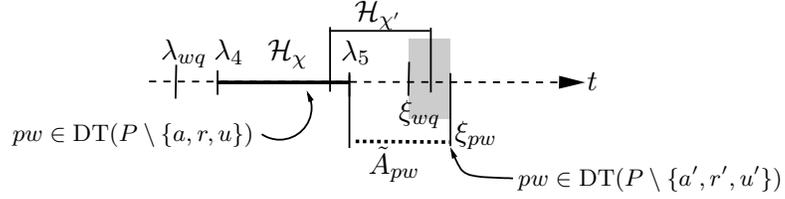}
\caption{\small Extending the almost-Delaunayhood of $pw$ to $[\t_5,\xi_{pw}]$. $\xi_{pw}$ is the first time 
in $[\xi_{wq},\infty)$ when $pw$ belongs to some reduced triangulation $\DT(P\setminus \{a',r',u'\})$, for $\{a',r',u'\}\cap \{q,r\}=\emptyset$. We apply Theorem \ref{Thm:RedBlue} within $\A_{pw}$ over the interval $(\t_5,\xi_{pw})$, noting that $(qp,w,\H_{\SQ})$ is among the last $O(\ell)$ such $(p,w)$-crossings $(q'p,w,\H_{\SQ'})$ to end before any charged event.}
\label{Fig:TimelinePwD1}
\vspace{-0.4cm}
\end{center}
\end{figure}

We are finally ready to apply Theorem \ref{Thm:RedBlue} in $\A_{pw}$ over the interval $(\t_5,\xi_{pw})$ (see Figure \ref{Fig:TimelinePwD1}). This is done with the third constant $h\gg \ell$ and with respect to the smaller set $P\setminus \{a',r',u'\}$. If at least one of the first two conditions of Theorem \ref{Thm:RedBlue} holds, we charge $\SQ$ within $\A_{pw}$ either to an $(h+3)$-shallow collinearity, or to $\Omega(h^2)$ $(h+3)$-shallow co-circularities (as in the previous chargings, these events are $h$-shallow in $P\setminus \{a',r',w'\}$, and $(h+3)$-shallow in $P$).
Clearly, each $(h+3)$-shallow event in $\A_{pw}$ is charged as above by at most $O(\ell)$ special quadruples $\SQ$, 
because $(qp,w,\H_{\SQ'})$ is among the last $O(\ell)$ such $(p,w)$-crossings to end before the event. 
Hence, the above charging accounts for $O\left(\ell h^2N(n/h)+\ell h n^2\beta(n)\right)$ special quadruples $\SQ$.

We can therefore assume that Condition (iii) of Theorem \ref{Thm:RedBlue} holds. Hence, there is a subset $\tilde{A}_{pw}$ of at most $3h+3$ points (including the above three points $a',r',u'\in P\setminus \{p,q,r,w\}$), whose removal restores the Delaunayhood of $pw$ throughout $(\t_5,\xi_{pw})$. Therefore, $pw$ belongs to $\DT(P\setminus (A_{pw}\cup \tilde{A}_{pw}))$ throughout the entire interval $[\t_{wq},\xi_{pw}]=[\t_{wq},\xi_{wq}]\cup (\t_5,\xi_{pw}]$ (where $A_{pw}$ denotes the set of at most $6\ell+3$ points, including $a,r,u$, whose removal restores the Delaunayhood of $pw$ throughout $[\t_{wq},\xi_{wq}]$).

\medskip
\noindent{\bf Case (d1)--Wrap up.} We again emphasize that the times of the various events discussed so far appear in the order

$$
\xi_{pq}<\t_{wq}<\xi_{-1}<\zeta_1<\delta_1<\t_q<\t_{pq}<\xi_{wq}<\xi_{pw},
$$ 
that $\delta_2\in (\delta_1,\t_{pq}]$, and that $w$ crosses $rp$ from $\L_{rp}^-$ to $\L_{rp}^+$ in the interval $(\delta_1,\t_q)$, as part of a single Delaunay crossing $(rp,w,\T=[\tau_0,\tau_1])$ (which occurs in $(P\setminus A_{pr})\cup \{w\}$). Refer to Figure \ref{Fig:SummaryD1}.

\begin{figure}[htbp]
\begin{center}
\input{SummaryD1.pstex_t}\hspace{3cm}\input{CaseD1Possible.pstex_t}
\caption{\small Case (d1): A (partial) summary of what we assume at the end of the analysis. Left: Various events occur in the depicted order (and $\delta_2$ lies in $(\delta_1,\t_{pq}]$). Right: A possible motion of $w$ after $r$ enters $\L_{pq}^+$ (during $I$).}
\label{Fig:SummaryD1}
\vspace{-0.4cm}
\end{center}
\end{figure} 

By the definition of $A_{pw}$ and $\tilde{A}_{pw}$ (of total cardinality $6\ell+3+3h+3=O(h)$), the edge $pw$ belongs to $\DT(P\setminus (A_{pw}\cup \tilde{A}_{pw}))$ throughout the interval $[\delta_1,\xi_{pw}]\subseteq [\t_{wq},\xi_{pw}]$. Furthermore, $pw$ belongs at time $\xi_{pw}$ to the triangulation $\DT(P\setminus \{a',r',u'\})$, where $a',r',u'\in \tilde{A}_{pw}\setminus \{q,r\}$.

Recall also that, since $\zeta_1$ belongs to both intervals $[t_0,\t_1]={\sf conv}(I\cup\I_r)$ and $(\xi_{-1},\xi_0)\subset [\xi_{pq},\t_{pq}]$, the combination of conditions (Q8) and (S6) (on, respectively, $\Q$ and $\SQ$) implies that the edge $pq$ belongs to $\DT(P\setminus \{a,w,r,u\})$ throughout the interval $[t_0,\t_{pq}]\subseteq [t_0,\t_1]\cup [\xi_{pq},\t_{pq}]$. 

Finally, we continue to assume that the edge $wq$ undergoes a Delaunay crossing by $r$ within $P\setminus \{a,p,u\}$. (The precise interval of this crossing is immaterial for our future analysis.)

In what follows, we use $A_{pq}^+$ to denote the set of all points of $P$ that appear in the cap $C_{pq}^+$ at some time in $(\xi_{-1},\t_q)$. By Lemma \ref{Prop:ShallowCup}, the cardinality of $A_{pq}^+$ does not exceed $8\ell+1$.

\medskip
\noindent{\bf Case (d1) -- charging terminal quadruples.}
To proceed, we draw a random sample $R$ of $\lceil n/h \rceil$ points of $P$. Notice that the following two events occur simultaneously with probability at least $\Omega(1/h^4)$:
(1) The four points $p,q,w,r$ belong to $R$, and (2) $R$ includes none of the points of 
$$
(A_{pq}^+\cup\tilde{A}_{pw}\cup A_{pw}\cup A_{pr}\cup\{a,u\})\setminus \{p,q,r,w\}.
$$

Suppose that the sample $R$ is indeed successful for the $3$-restricted right special quadruple $\SQ=(a,p,w,q)$ at hand, with respective two outer points $r$ and $u$. Then we can charge $\SQ$ to the quadruple $\varrho=(p,q,r,w)$, which satisfies the following conditions with respect to the sample $R$ (see Figure \ref{Fig:SetupA} in Section \ref{Subsec:TypeA} for a schematic summary, with $R$ replaced by $P$). 

\medskip
\noindent  {\bf (A1)} The edge $pq$ undergoes (in $R$) a Delaunay crossing $(pq,r,I=[t_0,t_1])$ and is crossed by $w$, from $\L_{pq}^+$ to $\L_{pq}^-$, at some later time $\t_q>t_1$.
In addition, $pq$ is again Delaunay at some time $\t_{pq}$ which is the first such time after $t_q$, and it belongs to $\DT(P\setminus \{r,w\})$ throughout $(t_1,\t_{pq})$. Hence, its reversely oriented copy $qp$ undergoes in $R\setminus \{r\}$ (and entirely within $(t_1,\t_{pq}]$) a Delaunay crossing by $w$.

\medskip
\noindent {\bf (A2)} The points $p,q,w,r$ are co-circular at times $\delta_0\in I$, $\delta_1\in [t_1,\t_q]$, and $\delta_2\in (\delta_1,\t_{pq}]$, and the following properties hold:

\noindent (i) The co-circularity at time $\delta_0$ is red-blue with respect to $pq$. 

\noindent (ii) The co-circularity at time $\delta_1$ is red-red with respect to $pq$ and red-blue with respect to the edge $pr$, whose Delaunayhood is violated right after time $\delta_1$ by $q$ and $w$. Furthermore, the open cap $C_{pq}^+=B[p,q,w]\cap \L_{pq}^+$ contains no points of $P$ at time $\delta_1$.

\noindent (iii) The co-circularity at time $\delta_2$ is again red-blue with respect to $pq$.
It arises during a single Delaunay crossing of $qp$ by $w$, which occurs in $\DT(\{p,q,r,w\})$ during some sub-interval of $(\delta_1,\t_{pq}]$.


\medskip
\noindent {\bf (A3)} The set $R\setminus \{q\}$ induces a (single) Delaunay crossing $(rp,w,\T=[\tau_0,\tau_1])$, where $w$ crosses $rp$ from $\L_{rp}^-$ to $\L_{rp}^+$ during $(\delta_1,\t_q)$. 


Similarly, the set $R\setminus \{p\}$ induces a Delaunay crossing of $wq$ by $r$, where $r$ crosses $wq$ before $\delta_1$, and from $\L_{wq}^-$ to $\L_{wq}^+$.

\medskip
\noindent {\bf (A4)} There exists a time $\xi_{pw}>\t_{pq}$ so that (i) the edge $pw$ is Delaunay (in $R$) at time $\xi_{pw}$, and (ii) $pw$ belongs to $\DT(R\setminus \{q,r\})$ throughout the interval $[\delta_1,\xi_{pw}]$.

\smallskip
In Section \ref{Subsec:TypeA} we show that $pw$ is Delaunay also at time $\delta_1$. In addition, Lemma \ref{Lemma:Crossing} implies that $pw$ belongs to $\DT(R\setminus \{q\})$ throughout the interval $\T=[\tau_0,\tau_1]$, which obviously intersects $[\delta_1,\xi_{pw}](\supset [\delta_1,\t_q])$.

\medskip
Notice that any such quadruple $\varrho=(p,q,r,w)$ in $R$ is charged as above by at most one $3$-restricted right special quadruple $\SQ=(a,p,w,q)$ in $\F$ (with outer points $r$ and $u$), because the latter quadruple is uniquely determined by each of the triples $(p,q,r)$ and $(p,q,w)$.

We say that a quadruple $\varrho=(p,q,r,w)$ is {\it terminal of type A} if it satisfies the above four conditions (A1)--(A4) with respect to the underlying set $R$. (In Section \ref{Sec:Terminal}, we shall again use $P$ to denote the underlying point set of our terminal quadruples. See Figure \ref{Fig:SetupA} in that section for a partial summary of the properties of terminal quadruples of type A.)

Let $\TF_R^A$ denote the resulting family of terminal quadruples $\varrho=(p,q,r,w)$ (of type A) in $R$ that are charged by $3$-restricted right special quadruples in $P$ through the above probabilistic argument.

\begin{lemma}
With the above assumptions, each terminal quadruple $\varrho=(p,q,r,w)$ in $\TF^A_R$ is uniquely determined by each of 
its sub-triples $(p,q,r), (p,q,w)$, $(p,r,w)$. Furthermore, any triple $(q,r,w)$ is shared by at most six terminal quadruples of $\TF^A_R$.
\end{lemma}
\begin{proof}
Clearly, the second part of the lemma is directly implied by Lemma \ref{Claim:UniqueQrw}, so it suffices to establish the first part of it.

By condition (A1), $w$ is the first point of $P$ to hit the edge $pq$ after its Delaunay crossing $(pq,r,I=[t_0,t_1])$ by $r$. Hence,  $\varrho=(p,q,r,w)$ is uniquely determined by the choice of $p,q$ and $r$.
A similar agrument implies that $\varrho=(p,q,r,w)$ is uniquely determined by the triple $(p,q,w)$.

To see that $\varrho$ is uniquely determined by $(p,r,w)$, let us assume for a contradiction that $\TF^A_R$ contains another such quadruple $\varrho'=(p,q',r,w)$ (of type A and with $q'\neq q$). Furthermore, assume with no loss of generality that 
the respective $(p,r)$-crossing $(pq',r,I')$ of $\varrho'$ ends after $I=[t_0,t_1]$. Note though that $I'$ must end before $w$ enters $\L_{pr}^-$ through $pr$ (as prescribed by condition (A3)).
However, in that case $I'$ would end in $(t_1,\delta_1)$, so $q'$ would have been included in the respective set $A_{pr}$ of $\varrho$, and, therefore, omitted\footnote{Clearly, we have $q'\neq w$ (i.e., there is no crossing $(pw,r,I')$), because $r$ can enter the halfplane $\L_{pw}^+$ only once, and it is already assumed to cross the line $\L_{pw}$, from $\L_{pw}^-$ to $\L_{pw}^+$, and {\it outside} $pw$ (as prescribed by Lemma \ref{Prop:HitQw} (ii)).} from $R$, contrary to the choice of $\varrho'\in \TF_R^A$.  
\end{proof}

To simplify the presentation, in what follows we only consider a subfamily $\TF^A=\TF^A_R$ of terminal quadruples of type A whose members $\varrho=(p,q,r,w)$ are {\it uniquely determined} by each one of their respective four sub-triples $(p,q,r),(p,q,w),(p,r,w)$, and $(q,r,w)$. This stronger uniqueness condition can be enforced by prunning $\TF_R^A$ (without affecting its asymptotic cardinality), so that, for each triple $(q,r,w)$, we keep in $\TF_R^A$ only one terminal quadruple $(p,q,r,w)$, if such quadruples exist at all in $\TF^A_R$.


Let $T^A(m)$ denote the maximum cardinality of a family $\TF^A$ of terminal quadruples of type A (with the above uniqueness property) that can be defined over a set of $m$ moving points.
The preceding analysis implies that the overall number of special quadruples that fall into Case (d1) is at most 
$$
O\left(h^4 T^A(n/h)+\ell h^2 N(n/h)+\ell h n^2\beta(n)\right).
$$

\paragraph{Case (d2).} The co-circularity at time $\delta_1$ is red-blue with respect to the edge $qr$, whose Delaunayhood is violated right after that by $p$ and $w$. We continue to assume that $r$ does not cross $pq$ again during $(t_1,\t_{pq}]$.

As in case (d1), we use $\upsilon_{pq}$ to denote the time in $I=[t_0,t_1]$ when $r$ enters the halfplane $\L_{pq}^+$.
We have the following lemma, whose proof is fully symmetric to that of Lemma \ref{Prop:HitQw}. 

\begin{lemma}\label{Prop:HitPw}
With the above notation, the following two properties hold in case (d2):

(i) The edge $pw$ is hit in $(\upsilon_{pq},\delta_1)$ by at least one of the points $q,r$. Namely, either $r$ crosses $pw$ from $\L_{pw}^-$ to $\L_{pw}^+$, or $q$ crosses $pw$ in the reverse direction. Moreover,
the Delaunayhood of $pw$ is violated by $q$ and $r$ after the last such crossing and until $\delta_1$.

(ii) The edge $rq$ is hit in $(\delta_1,\t_q)$ by at least one of the points $p,w$. Namely, either $w$ crosses $rq$ from $\L_{rq}^+$ to $\L_{rq}^-$, or $p$ crosses $rq$ in the reverse direction. Moreover, the Delaunayhood of $rq$ is violated by $w$ and $q$ after $\delta_1$ and until the first such crossing.
\end{lemma}

\begin{figure}[htbp]
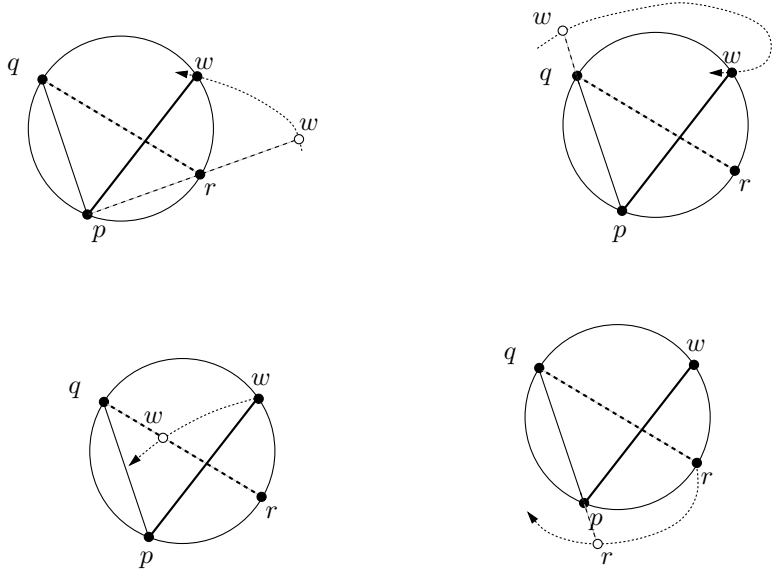

\begin{center}
\input{CaseD2HitR.pstex_t}\hspace{3cm}\input{CaseD2HitQ.pstex_t}\\
\vspace{1cm}
\input{CaseD2HitW.pstex_t}\hspace{3cm}\input{CaseD2HitP.pstex_t}
\caption{\small Lemma \ref{Prop:HitPw}. Top: Possible trajectory of $w$ during $(\upsilon_{pq},\delta_1)$, which realize the crossing of $pw$ by $r$ (left) or $q$ (right). Bottom: Possible trajectories of $w$ (left) and $r$ (right) during $(\upsilon_{pq},\delta_1)$, which realize the crossing of $rq$ by the respective point.}
\label{Fig:TrajectoriesD2}
\vspace{-0.6cm}
\end{center}
\end{figure}

Refer to Figure \ref{Fig:TrajectoriesD2}.
To prove part (i) of Lemma \ref{Prop:HitPw}, we note that, right before time $\delta_1$, the Delaunayhood of $pw$ is violated by $q\in \L_{pw}^-$ and $r\in \L_{pw}^+$, and that this violation does not hold either right before, or right after the time $\upsilon_{pq}$ when $r$ crosses $pq$. Hence, to obtain the desired crossing of $pw$, we can apply the time reversed variant Lemma \ref{Lemma:MustCross} for the triangulation $\DT(\{p,q,r,w\})$, over the interval $(\upsilon_{pq},\delta_1)$.

To prove part (ii) of Lemma \ref{Prop:HitPw}, we apply (the regular variant of) Lemma \ref{Lemma:MustCross} in $\DT(\{p,q,r,w\})$ over the interval $(\delta_1,\t_q)$, noting that the violation of $rq$ by $p\in \L_{rq}^-$ and $w\in \L_{rq}^+$, which holds right after time $\delta_1$, no longer exists either right before, or right after, the time $\t_q$ when $w$ hits $pq$.

\medskip
\noindent{\bf Case (d2) -- enforcing the crossing of $rq$ by $p$ or $w$.}
Our argument is fully symmetric to the one used in case (d1) to enforce a Delaunay crossing of $pr$ by $q$ or $w$.

\begin{figure}[htbp]
\begin{center}
\input{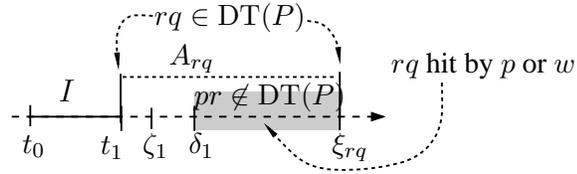}
\caption{\small Case (d2)--enforcing a crossing of $rq$ by at least one of the points $p,w$. The edge $rq$ is Delaunay throughout $I=[t_0,t_1]$ and at time $\xi_{rq}>\delta_1$, which is the first such time after $\delta_1$ when $rq$ re-enters $\DT(P)$.
The Delaunayhood of $rq$ is violated by $q$ and $w$ right after $\delta_1\in (t_1,\xi_{rq}]$, so it is hit by one of these points during $(\delta_1,\xi_{rq}]$.}
\label{Fig:CrossRq}
\vspace{-0.6cm}
\end{center}
\end{figure}

Recall that, according to Lemma \ref{Prop:FewPr}, at most $8\ell+1$ counterclockwise $(q,r)$-crossings can end in the interval $(t_1,\delta_0)$.
If $rq$ never re-enters $\DT(P)$ after time $\delta_1$, then $(pq,r,I)$ is among the last $8\ell+2$  counterclockwise Delaunay $(q,r)$-crossings in $\F$ (with respect to the standard order implied by Lemma \ref{Lemma: OrderOrdinaryCrossings}). Clearly, this scenario happens for at most $O(\ell n^2)$ special quadruples $\SQ$, because each of them is uniquely determined by the respective triple $(p,q,r)$ (according to Proposition \ref{Prop:TriplesSpecialQuad}).
Therefore, we may assume, in what follows, that $rq$ re-enters $\DT(P)$ at some future time $\xi_{rq}>\delta_1$ (which is the {\it first} such time when $rq$ is Delaunay); see Figure \ref{Fig:CrossRq}. By Lemma \ref{Prop:HitPw} (ii), $rq$ is hit during $(\delta_1,\xi_{rq}]\subset (t_1,\xi_{rq}]$ by $p$ or $w$. 
Furthermore, Lemma \ref{Prop:FewPr} (combined with Lemma \ref{Lemma:Crossing}) implies at most $8\ell+1$ counterclockwise (Delaunay) $(q,r)$-crossings in $\F$ can end during $(t_1,\xi_{rq}]$.

To enforce the desired crossing of $rq$, we apply Theorem \ref{Thm:RedBlue} in $\A_{rq}$ over the interval $(t_1,\xi_{rq})$, with the third threshold $h\gg \ell$. 

If one of the Conditions (i), (ii) holds, we charge $\SQ$ (via $(pq,r,I)$) either to an $h$-shallow collinearity or to $\Omega(h^2)$ $h$-shallow co-circularities. Clearly, each of these $h$-shallow events is charged at most $O(\ell)$ times in the above manner, because $(pq,r,I)$ is among the last $8\ell+2$ counterclockwise $(q,r)$-crossings (in $\F$) to end before the time of the event.
Hence, the above charging accounts for at most $O(\ell h^2N(n/h)+\ell hn^2\beta(n))$ special quadruples $\SQ$.

Assume, then, that Condition (iii) of Theorem \ref{Thm:RedBlue} holds, so we have a subset $A_{rq}$ of at most $3h$ points (possibly including $p$ or $w$, or both) whose removal restores the Delaunayhood of $rq$ throughout the entire interval $[t_0,\xi_{rq}]=I\cup [t_1,\xi_{rq}]$. To facilitate the subsequent analysis, we augment the set $A_{pr}$ as follows. For each crossing $(p'q,r,I')$ (in $\F$) that ends in the interval $(t_1,\delta_1)$ we add the respective point $p'$ to $A_{rq}$.

If $rq$ is hit during $(\delta_1,\xi_{rq}]$ by $p$, then the triple $(p,q,r)$ is involved in two Delaunay crossings, which occur within the smaller triangulation $\DT((P\setminus A_{rq})\cup \{p\})$. According to Lemma \ref{Lemma:TwiceCollin}, the overall number of such triples in $P$ does not exceed $O(hn^2)$. By Proposition \ref{Prop:TriplesSpecialQuad}, this also bounds the overall number of the respective special quadruples $\SQ$.

To conclude, we may assume, in what follows, that $rq$ is hit during $(\delta_1,\xi_{rq}]\subset (t_1,\xi_{rq})$ by $w$ (which crosses it from $\L_{rq}^+$ to $\L_{rq}^-$). Therefore, the reversely oriented copy $qr$ of $rq$ undergoes, within the reduced triangulation $\DT((P\setminus A_{rq})\cup \{w\})$, a Delaunay crossing $(qr,w,\T=[\tau_0,\tau_1])$. 

Notice that $(pq,r,I)$ is among the last $8\ell+2$ $(q,r)$-crossings in $\F$ to end before $w$ crosses $rq$ from $\L_{rq}^+$ to $\L_{rq}^-$, which implies 
the following symmetric analogue of Lemma \ref{Claim:UniquePrw}:

\begin{lemma}\label{Claim:LUniqueQrw}
Any triple $(q,r,w)$ is shared by at most $8\ell+2$ 3-restricted special quadruples $\SQ=(a,p,w,q)$ (with respective outer points $r$ and $u$) of the above kind.
\end{lemma}

If the above crossing $(qr,w,\T=[\tau_0,\tau_1])$ is a double Delaunay crossing, we apply Lemma \ref{Lemma:TwiceCollin} (in combination with the Clarkson-Shor argument) to establish an upper bound of $O(h n^2)$ on the overall number of such triples $(q,r,w)$ in $P$, which immediately yields an upper bound of $O(\ell hn^2)$ on the number of special quadruples $\SQ$ of this kind.
Hence, we may assume, in what follows, that the above crossing of $qr$ by $w$ in $\DT((P\setminus A_{rq})\cup \{w\})$ is a {\it single} Delaunay crossing.

\begin{figure}[htbp]
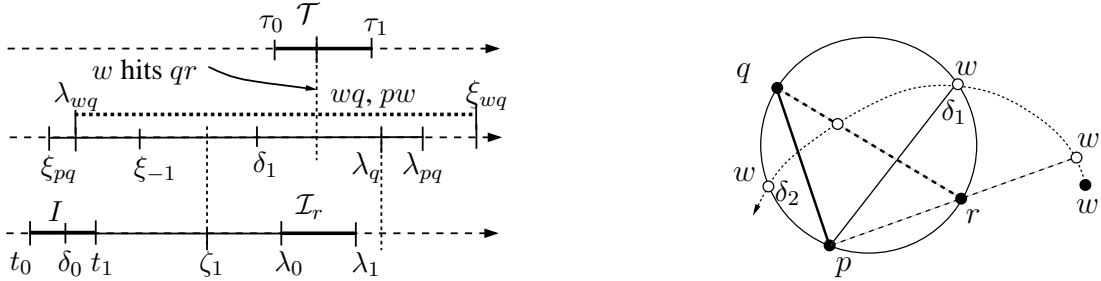

\begin{center}
\input{SummaryD2.pstex_t}\hspace{3cm}\input{D2Possible.pstex_t}
\caption{\small Case (d2): A (partial) summary of what we assume at the end of the analysis. Left: Various events occur in the depicted order (and $\delta_2$ lies in $(\delta_1,\t_{pq}]$). Right: A possible motion of $w$ after $r$ enters $\L_{pq}^+$ (during $I$).}
\label{Fig:SummaryD2}
\vspace{-0.4cm}
\end{center}
\end{figure}

We again emphasize that $\t_{wq}<\delta_1<\t_q<\t_{pq}<\xi_{wq}$ and $\delta_2\in (\delta_1,\t_{pq}]$, and that $w$ hits $qr$ (during $\T=[\tau_0,\tau_1]$) in the interval $(\delta_1,\t_q)$.
Furthermore, by condition (S6), $wq$ belongs to $\DT(P\setminus \{a,p,r,u\})$ throughout $[\t_{wq},\xi_{wq}]\subset (\delta_1,\xi_{wq})$ (and is Delaunay at times $\t_{wq}$ and $\xi_{wq}$).
See Figure \ref{Fig:SummaryD2}.

\medskip
\noindent{\bf Case (d2) -- charging terminal quadruples.}
As in case (d1), let $A_{pq}^+$ denote the set of at most $8\ell+1$ points that show up in the cap $C_{pq}^+=B[p,q,w]\cap \L_{pq}^+$ at some time in $(\xi_{-1},\t_q)$ (see Lemma \ref{Prop:ShallowCup}). 


To proceed, we draw a random sample of $R$ of $\lceil n/h \rceil$ points of $P$. Notice that the following two events occur simultaneously with probability at least $\Omega(1/h^4)$:
(1) The four points $p,q,w,r$ belong to $R$, and (2) $R$ includes none of the points of 
$$
(A_{pw}\cup A_{rq}\cup A_{pq}^+\cup\{a,u\})\setminus \{p,q,r,w\}
$$

Suppose that the sample $R$ is indeed successful for the $3$-restricted right special $\SQ=(a,p,w,q)$ at hand (with respective two outer points $r$ and $u$). Then we can charge $\SQ$ to the quadruple $\varrho=(p,q,r,w)$, which satisfies the following conditions with respect to the sample $R$:

\medskip
\noindent  {\bf (B1)} The edge $pq$ undergoes a Delaunay crossing $(pq,r,I=[t_0,t_1])$ and is crossed by $w$, from $\L_{pq}^+$ to $\L_{pq}^-$, at some later time $\t_q>t_1$.
In addition, $pq$ is again Delaunay at some time $\t_{pq}$ which is the first such time after $t_q$, and it belongs to $\DT(R\setminus \{r,w\})$ throughout $(t_1,\t_{pq})$. Hence, its reversely oriented copy $qp$ undergoes in $R\setminus \{r\}$ (and entirely within $(t_1,\t_{pq}]$) a Delaunay crossing by $w$.
Finally, $r$ does not cross $pq$ in $(t_1,\t_{pq}]$.

\medskip
\noindent {\bf (B2)} The points $p,q,w,r$ are co-circular at times $\delta_0\in I$, $\delta_1\in [t_1,\t_q]$, and $\delta_2\in (\delta_1,\t_{pq}]$, and the following properties hold: 

\noindent (i) The co-circularity at time $\delta_0$ is red-blue with respect to $pq$. 

\noindent (ii) The co-circularity at time $\delta_1$ is red-red with respect to $pq$ and red-blue with respect to the edge $rq$, whose Delaunayhood is violated right after time $\delta_1$ by $p$ and $w$. (In particular, this implies that $r$ remains in $\L_{pq}^+$ throughout $(t_1,\delta_1)$, after entering this halfplane during $I$.)
Furthermore, the open cap $C_{pq}^+=B[p,q,w]\cap \L_{pq}^+$ contains no points of $P$ at time $\delta_1$.

\noindent (iii) The co-circularity at time $\delta_2$ is again red-blue with respect to $pq$.
It arises during a single Delaunay crossing of $qp$ by $w$, which occurs in $\DT(\{p,q,r,w\})$ during some sub-interval of $(\delta_1,\t_{pq}]$. 


\medskip
\noindent {\bf (B3)} The set $R\setminus \{p\}$ induces a (single) Delaunay crossing $(qr,w,\T=[\tau_0,\tau_1])$, where $w$ crosses $rq$, from $\L_{rq}^+$ to $\L_{rq}^-$, during $(\delta_1,\t_q)$. 


\medskip
\noindent {\bf (B4)} There exists a time $\xi_{qw}>\t_{pq}$ so that (i) the edge $qw$ is Delaunay at time $\xi_{qw}$, and (ii) the edges $qw$ and $pw$ belong to, respectively, $\DT(R\setminus \{p,r\})$ and $\DT(R\setminus \{q.r\})$ throughout the interval $[\delta_1,\xi_{qw}]$.

\medskip
Notice that any such quadruple $\varrho=(p,q,r,w)$ in $R$ is charged as above by at most one $3$-restricted right special quadruple $\SQ=(a,p,w,q)$ in $\F$ (with respective outer points $r$ and $u$), because the latter quadruple is uniquely determined by each of the triples $(p,q,r)$ and $(p,q,w)$.

We say that a quadruple $\varrho=(p,q,r,w)$ is {\it terminal of type B} if it satisfies the above four conditions with respect to the underlying set $R$. (In Section \ref{Sec:Terminal}, we shall again use $P$ to denote the underlying set of our terminal quadruples.)

Let $\TF_R^B$ denote the resulting family of terminal quadruples $\varrho=(p,q,r,w)$ (of type B) in $R$ that are charged by $3$-restricted right special quadruples in $P$ through the above probabilistic argument.

\begin{lemma}\label{Claim:TypeBUniqueness}
With the above assumptions, each terminal quadruple $\varrho=(p,q,r,w)$ in $\TF^B_R$ is uniquely determined by each of 
its sub-triples $(p,q,r), (p,q,w)$, $(q,r,w)$. 
\end{lemma}
\begin{proof}
By condition (B1), $w$ is the first point of $P$ to hit the edge $pq$ after its Delaunay crossing $(pq,r,I=[t_0,t_1])$ by $r$. Hence,  $\varrho=(p,q,r,w)$ is uniquely determined by the choice of $p,q$ and $r$.
A similar agrument implies that $\varrho=(p,q,r,w)$ is uniquely determined by the triple $(p,q,w)$.

To see that $\varrho$ is uniquely determined by $(q,r,w)$, let us assume for a contradiction that $\TF^B_R$ contains another such quadruple $\varrho'=(p',q,r,w)$ (of type B and with $p'\neq p$). Furthermore, assume with no loss of generality that 
the respective counterclockwise $(q,r)$-crossing $(p'q,r,I')$ of $\varrho'$ ends after $I=[t_0,t_1]$. Note though that $I'$ must end before $w$ enters $\L_{pr}^-$ through $pr$ (as prescribed by condition (A3)).
However, in that case $I'$ would end in $(t_1,\delta_1)$, so $q'$ would be included in the respective set $A_{rq}$ of $\varrho$, and, thereby, omitted\footnote{Clearly, we have $p'\neq w$ (i.e., there is no crossing $(wq,r,I')$), because $r$ is already assumed to cross the line $\L_{wq}$, from $\L_{wq}^-$ to $\L_{wq}^+$, {\it outside} $pw$ (as prescribed by Lemma \ref{Prop:HitPw} (ii)).} from $R$, contrary to the choice of $\varrho'$ in $\TF_R^B$.  
\end{proof}

Let $T^B(m)$ denote the maximum cardinality of any family $\TF^B$ of terminal quadruples of type B (with the uniqueness property stated in Lemma \ref{Claim:TypeBUniqueness}) that can be defined over a set $P$ of $m$ moving points.
The preceding discussion implies that the number of special quadruples that fall into case (d2) is at most

$$
O\left(h^4 T^B(n/h)+\ell h^2 N(n/h)+\ell h n^2\beta(n)\right).
$$

We delegate the analysis of terminal quadruples of type B to Section \ref{Subsec:TypeB}. Note that the points of each such terminal quadruple $\varrho=(p,q,r,w)$ perform at least three Delaunay crossings (namely, the crossings of $pq$ by $r$ and $w$, and the crossing of $qr$ by $w$). Hence, it suffices to enforce two more crossings in order to ensure that some sub-triple of $\varrho$ be involved in {\it two} distinct Delaunay crossings.

As in the case of terminal quadruples of type A, we shall exploit the co-circularity at time $\delta_2$, which is red-blue with respect to $rw$, in order to enforce
a Delaunay crossing of that edge by at least one of the two points $p,q$. 
In addition, we shall enfore a Delaunay crossing of $pw$ by at least one of $r,q$ (during which $pw$ will be hit by $r$ or $q$, as suggested by Lemma \ref{Prop:HitPw} (i) and depicted in Figure \ref{Fig:TrajectoriesD2} (top)).

\medskip
\noindent{\bf $3$-restricted right special quadruples--wrap up.}
Putting together the previously established bounds on the maximum possible numbers of $3$-restricted right special quadruples that fall into cases (a), (b), (c), (d1) and (d2) yields the following recurrence:\\

\smallskip
\noindent$\Phi_3^R(n)=$
\begin{equation}
O\left(h^4 T^A(n/h)+h^4 T^B(n/h)+\ell h^2 N(n/h)+k\ell^2 N(n/\ell)+k^2N(n/k)+\ell h n^2\beta(n)\right).
\end{equation}

\medskip
\noindent{\bf Discussion.}
Notice that the roles of $p$ and $q$ in subcases (d1) and (d2) are largely symmetric, which enables us to enforce a Delaunay crossing of the respective edge $pr$ or $rq$ by at least one of the remaining two points of $p,q,r,w$.
In both scenarios, we first apply Theorem \ref{Thm:RedBlue} (with threshold $h\gg \ell$) in order to extend the (almost-)Delaunayhood of $pr$ or $qr$ from $I=[t_0,t_1]$ (where $pq$ undergoes the Delaunay crossing by $r$) to a larger interval. Lemma \ref{Prop:FewPr} implies that each event, that arises within the respective red-blue arrangement $\A_{pr}$ or $\A_{rq}$ during the gap interval, can be traced back to $\SQ$ (via $(pq,r,I)$) in only $O(\ell)$ possible ways.

The main difference between the two subcases stems from condition (S6), according to which $wq$ is almost-Delaunay in the interval $[\t_{wq},\xi_{wq}]$, and is {\it fully Delaunay} at the endpoints $\t_{wq},\xi_{wq}$.
Since the latter interval contains $\delta_1$, in subcase (d1) the corresponding Lemma \ref{Prop:HitQw} (i) immediately yields a Delaunay crossing of $wq$ by (at least) one of the points $p,r$.

In subcase (d2), however, we only know that $pw$ belongs throughout $[\t_{wq},\xi_{wq}]$ to some reduced triangulation $\DT(P\setminus A_{pw})$, where $A_{pw}$ is a subset of cardinality at most $6\ell+3$ which includes $a,r,u$, and perhaps also $q$.
That is, we are not necessarily able to restore the Delaunayhood of $pw$ at times $\t_{wq}$ and $\xi_{wq}$ without removing some of $r,q$, and thereby destroying $\varrho=(p,q,r,w)$.
In fact, it is not even known whether the collinearity mentioned in Lemma \ref{Prop:HitPw} (i) occurs in $[\t_{wq},\xi_{wq}]$ or {\it before} $\t_{wq}$.
In Section \ref{Subsec:TypeB} we use conditions (B1)--(B4) obtained above, to enforce the long-awaited crossing of $pw$ by $q$ or $r$.

\subsection{Stage 4: The number of left special quadruples}\label{Subsec:Stage4Left}
To bound the maximum possible number $\Phi_3^L(n)$ of $3$-restricted right special quadruples, we fix the underlying set $P$ of $n$ moving points, and a refined family $\F$. 

\medskip
\noindent{\bf Topological setup.} 
According to Proposition \ref{Prop:UniquePqw}, any $3$-restricted left special quadruple $\SQ=(a,p,w,q)$ shares its triple $(p,q,w)$ with at most two other such quadruples. (In other words, it suffices to bound the overall number of the corresponding triples $(p,q,w)$.) 
We strengthen the above property, by considering at most {\it one} $3$-restricted left quadruple for each triple $(p,q,w)$.
Therefore, in what follows every special quadruple $\SQ=(a,p,w,q)$ under our consideration will be uniquely determined by its triple $(p,q,w)$.

To proceed, we fix a $3$-restricted left special quadruple $\SQ=(a,p,w,q)$, with respect to $P$ and $\F$, whose two special $(a,q)$-crossings take place during the intervals $\I_r=[\t_0,\t_1]$ and $\J_u=[\t_2,\t_3]$ (in this order), where $r$ and $u$ are the respective outer points. 
Recall that the original ``regular" family $\F$ includes the quadruples $\Q_1=(p,q,a,r)$ and $\Q_2=(w,q,a,u)$.

By assumption, $\SQ$ satisfies the six conditions (S1)--(S2), (S3b), and (S4)--(S6). 
We emphasize that all these conditions, except for (S3b), are common to {\it all} $3$-restricted special quadruples, including the right special quadruples studied in Section \ref{Subsec:Stage4Right}. Moreover, one can switch the roles $p$ and $w$ by reversing the direction of the time axis, so our condition (S3b) of left special quadruples is fully symmetric to condition (S3a) on right special quadruples (which has been assumed throughout the analysis Section \ref{Subsec:Stage4Right}). See below for details. 

\begin{figure}[htbp]
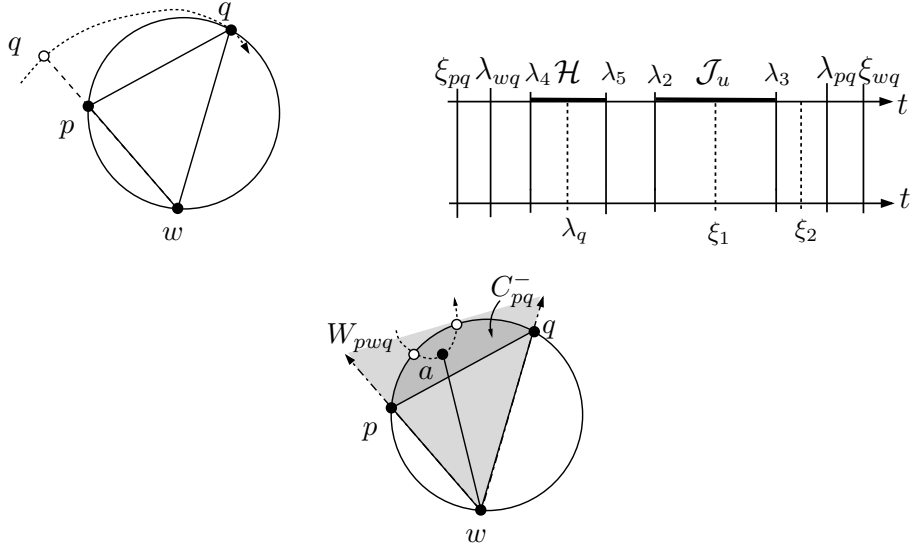

\begin{center}
\input{AlmostCrossingSpecialLeft.pstex_t}\hspace{2cm}\input{TimeLineLeft.pstex_t}\\
\vspace{0.3cm}
\input{TheWedgeLeft.pstex_t}
\caption{\small The topological setup during the interval $(\t_q,\xi_2)\subseteq [\xi_{pq},\xi_{wq}]$. 
Left: The edge $qw$ is hit at some time $\t_q\in [\t_{wq},\t_2)$ by $p$, so it undergoes a Delaunay crossing $(qw,p,\H=[\t_4,\t_5])$ within $\DT(P\setminus \{a,r,u\})$. Right: We have $\xi_{pq}\leq \t_4\leq \t_q<\t_5<\xi_1<\xi_2\leq\t_{pq}$. Bottom: The motion of $B[p,q,w]$ is continuous throughout $(\t_q,\xi_2]$ (the hollow circles represent the co-circularities at times $\xi_{1}$ and $\xi_2$).}
\label{Fig:TopSetupLeft}
\vspace{-0.4cm}
\end{center}
\end{figure}

Refer to Figure \ref{Fig:TopSetupLeft}.
As reviewed in the preceding Section \ref{Subsec:Stage4Left}, the $3$-restrictedness of $\SQ$ implies that there exist times $\t_{wq}\leq \t_0$, $\xi_{pq}\leq \t_{wq}$, $\t_{pq}\geq\t_3$ and $\xi_{wq}\geq\t_{pq}$, whose properties have been summarized in the beginning of that section.
In particular, $pq$ is Delaunay at times $\t_{pq}$ and $\xi_{pq}$, and $wq$
is Delaunay at the symmetric times $\t_{wq}$ and $\xi_{wq}$.
Furthermore, $pq$ and $wq$ are almost Delaunay during, respectively, $[\xi_{pq},\t_{pq}]$ and $[\t_{wq},\xi_{wq}]$.

Let us summarize what we know so far about the motion of $a,p,w,q$ if $\SQ=(a,p,w,q)$ is a $3$-restricted left special quadruple. 
By Condition (S3b), these points are co-circular at times $\xi_0\in \I_r\setminus \J_u$, and $\xi_1\in \J_u\setminus \I_r$, and $\xi_2\in (\t_3,\t_{pq}]$. 
Moreover, the Delaunayhood of $pq$ is violated, throughout $(\xi_1,\xi_2)$, by the points $a\in \L_{pq}^-$ and $w\in \L_{pq}^+$.
In particular, $a$ lies throughout that interval within the wedge $W_{pwq}=\L_{wp}^+\cap \L_{wq}^-$ and inside the cap $C^-_{pq}=B[p,q,w]\cap \L_{pq}^-$.
We emphasize that the order type of the quadruple $(q,p,w,a)$ remains unchanged during $(\xi_1,\xi_2)$.

In addition, by the same Condition (S3b), the smaller set $P\setminus \{a,r,u\}$ yields a (single) Delaunay crossing $(qw,p,\H_{\SQ})$, whose interval $\H=\H_{\SQ}=[\t_4,\t_5]$ is contained in $[\t_{wq},\t_2)$. 
Specifically, $w$ hits $pq$ at some moment\footnote{Recall from Section \ref{Subsec:Stage1Special} that $p$ can cross $qw$ either before or after $\xi_0$, depending on the location of $w$ when $q$ crosses $pa$. 
Our analysis only relies on the fact that $\t_q<\xi_1<\xi_{2}$.} $\t_{q}\in \H$, when $p$ crosses $\L_{wq}$ from $\L^+_{wq}$ to $\L^-_{wq}$. Since $p$ lies in $\L_{wq}^-$ at times $\xi_{1}$ and $\xi_{2}$, no further collinearities of $p,w,q$ can occur during 
$[\t_q,\xi_2)$. (Otherwise, the point $p$ would have to re-enter $\L_{wq}^+$ before $\xi_2$, and then the triple $p,q,w$ would be collinear three times, contrary to our assumptions.)  To conclude,
the disc $B[p,q,w]$ moves continuously throughout the interval $(\t_q,\xi_2]$, which is obviously contained in $[\xi_{pq},\t_{pq}]\cap [\t_{wq},\xi_{wq}]=[\t_{wq},\t_{pq}]$.

\paragraph{Overview.} We fix three constant parameters $k,\ell,h$, such that $12<k\ll \ell \ll h$, and distinguish between four possible cases. The first three cases (a)--(c) are fully symmetric to the cases (a)--(c) that we encountered 
in Section \ref{Subsec:Stage4Right} when handling right quadruples. (Moreover, the first two cases (a) and (b) are very similar to the the corresponding cases (a) and (b) in Section \ref{Subsec:Stage4Reg}.)

In the final, most involved, case (d), we re-introduce at last the outer point $u$. (The other outer point $r$ is not used in the analysis of left special quadruples.) The correspondence between $(wa,q,\J_u)$ and its ancestor quadruple 
$\Q_2=(w,q,a,u)$ in $\F$ implies that we have a single Delaunay crossing $(wq,u,I=[t_0,t_1])$ (which is the first among the two $(w,u)$-crossings of $\Q_2$). 
Since the points $u$ and $p$ cross the same edge $wq$ in opposite directions, $\SQ$ can again be charged to the resulting terminal quadruple $(w,q,u,p)$.

After ruling out cases (a)--(c), we may assume, in the last remaining case (d), that a total of at most $8\ell+1$ points of $P$ appear in the cap $C_{wq}^-=B[p,q,w]\cap \L_{wq}^-$ during $(\t_q,\xi_{2})$. (Notice that this condition is fully symmetric to the one in Lemma \ref{Prop:ShallowCup}.) 

As in Section \ref{Subsec:Stage4Right}, we use the interplay between $\SQ=(a,p,w,q)$ and $\Q_2=(w,q,a,u)$ to enforce as many Delaunay crossings as possible among $w,q,u,p$ before charging $\SQ$ to this terminal quadruple. 
Our analysis is largely simplified\footnote{In contrast, in the {\it almost-}symmetric case of right special quadruples we did not know whether the first crossing $(pq,r,I)$ of $\Q_1=(p,q,a,r)$ at all overlaps $(\xi_{-1},\t_q)$.} by the property that the interval $I=[t_0,t_1]$ of the first crossing of $\Q_2$
is entirely contained in the above interval $(\t_q,\xi_2)$; see below for details. 

We establish symmetric variants of Lemmas \ref{Prop:HitQw} and \ref{Prop:HitPw}.
Namely, we argue that (i) the edge $wu$ is hit in $(\t_q,t_0)$ by at least one of $p,q$, or else (ii) the edge $uq$ is hit in $(\t_q,t_0)$ by at least one of $p,r$.\footnote{These collinearities are fairly symmetric to the crossings of $pr$ and $rq$ that we enforced in Section \ref{Subsec:Stage4Right}.} In the first case (denoted as (d1)), we also show that $u$ hits $qp$ in 
$(\t_q,t_0)$. In the second case (denoted as (d2)) we similarly show that $u$  also hits $pw$ in $(\t_q,t_0)$. In both scenarios, we invoke Theorem \ref{Thm:RedBlue} to amplify the above two additional collinearities into full-fledged Delaunay crossings.
Therefore, by the time we charge $\SQ$ to the terminal quadruple $\TQ$, its various sub-triples among $w,q,u,p$ perform four Delaunay crossings (where some of these crossings occur in appropriately reduced subsets of $P$).

In Section \ref{Sec:Terminal} we express the number of such terminal quadruples, which arise in the analysis of left special quadruples, in terms of more elementary quantities, that were introduced in Section \ref{Sec:Prelim}. To do so, we enforce an additional, fifth crossing among $w,q,u,p$ (namely, the crossing of $pu$ by $w$ or $q$). As a result, some sub-triple among $w,q,u,p$ is involved in two Delaunay crossings, so our analysis bottoms out via Lemma \ref{Lemma:TwiceCollin}.

\begin{figure}[htbp]
\begin{center}
\input{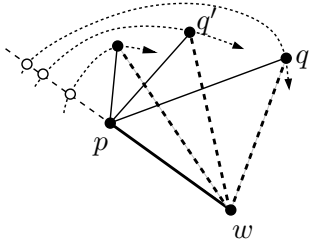}
\caption{\small Each left special quadruple $\SQ'=(a',p,w,q')\in G_{pw}^L$ (with respective outer points $r'$ and $u'$) comes with a counterclockwise $(w,p)$-crossing $(q'w,p,\H_{\SQ'})$, which occurs within $\DT(P\setminus \{a',r',u'\})$.}
\label{Fig:PwCrossingsLeft}
\vspace{-0.4cm}
\end{center}
\end{figure}

\bigskip
In what follows, we consider the family $\G^L_{pw}$ of all $3$-restricted left special quadruples of the form $\SQ'=(a',p,w,q')$, which share their middle pair with $\SQ$. 
We may assume that each $\SQ'=(a',p,w,q')\in \G^L_{pw}$ is uniquely determined by the choice of $q'$ (as the only ``free" point in the triple $(p,q',w)$). 
Note that the set $P_{\SQ'}$ of each $\SQ'$ includes, in addition to the four points $a',p,w,q'$ of $\SQ'$, the respective outer points $r'$ and $u'$ of its special crossings $(pa',q',\I_{r'})$ and $(wa',q',\J_{u'})$.
Furthermore, each of these quadruples $\SQ'\in \G^L_{pw}$ is accompanied by a counterclockwise $(w,p)$-crossing $(q'w,p,\H_{\SQ'}=\H')$, which occurs within the smaller triangulation $\DT(P\setminus\{a',r',u'\})$. See Figure \ref{Fig:PwCrossingsLeft}.
We use $\t_{q'}$ to denote the time in $\H'$ when the respective point $q'$ of $\SQ'$ enters the halfplane $\L_{wp}^+$ (or, equivalently, when $p$ crosses $q'w$ from $\L_{wq'}^+=\L_{q'w}^-$ to $\L_{q'w}^+$).

Notice that Lemma \ref{Lemma:OrderRelaxedCrossings} readily generalizes to the above $(w,p)$-crossings. Namely, a pair of such crossings $(qw,p,\H_{\SQ})$ and $(q'w,p,\H_{\SQ'})$, which occur within the respective triangulations $\DT(P\setminus \{a,r,u\})$ and $\DT(P\setminus \{a',r',u'\})$, are {\it compatible}, provided that $q'\neq a,r,u$ and $q\neq a',r',u'$, in the sense that the orders in which the intervals $\H_{\SQ}$ and $\H_{\SQ'}$ begin or end 
are both consistent with the time stamps $\t_q$ and $\t_{q'}$.

Clearly, for any special quadruple $\SQ=(a,p,w,q)\in \G^L_{pw}$ (with outer points $r$ and $u$) the family $\G^R_{pw}$ includes at most three other quadruples $\SQ'=(a',p,w,q')$ whose respective points $q'$ are equal to  
one of $a,r$ or $u$. The pigeonhole principle then implies that at least {\it one quarter} of all quadruples $\SQ=(a,p,w,q)$ in $\G^L_{pw}$ satisfy the following condition:

\smallskip
\noindent {\bf (PHL1)} \noindent{\it There exist at most three quadruples $\SQ'\in \G^L_{pw}$ with $q\in \{a',r',u'\}$.}

\smallskip
Since $p$ and $w$ are arbitrary points of $P$, (PHL1) holds for at least a quarter of all $3$-restricted left special quadruples under consideration; hence we may assume that it holds for the special quadruple $\SQ$ at hand.
Therefore, for all but $6$ quadruples $\SQ'=(a',p,w,q')\in \G^L_{pw}\setminus \{\SQ\}$ (with respective outer points $r'$ and $u'$) their respective $(w,p)$-crossings $(q'w,p,\H_{\SQ'})$ are compatible with $(qw,p,\H)$ via a suitable extension of Lemma \ref{Lemma:OrderRelaxedCrossings}.

\medskip
With the above preparations, we can now proceed with our case analysis.

\begin{figure}[htbp]
\begin{center}
\input{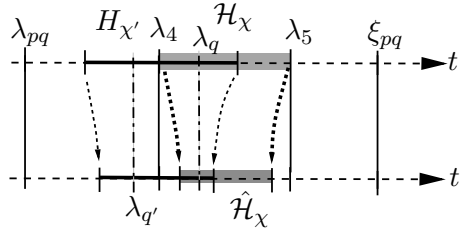}
\caption{\small Case (a): At least $k$ counterclockwise $(w,p)$-crossings $(q'w,p,\H_{\SQ'})$ either begin in $[\xi_{pq},\t_4)$ or end in $(\t_5,\t_{pq}]$ (one such crossing of the former type is depicted). Then, with some fixed and positive probability, the sample $\hat{P}$ yields a Delaunay crossing $(qw,p,\hat{\H}_{\SQ})$ that is $(q,p,\Theta(k))$-chargeable with respect to $[\xi_{pq},\t_{pq}]$.}
\label{Fig:TimelinePwLeft}
\vspace{-0.4cm}
\end{center}
\end{figure}

\medskip
\noindent{\bf Case (a).} For at least $k$ of the above quadruples $\SQ'=(a',p,w,q')\in \G^L_{pw}$, their respective $(w,p)$-crossings $(q'w,p,\H')$ either begin in $[\xi_{pq},\t_4)$, or end in $(\t_5,\t_{pq}]$. Refer to Figure \ref{Fig:TimelinePwLeft}. 
Recall that, by condition (S5), the edge $pq$ is Delaunay at each of the times $\xi_{pq}$ and $\t_{pq}$, and that it is almost Delaunay during the entire interval $[\xi_{pq},\t_{pq}]$.

To bound the number of such quadruples $\SQ$ that fall into case (a), we pass to a random sub-sample $\hat{P}$ of $n/4$ points in $P$, and
argue that, with some fixed positive probability,
the crossing $(qw,p,\H)$ becomes $(q,p,\Theta(k))$-chargeable there, for the reference interval $[\xi_{pq},\t_{pq}]$.
Therefore, Theorem \ref{Thm:Balanced} implies that the overall number of such triples $(p,q,w)$ in $P$ does not exceed
$$
O\left(k^2N(n/k)+kn^2\beta(n)\right),
$$ 
which also bounds the overall number of the corresponding $3$-restricted left special quadruples $\SQ$.

\paragraph{Preparing for cases (b) and (c): Charging events in $\A_{pw}$.}
We may assume, from now on, that there exist at most $k$ special quadruples $\SQ'\in \G_{pw}^L$ whose respective $(w,p)$-crossings $(q'w,p,\H')$ either begin in $[\xi_{pq},\t_4)$, or end in $(\t_5,\t_{pq}]$.

Before proceeding to the following cases, we apply Theorem \ref{Thm:RedBlue} in $\A_{pw}$ in order to extend the almost-Delaunayhood of $pw$ from $\H=[\t_4,\t_5]$ to $[\xi_{pq},\t_{pq}]$. We emphasize that $[\xi_{pq},\t_{pq}]\setminus \H$
consists of two intervals $[\xi_{pq},\t_4)$ and $(\t_5,\t_{pq}]$ (where the former interval can be empty), which we consider separately.
Note also that the edge $pw$ belongs during $\H$ to the reduced triangulation $\DT(P\setminus \{a,r,u\})$ (but not necessarily to $\DT(P)$), so Theorem \ref{Thm:RedBlue} must be applied, for each of these two intervals, with respect to $P\setminus \{a,r,u\}$.

In each of these applications, in cases (i) and (ii) we charge $\SQ$ (via its respective $(w,p)$-crossing $(qw,p,\H)$)
to $(\ell+3)$-shallow collinearities and co-circularities that occur in the full red-blue arrangement $\A_{pw}$. Since case (a) has been ruled out, the charging is almost unique, and accounts for at most $O\left(k\ell^2N(n/\ell)+k\ell n^2\beta(n)\right)$ left special quadruples.

At the end, we have either disposed of $\SQ$ through (conditions (i), (ii) of) Theorem \ref{Thm:RedBlue} or ended up with a set $A_{pw}$ of at most $6\ell+3$ points whose removal restores the Delaunayhood of $pw$ throughout $[\xi_{pq},\t_{pq}]$. Namely, $A_{pw}$ is composed of $a,r,u$, and of the two sets of at most $3\ell$ points each, which are obtained by separately applying Theorem \ref{Thm:RedBlue}, within $\A_{pw}$, over the intervals $(\xi_{pq},\t_4)$ and $(\t_5,\t_{pq})$.
Hence, we may assume, in what follows, that the above set $A_{pw}$ exists.

\begin{figure}[htbp]
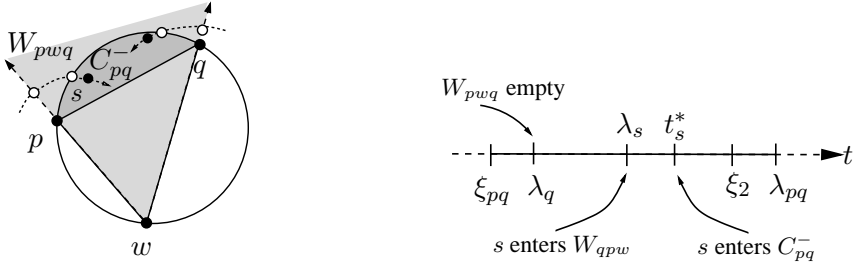

\begin{center}
\input{EnterTheWedgeLeft.pstex_t}\hspace{2.5cm}\input{TimelineEnterLeft.pstex_t}
\caption{\small Case (b). At least $\ell$ points $s\neq a,r,u$ visit the cap $C_{pq}^-$ during $(\t_q,\xi_2)$. 
Each of them must enter the wedge $W_{pwq}$ (through one of the rays $\vec{wp},\vec{wq}$, outside the respective edges $pw$ and $wq$) after time $\t_q$ and then enter the cap $C_{pq}^-$ (through the boundary of $B[p,q,w]$).}
\label{Fig:EnterTheWedgeLeft}
\vspace{-0.4cm}
\end{center}
\end{figure} 

\medskip
\noindent{\bf Case (b).} A total of at least $\ell$ points of $P$, distinct from $a,r,u$, appear in the cap $C^-_{pq}=B[p,q,w]\cap \L_{pq}^-$ at some time during the interval $(\t_q,\xi_2)$. (Note that some of these points $s$ may belong to $A_{pw}$.) Recall that $\t_q$ denotes the time in $\H$ when $p$ enters $\L_{wq}^-$, through $wq$, and that no additional collinearities of $p,q,w$ can occur during $(\t_q,\xi_2)$, so the motion of $B[p,q,w]$ is fully continuous in that interval.

Refer to Figure \ref{Fig:EnterTheWedgeLeft}.
Let $s\in P\setminus \{a,r,u\}$ be one of the points that visit $C^-_{pq}$ during $(\t_q,\xi_2)$. Since the above cap $C^-_{pq}$ is fully contained there in the wedge $W_{pwq}=\L_{wp}^+\cap \L_{wq}^-$, $s$ must enter $W_{pwq}$ after time $\t_q$ (when $W_{pwq}$ co-incides with the single ray $\vec{wp}=\vec{wq}$) through one of the rays $\vec{wp},\vec{wq}$. We also note that, by condition (S5) (and since $(\t_q,\xi_2)\subseteq [\xi_{pq},\t_{pq}]$), the edge $pq$ is Delaunay in $P\setminus \{a,w,r,u\}$ throughout $(\t_q,\xi_2)$, so $s$, which has to enter $C^-_{pq}$ before it enters $W_{pwq}$, can do so only through the boundary of $B[p,q,w]$. This results in a co-circularity of $p,q,w,s$, and is easily seen to imply that $s$ enters $W_{pwq}$ by crossing one of the rays $\vec{wp}$ or $\vec{wq}$ {\it outside} the respective edges $wp$ or $wq$. 

In what follows, we assume that $s$ is among the last $\ell$ points to leave $C_{pq}^-$ during $(\t_q,\xi_2)$.
Let $t_{s}^*$ denote the time of the corresponding co-circularity of $p,q,w,s$, which occurs when $s$ leaves $C^-_{pq}$ through the boundary of $B[p,q,w]$. Since $\SQ$ satisfies condition (S5), the opposite cap $C^+_{pq}=B[p,q,w]\cap \L_{qp}^+$ 
contains no points of $P\setminus \{a,r,u\}$ at time $t_s^*$. (Otherwise, the Delaunayhood of $wq$ would be violated by $s$ and any of these points.) Therefore, the co-circularity at time $t_s^*$ has to be $(\ell-1)$-shallow in $P\setminus \{a,r,u\}$, and thus $(\ell+2)$-shallow in $P$.

Note also that the co-circularity at time $t_s^*$ is red-blue with respect to the edge $pq$, which is violated right before it by $w$ and $s$. Lemma \ref{Lemma:Crossing}, together with the choice of $s\neq a,r,u$, imply 
that this co-circularity cannot occur during the crossing $(qw,p,\H_{\SQ}=[\t_4,\t_5])$ (which occurs in $P\setminus \{a,r,u\}$), so $t_s^*>\t_5$. 


As in the symmetric case (b) of Section \ref{Subsec:Stage4Right}, we distinguish between two possible subcases. In each of them we manage to dispose of $\SQ$ by charging it, within one of the arrangements $\A_{wq}, \A_{pw}$, either to $\Omega(\ell^2)$ $(2\ell)$-shallow co-circularities, or to a $(2\ell)$-shallow collinearity.

\medskip
\noindent {\bf Case (b1).} At least half of the above points $s$ cross the line $\L_{wq}$, from $\L_{wq}^+$ to $\L_{wq}^-$, during $(\t_q,t_s^*)$. (This also includes points $s$ that possibly cross $\L_{wq}$ outside the ray $\vec{wq}$, before entering $W_{pwq}$ through the other ray $\vec{wp}$.) By Condition (S6) (and since $(\t_q,t_s^*)\subseteq (\t_q,\xi_2)\subseteq [\t_{wq},\xi_{wq}]$), each of these crossings occurs outside $wq$, within one of the outer rays of $\L_{wq}$.

For each $s$ we argue, exactly as in Section \ref{Subsec:Stage4Reg}, that the points $w,q,s$ are involved during $(\t_q,t_s^*)\subseteq (\t_q,\xi_2)$ either in a $(2\ell)$-shallow collinearity, or in $\Omega(\ell)$ $(2\ell)$-shallow co-circularities. That is, right after $s$ enters $\L_{wq}^-$ at time $\t_q$ (outside $wq$), the disc $B[w,q,s]$ ``swallows" the entire halfplane $\L_{wq}^+$. (In addition, $s$ must remain in $\L_{wq}^-$ until time $t_s^*$, for otherwise the points $w,q,s$ would be collinear more than twice.) If this disc, which contains at most $\ell+2$ points at the end of the process, contains at least $2\ell$ points at time $\t_q$, then each of the last $\ell-2$ resulting co-circularities are $(2\ell)$-shallow (in $P$). Otherwise, the collinearity of $q,p,s$ is $(2\ell)$-shallow.

Since $s$ can be chosen in at least $\Omega(\ell)$ different ways, the points $w$ and $q$ are involved during $(\t_q,\xi_2)$ either in $\Omega(\ell^2)$ $(2\ell)$-shallow co-circularities, or in a $(2\ell)$-shallow collinearity.
In both cases, we charge $\SQ$ to these events. 

Note that each $(2\ell)$-shallow event, which occurs in $\A_{pq}$ at some time $t^*\in (\t_q,\xi_2)$, can be traced back to $(qw,p,\H)$ (and, by Proposition \ref{Prop:UniquePqw}, also to $\SQ$) in at most $O(1)$ possible ways because $p$ is among the last four points to hit the edge $wq$ before time $t^*$, according to condition (S6). 
Hence, the above scenario happens for at most $O(\ell^2N(n/\ell)+\ell n^2\beta(n))$ special quadruples $\SQ$.

\begin{figure}[htbp]
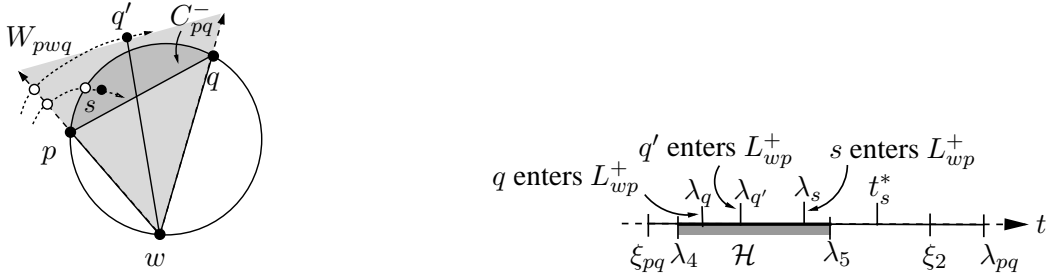

\begin{center}
\input{CandidatePropLeft.pstex_t}\hspace{3cm}\input{TimeLineLeftProp.pstex_t}
\caption{\small Proposition \ref{Prop:kUniqueSpecialLeft}. Left: $q$ is among the last $k+7$ candidates $q'$ to enter $\L_{wp}^+$ before time $\t_s$. 
Right: The various critical events occur in the depicted order. Note that $\t_s$ may occur in (the second part of) $\H=[\t_4,\t_5]$.
}
\label{Fig:PropositionCandidatesLeft}
\vspace{-0.5cm}
\end{center}
\end{figure} 

\medskip
\noindent {\bf Case (b2).} At least half of the above points $s\neq a,r,u$ remain in $\L_{wq}^-$ throughout the respective intervals $(\t_q,t_s^*)$. Each of these points must enter $W_{pwq}$, also during $(\t_q,t_s^*)$, through the ray emanating from $p$ in direction $\vec{wp}$, thereby crossing $\L_{pw}$ from $\L_{wp}^-$ to $\L_{wp}^+$. (See Figure \ref{Fig:PropositionCandidatesLeft} (left). Recall that such a collinearity can occur at most once, because the triple $p,w,s$ can be collinear at most twice.)

We again fix one of these points $s$, and use $\t_s$ to denote the corresponding time in $(\t_q,t_s^*)$ when $s$ enters $W_{pwq}$ through the ray emanating from $w$ in direction $pw$. As in the previous case, we conclude that either the collinearity of $p,w,s$ at time $t_s$ is $(2\ell)$-shallow, or the points $p,w,s$ are involved in $\Omega(\ell)$ $(2\ell)$-shallow co-circularities during the preceding interval $(\t_s,t_s^*)$. 
As in the matching scenarios (b2) in Sections \ref{Subsec:Stage4Reg} and \ref{Subsec:Stage4Right}, the main challenge is to argue that each of the above $(2\ell)$-shallow events, which occur in $\A_{pw}$ during $(\t_s,t_s^*]\subseteq (\t_q,\xi_2)$, can be traced back to $\SQ$ in at most $O(k)$ ways.

To show this, let $t^*\in (\t_q,\xi_2)$ be the time of a $(2\ell)$-shallow collinearity or co-circularity that occurs in $\A_{pw}$. 
First, we guess the points $p$ and $w$ of $\SQ$ in $O(1)$ possible ways among the three or four points involved in the event. 
We next recall that, in the charging scheme of case (b2), each $(2\ell)$-shallow co-circularity or collinearity or collinearity that we charge in $\A_{pw}$ is obtained via some point $s$, which is also involved in the event, that enters $\L_{wp}^+$ at the respective time $\t_s$. We, therefore, guess $s$ among the remaining one or two points involved in the event at time $t^*$.
To guess the remaining points $a$ and $q$ of $\SQ$, we examine all ``candidate" special quadruples $\SQ'\in \G_{pw}^L$ whose two ``middle" points $(p,w)$ are shared with $\SQ$. 
Recall that each of these quadruples is accompanied by the $(w,p)$-crossing $(q'w,p,\H'=\H_{\SQ'})$, where $q'$ enters $\L_{wp}^+$ at the respective time $\t_{q'}\in \H'$. 
Recall also that $\SQ'$ is uniquely determined by the choice of $q'$ (as long as $p$ and $w$ remain fixed).

Clearly, it suffices to consider only special quadruples $\SQ'=(a',p,w,q')$ in $\G_{pw}^L$ with the following properties: 
(1) $s\neq a',r',u'$, where $r'$ and $u'$ are the outer points of $\SQ'$, (2) $\t_{q'}<\t_s$, and (3) $s$ lies in $\L_{wq'}^-$ during the second portion of $\H_{\SQ'}$ (after $\t_{q'}$). This is because each of these conditions holds for $\SQ$ and $s$ in the charging scheme of case (b2). For example, (3) follows because case (b1) does not occur for $s$ (and $t_s^*>\t_5$). 

If a special quadruple $\SQ'=(a',p,w,q')\in \G_{pw}^L$ satisfies the above three conditions (1)--(3), we say that the respective point $q'$ (which uniquely determines $\SQ'$) is a candidate (for $q$).



The following symmetric variant of Proposition \ref{Prop:kUniqueSpecial} guarantees that each $(2\ell)$-shallow event, which occurs in $\A_{pw}$ at some fixed time $t^*\in (\t_q,\xi_2)$, is charged by at most $k+7$ quadruples in $\SQ'\in \G_{pw}^L$, because its points $q$ is among the last $k+7$ similar candidates $q'$ to enter $\L_{wp}^+$ before time $\t_s$. See Figure \ref{Fig:PropositionCandidatesLeft}.

\begin{proposition}\label{Prop:kUniqueSpecialLeft}
With the above assumptions, the point $q$ is among the last $k+7$ candidates $q'$ to enter the halfplane $\L_{wp}^+$ before $\t_s$.
\end{proposition}

We omit the fairly technical proof of Proposition \ref{Prop:kUniqueSpecialLeft}, noting that it is fully symmetric to the proof of Proposition \ref{Prop:kUniqueSpecial}, and very similar to the proof of Proposition \ref{Prop:kUnique}.

Repeating the same charging argument for each of the $\Omega(\ell)$ possible choice of $s$ shows that at most\\
$O\left(k\ell^2N(n/\ell)+k\ell n^2\beta(n)\right)$
special quadruples can fall into case (b2). 



\bigskip
\noindent{\bf Case (c).} A total of at least $\ell$ points $s\in P\setminus A_{pw}$ appear in the cap $C^-_{wp}=B[p,q,w]\cap \L_{wp}^-$ at some time during $(\t_q,\xi_2)$. Here $A_{pw}$ continues to denote the subset of at most $6\ell+3$ points, including $a,r$ and $u$, whose removal restores the Delaunayhood of $pw$ throughout the interval $[\xi_{pq},\t_{pq}]$. (Recall that $A_{pw}$ was obtained by applying Theorem \ref{Thm:RedBlue} in $\A_{pw}$, after ruling out case (a).)

\begin{figure}[htbp]
\begin{center}
\input{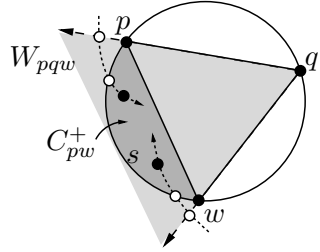}
\caption{\small Case (c). A total of at least $\ell$ points $s\in P\setminus A_{pw}$ enter in the cap $C^-_{wp}$ during $(\t_q,\xi_2)$. 
Each of them must enter the wedge $W_{pqw}$ (through one of the rays $\vec{qp},\vec{qw}$, outside the respective edges $pq$ and $wq$), and only then cap $C_{wp}^-$ (through the boundary of $B[p,q,w]$).}
\label{Fig:EnterTheWedgeWp}
\vspace{-0.4cm}
\end{center}
\end{figure} 

Clearly, $C^-_{wp}$ is contained in the wedge $W_{pqw}=\L_{pq}^+\cap \L_{wq}^-$, which shrinks at time $\t_q$ to the ray $\vec{qp}=\vec{qw}$. Hence, each of these points $s$ has to enter $W_{pqw}$ and $C^-_{wp}$ (in this order) before time $\t_q$. Furthermore, $s$ can leave $C^+_{pw}$ only through the boundary of $B[p,q,w]$, at a co-circularity of $p,q,w,s$. (Otherwise $s$ would have to hit $pw$ and, therefore, belong to $A_{pw}$.)
In addition, $s$ can leave $W_{pqw}$ only through one of the rays $\vec{qp}$ and $\vec{qw}$ (outside the respective segments $qp,qw$). 
See Figure \ref{Fig:EnterTheWedgeWp}.

As in the previous case (b), we may assume that each $s$ under consideration is among the first $\ell$ such points of $P\setminus A_{pw}$ to enter $C_{wp}^-$ during $(\t_q,\xi_2)$, and use $t_s^*$ to denote the time of the respective co-circularity.
Clearly, the opposite cap $C^+_{wp}=B[p,q,w]\cap \L_{wp}^+$ contains then no points of $P\setminus A_{pw}$. Indeed, otherwise the Delaunayhood of $pw$ would be violated by $s$ and any one of these points (contrary to our assumption that $pw\in \DT(P\setminus A_{pw})$ throughout $[\xi_{pq},\t_{pq}]\supset (\t_q,\xi_2)$).
Hence, the resulting co-circularity of $p,q,w,s$ at time $t_s^*$ is $(7\ell+2)$-shallow in $P$, because, at the time of co-circularity, the circumdisc $B[p,q,w]=B[p,s,w]$ can contain in its interior at most $6\ell+3$ points of $A_{pw}$ and at most $\ell-1$ points of $P\setminus A_{pw}$.

\medskip
\noindent {\bf Case (c1).} If at least half of the above points $s$ cross the line $\L_{wq}$ (from $\L_{wq}^+$ to $\L_{wq}^-$) during their respective intervals $(\t_q,t_s^*)$,
then we argue exactly as in subcase (b1). 

Namely, we fix one of the these points $s$ and notice that, right after $s$ enters $\L_{wq}^-$ outside $wq$, the disc $B[w,q,s]$ contains the entire halfplane $\L_{wq}^-$.
Therefore, the points $p,q,s$ are involved, during $(\t_{q},t_s^*)$, either in an $(8\ell)$-shallow common collinearity (which occurs when $s$ enters $\L_{wq}^-$), or in $\Omega(\ell)$ $(8\ell)$-shallow co-circularities. 

We repeat the above argument for each of the $\ell/2$ possible choices of $s$ and charge $\SQ$ within $\A_{wq}$ (via $(qw,p,\H)$) to the above $(8\ell)$-shallow events.
As in case (b1), each $(8\ell)$-shallow collinearity or co-circularity occurs during $(\t_q,\xi_2)$, and involves $w$ and $q$, so it is charged by at most $O(1)$ special quadruples $\SQ$ (because $\SQ$ is uniquely determined by $(p,q,w)$ and $p$ is among the last four points to hit $wq$ before the respective time $t^*$ of the event). 


\medskip
\noindent {\bf Case (c2).} We may assume, then, that at least half of the above points $s$ enter $W_{pqw}$ through the ray $\vec{qp}$.
For each of these points $s$, the triple $q,p,s$ are involved during $(\t_q,t_s^*)$ either in an $(8\ell)$-shallow collinearity, or in $\Omega(\ell)$ $(8\ell)$-shallow co-circularities. As before, we repeat the above argument for the $\ell/2$ eligible choices of $s$ and charge $\SQ$, within $\A_{pq}$, either to $\Omega(\ell^2)$ $(8\ell)$-shallow co-circularities or to an $(8\ell)$-shallow collinearity. 

We claim that each of the resulting $(8\ell)$-shallow events, which occur in $\A_{pq}$ during $(\t_q,\xi_2)$, can be traced back to $\SQ$ in at most $O(1)$ possible ways.
Indeed, fix any of the above events, at some time $t^*\in (\t_q,\xi_2)$.
We first guess $p$ and $q$ in $O(1)$ possible ways among the three or four points involved in the event.

To guess the point $a$ (which would immediately determine $(pa,q,\I_r)$ and thereby also $\SQ$), we consider all special $(p,q)$-crossings $(pa',q,\I_{r'})$ (in $\F$) and recall that, according to conditions (S1) and (S5), at most $O(1)$ such crossings can begin during $[\xi_{pq},\t_{0})$ or end during $(\t_1,\t_{pq}]$. 
Notice also that the interval $[\t_{pq},\xi_{pq}]$, which covers $(\t_q,\xi_2)$, is contained in the union of $[\xi_{pq},\t_{0})$, $\I_r=[\t_0,\t_1]$, and $(\t_1,\t_{pq}]$. 

To guess $a$ (based on $t^*,q$ and $p$), we distinguish between two possible situations. As before, our analysis is 
fully symmetric to that given in case (c2) of Section \ref{Subsec:Stage4Right}, so we only briefly review it.

\begin{figure}[htbp]
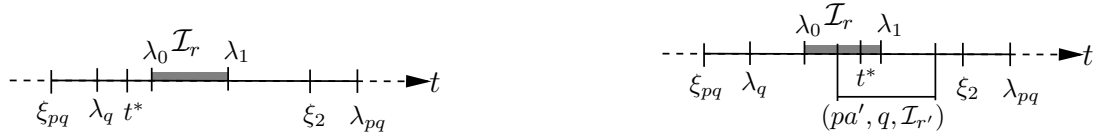

\begin{center}
\input{CaseCLeft1.pstex_t}\hspace{3cm}\input{CaseCLeft2.pstex_t}
\caption{\small Case (c2): Guessing $a$ based on $t^*$, $p$ and $q$. Left: If $t^*\in (\t_q,\t_0)$, then $(pa,q,\I_r=[\t_0,\t_1])$ is among the first $O(1)$ special clockwise $(p,q)$-crossings to 
begin after $t^*$. Right: If $t^*\in [\t_0,\xi_2)$, then $(pa,q,\I_r)$ is among the last $O(1)$ special clockwise $(p,q)$-crossings to begin before (or at) $t^*$.} 
\label{Fig:CaseCLeft}
\vspace{-0.4cm}
\end{center}
\end{figure}

\noindent(i) If $t^*$ belongs to $(\t_q,\t_0)\subseteq [\xi_{pq},\t_0)$ then $(pa,q,\I_r=[\t_0,\t_1])$ is among the last $O(1)$ special clockwise $(p,q)$-crossings to 
begin after $t^*$, because $\SQ$ satisfies condition (S5). See Figure \ref{Fig:CaseCLeft} (left).

\noindent(ii) If $t^*$ belongs to the interval $[\t_0,\xi_{2}]$, which is contained in $\I_r\cup (\t_1,\t_{pq}]$,  
then we resort to a more subtle argument (which is fully symmetic to the one given in case (c1) of Section \ref{Subsec:Stage4Right}) to show that $(pa,q,\I_r)$ is among the last $O(1)$ special clockwise $(p,q)$-crossings to 
begin before $t^*$. See Figure \ref{Fig:CaseCLeft} (right).

\medskip
To recap, in each of the cases (c1) and (c2) we charge $\SQ$ (via $(pa,q,\I_r)$) either to $\Omega(\ell^2)$ $(8\ell)$-shallow co-circularities, or to an $(8\ell)$-shallow collinearity, which occur in one of the arrangements $\A_{pq}, A_{wq}$ during the interval $(\t_q,\xi_2)$. Furthermore, each $(8\ell)$-shallow event is charged by at most $O(1)$ special quadruples. Hence, at most 
$O\left(\ell^2N(n/\ell)+\ell n^2\beta(n)\right)$ 
special quadruples $\SQ$ fall into case (c).


\bigskip
\noindent{\bf Case (d).} Assume that none of the preceding cases occurs. 
In particular, there is a subset $A_{pw}$ of at most $6\ell+3$ points (including $a$, $r$ and $u$) whose removal restores the Delaunayhood of $pw$ throughout the interval $[\xi_{pq},\t_{pq}]$. Furthermore, a total of fewer than $\ell$ points of $P\setminus \{a,r,u\}$ appear in the cap $C^-_{pq}$ during $(\t_q,\xi_2)$, and a total of fewer than $\ell$ points of $P\setminus A_{pw}$ points appear in the cap $C^-_{pw}$ during that interval.

\begin{figure}[htbp]
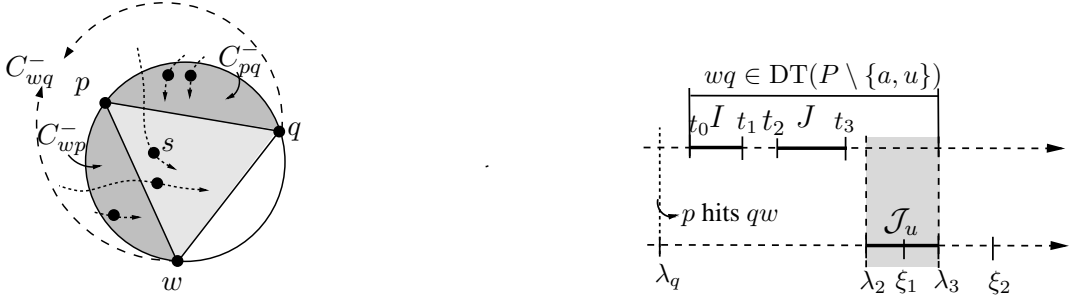

\begin{center}
\input{WqCap.pstex_t}\hspace{2cm}\input{TwoQuadsInterplayLeft.pstex_t}
\caption{\small Left: Lemma \ref{Prop:ShallowCupLeft}. A total of at most $8\ell+1$ points $s$ of $P$ appear in the cap $C_{wq}^-=B[p,q,w]\cap \L_{wq}^-$ (consisting of all the shaded portions) during $(\t_q,\xi_2)$. 
All of these points must enter $C_{wq}^-$ after $\t_q$, and none of them can enter
$C_{wq}^-$ through $wq$, unless it is one of $a,r,u$. Right: The regular quadruple $\Q_2$ of $(wa,q,\J_u)$ is composed of two $(w,u)$-crossings $(wq,u,I)$, $(wa,u,J)$, which end before the beginning time $\t_2$ of $\J_u$. By condition (Q8), the edge $wq$ belongs to $\DT(P\setminus \{a,u\})$ throughout $[t_0,\t_3]$, implying that $\t_q<t_0$. }
\label{Fig:WqCapShallow}\label{Fig:TwoQuadsInterplayLeft}
\vspace{-0.4cm}
\end{center}
\end{figure}

The above assumptions imply the following symmetric variant of Lemma \ref{Prop:ShallowCup}, whose proof is also fully symmetric to its predecessor (see Figure \ref{Fig:WqCapShallow} (left)).

\begin{lemma}\label{Prop:ShallowCupLeft}
With the above assumptions, a total of at most $8\ell+1$ points of $P$ appear in the cap $C_{wq}^-=C[p,q,w]\cap \L_{wq}^-$ during $(\t_q,\xi_2)$.
\end{lemma}

With the above preparations, we can finally describe the interplay between the special quadruple $\SQ$ under consideration and the ordinary Delaunay quadruple $\Q_2=(w,q,a,u)$ in $\F$, which corresponds to the {\it second} special $(a,q)$-crossing $(wa,q,\J_u=[\t_2,\t_3])$ of $\SQ$. At the end of this section, we shall charge $\SQ$ to the terminal quadruple $\varrho=(w,q,u,p)$, which is composed of the edge $wq$, and of the two points $u$ and $p$ that cross $pq$ in opposite directions. 
As in the case of right special quadruples, we first try to enforce as many Delaunay crossings as possible among $w,q,u,p$, before charging $\SQ$ to this terminal quadruple. 

\begin{figure}[htbp]
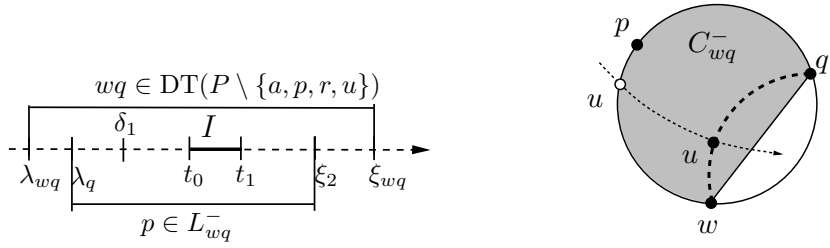

\begin{center}
\input{SchemeDelta1Left.pstex_t}\hspace{2cm}\input{Delta1Left.pstex_t}
\caption{\small Left: Proposition \ref{Prop:EntirelyWithin}: The interval $I=[t_0,t_1]$ (where $wq$ undergoes a crossing by $u$) is fully contained in the interval $(\t_q,\xi_2)$, during which $p$ lies in $\L_{wq}^-$. Right: The cap $C_{wq}^-$ is empty right after time $\t_q$, so $u$ must enter $C_{wq}^-$ before the beginning $t_0$ of $I$. Unless $u$ crosses $qw$ in $(\t_{wq},t_0)\supset (\t_q,t_0)$, $u$ must enter $C_{wq}^-$ at a blue-blue co-circularity of $w,q,u,p$ with respect to $wq$, at some time $\delta_1\in (\t_q,t_1)$.}
\label{Fig:Delta1Left}
\vspace{-0.4cm}
\end{center}
\end{figure}

Recall that the quadruple $\Q_2=(w,q,a,u)$ belongs to the refined family $\F$, so it satisfies the eight properties (Q1)-(Q8). (Refer to Figure \ref{Fig:TwoQuadsInterplayLeft} (right).) Specifically, $\Q_2$ is composed of two clockwise $(w,u)$-crossings $(wq,u,I=[t_0,t_1])$ and $(wa,u,J=[t_2,t_3])$, where $I$ ends before the end $t_3$ of $J$, and $J$ ends before the beginning $\t_2$ of $\J_u$.
(In particular, $\J_u$ is disjoint from both of $I,J$.)
Since $\Q_2$ satisfies condition (Q8), the edge $wq$ belongs to $\DT(P\setminus \{a,u\})$ throughout the interval $\conn{I}{\J_u}={\sf conv}(I\cup\J_u)=[t_0,\t_3]$.
Therefore (and since $\t_q<\t_2<\t_3$), the point $p$ can cross $wq$ (at time $\t_q$, and from $\L_{wq}^+$ to $\L_{wq}^-$) only before the beginning $t_0$ of $I$, and the entire crossing $(qw,p,\H=[\t_4,\t_5])$ occurs in $P\setminus \{a,r,u\}$ before $I$. 
We thus obtain the following important property of $3$-restricted left special quadruples (see Figure \ref{Fig:Delta1Left} (left)):\footnote{Though it is not necessary for our analysis, Proposition \ref{Prop:EntirelyWithin} holds for {\it all} $1$-restricted left special quadruples $\SQ=(a,p,w,q)$, with respective outer points $r$ and $u$.}

\begin{proposition}\label{Prop:EntirelyWithin}
With the above assumptions, the first Delaunay crossing $(wq,u,I=[t_0,t_1])$ occurs entirely within $(\t_q,\t_2)\subset (\t_q,\xi_1)\subset (\t_q,\xi_2)$. In particular, $p$ crosses $wq$ at time $\t_q$ (from $\L_{wq}^+$ to $\L_{wq}^-$) before $u$ does so in the opposite direction (during $I$, from $\L_{wq}^-$ to $\L_{wq}^+$).
\end{proposition}

Recall that $p$ remains in $\L_{wq}^-$ throughout the interval $(\t_q,\xi_2)$, which contains $I$; see Figure \ref{Fig:Delta1Left} (left). Note that the open cap $B[w,q,u]\cap \L_{wq}^-$ contains no points of $P$ at time $t_0$ (when the Delaunay crossing of $wq$ by $u$ begins). Hence, $u$ lies at that moment within the cap $C_{wq}^-=B[p,q,w]\cap \L_{wq}^-$; see Figure \ref{Fig:Delta1Left} (right).
Since $C_{wq}^-$ is empty right after time $\t_q$, the point $u$ has to enter $C_{wq}^-$ in the interval $(\t_q,t_0)$.

Assume first that $u$ hits $wq$ during $(\t_{wq},t_0)$. (In particular, this includes the scenario where 
$u$ enters $C_{wq}^-$ during $(\t_q,t_0)$ through the relative interior of $wq$.) Recall that $wq$ is Delaunay in $P\setminus \{a,p,r,u\}$ throughout $[\t_{wq},t_0]\subset [\t_{wq},\xi_{wq}]$ (in addition to its Delaunayhood in $P$ at times $\t_{wq}$ and $t_0$). Hence, in the reduced set $P\setminus \{a,p,r\}$, the edge $wq$ or, more, precisely, its reversely oriented copy $qw$, undergoes a  Delaunay crossing by $u$ during some sub-interval of $[\t_{wq},t_0)$. Therefore, together with the crossing $(wq,u,I)$, the triple $w,q,u$ performs two single Delaunay crossings in $P\setminus \{a,p,r\}$. Combining Lemma \ref{Lemma:TwiceCollin} with the probabilistic argument of Clarkson and Shor, we obtain that the number of such triples $q,w,u$ in $P$ cannot exceed $O(n^2)$. By Proposition \ref{Prop:TriplesSpecialQuad}, the same quadratic bound must also hold for the overall number of such left special quadruples $\SQ$.

\begin{figure}[htbp]
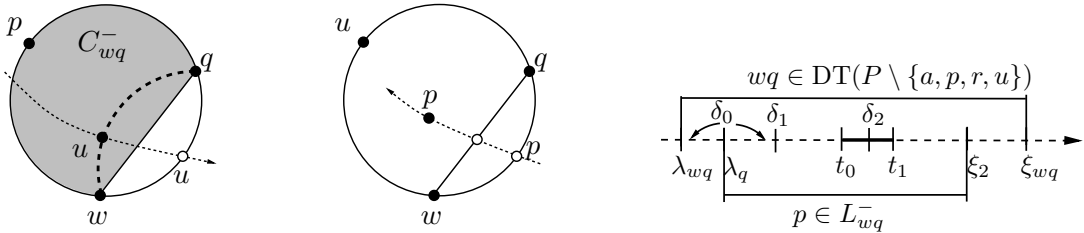

\begin{center}
\input{Delta2Left.pstex_t}\hspace{1.5cm}\input{Delta0Left.pstex_t}\hspace{1.5cm}\input{SchemeThreeCocircsLeft.pstex_t}
\caption{\small Left: The red-blue co-circularity of $w,q,u,p$ with respect to $wq$, which must occur at some time $\delta_2\in I$. Center: The points $w,q,u,p$ are involved at some time $\delta_0\in [\t_{wq},\delta_1)$ in their first co-circularity, which is also red-blue with respect to $wq$. Right: A schematic summary of the motion of $w,q,u,p$ (assuming that $u$ does not cross $wq$ during $(\t_q,t_0)$).}
\label{Fig:Delta2Left}
\vspace{-0.4cm}
\end{center}
\end{figure}

\paragraph{Case (d): The three co-circularities of $w,q,u,p$.}
Assume, then, that $u$ does not cross $wq$ in $[\t_{wq},t_0)$.
In particular, $u$ enters $B[p,q,w]$ in $(\t_q,t_0)$ through the boundary of $B[p,q,w]$, at a blue-blue co-circularity of $w,q,u,p$ with respect to $wq$ (as depicted in Figure \ref{Fig:Delta2Left} (right)). 
We claim that this is the second co-circularity of $w,q,u,p$, denoting its time by $\delta_1$.

Indeed, by Lemma \ref{Lemma:OnceCollin}, another co-circularity of $w,q,u,p$ occurs at some time $\delta_2\in I=[t_0,t_1]$ (where $wq$ undergoes a single Delaunay crossing by $u$), and is red-blue with respect to $wq$. Refer to Figure \ref{Fig:Delta2Left} (left).
Furthermore, since $u$ does not hit $wq$ during $[\t_{wq},\delta_1)\subset [\t_{wq},t_0)$ (and $wq$ belongs to $\DT(\{w,q,u,p\})$ at times $\t_{wq}$ and $\delta_1$), 
the edge $qw$ undergoes a Delaunay crossing by $p$ in the triangulation of $\{w,q,u,p\}$ too. This crossing occurs during some sub-interval of $[\t_{wq},\delta_1)$ so, by Lemma \ref{Lemma:OnceCollin}, $w,q,u,p$ are involved  in another co-circularity at some time $\delta_0\in [\t_{wq},\delta_1)$; see Figure \ref{Fig:Delta2Left} (center).

To conclude, the four points $w,q,u,p$ are co-circular at times $\delta_0\in [\t_{wq},t_0)$, $\delta_1\in (\delta_0,t_0)$, and $\delta_2\in I=[t_0,t_1]$. (See Figure \ref{Fig:Delta2Left} (right) for a schematic summary.) Here the two extremal co-circularities, which occur at times $\delta_0$ and $\delta_1$, are red-blue with respect to the edges, and the middle co-circularity at time $\delta_1$, is blue-blue with respect to $wq$ (and occurs when $u$ enters the cap $C^-_{wq}$). 
We emphasize that $u$ remains in $C^-_{wq}$ throughout $(\delta_1,t_0)$.

Furthermore, the order type of the third co-circularity (at time $\delta_2\in I$) is completely determined by Proposition \ref{Prop:EntirelyWithin} and the fact that $p$ lies in $\L_{wq}^-$ throughout $(\t_q,\xi_2)$. Hence, this co-circularity occurs during the second portion of $I$ (i.e., after $u$ enters $\L_{wq}^+$), when $p$ leaves the cap $B[w,q,u]\cap \L_{wq}^-$.

Notice that the caps $B[w,q,u]\cap \L_{wq}^-$ and $C_{wq}^-$ coincide at time $\delta_2\in I\subset (\t_q,\xi_2)$. Therefore, Lemma \ref{Prop:ShallowCupLeft}, together with $P$-emptiness of $B[w,q,u]\cap \L_{wq}^+$ during the second portion of $I$, imply that the co-circularity at time $\delta_2$ is $(8\ell+1)$-shallow. 

Recall that $(wq,u,I)$ is a clockwise $(w,q)$-crossing, and a counterclockwise $(q,u)$-crossing. 
Lemma \ref{Prop:ShallowCupLeft} yields the following symmetric analogue of Lemma \ref{Prop:FewPr} (with somewhat simpler proof, due to Proposition \ref{Prop:EntirelyWithin}).

\begin{lemma}\label{Prop:FewWu}
With the above assumptions, at most $8\ell+1$ clockwise $(w,u)$-crossings $(wq',u,I')$, and at most $8\ell+1$ counterclockwise $(q,u)$-crossings $(w'q,u,I')$, can begin in the interval $(\delta_{1},t_0)$.
\end{lemma}
\begin{proof}
Let $(wq',u,I')$ be a clockwise $(w,u)$-crossing that begins in $(\delta_1,t_0)$. 
By Lemma \ref{Lemma:OnceCollin}, the four points $w,q,u,q'$ are co-circular at some moment $\zeta'\in I'\setminus I\subset (\delta_1,t_0)$, and this co-circularity is red-blue with respect to the edges $wq',uq$, and monochromatic with respect to $wq$. Furthermore, since $p$ remains in $C_{wq}^-$ throughout $(\delta_1,t_0)\subset (\t_q,\xi_2)$, the above co-circularity is, in fact, blue-blue with respect to $wq$ (see Figure \ref{Fig:FewWu}). Hence, both points $u,q'$ lie at time $\zeta'$ inside the cap $C_{wq}^-$.
Lemma \ref{Prop:ShallowCupLeft} now implies that the overall number of such points $q'$ (and, therefore, of their respective $(w,u)$-crossings $(wq',u,I')$) cannot exceed $8\ell+1$.

\begin{figure}[htbp]
\begin{center}
\input{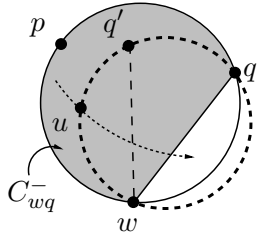}
\caption{\small Lemma \ref{Prop:FewWu}: Proving that at most $8\ell+1$ clockwise $(w,u)$-crossings $(wq',u,I')$ begin in $(\delta_1,t_0)$. For each of these crossings, the four points $w,q,u,q'$ are involved in a blue-blue co-circularity with respect to $wq$ at some time $\zeta\in I'\setminus I\subset (\delta_1,t_0)$, so their respective points $q'$ enter $C_{wq}^-$ during $(\t_q,\xi_2)$.}
\label{Fig:FewWu}
\vspace{-0.4cm}
\end{center}
\end{figure} 

A fully symmetric argument shows that at most $8\ell+1$ counterclockwise $(q,u)$-crossings $(w'q,u,I')$ can begin in the interval $(\delta_{1},t_0)$, because their respective points $w'$ must appear in $C_{wq}^-$ at some moment during $(\delta_1,t_0)\subset (\t_q,\xi_2)$.
\end{proof}

To proceed, we distinguish between two possible cases depicted in Figure \ref{Fig:CasesD12Left}.

\begin{figure}[htbp]
\begin{center}
\input{CaseD1Left.pstex_t}\hspace{3cm}\input{CaseD2Left.pstex_t}
\caption{\small Left: Case (d1). The co-circularity at time $\delta_1$ is red-blue with respect to the edges $wu$ and $pq$. Right before time $\delta_1$, the Delaunayhood of $wu$ is violated by $p$ and $q$. Right: Case (d2). The co-circularity at time $\delta_1$ is red-blue with respect to the edges $uq$ and $wp$. Right before time $\delta_1$, the Delaunayhood of $uq$ is violated by $p$ and $w$.}
\label{Fig:CasesD12Left}
\vspace{-0.6cm}
\end{center}
\end{figure}

\medskip
\noindent{\bf Case (d1).} The co-circularity at time $\delta_{1}$ is red-blue with respect to the edge $wu$ whose Delaunayhood is violated right before $\delta_1$ by $p\in \L_{wu}^-$ and $q\in \L_{wu}^+$ (see Figure \ref{Fig:CasesD12Left} (left)). 

Note that the above violation of $wu$ does not hold at time $\t_q$, when the segments $pq$ and $wu$ do not even intersect. Therefore, and since $\delta_1$ is the {\it only} blue-blue co-circularity of $w,q,u,p$ with respect to $wq$, applying (the time-reversed variant of) Lemma \ref{Lemma:MustCross} in $\DT(\{w,q,u,p\})$ over the interval $(\t_q,\delta_1)$ shows that $wu$ is hit in that interval by at least one of $p$ or $q$ (see Figure \ref{Fig:TrajectoriesD1Left}).

A very similar argument shows that the edge $pq$, whose Delaunayhood is violated right after time $\delta_1$ by $u\in \L_{pq}^-$ and $w\in \L_{pq}^+$, is hit by $u$ after $\delta_1$ and before $u$ enters $\L_{wq}^+$ (during $I$).
Indeed, let $\upsilon_{wq}$ denote the time in $I$ when $u$ hits $wq$.
Note that the above violation of $pq$ does not hold at time $\upsilon_{wq}$. Therefore, another application of Lemma \ref{Lemma:MustCross} in $\DT(\{p,q,w,u\})$ shows that the edge $pq$ is hit during $(\delta_1,\upsilon_{wq})$ by at least one of the two points $w$ or $u$.
Recall, however, that $I\subset (\t_q,\xi_2)$ (by Proposition \ref{Prop:EntirelyWithin}). Hence, both times $\delta_1\in (\t_q,t_0)$ and $\upsilon_{wq}\in I=[t_0,t_1]$ belong to the interval $(\t_q,\xi_{2})$ (during which $p$ lies in $\L_{wq}^-$), ruling out the crossing of $pq$ by $w$ in $(\delta_1,\upsilon_{wq})$. Hence, it must be the case that $pq$ is by $u$, as depicted in Figure \ref{Fig:TrajectoriesD1Left}.

\begin{figure}[htbp]
\begin{center}
\input{CaseD1LeftHitP.pstex_t}\hspace{3cm}\input{CaseD1LeftHitQ.pstex_t}\\
\caption{\small The two possible trajectories of $u$ according to Lemma \ref{Prop:HitQw}. The edge $uw$ is hit in $(t_q,\delta_1)$ by $p$ (left) or $q$ (right). In both scenarios, $u$ hits the edge $pq$ after $\delta_1$ and before the time $\upsilon_{wq}\in I$ when $u$ hits $wq$.}
\label{Fig:TrajectoriesD1Left}
\vspace{-0.6cm}
\end{center}
\end{figure}

To conclude, we have established the following lemma.

\begin{lemma}\label{Prop:HitPq}
With the above notation, the following two claims hold in case (d1):

(i) The edge $pq$ is hit in $(\delta_1,\upsilon_{wq})$ by $u$, which crosses $pq$ from $\L_{pq}^-$ to $\L_{pq}^+$. 

(ii) The edge $wu$ is hit in $(\t_q,\delta_1)$ by at least one of the points $p,q$. Namely, either $p$ crosses $wu$ from $\L_{wu}^+$ to $\L_{wu}^-$, or $q$ crosses $wu$ in the reverse direction. Moreover, the Delaunayhood of $wu$ is violated by $p$ and $q$ right after the last such crossing and until $\delta_1$.
\end{lemma}

\begin{figure}[htbp]
\begin{center}
\input{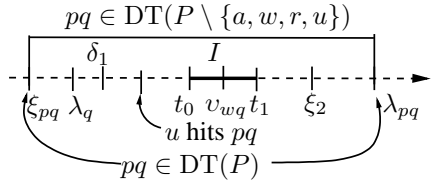}
\caption{\small Case (d1)--obtaining a Delaunay crossing of $pq$ by $u$. The edge $pq$ is Delaunay at times $\xi_{pq}$ and $\t_{pq}$, and almost Delaunay in $(\xi_{pq},\t_{pq})$. 
Since $u$ hits $pq$ in $(\delta_1,\upsilon_{wq})\subset (\xi_{pq},\t_{pq})$, $pq$ undergoes a Delaunay crossing by $u$ in $P\setminus\{a,w,r\}$.}
\label{Fig:CrossPq}
\vspace{-0.6cm}
\end{center}
\end{figure}

\noindent {\bf Case (d1) -- the crossing of $pq$ by $u$.} Refer to Figure \ref{Fig:CrossPq}. Recall that both $\t_q$ and $\delta_1$ belong to the interval $(\t_q,\xi_2)\subset(\xi_{pq},\t_{pq})$ where, by condition (S5), $pq$ belongs to $\DT(P\setminus \{a,w,r,u\})$ (in addition to its Delaunayhood in $P$ at times $\xi_{pq},\t_{pq}$).

By Lemma \ref{Prop:HitPq} (i), $pq$ is hit by $u$ in $(\t_q,\delta_1)\subset[\xi_{pq},\t_{pq}]$.
Therefore, and since $pq$ is Delaunay at times $\xi_{pq}$ and $\t_{pq}$, this edge (or its reversely oriented copy $qp$) undergoes a Delaunay crossing by $u$ within a suitably reduced triangulation $\DT(P\setminus \{a,w,r\})$. 

\medskip
\noindent{\bf Case (d1)--enforcing the crossing of $wu$ by $p$ or $q$.} 
If the edge $wq$ is never Delaunay in $P$ before time $\delta_1$ then, by Lemma \ref{Prop:FewWu}, $(wq,u,I)$ is among the first $O(\ell)$ clockwise $(w,u)$-crossings (because $wu$ is Delaunay during each of these crossings). Proposition \ref{Prop:TriplesSpecialQuad} implies that this can occur for at most $O(\ell n^2)$ special quadruples $\SQ$.
Therefore, we may assume that $wu$ has appeared in $\DT(P)$ also before $\delta_1$.

\begin{figure}[htbp]
\begin{center}
\input{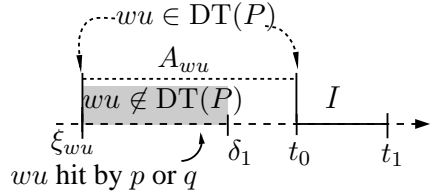}
\caption{\small Case (d1)--enforcing a crossing of $wu$ by at least one of the points $p,q$. The edge $wu$ is Delaunay throughout $I=[t_0,t_1]$ and at time $\xi_{wu}<\delta_1$ (which is the last such time before $\delta_1$).
The Delaunayhood of $wu$ is violated by $p$ and $q$ right before $\delta_1\in (\xi_{wu},t_0]$, so the promised crossing of $pq$, by at least one of $p,q$, must occur in $[\xi_{wu},\delta_1)$.}
\label{Fig:CrossWu}
\vspace{-0.6cm}
\end{center}
\end{figure}

Let $\xi_{wu}$ denote the last time in $(-\infty,\delta_1)$ when the edge $wu$ belongs to $\DT(P)$; see Figure \ref{Fig:CrossWu}. 
Notice that the time when $wu$ is hit by one of $p,q$, as prescribed by Lemma \ref{Prop:HitPq} (ii), must belong to
the interval $[\xi_{wu},\delta_1)$, which is contained in $[\xi_{wu},t_0)$.
To enforce the desired Delaunay crossing of $wu$, we apply Theorem  \ref{Thm:RedBlue} in $\A_{wu}$ over the interval $(\xi_{wu},t_0)$, with the third constant $h\gg \ell$. 

If at least one of the Conditions (i), (ii) holds, we can charge $\SQ$, within $\A_{wu}$, either to an $h$-shallow collinearity or to $\Omega(h^2)$ $h$-shallow co-circularities. Lemma \ref{Prop:FewPr} ensures that each $h$-shallow event, that occurs in $\A_{wu}$ at some time $t^*\in (\xi_{wu},t_0)$, is charged in this manner by at most $O(\ell)$ left special quadruples. Indeed, the corresponding points $w$ and $u$ are involved in the event, so we can guess them in $O(1)$ possible ways, and $(wq,u,I)$ is among the first $8\ell+2$ clockwise $(w,u)$-crossings to begin after time $t^*$.
Therefore, the above charging accounts for at most $O\left(\ell h^2N(n/h)+\ell hn^2\beta(n)\right)$ special quadruples $\SQ$.

We may assume, then, that Condition (iii) of Theorem \ref{Thm:RedBlue} holds. That is, there is a subset $A_{wu}$ of at most $3h$ points (perhaps including some of $p,q,a$, and $r$) whose removal restores the Delaunayhood of $wu$ throughout the interval $[\xi_{wu},t_0]$. 

If $wu$ is crossed during $[\xi_{wu},t_0)$ by $q$ (from $\L_{wu}^-$ to $\L_{wu}^+$), then, together with $(wq,u,I)$, the triple $w,q,u$ performs two Delaunay crossings in $(P\setminus A_{wu})\cup \{q\}$. A routine combination of Lemma \ref{Lemma:TwiceCollin} with the probabilistic argument of Clarkson and Shor implies that $P$ contains at most $O(hn^2)$ triples $w,q,u$ of this kind. By Proposition \ref{Prop:TriplesSpecialQuad}, this also bounds the overall number of such left special quadruples $\SQ$. 

To conclude, we may assume that the edge $wu$ (or its reversely oriented copy $uw$) undergoes a Delaunay crossing by $p$ in the smaller set $(P\setminus A_{wu})\cup \{p\}$. In addition, we have shown that the edge $pq$ (or its reversely oriented copy $qp$) undergoes a Delaunay crossing by $u$ in $P\setminus \{a,r,w\}$. (Note that one, or both of these crossings can be a double Delaunay crossing.)
Therefore, together with the crossings $(wq,u,I)$ and $(qw,p,\H)$, each of the four possible sub-triples of $w,q,u,p$ performs a Delaunay crossing within a suitably refined triangulation.

Finally, recall that the four points $w,q,u,p$ are involved at some time $\delta_2\in I\subset (\t_q,\xi_2)$ in their third (and last) co-circularity, which is red-blue with respect to the edges $wq$ and $pu$. Moreover, this co-circularity is $(8\ell+1)$-shallow in $P$ (because of Lemma \ref{Prop:ShallowCupLeft}), and the Delaunayhood of $up$ is violated right after it by $q\in \L_{pu}^-$ and $p\in \L_{pu}^+$. Let $A_{\delta_2}$ be the set of at most $8\ell+1$ points that lie at time $\delta_2$ within the circumdisc of $w,q,u,p$.

\paragraph{Case (d1): Charging terminal quadruples.} We consider a subset $R$ of $\lceil n/h\rceil$ points chosen at random from $P$. Notice that the following two events occur simultaneously, with probability at least $\Omega(1/h^4)$:
(1) $R$ contains the four points $w,q,u,p$, and (2) none of the points of $(A_{wu}\cup A_{\delta_2}\cup \{a,r\})\setminus \{w,q,u,p\}$ belong to $R$.

\begin{figure}[htbp]
\begin{center}
\input{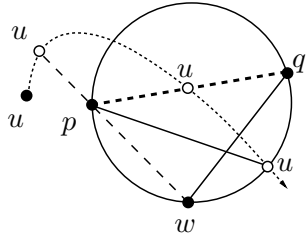}\\
\caption{\small A possible trajectory of $u$ if $\TQ=(w,q,u,p)$ is a terminal quadruple of type C. The points of $\TQ$ are involved in an extremal (last) Delaunay co-circularity, right after which the Delaunayhood of $pu$ is violated by $q\in \L_{pu}^-$ and $w\in \L_{pu}^+$.}
\label{Fig:TypeC}
\vspace{-0.6cm}
\end{center}
\end{figure}

In the case of success, we charge $\SQ$ to the quadruple $\TQ=(w,q,u,p)$, which satisfies the following two conditions with respect to $R$ (see Figure \ref{Fig:TypeC}):

\medskip
\noindent {\bf (C1)} The edge $pq$ (or $qp$) undergoes a Delaunay crossing by $u$ in $R\setminus \{w\}$. Similarly, the edge $wu$ (or $uw$) undergoes a Delaunay crossing by $p$ in $R\setminus \{q\}$.

\smallskip
\noindent {\bf (C2)} The four points of $\TQ$ are involved in a Delaunay co-circularity, right after which the Delaunayhood of $pu$ is violated by $q\in \L_{pu}^-$ and $w\in \L_{pu}^+$. Furthermore, this is the last co-circularity of $w,q,u,p$.

Note that $\SQ$ is uniquely determined by $\TQ$.

\medskip
\noindent{\it Definition.} Let $P$ be a finite set of moving points in $\reals^2$. We say that a quadruple $\TQ=(w,q,u,p)$ in $P$ is {\it terminal of type C} if it satisfies the above conditions (C1) and (C2), with $R$ replaced by $P$.

\smallskip
Let $T^C(m)$ denote the maximum possible number of terminal quadruples of type C that can arise in an underlying set of $m$ moving points. Then the overall number of 3-restricted left special quadruples that fall into case (d1)
is at most
$$
O\left(h^4 T^C(n/h)+\ell h^2 N(n/h)+\ell h n^2\beta(n)\right).
$$

In Section \ref{Subsec:TerminalCD} we will use the corresponding extremal Delaunay co-circularity of $w,q,u,p$ of each terminal quadruple $\TQ$ to enforce a Delaunay crossing of $pu$ by at least one of the remaining two points $w,q$ of $\TQ$. Together with the Delaunay crossings in condition (C1), at least one of the triples $p,u,w$ or $p,u,q$ will perform two (single) Delaunay crossings. Therefore, our analysis will again bottom up via Lemma \ref{Lemma:TwiceCollin}.

\smallskip
\noindent{\bf Remark.} Notice that in condition (C1) we omit the crossings $(wq,u,I)$ and $(qw,p,\H)$ which gave rise to the terminal quadruple $\TQ=(w,q,u,p)$, after having used them to enforce the crossings of $p,u,w$ and $p,u,q$.

\paragraph{Case (d2).} The co-circularity at time $\delta_{1}$ is red-blue with respect to the edge $uq$ whose Delaunayhood is violated right before $\delta_1$ by $p\in \L_{uq}^-$ and $w\in \L_{uq}^+$ (see Figure \ref{Fig:CasesD12Left} (right)). 

Using $\upsilon_{wq}$ as before to denote the unique time in $I=[t_0,t_1]$ when $u$ hits $wq$, we have the following symmetric variant of Lemma \ref{Prop:HitPq}, which can be established by switching the roles of $w$ and $q$ in the argument that implied Lemma \ref{Prop:HitPq}. 
 
\begin{lemma}\label{Prop:HitWp}
With the above notation, the following two properties hold in case (d2):

(i) The edge $wp$ is hit in $(\delta_1,\upsilon_{wq})$ by $u$, which crosses $wp$ from $\L_{wp}^-$ to $\L_{wp}^+$. 

(ii) The edge $uq$ is hit in $(\t_q,\delta_1)$ by at least one of the points $p,w$. Namely, either $p$ crosses $uq$ from $\L_{uq}^+$ to $\L_{uq}^-$, or $w$ crosses $uq$ in the reverse direction. Moreover, the Delaunayhood of $uq$ is violated by $p$ and $w$ right after the last such crossing and until $\delta_1$.
\end{lemma}

\begin{figure}[htbp]
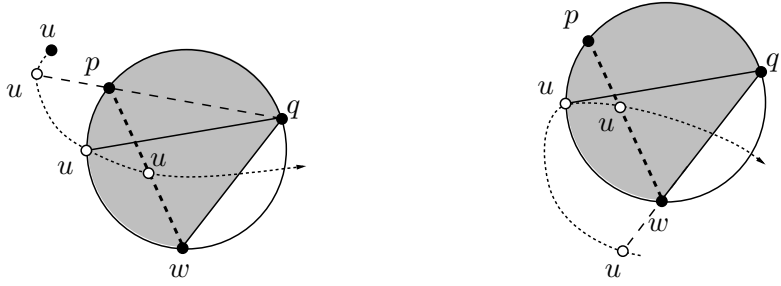

\begin{center}
\input{CaseD2LeftHitP.pstex_t}\hspace{3cm}\input{CaseD2LeftHitW.pstex_t}\\
\caption{\small The two possible trajectories of $u$ according to Lemma \ref{Prop:HitWp}. The edge $uq$ is hit in $(t_q,\delta_1)$ by $p$ (left) or $w$ (right). In both scenarios, $u$ hits the edge $wp$ after $\delta_1$ and before the time $\upsilon_{wq}\in I$ when $u$ hits $wq$.}
\label{Fig:TrajectoriesD2Left}
\vspace{-0.6cm}
\end{center}
\end{figure}

We next amplify the collinearities in Lemma \ref{Prop:HitWp} into full-fledged Delaunay crossings.
We again emphasize that $(wq,u,I)$ is a clockwise $(w,u)$-crossing, and a counterclockwise $(u,q)$-crossing, so the role of $uq$ in the present case (d2) is fully symmetric to the role of $wu$ in case (d1).
In particular, the crossing of $uq$ (or of its reversed copy $qu$) by $p$ or $w$ will be enforced using essentially the same argument as was used in case (d1) to enforce the Delaunay crossing of $wu$ by $p$ or $q$.

In contrast, the properties of $pq$ (in case (d1)) and $wp$ (in case (d2)) are {\it not} symmetric.
Indeed, the edge $pq$ (which was crossed by $u$ in case (d1)) is almost Delaunay throughout the interval $[\xi_{pq},\t_{pq}]$ (which covers $(\t_q,\xi_2)\supset (\delta_1,\upsilon_{wq})$, where $wp$ is hit by $u$ or $q$), and Delaunay at both times $\xi_{pq},\t_{pq}$. However, the edge $wp$ (which is crossed by $u$ in the present case (d2)) becomes Delaunay in $[\xi_{pq},\t_{pq}]$ only after removal of a subset $A_{pw}$ of at most $6\ell+3$ points (including $u$), which is not enough to obtain a Delaunay crossing of $wp$ by $u$.

\smallskip

\noindent{\bf Case (d2): Enforcing a Delaunay crossing of $wp$ by $u$.}
We emphasize that the third co-circularity of $w,q,u,p$ is $(8\ell+1)$-shallow and occurs at some time $\delta_2$ during the second portion of $I$, starting right after the unique time $\upsilon_{wq}$ in $I$ when $u$ hits $wq$.
Recall also that $I$ begins after $\delta_1$ and is contained in the nested intervals $(\t_q,\xi_{2})$ and $(\xi_{pq},\t_{pq})$ (where the Delaunayhood of $pw$ can be restored by removing the above set $A_{pw}$ of at most $6\ell+3$ points).

\begin{figure}[htbp]
\begin{center}
\input{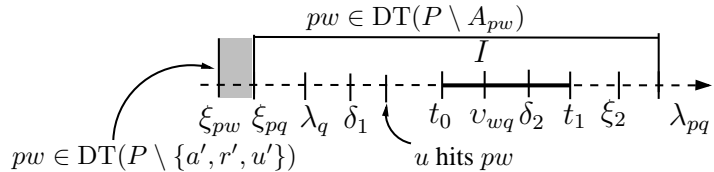}
\caption{\small Case (d2)--enforcing the crossing of $wp$ by $u$. The edge $pw$ is Delaunay in $P\setminus A_{pw}$ throughout the interval $[\xi_{pq},\t_{pq}]$, which contains $\delta_1$ and $I$ (inclduing $\upsilon_{wq}$ and $\delta_2$). We first obtain a time $\xi_{pw}\leq \xi_{pq}$ when $pw$ belongs to some reduced triangulation $\DT(P\setminus \{a',r',u'\})$, so that none of the obstruction points $a',r',u'$ is equal to $u$. Note that $u$ hits $pw$ in the interval $(\xi_{pw},\delta_2)$.}
\label{Fig:SetupCrossWp}
\vspace{-0.6cm}
\end{center}
\end{figure}

Notice that the Delaunayhood of $pw$ at time $\delta_2$ can be enforced by removing the subset $A_{\delta_2}$ of at most $8\ell+1$ points that lie at time $\delta_2$ in the interior of the circumdisc of $w,q,u,p$. Since $A_{\delta_2}$ does not include $u$, its removal does not destroy the crossing triple $w,p,u$.

We first obtain a time $\xi_{pw}\leq\xi_{pq}$ when the edge $pw$ belongs to some reduced triangulation $\DT(P\setminus \{a',r',u'\})$, for some $a',r',u'\in P\setminus \{w,p,u\}$.
In particular, $(\xi_{pw},\delta_2)$ contains the above time in $(\delta_1,\upsilon_{wq})$ when $u$ crosses $wp$ from $\L_{wp}^-$ to $\L_{wp}^+$.
We then use Theorem \ref{Thm:RedBlue} to extend the almost-Delaunayhood of $pw$ to $[\xi_{pw},\xi_{pq})$, so as to cover the entire $[\xi_{pw},\t_{pq}]$. As a result, $wp$ will undergo a Delaunay crossing by $u$ during some sub-interval of $[\xi_{pw},\delta_2]$ (and in an appropriately reduced subset of $P$).

To obtain the above time $\xi_{pw}\leq\xi_{pq}$, we return to the subfamily $\G^L_{pw}$ of all $3$-restricted left special quadruples $\SQ'=(a',p,w,q')$ (each coming with respective outer points $r'$ and $u'$) whose two middle points are equal to $p$ and $w$, respectively. Recall that each quadruple in $\G^L_{pw}$ is uniquely determined by its respective point $q'$. In addition, we can assume that all quadruples in $\G_{pw}^L$ fall into case (d2) of the present analysis (because the remaining quadruples in $\G_{pw}^L$ are handled using previous charging arguments). In particular, $\G^L_{pw}$ contains the quadruple $\SQ=(a,p,w,q)$ under consideration.

Our analysis relies on the following uniqueness property:
\begin{lemma}\label{Lemma:UniqueWpu}
With the above assumptions, the family $\G_{pw}^L$ contains at most $3\ell+1$ other 3-restricted left special quadruples $\SQ'=(a',p,w,q')$, with respective outer points $r'$ and $u'$, that fall into case (d2) and satisfy $u'=u$.
\end{lemma}
In other words, any triple $w,p,u$ can be shared by at most $3\ell+2$ 3-restricted left special quadruples $\SQ$ under consideration.
\begin{proof}
Notice that, for each terminal quadruple $\SQ'=(a',p,w,q')\in \G_{pw}^L$ under consideration, with respective outer points $r'$ and $u'=u$, the four points $w,u,p,q'$ are involved in their third co-circularity at some time $\delta'_2$ during the respective regular crossing of $wq'$ by $u$. Right after time $\delta'_2$, the Delaunayhood of $pu$ is violated by $q'\in\L_{pu}^-$ and $w\in \L_{pu}^+$. Clearly, the lemma will follow if we show that $\delta_2$ is among the first $8\ell+2$ such times $\delta'_2$ to occur after $u$ crosses $wp$ from $\L_{wp}^-$ to $\L_{wp}^+$ (as prescribed in Lemma \ref{Prop:HitWp} (i)). (See Figure \ref{Fig:UniqueWpu}.)

\begin{figure}[htbp]
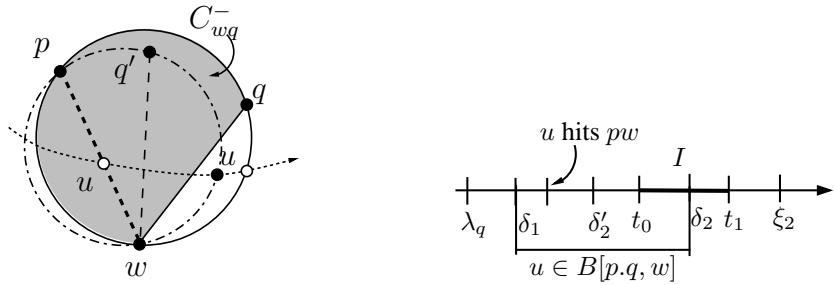

\begin{center}
\input{UniqueWpu.pstex_t}\hspace{2cm}\input{SetupUniqueWpu.pstex_t}
\caption{\small Proof of Lemma \ref{Lemma:UniqueWpu}. We fix a quadruple $\SQ'=(a',p,w,q')\in \G_{pw}^L$ whose second outer point $u'$ equal to $u$, so that the third co-circularity of $w,p,u,q'$ occurs at some time $\delta'_2$ after $u$ crosses $wp$ (from $\L_{wp}^-$ to $\L_{wp}^+$) and before $\delta_2$. 
We claim that $q'$ lies at time $\delta'_2$ in the cap $C^-_{wq}$. The two hollow circles in the left figure represent the location of $u$ when it hits $wp$, and at time $\delta_2>\delta'_2$ (when $u$ leaves $B[p,q,w]$).}
\label{Fig:UniqueWpu}
\vspace{-0.6cm}
\end{center}
\end{figure}

To establish the last claim, let $\SQ'=(a',p,w,q')$ be a $3$-restricted left quadruple, with respective outer points $r'$ and $u'=u$, and such that the corresponding third co-circularity of $w,u,p,q'$ occurs at some time $\delta'_2$ after $w$ enters $\L_{wp}^+$ through $wp$ and before $\delta_2$. 
We claim that $q'$ lies at time $\delta_2$ within the cap $C^-_{wq}=B[p,q,w]\cap \L_{wq}^-$ (as depicted in Figure \ref{Fig:UniqueWpu} (left)), so, by Lemma \ref{Prop:ShallowCupLeft}, the overall number of such points $q'$ (and, therefore, also of their respective quadruples $\SQ'$) cannot exceed $8\ell+1$.

Indeed, recall that the motion of $B[p,q,w]$ is continuous in the interval $(\t_q,\xi_2)$, which contains $\delta_1\in (\t_q,t_0)$ and  $\delta_2\in I=[t_0,t_1]$. Therefore, and since $\delta_2$ is (the time of) the last co-circularity of $w,q,u,p$, the point $u$ must remain in $B[p,w,q]$ after time $\delta_1$, when $u$ enters that disk, and until the time $\delta_2$, when $u$ leaves $B[p,q,w]$.
Therefore, both $u$ and $q'$ lie in $B[p,q,w]\cap \L_{wp}^+$ at time $\delta'_2$, when we encounter a red-red co-circularity of $p,w,u,q'$ with respect to $wp$.
It hence suffices to show that $q'$ lies in $\L_{wq}^-$ at time $\delta_2$. 

Assume for a contradiction that $q'$ lies at time $\delta'_2$ in the opposite cap $C_{wq}^+=B[p,q,w]\cap \L_{wq}^+$. This readily implies that the four edges $wp,wq,wq'$, and $wu$, appear around $w$ in this clockwise order at time $\delta_2$; see Figure \ref{Fig:UniqueWpuContradict} (left). In particular, the point $u$ too lies at time $\delta'_2$ in $C_{wq}^+$, so $\delta'_2$ belongs to the second portion of $I$ (which starts at time $\upsilon_{wq}$, when $u$ hits $wq$); see Figure \ref{Fig:UniqueWpuContradict} (right). Notice that the Delaunayhood of $wq$ is violated in $(\upsilon_{wq},\delta_2)$ by $p\in \L_{wq}^-$ and $u\in \L_{wq}^+$, so $p$ must lie in the cap $B[w,q,u]\cap \L_{wu}^-$ at time $\delta'_2\in (\upsilon_{wq},\delta_2)$.
However, since the co-circularity of $w,u,p,q'$ at time $\delta'_2$ is blue-blue with respect to $wu$, the above cap $B[w,q,u]\cap \L_{wu}^-$ must then contain also the point $q'$. In particular, $q'$ lies at time $\delta'_2$ within the disk $B[w,q,u]$, contrary to the $P$-emptiness of $B[w,q,u]\cap \L_{wq}^+$ during the second portion of $I$.

\begin{figure}[htbp]
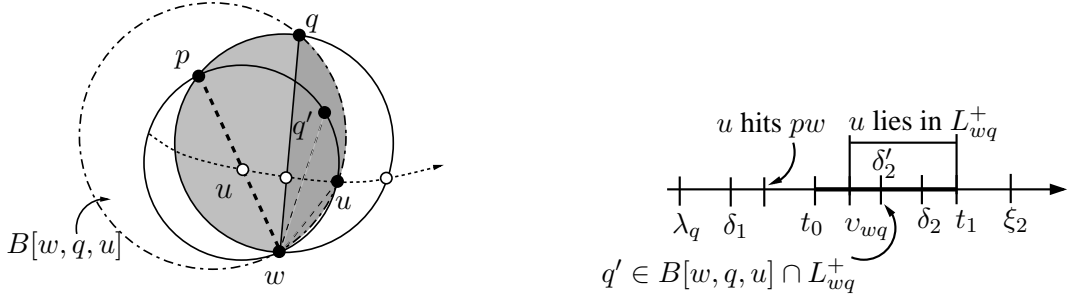

\begin{center}
\input{UniqueWpuOppositeCap.pstex_t}\hspace{2cm}\input{ContradictUniqueWpu.pstex_t}
\caption{\small If $q'$ lies at time $\delta'_2$ in the opposite cap $C_{wq}^+$, then this co-circularity occurs during the second portion $(\upsilon_{wq},t_1]$ of $I$. In this hypothetic case, $q'$ lies at time $\delta_2$ within the disk $B[w,q,u]$, contrary to the $P$-emptiness of $B[w,q,u]\cap \L_{wq}^+$ during $(\upsilon_{wq},t_1]$.}
\label{Fig:UniqueWpuContradict}
\vspace{-0.6cm}
\end{center}
\end{figure}

To conclude, the above contradiction implies that $q'$ lies at time $\delta'_2$ in the cap $B[p,q,w]\cap \L_{wq}^-$. Hence, Lemma \ref{Prop:ShallowCupLeft} implies the overall number of such points $q'$ cannot exceed $8\ell+1$. Therefore, the family $\G_{pw}^L$ contains at most $8\ell+1$ 3-restricted left special quadruples $\SQ'=(a',p,w,q')$, with respective outer points $r'$ and $u'=u$, that fall into case (d2), and whose respective third co-circularities $\delta'_2$ occur after $u$ crosses $wp$ from $\L_{wp}^-$ to $\L_{wp}^+$ and before $\delta_2$. In other words, $\delta_2$ is among the first $8\ell+2$ such times $\delta'_2$ to occur after $u$ crosses $wp$ as above.
\end{proof}

Lemma \ref{Lemma:UniqueWpu} implies, through the standard pigeonhole argument, that at least some constant positive fraction of all 3-restricted left special quadruples $\SQ=(a,p,w,q)$ under consideration (with respective outer points $r$ and $u$) satisfy the following condition:

\smallskip
\noindent {\bf (PHL2)} {\it There exist at most $O(\ell)$ quadruples $\SQ'\in \G^L_{pw}$, with respective outer points $r'$ and $u'$, so that $u\in \{a',r',u'\}$.}

We may assume, with no loss of generality, that (PHL2) holds for $\SQ$ under consideration.
With these preparations, we can proceed to the main argument in $\A_{pw}$.
Recall that each such quadruple $\SQ'\in \G_{pw}^L$ is uniquely determined by the respective point $q'$, and is accompanied by a counterclockwise $(w,p)$-crossing $(q'w,p,\H_{\SQ'})$ which occurs in the reduced triangulation $\DT(P\setminus \{a',r',u'\})$.

\begin{figure}[htbp]
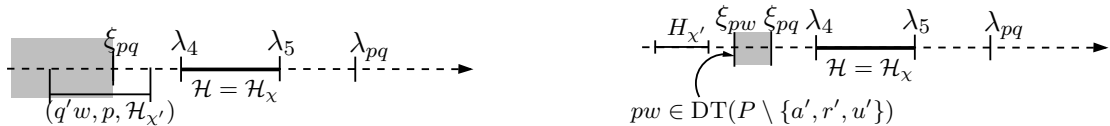

\begin{center}
\input{NoTimeBeforeWp.pstex_t}\hspace{2cm}\input{PwExistLeft.pstex_t}
\caption{\small Left: If there exists no quadruple $\SQ'$ (with respective outer points $r'$ and $u'$) in $\G_{pw}^L$ that satisfies $a',r',u'\neq u$, and whose respective $(w,p)$-crossing $(q'w,p,\H_{\SQ'})$ begins before $\xi_{pq}$, then $\G_{pw}^L$ contains a total of at most $O(\ell)$ quadruples $\SQ'$ whose respective $(w,p)$-crossings $(q'w,p,\H_{\SQ'})$ begin before the starting time $\t_4$ of $\H=\H_{\SQ}$. Right: Otherwise, there is a time $\xi_{pw}\leq \xi_{pq}$ which is the last such time when $pw$ belongs to some reduced triangulation $\DT(P\setminus \{a',r',u'\})$, for $a',r',u'\neq u$.}
\label{Fig:ChooseTimePwLeft}
\vspace{-0.6cm}
\end{center}
\end{figure}

Refer to Figure \ref{Fig:ChooseTimePwLeft}.
Assume first that there is no quadruple $\SQ=(a',p,w,q')\in \G_{pw}^L$ (with respective outer points $r'$ and $u'$) such that $a',r',u'\neq u$, and whose respective $(w,p)$-crossing $(q'w,p,\H_{\SQ'})$ begins before $\xi_{pq}$. (See Figure \ref{Fig:ChooseTimePwLeft} (left).) Since case (a) has been ruled out, $\G_{pw}^L$ contains at most $k$ special quadruples whose respective $(w,p)$-crossings begin in $[\xi_{pq},\t_4)$.
Thus, $G_{pw}^L$ contains a total of at most $O(k+\ell)$ quadruples whose respective $(w,p)$-crossing $(q'w,p,\H_{\SQ'})$ begin before the starting $\t_4$ of $\H=\H_{\SQ}$ (including the at most $O(\ell)$ such $(w,p)$-crossings that begin before $\xi_{pq}$ and have one of their respective obstruction points $a',r',u'$ equal to $u$). We charge $\SQ$ to the edge $pw$, noting that the above scenario can occur for at most $O(\ell n^2)$ $3$-restricted left special quadruples under consideration.

We thus can assume, in what follows, that there is at least one quadruple $\SQ'=(a',p,w,q')$, with respective outer points $r'$ and $u'$, that satisfies $a',r',u'\neq u$, and whose respective $(w,p)$-crossing $(q'w,p,\H')$ in $P\setminus \{a',r',u'\}$ begins before (or at) $\xi_{pq}$. (See Figure \ref{Fig:ChooseTimePwLeft} (right).) In particular, Lemma \ref{Lemma:Crossing} implies that there is a time before (or at) $\xi_{pq}$ when $pw$ belongs to some reduced triangulation $\DT(P\setminus \{a',r',u'\})$, for some three points $a',r',u'$ distinct from $u$. We choose $\xi_{pw}$ as the {\it last} such time in $(\infty,\xi_{pq}]$.

Notice that the above choice of $\xi_{pw}$ guarantees that there exist at most $O(\ell)$ quadruples $\SQ'\in \G_{pw}^L$ whose respective $(w,p)$-crossings begin in $[\xi_{pw},\t_4)$. In what follows, we will use $a',r',u'$  to denote some three fixed points whose removal restores the Delaunayhood of $pw$ at time $\xi_{pw}$.

We next apply Theorem \ref{Thm:RedBlue} in $\A_{pw}$ over the interval $(\xi_{pw},\t_4)$. This is done with respect to the reduced set $P\setminus \{a',r',u'\}$ (which ensures the Delaunayhood of $pw$ at the endpoint $\xi_{pw}$), and with the third constant $h\gg \ell$. 

In cases (i), (ii) of Theorem \ref{Thm:RedBlue} we charge $\SQ$ within the reduced arrangement $\A_{pw}$ either to $\Omega(h^2)$ $h$-shallow co-circularities, or to an $h$-shallow collinearity. Notice that each of the charged events is $(h+3)$-shallow with respect to the original set $P$, and is charged by at most $O(\ell)$ left special quadruples $\SQ$.
(The latter holds because the respective $(w,p)$-crossing $(qw,p,\H=[\t_4,\t_5])$ of $\SQ$ is among the first $O(\ell)$ such $(p,w)$-crossings to begin after the time of the event.) Therefore, the above charging accounts for at most $O\left(\ell h^2N(n/h)+\ell hn^2\beta(n)\right)$ special quadruples $\SQ$.

\begin{figure}[htbp]
\begin{center}
\input{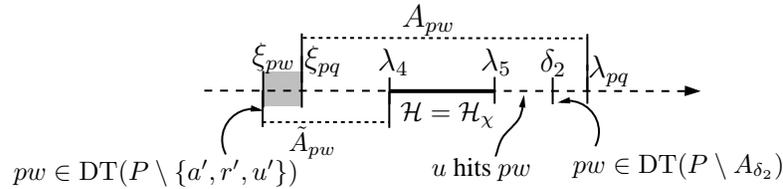}
\caption{\small In case (iii) of Theorem \ref{Thm:RedBlue} we end up with a subset $\tilde{A}_{pw}$ of at most $3h+3$ points (including $a',r',u'$) whose removal restores the Delaunayhood of $pw$ throughout $[\xi_{pw},\t_4]$. In addition, $pw$ is Delaunay in $P\setminus A_{pw}$ throughout the interval $[\xi_{pq},\t_{pq}]$ (which contains $\H=[\t_4,\t_5]$, $\delta_2$, and the time before $\delta_2$ when $u$ crosses $wp$ from $\L_{wp}^-$ to $\L_{wp}^+$), and it is Delaunay in $P\setminus A_{\delta_2}$ at time $\delta_2$. Hence, if we omit the $O(h)$ points of $(\tilde{A}_{pw}\cup A_{pw}\cup A_{\delta_2})\setminus \{u\}$, the edge $wp$ (or $pw$) undergoes a Delaunay crossing by $u$.}
\label{Fig:PwExtendD2Left}
\vspace{-0.6cm}
\end{center}
\end{figure}

Finally, in case (iii) of Theorem \ref{Thm:RedBlue} we end up with a subset $\tilde{A}_{pw}$ of at most $3h+3$ points
(including the three points $a',r',u'$ which were put aside) whose removal restores the Delaunayhood of $pw$ throughout $[\xi_{pw},\t_4]$; see Figure \ref{Fig:PwExtendD2Left}.
In particular, $pw$ is Delaunay in $P\setminus (A_{pw}\cup \tilde{A}_{pw})$ 
throughout the interval $[\xi_{pq},\t_{pq}]=[\xi_{pw},\t_4)\cup [\xi_{pq},\t_{pq}]$, which contains $\delta_2$ and the above time in $(\delta_1,\upsilon_{wq})\subset (\delta_1,\delta_2)$ when $u$ crosses $wp$ from $\L_{wp}^-$ to $\L_{wp}^+$.
Furthermore, recall that the co-circularity of $p,q,w,u$ at time $\delta_2$ is a Delaunay co-circularity in $P\setminus {A_{\delta_2}}$, where $A_{\delta_2}\subset P$ is a subset of cardinality at most $8\ell+1$. In particular, $pw$ is Delaunay in $P\setminus {A_{\delta_2}}$ at time $\delta_2$.
Hence, in the even more reduced set
$(P\setminus (A_{\delta_2}\cup A_{pw}\cup \tilde{A}_{pw}))\cup \{u\}$, the edge $wp$ (or its reversely oriented copy $pw$) undergoes a Delaunay crossing by $u$ during some sub-interval of $[\xi_{pw},\delta_2)$. (Specifically, the Delaunayhood of $wp$ at times $\xi_{pw}$ and $\delta_2$ is guaranteed by removal of $a',r',u'\in \tilde{A}_{pw}\setminus \{u\}$ and $A_{\delta_2}\subset P\setminus \{p,q,w,u\}$.)

\medskip
\noindent{\bf Case (d2): enforcing the crossing of $qu$ by $p$ or $w$.} 
If the edge $uq$ is never Delaunay in $P$ before time $\delta_1$, Lemma \ref{Prop:FewWu} implies that $(wq,u,I)$ is among the first $O(\ell)$ counterclockwise $(q,u)$-crossings (because $uq$ is Delaunay during each of these crossings). Proposition \ref{Prop:TriplesSpecialQuad} implies that this can occur for at most $O(\ell n^2)$ special quadruples $\SQ$.
Therefore, we may assume that $uq$ appears in $\DT(P)$ also before $\delta_1$.

\begin{figure}[htbp]
\begin{center}
\input{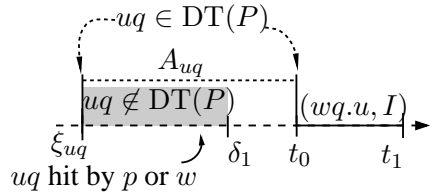}
\caption{\small Case (d2)--enforcing a Delaunay crossing of $qu$ by at least one of the points $p,w$. The edge $qu$ is Delaunay throughout $I=[t_0,t_1]$ and at time $\xi_{uq}<\delta_1$ (which is the last such time before $\delta_1$).
The Delaunayhood of $uq$ is violated by $p$ and $w$ right before $\delta_1\in (\xi_{uq},t_0)$, so it is hit in $[\xi_{uq},\delta_1)$ by at least one of $p,w$.}
\label{Fig:CrossUq}
\vspace{-0.6cm}
\end{center}
\end{figure}

Let $\xi_{uq}$ denote the last time before $\delta_1$ when the edge $uq$ belongs to $\DT(P)$; see Figure \ref{Fig:CrossUq}. 
Notice that the time when $uq$ is hit by one of $p,w$, as prescribed by Lemma \ref{Prop:HitPq} (ii), must belong to
the interval $[\xi_{uq},\delta_1)$, which is contained in $[\xi_{uq},t_0)$.
To enforce the desired Delaunay crossing of $qu$, we apply Theorem \ref{Thm:RedBlue} in $\A_{qu}$ over the interval $(\xi_{qu},t_0)$, with the third constant $h\gg \ell$. 

If at least one of the Conditions (i), (ii) holds, we can charge $\SQ$, within $\A_{uq}$, either to an $h$-shallow collinearity or to $\Omega(h^2)$ $h$-shallow co-circularities. Lemma \ref{Prop:FewPr} ensures that each $h$-shallow event, that occurs in $\A_{uq}$ at some time $t^*\in (\xi_{uq},t_0)$, is charged in this manner by at most $O(\ell)$ left special quadruples. Indeed, the corresponding points $u$ and $q$ are involved in the event, so we can guess them in $O(1)$ possible ways, and $(wq,u,I)$ is among the first $8\ell+2$ (regular) counterclockwise $(q,u)$-crossings to begin after time $t^*$.
Therefore, and since $\SQ$ is uniquely determined by $(wq,u,I)$ (see Proposition \ref{Prop:TriplesSpecialQuad}), the above charging accounts for at most $O\left(\ell h^2N(n/h)+\ell hn^2\beta(n)\right)$ special quadruples $\SQ$.

We may assume, then, that Condition (iii) of Theorem \ref{Thm:RedBlue} holds. That is, there is a subset $A_{uq}$ of at most $3h$ points (perhaps including some of $p,q,a$, and $r$) whose removal restores the Delaunayhood of $uq$ throughout the interval $[\xi_{uq},t_0]$. 

If $uq$ is crossed during $[\xi_{uq},t_0)$ by $w$ (from $\L_{uq}^-$ to $\L_{uq}^+$), then the triple $w,q,u$ performs two Delaunay crossings in $(P\setminus A_{uq})\cup \{w\}$. A routine combination of Lemma \ref{Lemma:TwiceCollin} with the probabilistic argument of Clarkson and Shor implies that $P$ contains at most $O(hn^2)$ triples $w,q,u$ of this kind. By Proposition \ref{Prop:TriplesSpecialQuad}, this also bounds the overall number of such left special quadruples $\SQ$. 

\paragraph{Case (d2): Converging.} To recap, after excluding $O\left(\ell h^2N(n/h)+\ell hn^2\beta(n)\right)$ special quadruples $\SQ$, we may assume that the edge $uq$ is hit in $(\xi_{uq},t_0)$ by $p$, so it (or its reversely oriented copy $qu$) undergoes a Delaunay crossing by $p$ in the smaller set $(P\setminus A_{uq})\cup \{p\}$. 

In addition, the four points $w,q,u,p$ are involved at some time $\delta_2\in I$ in their third (and last) co-circularity, which is red-blue with respect to the edges $wq$ and $pu$. Moreover, this co-circularity is $(8\ell+1)$-shallow in $P$, and the Delaunayhood of $up$ is violated right after it by $q\in \L_{pu}^-$ and $p\in \L_{pu}^+$. 
As before, we use $A_{\delta_2}$ to denote the set of at most $8\ell+1$ points that lie at time $\delta_2$ within the circumdisc of $w,q,u,p$.

Finally, there exist sets $A_{pw}$ and $\tilde{A}_{pw}$ that contain at most $O(\ell+h)=O(h)$ points in total, so that $wp$ (or its reversely oriented copy $pw$) undergoes a Delaunay crossing by $u$ in $(P\setminus (A_{\delta_2}\cup A_{pw}\cup \tilde{A}_{pw}))\cup \{u\}$. (Note that one, or both of these crossings can be a double Delaunay crossing.)

\paragraph{Case (d2): Charging terminal quadruples.} We consider a subset $R$ of $\lceil n/h\rceil$ points chosen at random from $P$. Notice that the following two events occur simultaneously, with probability at least $\Omega(1/h^4)$:
(1) $R$ contains the four points $w,q,u,p$, and (2) none of the points of $(A_{uq}\cup A_{pw}\cup \tilde{A}_{pw}\cup A_{\delta_2})\setminus \{w,q,u,p\}$ belong to $R$.

\begin{figure}[htbp]
\begin{center}
\input{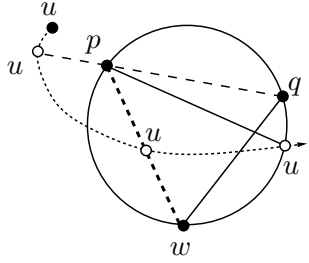}\\
\caption{\small A possible trajectory of $u$ if $\TQ=(w,q,u,p)$ is a terminal quadruple of type D. The points of $\TQ$ are involved in an extremal (last) Delaunay co-circularity, right after which the Delaunayhood of $pu$ is violated by $q\in \L_{pu}^-$ and $w\in \L_{pu}^+$.}
\label{Fig:TypeD}
\vspace{-0.6cm}
\end{center}
\end{figure}

In the case of success, we charge $\SQ$ to the quadruple $\TQ=(w,q,u,p)$, which satisfies the following two conditions with respect to $R$ (noting that $\SQ$ is uniquely determined by $\TQ$); see Figure \ref{Fig:TypeD}:

\medskip
\noindent {\bf (D1)} The edge $wp$ (or $wp$) undergoes a Delaunay crossing by $u$ in $R\setminus \{q\}$. Similarly, the edge $qu$ (or $uq$) undergoes a Delaunay crossing by $p$ in $R\setminus \{q\}$.

\smallskip
\noindent {\bf (D2)} The four points of $\TQ$ are involved in a Delaunay co-circularity, right after which the Delaunayhood of $pu$ is violated by $q\in \L_{pu}^-$ and $w\in \L_{pu}^+$. Furthermore, this is the last co-circularity of $w,q,u,p$.

\medskip
\noindent{\it Definition.} Let $P$ be a finite set of moving points in $\reals^2$. We say that a quadruple $\TQ=(w,q,u,p)$ in $P$ is {\it terminal of type D} if it satisfies the above conditions (D1) and (D2), with $R$ replaced by $P$.

\smallskip
Let $T^D(m)$ denote the maximum possible number of terminal quadruples of type D that can arise in an underlying set of $m$ moving points. Then the overall number of 3-restricted left special quadruples that fall into case (d1)
is at most
$$
O\left(h^4 T^D(n/h)+\ell h^2 N(n/h)+\ell h n^2\beta(n)\right).
$$

In Section \ref{Subsec:TerminalCD} we will use the corresponding extremal Delaunay co-circularity of $w,q,u,p$ of each terminal quadruple $\TQ$ to enforce a Delaunay crossing of $pu$ by at least one of the remaining two points $w,q$ of $\TQ$. Together with the Delaunay crossings in condition (D1), at least one of the triples $p,u,w$ or $p,u,q$ will perform two (single) Delaunay crossings. Therefore, our analysis will again bottom up via Lemma \ref{Lemma:TwiceCollin}.


\paragraph{$3$-restricted left special quadruples--wrap up.}
Putting together the previously established bounds on the maximum possible numbers of $3$-restricted left special quadruples that fall into cases (a), (b), (c), (d1) and (d2) yields the following recurrence:
\begin{equation}
\Phi_3^L(n)=O\left(h^4 T^C(n/h)+h^4 T^D(n/h)+\ell h^2 N(n/h)+k\ell^2 N(n/\ell)+k^2N(n/k)+\ell h n^2\beta(n)\right).
\end{equation}

\section{The number of terminal quadruples}\label{Sec:Terminal}
In this section we obtain ``quadratic" recurrences for the maximum numbers $T^A(n),T^B(n),T^C(n)$, and $T^D(n)$, of terminal quadruples of the respective types A, B, C, and D, which arise at the last stage of the analysis in Section \ref{Sec:Special}. Each of these four quantities is expressed only in terms of the maximum number of Delaunay co-circularities in smaller-size sets, plus a nearly quadratic additive term. In other words, our analysis bottoms out.
Combining these four new recurrences with the ones, obtained in Sections \ref{Sec:ReduceToCrossings}, \ref{Sec:CountQuad}, and \ref{Sec:Special}, we finally get a complete system of ``quadratic" recurrences, whose 
solution is $N(n)=O(n^{2+\eps})$, for any $\eps>0$. This completes the proof of Theorem \ref{Thm:OverallBound}. 

\subsection{Terminal quadruples of type A}\label{Subsec:TypeA}
In this section we finally express the maximum possible cardinality $T^A(n)$ of a family $\TF^A$ of terminal quadruples of type A (where each quadruple in $\TF^A$ is uniquely determined by each of its four sub-triples) in terms of more elementary quanitities that were introduced in Section \ref{Sec:Prelim}. 

To do so, we fix the underlying set $P$ of $n$ moving points, a family $\TF^A$ as above, and a terminal quadruple $\varrho=(p,q,r,w)$ in $\TF_A$. We emphasize that $\TQ$, as well as any other quadruple $(p,q,r,w)\in \TF^A$, is uniquely determined by each of its four sub-triples $(p,q,r),(p,q,w),(p,r,w), (q,r,w)$.

Recall that the four points of $\varrho$ perform {\it four} Delaunay crossings, namely the crossing of $pq$ by $r$, the crossing of $qp$ by $w$, the crossing of $rp$ by $w$, and the crossing of $wq$ by $r$.
Here only the first crossing, $(pq,r,I=[t_0,t_1])$, is defined with respect to the compelete point set $P$. Each of the remaining three crossings of $\varrho$ occurs within a reduced point set, which is obtained by omitting from $P$ the fourth point of $\varrho$ (not directly involved in the crossing).

In this section, we shall enforce on the points of $\varrho$ an additional {\it fifth} crossing, namely the crossing of $rw$ (or its reversely oriented copy $wr$) by one of $p,q$.
As a result, at least one of the triples $p,r,w$ or $q,r,w$ will perform two Delaunay crossings (within an appropriately reduced triangulation).
We thus shall charge $\varrho$ to that triple and complete our analysis by invoking Lemma \ref{Lemma:TwiceCollin}.

\begin{figure}[htbp]
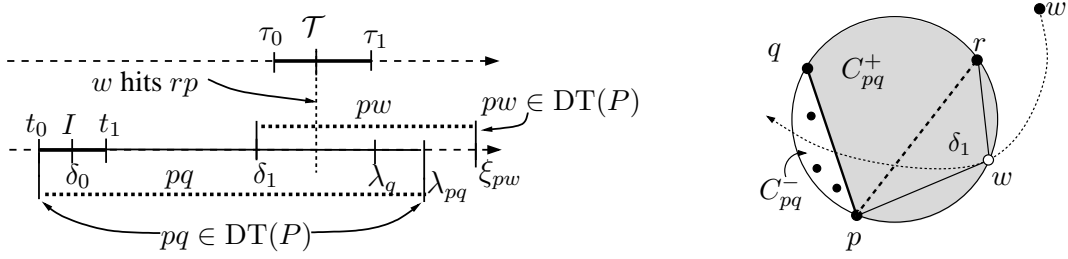

\begin{center}
\input{SetupA.pstex_t}\hspace{3cm}\input{BothDelaunay.pstex_t}
\caption{\small A partial summary of the properties of a terminal quadruple $\varrho=(p,q,r,w)$ of type A.
Left: Various events occur in the depicted order (and $\delta_2$ occurs in $(\delta_1,\t_{pq}]$). Notice that $w$ hits $rp$ in $[\tau_0,\tau_1]\cap (\delta_1,\t_q)$. Right: The edges $pw$ and $rw$ are Delaunay at time $\delta_1$, because the open cap $C_{pq}^+$ contains then no points of $P$.}
\label{Fig:SetupA}
\vspace{-0.4cm}
\end{center}
\end{figure}

\paragraph{Topological setup.} Refer to Figure \ref{Fig:SetupA}. By condition (A1), the edge $pq$ 
is crossed by $r$ (during $I=[t_0,t_1]$, as part of the corresponding Delaunay crossing) and $w$ (at some later time $\t_q>t_1$), in opposite directions.
Furthermore, $pq$ re-enters $\DT(P)$ at some later time $\t_{pq}$ after $\t_q$, and it belongs to $\DT(P\setminus \{r,w\})$ throughout $[t_0,\t_{pq}]$.

By condition (A2), the four points $p,q,r$ and $w$ are co-circular at some times $\delta_0\in I$, $\delta_1\in (t_1,\t_{pq}]$ and $\delta_2\in (\delta_1,\t_{pq}]$, where the two extremal co-circularities (at times $\delta_0$ and $\delta_2$) are red-blue with respect to $pq$, and the middle co-circularity (at time $\delta_1$) is red-red with respect to $pq$ (and red-blue with respect to $pr$). 

As a matter of fact, $\delta_2$ arises as part of a single Delaunay of $qp$ by $w$, which occurs in the triangulation $\DT(\{p,q,r,w\})$ within the interval $(\delta_1,\t_{pq}]$. Therefore, if $w$ lies at that moment in $\L_{pq}^-$ (so $r$ lies then in $\L_{pq}^+$), the Delaunayhood of $rw$ is violated right after $\delta_2$ by $p$ and $q$, and otherwise the Delaunayhood of $pq$ is violated right after $\delta_2$ by $r$ and $w$.

Furthermore, the open cap $C_{pq}^+=B[p,q,w]\cap \L_{pq}^+$ contains no points of $P$ at time $\delta_1$, which is easily seen to imply the following property:

\begin{claim}\label{Claim:FullyDelaunayDelta1}
With the above assumptions, both edges $pw$ and $rw$ are Delaunay at time $\delta_1$.
\end{claim}
\begin{proof}
If the the opposite cap $C_{pq}^-=B[p,q,w]\cap \L_{pq}^-$ contains no points of $P$ at time $\delta_1$, then this co-circularity of $p,q,r,w$ is Delaunay, and we are done.
Otherwise, $pq$ is not Delaunay even in $P\setminus \{w\}$, and each of its violating pairs in $P\setminus \{w\}$ must involve $r$ (because $\delta_1\in (t_0,\t_{pq})$). 
Hence, applying Lemma \ref{Lemma:Incremental} to $pq$ and $r$ in $P\setminus\{w\}$ shows that both edges $pr$ and $rq$ belong at that moment to the triangulation $\DT(P\setminus \{w\})$. Furthermore, since $pr$ does not belong to $\DT(P)$ (as $C_{pq}^-\subseteq B[p,q,r]\cap \L_{pr}^-$ is not empty), the claim now follows by another application of Lemma \ref{Lemma:Incremental}, this time to $pr$ and $w$.
\end{proof}

By condition (A4), we have a time $\xi_{pw}>\t_{pq}>\t_q$ so that $pw$ belongs to $\DT(P\setminus \{r,q\})$ throughout the interval $(\delta_1,\xi_{pw})$, and it is Delaunay at time $\xi_{pw}$ (in addition to its Delaunayhood at time $\delta_1$).

Finally, by condition (A3), the edge $rp$ undergoes in $P\setminus \{q\}$ a single Delaunay crossing $(rp,w,\T=[\tau_0,\tau_1])$, where $w$ enters $\L_{rp}^+=\L_{pr}^-$ in the interval $(\delta_1,\t_q)$. 
Hence, Lemma \ref{Lemma:Crossing} implies that $pw$ belongs $\DT(P\setminus \{q\})$ throughout the interval $\T=[\tau_0,\tau_1]$, which clearly intersects $(\delta_1,\xi_{pw})\supset (\delta_1,\t_q)$.
Similarly, the edge $wq$ undergoes in $P\setminus \{p\}$ a Delaunay crossing by $r$. 

In what follows, we consider a subfamily $\TF^A_{rw}$ of all terminal quadruples $\varrho'=(p',q',r,w)$ in $\TF^A$ whose last two points are equal to, respectively, $r$ and $w$. In particular, $\TF^A_{rw}$ includes the terminal quadruple $\varrho=(p,q,r,w)$ under consideration.
Note that each $\varrho'=(p',q',r,w)\in \TF^A_{rw}$ is accompanied by a clockwise $(r,w)$-crossing $(rp',w,\T')$
(which occurs within an appropriately reduced triangulation $\DT(P\setminus \{q'\})$.

To enforce a Delaunay crossing of by $rw$ by $p$ or $q$, we fix a pair of constants $k\ll \ell$ and distinguish between three possible cases, treating each in turn. 

\medskip
\noindent{\bf Case (a)} The crossing $(rp,w,\T=[\tau_0,\tau_1])$ begins after $\delta_1$ and $\TF^A_{rw}$ contains at least $k$ terminal quadruples $\varrho'=(p',q',r,w)$ 
whose respective clockwise $(r,w)$-crossings $(rp',w,\T')$
begin in $[\delta_1,\tau_0)$, or $[\tau_0,\tau_1]$ ends before $\xi_{pw}$ and $\TF^A_{rw}$ contains at least $k$ terminal quadruples $\varrho'=(p',q',r,w)$ 
whose respective clockwise $(r,w)$-crossings $(rp',w,\T')$ end in $(\tau_1,\xi_{pw}]$.

\begin{figure}[htbp]
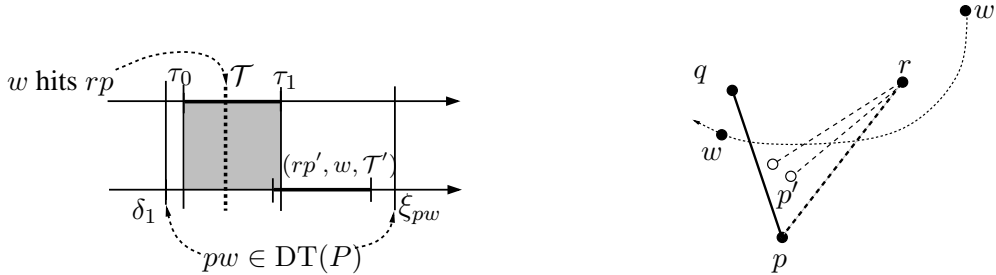

\begin{center}
\input{TerminalACaseA.pstex_t}\hspace{3cm}\input{TypeACaseAEdges.pstex_t}
\caption{\small Case (a): The scenario where $(rp,w,\T=[\tau_0,\tau_1])$ ends before $\xi_{pw}$, and the family $\TF^A_{rw}$ contains at least $k$ terminal quadruples $\varrho'=(p',q',r,w)$ whose respective $(r,w)$-crossings $(rp',w,\T')$ end in $(\tau_1,\xi_{pw}]$. At at least $k-2$ of these quadruples satisfy $p'\neq q$ and $q'\neq p$, so their respective intervals $\T'$ are entirely contained in $[\tau_0,\xi_{pw}]$.}
\label{Fig:TypeACaseA}
\vspace{-0.4cm}
\end{center}
\end{figure} 

Assume without loss of generality that the latter scenario occurs, so at least $k$ clockwise $(r,w)$-crossings $(rp',w,\T')$ end in $(\tau_1,\xi_{pw}]$; see Figure \ref{Fig:TypeACaseA}.
Notice that each of them occurs within a smaller triangulation $\DT(P\setminus \{q'\})$ which is, in general, distinct from the ambient triangulation $\DT(P\setminus \{q\})$ of $(rp,w,\T)$.
Fortunately, any terminal quadruple $\varrho'=(p',q',r,w)\in \TF^A_{rw}$ is uniquely determined by each of its respective points $p'$ and $q'$. Hence, at least $k-2$ of the above quadruples $\varrho'$ satisfy $p'\neq q$ and $q'\neq p$, in which case their respective $(r,w)$-crossings are compatible with $(rp,w,\T)$ (through Lemma \ref{Lemma:OrderRelaxedCrossings}) and, therefore, occur entirely within $[\tau_0,\xi_{pw}]=\T\cup (\tau_1,\xi_{pw}]$. 

We sample a subset $\hat{P}$ of $n/4$ points and argue that, with some positive fixed probability, $(rp,w,\T)$ becomes a $(p,w,\Theta(k))$-chargeable Delaunay crossing within $\DT(\hat{P})$. 
Namely, we notice that the following two events occur simultaneously with some fixed positive probability:
(1) $\hat{P}$ includes the three points $p,r,w$, but not $q$, and (2) $\hat{P}$ includes $p'$ but not $q'$ for at least some constant fraction of the above quadruples $\varrho'=(p',q',r,w)\in \TF^A_{rw}$ (whose respective $(r,w)$-crossings $(rp',w,\T')$ end in $(\tau_1,\xi_{pw})$).
In the case of success, condition (1) implies that $rp$ still undergoes a single Delaunay crossing by $w$ in $\hat{P}$, which occurs in some sub-interval of $\T=[\tau_0,\tau_1]\subset [\tau_0,\xi_{pw}]$. Similarly, condition (2) implies that at least $\Omega(k)$ clockwise $(r,w)$-crossings in $R$ occur within $[\tau_0,\xi_{pw}]$.

By Theorem \ref{Thm:Balanced}, the overall number of such triples $(p,w,r)$ in $\hat{P}$ (and, thereby, in $P$) cannot exceed $O\left(k^2N(n/k)+k n^2\beta(n)\right)$.
Clearly, this also bounds the overall number of the corresponding terminal quadruples $\varrho=(p,q,r,w)$ in $P$.
If $(rp,w,\T)$ ends before $\xi_{pw}$, and $\Sigma_{rw}^A$ contains at least $k$ terminal quadruples $\TQ'$ whose respective $(r,w)$-crossings $(rp',w,\T')$ end in $(\tau_1,\xi_{pw}]$, we argue in a fully symmetrical manner, for the same upper bound on the number of such terminal quadruples $\varrho$.

We thus can assume, in what follows, that either the crossing $(rp,w,\T=[\tau_0,\tau_1])$ ends after $\xi_{pw}$, or the sub-family $\TF_{rw}^A$ contains at most $k$ other quadruples $\varrho'=(p,q,r,w)$ whose respective $(r,w)$-crossings $(rp',w,\T')$ end in $(\tau_1,\xi_{pw}]$. Similarly, we can assume that either $[\tau_0,\tau_1])$ begins before $\delta_1$, or the sub-family $\TF_{rw}^A$ contains at most $k$ other quadruples $\varrho'=(p',q',r,w)$ whose respective $(r,w)$-crossings $(rp',w,\T')$ begin in $[\delta_1,\tau_0)$.

\medskip
\noindent{\bf Case (b)} The family $\TF_{rw}^A$ contains no terminal quadruple $\varrho'=(p',q',w,r)\neq \varrho$ that satisfies $q'\neq p$, and whose respective $(r,w)$-crossing $(p'r,w,\T')$ ends in $[\xi_{pw},\infty)$.

Since case (a) has been ruled out (and $\TF_{rw}^A$ contains at most one quadruple $\varrho'=(p',q',r,w)$ with $q'=p$), we conclude that there exist at most $k+1$ terminal quadruples $\varrho'\in \TF^A_{rw}$ whose respective $(r,w)$-crossings $(p'r,w,\T')$ end after $\T=[\tau_0,\tau_1]$.
Hence, we charge $(p,q,r,w)$ (via its respective $(r,w)$-crossing $(pr,w,\T=[\tau_0,\tau_1])$) to the edge $rw$ and notice that any edge can be charged in this manner by at most $k+2$ terminal quadruples. 

To conclude, the above scenario is encountered for at most $O(kn^2)$ terminal quadruples $\varrho$.


\medskip
\noindent{\bf Case (c)} None of the previous cases occurs. 
In particular, since case (b) has been ruled out,
the family $\TF^A_{rw}$ contains at least one quadruple $\varrho'=(p',q',r,w)\neq \varrho$, with $q'\neq p$, and whose respective $(r,w)$-crossing $(rp',w,\T')$ ends in $[\xi_{pw},\infty)$. (Clearly, we have $q'\neq q$, for otherwise $\varrho$ would coincide with $\varrho'$.)

\begin{figure}[htbp]
\begin{center}
\input{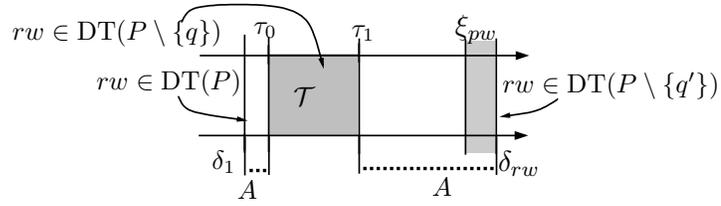}
\caption{\small Case (c): Extending the almost-Delaunayhood of $rw$ to $[\delta_1,\delta_{rw}]$. Here $\delta_{rw}$ is the first time in $[\xi_{pw},\infty)$ when $rw$ belongs to some reduced triangulation $\DT(P\setminus \{q'\})$, for some $q'\neq p,q$.}
\label{Fig:ExtendRw}
\vspace{-0.4cm}
\end{center}
\end{figure}

Applying Lemma \ref{Lemma:Crossing} to the crossing $(rp',w,\T')$ (in its ambient set $P\setminus \{q'\}$) implies, then, there is a time $\delta_{rw}\geq \xi_{pw}$ which is the first such time when the edge $rw$ belongs to some reduced triangulation $\DT(P\setminus \{q'\})$, where $q'\neq p,q$.
In what follows, we use $q'$ to denote a fixed point in $P\setminus \{p,q,r,w\}$ whose removal restores the Delaunayhood at time $\delta_{rw}$; see Figure \ref{Fig:ExtendRw}.

Note that we have $\delta_{rw}>\t_{pq}>\delta_1$.
Since case (a) has been ruled out, the choice of $\delta_{rw}$ guarantees that, unless $\delta_{rw}$ belongs to  $\T=[\tau_0,\tau_1]$, there exist at most $k+1$ quadruples $\varrho'\in \TF^A_{rw}$ whose respective $(r,w)$-crossings $(rp',w,\T')$ end in $(\tau_1,\delta_{rw}]$. (In particular, by the choice of $\delta_{rw}$, there is at most one quadruple $\TQ'=(p',q',r,w)$ whose respective $(r,w)$-crossing ends in $(\xi_{pw},\delta_{rw}]$, and it must satisfy $q'=p$.)

\paragraph{Charging events in $\A_{rw}$.} We next invoke Theorem \ref{Thm:RedBlue} in order to extend the almost-Delaunayhood of $rw$, which already belongs to $\DT(P\setminus \{q\})$ throughout $\T=[\tau_0,\tau_1]$ (by Lemma \ref{Lemma:Crossing}), to the interval $[\delta_1,\delta_{rw}]$, which clearly intersects $\T$.

Note that $[\delta_1,\delta_{rw}]\setminus \T$ is composed of two disjoint (and possibly empty) sub-intervals $[\delta_1,\tau_0)$ and $(\tau_1,\delta_{rw}]$. We apply Theorem \ref{Thm:RedBlue} separately over each of these sub-intervals (and only if they are non-empty). In both cases, we use the second threshold parameter $\ell\gg k$.

The first application of Theorem \ref{Thm:RedBlue} in $\A_{rw}$, over $(\delta_1,\tau_0)$, can be done with respect to the complete point set $P$ (using the Delaunayhood of $rw$ at time $\delta_1$, given in Claim \ref{Claim:FullyDelaunayDelta1}). It is necessary only if $\delta_1<\tau_0$.

If at least one of the conditions (i), (ii) of that theorem is satisfied, we charge $\varrho$ within $\A_{rw}$, via $(rp,w,\T)$, either to an $\ell$-shallow collinearity, or to $\Omega(\ell^2)$ $\ell$-shallow co-circularities during $(\delta_1,\tau_0)$.
Since case (a) has been ruled out, $(rp,w,\T)$ is among the first $k+1$ such $(r,w)$-crossings to begin after any event that we charge. Hence, the above charging accounts for at most $O(k\ell^2N(n/\ell)+k\ell n^2\beta(n))$ quadruples $\varrho\in \TF^A$.
Otherwise, we end up with a subset of at most $3\ell$ points (perhaps including $p$ or $q$, or both) whose removal restores the Delaunayhood of $rw$ throughout $[\delta_1,\tau_0]$.

The similar second application of Theorem \ref{Thm:RedBlue} (over $(\tau_1,\delta_{rw})$) is done with respect to the reduced point set $P\setminus \{q'\}$ (where $q'$ denotes the point whose removal restores the Delaunayhood of $rw$ at time $\delta_{rw}$). It is necessary only if $\tau_1<\delta_{rw}$.

If at least one of the conditions (i), (ii) of that theorem holds, we charge $\varrho$ (via $(rp,w,\T)$) within $\A_{rw}$ either to an $(\ell+1)$-shallow collinearity, or to $\Omega(\ell^2)$ $(\ell+1)$-shallow co-circularities (which are $\ell$-shallow with respect to $P\setminus \{q'\}$).
By the choice of $\delta_{rw}$, $(rp,w,\T)$ is among the last $k+2$ such $(r,w)$-crossings to end after the event, so any $(\ell+1)$-shallow event in $\A_{rw}$ is charged by at most $O(k)$ quadruples $\varrho$.
Otherwise, we end up with a subset of at most $3\ell+1$ points (inclding $q'$, and perhaps also some of $p,q$) whose removal restores the Delaunayhood of $rw$ throughout $[\tau_1,\delta_{rw}]$.

\begin{figure}[htbp]
\begin{center}
\input{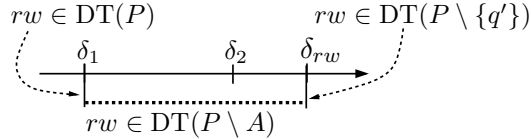}
\caption{\small Case (c): The edge $rw$ belongs to $\DT(P\setminus A)$ throughout the interval $[\delta_1,\delta_{rw}]$, which contains the last co-circularity $\delta_2$ of $p,q,r,w$. In addition, $rw$ belongs to $\DT(P)$ and $\DT(P\setminus \{q'\})$ at times $\delta_1$ and $\delta_{rw}$, respectively.}
\label{Fig:CrossRw}
\vspace{-0.4cm}
\end{center}
\end{figure}

To conclude, we may assume that there is a subset $A_{rw}$ of at most $6\ell+1$ points (including $q'$) whose removal restores the Delaunayhood of $rw$ throughout $[\delta_1,\delta_{rw}]$. To obtain the  crossing of $rw$ by $p$ or $q$ (which would occur in, respectively, $\DT((P\setminus A_{rw})\cup \{p\})$ or $\DT((P\setminus A_{rw})\cup \{q\})$), it suffices to show that $rw$ is hit by one of these two points during the interval $[\delta_1,\delta_{rw}]$. Notice that the latter interval contains $\delta_2\in (\delta_1,\t_{pq})\subset (\delta_1,\xi_{pw}]$. See Figure \ref{Fig:CrossRw}.
To do so, we distinguish between two possible sub-scenarios, depending on the precise order type of the co-circularity (at time) $\delta_2$, which is red-blue with respect to $pq$ and $rw$.

\begin{figure}[htbp]
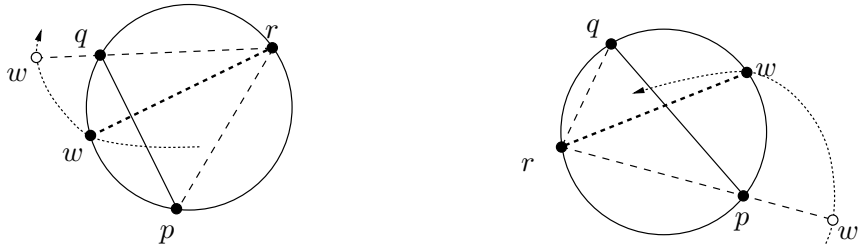

\begin{center}
\input{ViolateRw.pstex_t}\hspace{3cm}\input{ViolateRwBefore.pstex_t}
\caption{\small Case (c): Left: A possible motion of $w$ if it lies at time $\delta_2$ in the halfplane $\L_{pq}^-$ (so $r$ lies then in $\L_{pq}^+$). The Delaunayhood of $rw$ is violated right after this event by $p$ and $q$, so at least one of them must cross $rw$ during $(\delta_2,\delta_{rw})$. Right: A possible motion of $w$ if it lies at time $\delta_2$ in the halfplane $\L_{pq}^-$ (so $r$ lies then in $\L_{pq}^+$). The Delaunayhood of $rw$ is violated right after this event by $p$ and $q$, so at least one of them must cross $rw$ during $(\delta_1,\delta_2)$.}
\label{Fig:CrossRwHit}
\vspace{-0.4cm}
\end{center}
\end{figure}

If $r$ lies in $\L_{pq}^-$ when $w$ enters $\L_{pq}^-$ (through $pq$), then the Delaunayhood of $rw$ is violated right {\it after} $\delta_2$ by $p\in \L_{rw}^-$ and $q\in \L_{rw}^+$, as depicted in Figure \ref{Fig:CrossRwHit} (left).
Since $\delta_2$ is the {\it last} co-circularity of $p,q,r,w$, Lemma \ref{Lemma:MustCross} implies that $rw$ is hit during $(\delta_2,\delta_{rw}]$ by at least one of $p,q$ (because $q'\neq p,q,r,w$), so we are done.

Assume, then, that $r$ lies in $\L_{pq}^+$ when $w$ enters $\L_{pq}^-$, so the Delaunayhood of $rw$ is violated right {\it before} $\delta_2$ by $p\in \L_{rw}^+$ and $q\in \L_{rw}^-$, as depicted in Figure \ref{Fig:CrossRwHit} (right). Notice that this violation does not hold at time $\delta_1$. Hence, we can obtain the desired crossing of $rw$ in $(\delta_1,\delta_2)$ by applying the time-reversed variant of Lemma \ref{Lemma:MustCross} (from $\delta_2$).
The crucial observation is that $\delta_1,\delta_2$ have different order types, which rules out the last case in Lemma \ref{Lemma:MustCross}.

If $rw$ is hit during $(\delta_1,\delta_{rw}]$ by the point $p$, then the triple $p,r,w$ performs two Delaunay crosings within the triangulation $\DT((P\setminus A_{rw})\cup \{p\})$, namely, $(rp,w,\T)$ and the just established crossing of $wr$ by $p$.
Otherwise, if $rw$ is hit during $(\delta_1,\delta_{rw}]$ by $q$, the other triple $q,r,w$ performs two Delaunay crossings within the triangulation $\DT((P\setminus A)\cup \{q\})$, namely, the crossing of $wq$ by $r$ (as prescribed by condition (A3)) and the just established crossing of $rw$ by $q$.

In both cases, a standard combination of Lemma \ref{Lemma:TwiceCollin} with the probabilistic argument of Clarkson and Shor implies that the overall number of the corresponding triples $(p,r,w)$ or $(q,r,w)$ in $P$ cannot exceed $O(\ell n^2)$. Since the quadruple $\varrho$ at hand is uniquely determined by each of its four sub-triples, this also bounds the overall number of such quadruples in $\TF^A$.

To conclude, we have established the following bound on the maximum possible cardinality of $\TF^A$:
\begin{equation}
T^A(n)=O\left(k\ell^2 N(n/\ell)+k^2N(n/k)+k\ell n^2\beta(n)\right).
\end{equation}

Notice that we have expressed the maximum possible number of terminal quadruples of type A in terms of more elementary quantities which were introduced in Section \ref{Sec:Prelim}.

\subsection{Terminal quadruples of type B}\label{Subsec:TypeB}
In this section we at last express the maximum possible cardinality $T^B(n)$ of a family $\TF^B$ of terminal quadruples of type B (where each quadruple $(p,q,r,w)\in \TF^B$ is uniquely determined by each of the respective sub-triples $(p,q,r)$, $(p,q,w)$ and $(q,r,w)$) in terms of more elementary quanitities that were introduced in Section \ref{Sec:Prelim}.
To do so, we fix the underlying set $P$ of $n$ moving points, a family $\TF^B$ as above, and a terminal quadruple  $\varrho=(p,q,r,w)$ of type B in $\TF^B$.

Recall that the four points of $\varrho$ perform (at least) three Delaunay crossings, namely the crossing of $pq$ by $r$, the crossing of $qp$ by $w$, and the crossing of $qr$ by $w$.
Here only the first crossing, namely, $(pq,r,I=[t_0,t_1])$, is defined with respect to the compelete point set $P$. Each of the remaining three crossings of $\varrho$ occurs within a reduced point set, which is obtained from $P$ by removing the fourth point of $\varrho$ (not directly involved in the crossing).

In the course of this section, we will enforce on the points of $\varrho$ two additional crossings, namely the crossing of $pw$ by one of $q,r$, and, finally, the crossing of $rw$ by one of $p,q$.
As a result, at least one of the triples $(p,q,w)$, $(p,r,w)$ or $(q,r,w)$ will perform two Delaunay crossings (within an appropriately reduced triangulation).
We will thus charge $\varrho$ to that triple and bottom out by invoking Lemma \ref{Lemma:TwiceCollin}.

\begin{figure}[htbp]
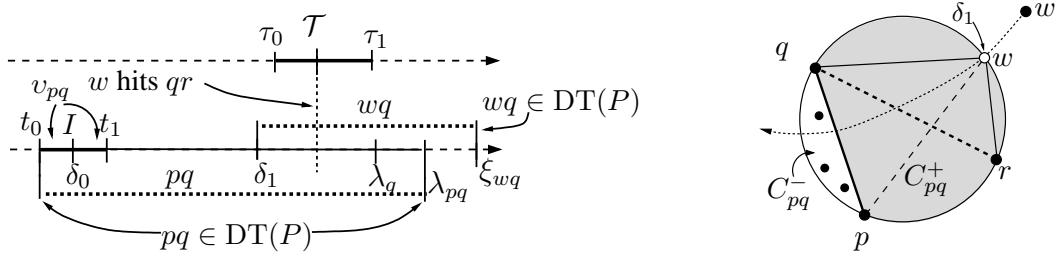

\begin{center}
\input{SetupB.pstex_t}\hspace{3cm}\input{BothDelaunayB.pstex_t}
\caption{\small A partial summary of the properties of a terminal quadruple $\varrho=(p,q,r,w)$ of type B.
Left: Various events occur in the depicted order (and $\delta_2$ occurs in $(\delta_1,\t_{pq}]$). $\upsilon_{pq}$ is the time in $I$ at which $r$ hits $pq$. Notice that $w$ hits $qr$ in $[\tau_0,\tau_1]\cap (\delta_1,\t_q)$. Right: The edges $wq$ and $rw$ are Delaunay at time $\delta_1$, and $pw$ belongs to $\DT(P\setminus \{r\})$, because the open cap $C_{pq}^+$ contains then no points of $P$.}
\label{Fig:SetupB}
\vspace{-0.4cm}
\end{center}
\end{figure}

\paragraph{Topological setup.} Refer to Figure \ref{Fig:SetupB}. By condition (B1), the edge $pq$ 
is crossed by $r$ (during $I=[t_0,t_1]$, as part of the corresponding Delaunay crossing) and $w$ (at some later time $\t_q>t_1$), in opposite directions.
Furthermore, $pq$ re-enters $\DT(P)$ at some later time $\t_{pq}\geq \t_q$, and it belongs to $\DT(P\setminus \{r,w\})$ throughout $[t_0,\t_{pq}]$.

By condition (B2), the four points of $p,q,r$ and $w$ are co-circular at some three times $\delta_0\in I$, $\delta_1\in (t_1,\t_{pq}]$ and $\delta_2\in (\delta_1,\t_{pq}]$, where the two extremal co-circularities (at times $\delta_0$ and $\delta_2$) are red-blue with respect to $pq$, and the middle co-circularity (at time $\delta_1$) is red-red with respect to $pq$ (and red-blue with respect to $rq$). 
Clearly, $r$ remains in $\L_{pq}^+$ throughout $(t_1,\delta_1)$ after entering this halfplane during $I$ (for otherwise $r$ would have to cross $\L_{pq}$ three times).

As a matter of fact, the last co-circularity at time $\delta_2$ arises as part of a single Delaunay of $qp$ by $w$, which occurs in the triangulation $\DT(\{p,q,r,w\})$ within the interval $(\delta_1,\t_{pq}]$. Therefore, if $w$ lies at that moment in $\L_{pq}^-$ (so $r$ lies then in $\L_{pq}^+$), the Delaunayhood of $rw$ is violated right after $\delta_2$ by $p$ and $q$, and otherwise the Delaunayhood of $pq$ is violated right after $\delta_2$ by $r$ and $w$.

Furthermore, the open cap $C_{pq}^+=B[p,q,w]\cap \L_{pq}^+$ contains no points of $P$ at time $\delta_1$.
Using Lemma \ref{Lemma:Incremental}, we obtain the following property: 

\begin{claim}\label{Claim:FullyDelaunayDelta1B}
With the above assumptions, both edges $wq$ and $rw$ belong to $\DT(P)$ at time $\delta_1$.  Furthermore, the edge $pw$ belongs then to $\DT(P\setminus \{r\})$. 
\end{claim}
\begin{proof}
The first part of the claim is fully symmetric to Claim \ref{Claim:FullyDelaunayDelta1}, and can be established using a fully symmetric argument (switching the roles of $p$ and $q$). We thus proceed to proving the Delaunayhood of $pw$ in $P\setminus \{w\}$.
Indeed, if the opposite cap $C_{pq}^-=B[p,q,w]\cap \L_{pq}^-$ contains no points of $P$ at time $\delta_1$, then this co-circularity of $p,q,r,w$ is Delaunay, and we are done.
Otherwise, $pq$ is not Delaunay even in $P\setminus \{r\}$, and each of its violating pairs in $P\setminus \{r\}$ must involve $w$ (because $\delta_1\in (t_0,\t_{pq})$). 
Hence, Lemma \ref{Lemma:Incremental} implies that $pw$ belongs at that moment to the triangulation $\DT(P\setminus \{r\})$. 
\end{proof}

By condition (B4), we have a time $\xi_{wq}>\t_{pq}>\t_q$ so that $wq$ belongs to $\DT(P\setminus \{p,r\})$ throughout the interval $[\delta_1,\xi_{wq}]$, and it is Delaunay at time $\xi_{wq}$ (in addition to its Delaunayhood at time $\delta_1$).

Finally, by condition (B3), the edge $qr$ undergoes in $P\setminus \{p\}$ a single Delaunay crossing $(qr,w,\T=[\tau_0,\tau_1])$, where $w$ enters $\L_{qr}^+=\L_{rq}^-$ in the interval $(\delta_1,\t_q)$. 
Hence, Lemma \ref{Lemma:Crossing} implies that $wq$ belongs $\DT(P\setminus \{p\})$ throughout the interval $\T=[\tau_0,\tau_1]$, which clearly intersects $[\delta_1,\xi_{wq}]\supset [\delta_1,\t_q]$.

\paragraph{Overview.}
Clearly, the motion of $p,q,r$ and $w$ still obeys Proposition \ref{Prop:HitPw}. In particular, using $\upsilon_{pq}$ to denote the time\footnote{Note that the order of $\delta_0$ and $\upsilon_{pr}$ is unknown, and it is determined by the location of $w$ at time $\upsilon_{pq}$.} in $I$ when $r$ enters $\L_{pq}^+$ through $pq$, the edge $pw$ is hit in $(\upsilon_{pq},
\delta_1)$ by at least one of the points $q,r$. Namely, $q$ crosses $pw$ from $\L_{pw}^+$ to $\L_{pw}^-$, or $r$ crosses $pw$ in the reverse direction.
Our analysis proceeds in two steps. 
At the first step, we refine this collinearity into a full-fledged Delaunay crossing of $pw$. If $pw$ (or, more precisely, its reversely oriented copy $wp$) is crossed by $q$, then our analysis bottoms out through Lemma \ref{Lemma:TwiceCollin}. 
If $pw$ is crossed by $r$, we proceed to the second step, at which we enforce a Delaunay crossing of $rw$ by at least one of $p,q$. At this step, our analysis is fully symmetric to the one that was used in Section \ref{Subsec:TypeA} to enforce the same type of crossing.

\paragraph{Part 1: Enforcing the crossing of $pw$ by $q$ or $r$.}

We consider the subfamily $\TF^B_{pw}$ of all terminal quadruples $\varrho'=(p,q',r',w)\in \TF^B$ of type B whose first and last points are equal to, respectively, $p$ and $w$. (By the definition of $\TF^B$, each $\varrho'\in \TF^B_{pw}$ is uniquely determined by its respective point $q'$.)
In particular, $\TF^B_{pw}$ includes the quadruple $\TQ$ under consideration.

For each $\TQ'\in \TF^B_{pw}$ we use $\delta'_1$ to denote the respective time when $p,q',r',w$ are involved in a red-red co-circularity with respect to $pq'$, as prescribed by Condition (B2).
We emphasize that, by condition (B2), the open cap $C_{pq'}^+=B[p,q',w]\cap \L_{pq}^+$ contains no points of $P$ at time $\delta'_1$, so the edge $pw$ belongs at that moment to the triangulation $\DT(P\setminus \{r'\})$.

Let $\upsilon_{pw}$ denote the time when $pw$ is hit by $r$ or $q$, as prescribed in Proposition \ref{Prop:HitPw}.
Namely, we assume that $\upsilon_{pw}<\delta_1$, and that the Delaunayhood of $pw$ is violated by $r\in \L_{pw}^-$ and $q\in \L_{pw}^+$ throughout the interval $(\upsilon_{pw},\delta_1)$.


\begin{proposition}\label{Prop:Prw}
With the above notation, there exist no terminal quadruples $\varrho'\in \TF_{pw}^B$ whose respective second co-circularities $\delta'_1$ occur in $(\upsilon_{pw},\delta_1)$. 
\end{proposition}
\begin{proof}
Assume for a contradiction that there is a terminal quadruple $\varrho'=(p,q',r',w)$ whose respective time $\delta'_1$ belongs to $(\upsilon_{pw},\delta_1)$, where the Delaunayhood of $pw$ is violated by $q\in \L_{pw}^-$ and $r\in \L_{pw}^+$. Note that $q\neq q'$. By Claim \ref{Claim:FullyDelaunayDelta1B}, $pw$ belongs to $\DT(P\setminus \{r'\})$ at time $\delta'_1$. Therefore, and since both $r$ and $r'$ lie then in $\L_{pw}^+$, we obtain $r=r'$. (Otherwise, the Delaunayhood of $pw$ would be violated at time $\delta'_1$ by the points $r$ and $q$, none of them equal to $r'$.) In other words, $\TQ$ and $\TQ'$ differ only in their second points. Hence, $q$ lies at time $\delta'_1$ within the disc $B[p,q',r]=B[p,q',w]$.

Since $q$ cannot lie at time $\delta'_1$ inside the cap $C_{pw}^+=B[p,q',w]\cap \L_{pq'}^+$, it has to lie inside the complementary cap $C_{pq}^-=B[p,q',w]\cap \L_{pq'}^-$, which coincides with $B[p,q',r]\cap \L_{pq'}^-$.
In other words, the Delaunayhood of both $pq'$ and $pw$ is violated at time $\delta'_1$ by $q\in \L_{pq'}^-$ and $r\in \L_{pq'}^+$. See Figure \ref{Fig:ContradictionPw}.

\begin{figure}[htbp]
\begin{center}
\input{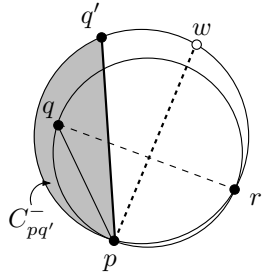}
\caption{\small Proof of Proposition \ref{Prop:Prw}. We assume that $\TQ'=(p,q',r,w)$ is a terminal quadruple in $\TF_{pw}^B$, whose second co-circularity occurs at time $\delta'_1\in (\upsilon_{pw},\delta_1)$. The point $q$ must lie at time $\delta'_1$ in the cap  $C_{pq}^-=B[p,q',w]\cap \L_{pq'}^-$, which coincides with $B[p,q',r]\cap \L_{pq'}^-$.}
\label{Fig:ContradictionPw}
\vspace{-0.2cm}
\end{center}
\end{figure} 

Recall that $\delta'_1$ occurs after the end of the respective $(p,r)$-crossing $(pq',r,I'=[t'_0,t'_1])$ of $\TQ'$ (which is prescribed by condition (B1)). Since the disc $B[p,q',r]$ 
contains no points of $P$ right after time $t'_1$ (and the motion of $B[p,q',r]$ is continuous throughout $(t'_1,\delta'_1)$), the point $q$ must enter the cap $B[p,q',r]\cap \L_{pq'}^-$ in $(t'_1,\delta'_1)$. Furthermore, conditions (B1) and (B2) imply that $q$ cannot hit $pq'$ in $(t'_1,\delta'_1)$, so $q$ can enter $B[p,q',r]\cap \L_{pq'}^-$ only through the boundary of $B[p,q',r]$, at a common co-circularity of $p,q,q',r$. See Figure \ref{Fig:ContradictionPwIntervals} (left). In what follows, we use $\delta'$ to denote the time of (the last) such co-circularity in $(t'_1,\delta'_1)$, noting that $q$ remains in $B[p,q',r]\cap \L_{pq'}^-$ throughout $(\delta',\delta'_1)$.

\begin{figure}[htbp]
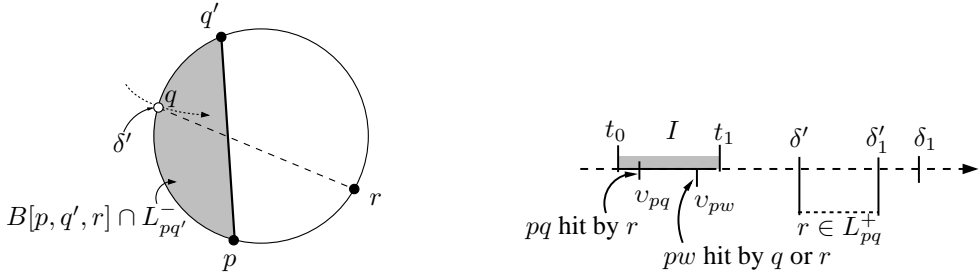

\begin{center}
\input{ContradictionPwEnter.pstex_t}\hspace{2cm}\input{TimelinePrw.pstex_t}
\caption{\small Proof of Proposition \ref{Prop:Prw}. Left: $q$ must enter $B[p,q',r]\cap \L_{pq'}^-$ at some time $\delta'\in (t'_1,\delta'_1)$. The Delaunayhood of $pq'$ is violated by $q$ and $r$ throughout $(\delta',\delta'_1)$. Right: Arguing that $\delta'$ occurs after $I$. Various events occur in the depicted order. The co-circularity at time $\delta'$ occurs after $I$, so it is the third co-circularity of $p,q,r,q'$.}
\label{Fig:ContradictionPwIntervals}
\vspace{-0.3cm}
\end{center}
\end{figure} 

We claim that $\delta'$ occurs after $I=[t_0,t_1]$; see Figure \ref{Fig:ContradictionPwIntervals} (right).
Indeed, since $q$ lies in $\L_{pr}^-(\supset B[p,q',r]\cap \L_{pq'}^-)$ throughout $(\delta',\delta'_1)$, and since 
$\delta'_1>\upsilon_{pw}>\upsilon_{pq}$, we obtain that $\upsilon_{pq}<\delta'$ (for, otherwise, $\upsilon_{pq}$ would belong to $(\delta',\delta'_1)$). 
Furthermore, by Lemma \ref{Lemma:Crossing}, $\delta'$ cannot occur during $I=[t_0,t_1]$, because it is (the time of) a red-blue co-circularity with respect to $rq$. 
Therefore, we have $\delta'>t_1$.

To conclude, $q$ enters $B[p,q',r]\cap \L_{pq'}^-$ at a common co-circularity of $p,q,q',r$, and only after the ends of $I$ and $I'$. According to Lemma \ref{Lemma:OnceCollin}, the points $p,q,q',r$ are involved in at least two previous co-circularities in the intervals $I\setminus I'$ and $I'\setminus I$. Hence, the co-circularity at time $\delta'$ has index $3$.
Note that the Delaunayhood of $pq'$ is violated by $q\in \L_{pq'}^-$ and $r\in \L_{pq'}^+$ throughout the interval $(\delta',\delta'_1)$.
Moreover, since $\TQ'$ satisfies condition (B1), the edge $pq'$ re-enters $\DT(P)$ at some time $\t_{pq'}>\delta'_1$.
Since $\delta'_1$ is the last co-circularity of $p,q',r,w$, Lemma \ref{Lemma:MustCross} implies that the edge $pq'$ is hit during $(\delta'_1,\t_{pq'}]$ by at least one of the points $q,r$, contrary to Condition (B1) on $\TQ'$. This last contradiction completes the proof of Proposition \ref{Prop:Prw}.
\end{proof}

Note that the subfamily $\TF^B_{pw}$ can contain at most one quadruple $\TQ'=(p,q',r',w)$ with $q'=r$. Applying the pigeonhole principle (as this was done in Section \ref{Subsec:Stage4Reg}) we get that at least half of all terminal quadruples $\TQ=(p,q,r,w)\in \TF_{pw}^B$ satisfy the following condition: 

\smallskip
\noindent {\bf (PHT)} {\it There is at most one quadruple $\varrho'=(p,q',r',w)\in \TF_{pw}^B$ that satisfies $r'=q$.} 

\smallskip

With no loss of generality, we can assume, in what follows, that (PHT) holds for the terminal quadruple $\varrho=(p,q,r,w)$ under consideration.
To proceed, we distinguish between two possible cases.

\medskip
\noindent{\bf Case (1a).} The edge $pw$ is hit at time $\upsilon_{pw}$ by $q$, which crosses $pw$ from $\L_{pw}^+$ to $\L_{pw}^-$. 

Assume first that there exist no terminal quadruples $\TQ'=(p,q',r',w)$ in $\TF_{pw}^B$, with $r'\neq q$, and whose respective second co-circularities $\delta'_1$ occur before $\upsilon_{pw}$. In this scenario, Proposition \ref{Prop:Prw} together with condition (PHT) imply that $\delta_1$ is among the first two such second co-circularities $\delta'_1$ of terminal quadruples $\TQ'\in \TF_{pw}^B$, so we can charge $\TQ$ (via $\delta_1$) to the edge $pw$. Clearly, this can happen for $O(n^2)$ terminal quadruples $\TQ\in \TF_{pw}^B$.

\begin{figure}[htbp]
\begin{center}
\input{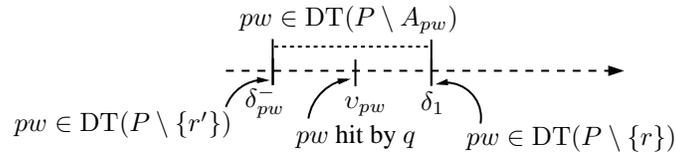}
\caption{\small Case (1a): $pw$ is hit by $q$ at time $\upsilon_{pw}$. We choose $\delta_{pw}^-$ as the last time before $\upsilon_{pw}$ when $pw$ belongs to a reduced triangulation $\DT(P\setminus \{r'\})$, for some $r'\neq q$, and apply Theorem \ref{Thm:RedBlue} over $(\delta_{pw}^-,\delta_1)$.}
\label{Fig:TimelinePw1a}
\vspace{-0.3cm}
\end{center}
\end{figure} 

To conclude, we may assume in what follows that the above scenario does not occur. 
Recall that, for each $\TQ'\in \TF_{pw}^B$, the edge $pw$ belongs to $\DT(P\setminus \{r'\})$ at the respective time $\delta'_1$. Hence, there is a time $\delta^-_{pw}<\upsilon_{pw}$ which is the last such time when $pw$ belongs to some reduced triangulation $\DT(P\setminus \{r'\})$, for $r'\neq p,q,w$. 

We apply Theorem \ref{Thm:RedBlue} for $pw$ in the interval $(\delta^-_{pw},\delta_1)$, with the first threshold parameter $k$. This is done with respect to the reduced set $P\setminus \{r,r'\}$ (to ensure the Delaunayhood of $pw$ at times $\delta^-_{pw}$ and $\delta_1$). Refer to Figure \ref{Fig:TimelinePw1a}.

In cases (i), (ii) of Theorem \ref{Thm:RedBlue}, we encounter in the appropriately reduced red-blue arrangement $\A^{(r,r')}_{pw}$ of $pw$ (defined with respect to $P\setminus \{r,r'\}$) either a $k$-shallow collinearity or $\Omega(k^2)$ $k$-shallow co-circularities, and charge $\TQ$ to these events, which are $(k+2)$-shallow with respect to the original set $P$. Notice that each $(k+2)$-shallow event in the full arrangement $\A_{pw}$ is charged in this manner by at most $O(1)$ terminal quadruples $\TQ\in \TF^B_{pw}$, whose respective second co-circularities $\delta_1$ are among the first two such co-circularities to occur in $\A_{pw}$ after the time of the event. Hence, the above charging accounts for at most $O(k^2 N(n/k)+kn^2\beta(n))$ terminal quadruples $\TQ\in \TF^B$.

Assume, then, that condition (iii) of Theorem \ref{Thm:RedBlue} is satisfied.
That is, there is a subset $A_{pw}$ of at most $3k+2$ points (including $r$ and $r'$) whose removal restores the Delaunayhood of $pw$ in $[\delta^-_{pw},\delta_1]$.
Moreover, since $q\neq r,r'$, the edge $pw$ belongs to $\DT((P\setminus A_{pw})\cup \{q\})$ at both times $\delta_{pw}^-$ and $\delta_{pw}^+$.
Therefore, the triple $p,q,w$ performs two (single) Delaunay crossings in the reduced set $(P\setminus A_{pw})\cup \{q\}$, namely, the crossing of $qp$ by $w$, and the crossing of $wp$ by $q$. A rountine combination of Lemma \ref{Lemma:TwiceCollin} with the probabilistic argument of Clarkson and Shor shows that the overall number of such triples $(p,q,w)$ (and, therefore, of their corresponding terminal quadruples $\TQ\in \TF^B$) is at most $O(kn^2)$.

In conclusion, we have shown that at most $O\left(k^2N(n/k)+kn^2\beta(n)\right)$ terminal quadruples fall into case (1a).

\medskip
\noindent{\bf Case (1b).} The edge $pw$ is hit at time $\upsilon_{pw}$ by the point $r$, which crosses $pw$ from $\L_{pw}^-$ to $\L_{pw}^+$.

Notice that, by Proposition \ref{Prop:Prw}, each terminal quadruple $\varrho=(p,q,r,w)$ falling into case (1a) is uniquely determined by the choice of $(p,r,w)$, because the second co-circularity $\delta_1$ of $\TQ$ is the first co-circularity of this kind (over all $\TQ'=(p,q',r,w)\in \TF^B$) to occur after the {\it unique} time when $r$ enters the halfplane $\L_{pw}^+$ through $pw$.

If there exists no terminal quadruple $\TQ'=(p,q',r',w)\in \TF_{pw}^B$ whose respective second co-circularity $\delta'_1$ occurs before $\upsilon_{pw}$, Proposition \ref{Prop:Prw} implies that $\delta_1$ is the first such co-circularity, so $\TQ$ can be charged to the edge $pw$. Clearly, this accounts for at most $O(n^2)$ terminal quadruples $\TQ$. 

For each of the remaining quadruples $\TQ\in \TF_{pw}^B$ (that fall into case (1b)), $\TF^B_{pw}$ contains another quadruple $\TQ'=(p,q',r',w)$, necessarily with $r'\neq r$, so that the edge $pw$ is Delaunay in $P\setminus\{r'\}$ at the time $\delta'_1<\upsilon_{pw}$ of the respective second co-circularity of $\TQ'$.
In particular, we can choose a time $\delta^-_{pw}<\upsilon_{pw}$ which is the last such time when $pw$ belongs to a reduced triangulation $\DT(P\setminus \{r'\})$, for some $r'\neq p,w,r$. 

Similarly, if there exists no quadruple $\TQ'=(p,q',r',w)\in \TF_{pw}^B$ whose respective second co-circularity $\delta'_1$ occurs after $\delta_1$, we can charge $\TQ$ (via its respective time stamp $\delta_1$) to $pw$.
Otherwise, there is a time $\delta^+_{pw}$ which is the first such time when $pw$ belongs to a reduced triangulation $\DT(P\setminus \{r''\})$, for some $r''\neq p,w,r$.

\begin{figure}[htbp]
\begin{center}
\input{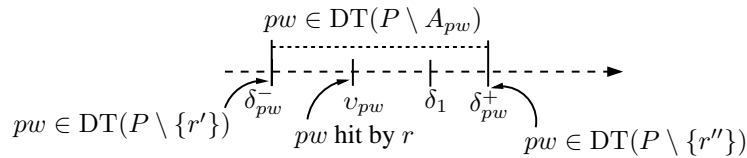}
\caption{\small Case (1b): $pw$ is hit by $r$ at time $\upsilon_{pw}$. We choose $\delta_{pw}^-$ as the last time before $\upsilon_{pw}$ when $pw$ belongs to a reduced triangulation $\DT(P\setminus \{r'\})$, for some $r'\neq r$, and apply Theorem \ref{Thm:RedBlue} over $(\delta_{pw}^-,\delta_1)$.
In addition, we choose $\delta_{pw}^+$ as the first time after $\delta_1$ when $pw$ belongs to a reduced triangulation $\DT(P\setminus \{r''\})$, for some $r''\neq r$, and apply Theorem \ref{Thm:RedBlue} over $(\delta_1,\delta_{pw}^+)$.}
\label{Fig:TimelinePw1b}
\vspace{-0.3cm}
\end{center}
\end{figure} 

For each of the remaining quadruples $\TQ\in \TF^B$ (that fall into case (1b)) there exist times $\delta_{pw}^-<\upsilon_{pw}<\delta_1$ and $\delta_{pw}^+>\delta_1$ as above, with respective obstruction points $r',r''\not\in \{p,w,r\}$; refer to Figure \ref{Fig:TimelinePw1b}.
We can now apply Theorem \ref{Thm:RedBlue} for the edge $pw$, over the interval $(\delta^-_{pw},\delta^+_{pw})$ (containing $\delta_1$).
This is done with the threshold $k$, and with respect to the reduced point set $P\setminus \{r',r''\}$.

In cases (i) and (ii) of Theorem \ref{Thm:RedBlue}, we charge $\varrho$ within $\A_{pw}$ (via $\delta_1$) either to a $(k+2)$-shallow collinearity, or to 
$\Omega(k^2)$ $(k+2)$-shallow co-circularities. 
Note that each $(k+2)$-shallow event, that occurs in $\A_{pw}$ during $(\delta^+_{pw},\delta^+_{pw})$, is charged by at at most $O(1)$ terminal quadruples in $\TF^B_{pw}$ (that fall into case (1b)), because the second co-circularity $\delta_1$ of $\TQ$ is either the last such co-circularity to occur before the time $t^*$ of the event, or the first such co-circularity to occur after $t^*$. Therefore, the above charging accounts for at most $O(k^2N(n/k)+kn^2\beta(n))$ terminal quadruples.

Now assume that Condition (iii) of Theorem \ref{Thm:RedBlue} holds. That is, we have a subset $A_{pw}$ of at most $3k+2$ points (including $r'$ and $r''$) whose removal restores the almost Delaunayhood of $pw$ throughout the interval $[\delta^-_{pw},\delta^+_{pw}]$. Moreover, since $r\neq r',r''$, the edge $pw$ belongs to $\DT((P\setminus A_{pw})\cup \{r\})$ at both times $\delta_{pw}^-$ and $\delta_{pw}^+$. Therefore, the edge $pw$ undergoes a Delaunay crossing by $r$ within the reduced triangulation
$\DT((P\setminus A_{pw})\cup \{r\})$.

\paragraph{Part 2: Enforcing a Delaunay crossing of $rw$.}
To conclude, we may assume, from now on, that each terminal quadruple $\varrho=(p,q,r,w)\in \TF^B$ under consideration is uniquely determined by each of its four sub-triples $(p,q,r), (p,q,w),(p,r,w)$ and $(q,r,w)$. Moreover, each of these triples defines a Delaunay crossing (in an appropriately reduced subset of $P$). 

We now exploit the last co-circularity of $p,q,r,w$ (at time $\delta_2\in [\delta_1,\t_{pq}]$) to enforce a fifth such crossing, namely the Delaunay crossing of $rw$ by one of $p,q$.
Here our argument is symmetric to the one that was used in Section \ref{Subsec:TypeA}. (Namely, we now switch the roles of $p$ and $q$).
In the case of success, at least one of the triples $(p,r,w),(q,r,w)$ performs two (single) Delaunay crossings, so Lemma \ref{Lemma:TwiceCollin} can be invoked. Otherwise, we dispose of $\TQ$ either through Theorem \ref{Thm:Balanced}, or by charging it within $\A_{rw}$.

Before proceeding with our case analysis, we emphasize that $wq$ is Delaunay at times $\delta_1$ and $\xi_{wq}>\t_{pq}(>\delta_2)$, and that the single Delaunay crossing $(qr,w,\T=[\tau_0,\tau_1])$ is defined with respect to a smaller point set $P\setminus \{p\}$. In addition, both $[\delta_1,\t_{wq}]$ and $[\tau_0,\tau_1]$ contain the time when $w$ crosses $rq$ from $\L_{rq}^+$ to $\L_{rq}^-$. 

We keep $\TQ=(p,q,r,w)\in \TF^B$ fixed and consider a subfamily $\TF_{rw}^B$ of all such terminal quadruples $\TQ'=(p',q',r,w)\in \TF^B$ whose last two points are equal to, respectively, to $r$ and $w$. (In particular, $\TF_{rw}^B$ includes the terminal quadruple $\TQ=(p,q,r,w)$ at hand.)
As in the symmetric setup of Section \ref{Subsec:TypeA}, we distinguish between three possible scenarios (a)--(c), ruling them out one by one.

\medskip
\noindent{\bf Case (a)} The crossing $(qr,w,\T=[\tau_0,\tau_1])$ begins after $\delta_1$ and $\TF^B_{rw}$ contains at least $k$ terminal quadruples $\varrho'=(p',q',r,w)$ 
whose respective counterclockwise $(r,w)$-crossings $(q'r,w,\T')$
begin in $[\delta_1,\tau_0)$, or $[\tau_0,\tau_1]$ ends before $\xi_{wq}$ and $\TF^B_{rw}$ contains at least $k$ terminal quadruples $\varrho'=(p',q',r,w)$ 
whose respective counterlclockwise $(r,w)$-crossings $(q'r,w,\T')$ end in $(\tau_1,\xi_{wq}]$.

\begin{figure}[htbp]
\begin{center}
\input{TerminalBCaseA.pstex_t}\hspace{3cm}\input{TypeBCaseAEdges.pstex_t}
\caption{\small Case (a): The scenario where $(qr,w,\T=[\tau_0,\tau_1])$ ends before $\xi_{wq}$, and the family $\TF^B_{rw}$ contains at least $k$ terminal quadruples $\varrho'=(p',q',r,w)$ whose respective $(r,w)$-crossings $(q'r,w,\T')$ end in $(\tau_1,\xi_{wq}]$. At at least $k-2$ of these quadruples satisfy $p'\neq q$ and $q'\neq p$, so their respective intervals $\T'$ are entirely contained in $[\tau_0,\xi_{wq}]$.}
\label{Fig:TypeBCaseA}
\vspace{-0.4cm}
\end{center}
\end{figure} 

Assume without loss of generality that the latter scenario occurs, so at least $k$ counterclockwise $(r,w)$-crossings $(q'r,w,\T')$ end in $(\tau_1,\xi_{wq}]$; see Figure \ref{Fig:TypeBCaseA}.
Notice that each of them occurs within a smaller triangulation $\DT(P\setminus \{p'\})$ which, in general, is distinct from the ambient triangulation $\DT(P\setminus \{p\})$ of $(qr,w,\T)$.
Fortunately, any terminal quadruple $\varrho'=(p',q',r,w)\in \TF^B_{rw}$ is uniquely determined by each of its respective points $p'$ and $q'$. Hence, at least $k-2$ of the above quadruples $\varrho'$ satisfy $p'\neq q$ and $q'\neq p$, in which case their respective $(r,w)$-crossings are compatible with $(qr,w,\T)$ through Lemma \ref{Lemma:OrderRelaxedCrossings}, and, therefore, occur entirely within $[\tau_0,\xi_{wq}]=\T\cup (\tau_1,\xi_{wq}]$. 

We sample a subset $\hat{P}$ of $n/4$ points and argue that, with some positive fixed probability, $(qr,w,\T)$ becomes a $(q,w,\Theta(k))$-chargeable Delaunay crossing within $\DT(\hat{P})$. 
Namely, we notice that the following two events occur simultaneously with some fixed positive probability:
(1) $\hat{P}$ includes the three points $q,r,w$, but not $p$, and (2) $\hat{P}$ includes $q'$ but not $p'$ for at least some constant fraction of the above quadruples $\varrho'=(p',q',r,w)\in \TF^B_{rw}$ (whose respective $(r,w)$-crossings $(q'r,w,\T')$ end in $(\tau_1,\xi_{wq}]$).
In the case of success, condition (1) implies that $qr$ still undergoes a single Delaunay crossing by $w$ in $\hat{P}$, which occurs in some sub-interval of $\T=[\tau_0,\tau_1]\subset [\tau_0,\xi_{wq}]$. Similarly, condition (2) implies that at least $\Omega(k)$ counterclockwise $(r,w)$-crossings in $R$ occur within $[\tau_0,\xi_{wq}]$.

By Theorem \ref{Thm:Balanced}, the overall number of such triples $(q,r,w)$ in $\hat{P}$ (and, thereby, in $P$) cannot exceed $O\left(k^2N(n/k)+k n^2\beta(n)\right)$, which also bounds the overall number of the corresponding terminal quadruples $\varrho=(p,q,r,w)$ in $P$.

We thus can assume, in what follows, that either the crossing $(qr,w,\T=[\tau_0,\tau_1])$ ends after $\xi_{pw}$, or the sub-family $\TF_{rw}^A$ contains at most $k$ other quadruples $\varrho'=(p,q,r,w)$ whose respective $(r,w)$-crossings $(q'r,w,\T')$ end in $(\tau_1,\xi_{wq}]$. Similarly, we can assume that either $[\tau_0,\tau_1])$ begins before $\delta_1$, or the sub-family $\TF_{rw}^B$ contains at most $k$ other quadruples $\varrho'=(p',q',r,w)$ whose respective $(r,w)$-crossings $(q'r,w,\T')$ begin in $[\delta_1,\tau_0)$.

\medskip
\noindent{\bf Case (b)} The family $\TF_{rw}^B$ contains no terminal quadruple $\varrho'=(p',q',w,r)\neq \varrho$ that satisfies $p'\neq q$, and whose respective $(r,w)$-crossing $(rq',w,\T')$ ends in $[\xi_{wq},\infty)$.

Since case (a) has been ruled out (and $\TF_{rw}^B$ contains at most one quadruple $\varrho'=(p',q',r,w)$ with $q'=p$), we conclude that there exist at most $k+1$ terminal quadruples $\varrho'\in \TF^B_{rw}$ whose respective $(r,w)$-crossings $(rq',w,\T')$ end after $\T=[\tau_0,\tau_1]$.
Hence, we charge $(p,q,r,w)$ (via its respective $(r,w)$-crossing $(pr,w,\T=[\tau_0,\tau_1])$) to the edge $rw$ and notice that any edge can be charged in this manner by at most $k+2$ terminal quadruples. 

To conclude, the above scenario happens for at most $O(kn^2)$ terminal quadruples $\varrho$.


\medskip
\noindent{\bf Case (c)} None of the previous cases occurs. 
In particular, since case (b) has been ruled out,
the family $\TF^B_{rw}$ contains at least one quadruple $\varrho'=(p',q',r,w)\neq \varrho$, with $p'\neq q$, and whose respective $(r,w)$-crossing $(q'r,w,\T')$ ends in $[\t_{wq},\infty)$. (Clearly, we have $p'\neq p$, for otherwise $\varrho$ would coincide with $\varrho'$.)

\begin{figure}[htbp]
\begin{center}
\input{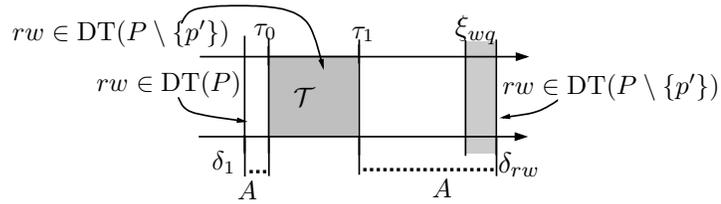}
\caption{\small Case (c): Extending the almost-Delaunayhood of $rw$ to $[\delta_1,\delta_{rw}]$. Here $\delta_{rw}$ is the first time in $[\xi_{wq},\infty)$ when $rw$ belongs to some reduced triangulation $\DT(P\setminus \{p'\})$, for some $p'\neq p,q$.}
\label{Fig:ExtendRwB}
\vspace{-0.4cm}
\end{center}
\end{figure}

Lemma \ref{Lemma:Crossing} implies, then, there is a time $\delta_{rw}\geq \xi_{wq}$ which is the first such time when the edge $rw$ belongs to some reduced triangulation $\DT(P\setminus \{p'\})$, for $p'\neq p,q$.
In what follows, we use $p'$ to denote a fixed point in $P\setminus \{p,q,r,w\}$ whose removal restores the Delaunayhood at time $\delta_{rw}$; see Figure \ref{Fig:ExtendRwB}.

Note that we have $\delta_{rw}>\xi_{wq}>\t_{pq}>\delta_1$.
Since case (a) has been ruled out, the choice of $\delta_{rw}$ guarantees that, unless $\delta_{rw}$ belongs to  $\T=[\tau_0,\tau_1]$, there exist at most $k+1$ quadruples $\varrho'\in \TF^B_{rw}$ whose respective $(r,w)$-crossings $(q'p,w,\T')$ end in $(\tau_1,\delta_{rw}]$.

\paragraph{Charging events in $\A_{rw}$.} We are now ready to invoke Theorem \ref{Thm:RedBlue} in order to extend the almost-Delaunayhood of $rw$, which already belongs to $\DT(P\setminus \{p\})$ throughout $\T=[\tau_0,\tau_1]$ (by Lemma \ref{Lemma:Crossing}), to the interval $[\delta_1,\delta_{rw}]$, which clearly intersects $\T$.

Note that $[\delta_1,\delta_{rw}]\setminus \T$ is composed of two disjoint (and possibly empty) sub-intervals $[\delta_1,\tau_0)$ and $(\tau_1,\delta_{rw}]$. We apply Theorem \ref{Thm:RedBlue} separately over each of these sub-intervals (and only if they are non-empty). In both cases, we use the second threshold parameter $\ell\gg k$.

The first application of Theorem \ref{Thm:RedBlue}, over $(\delta_1,\tau_0)$, is done with respect to the complete point set $P$ (using the Delaunayhood of $rw$ at time $\delta_1$). It is necessary only if $\delta_1<\tau_0$.

If at least one of the conditions (i), (ii) of that theorem is satisfied, we charge $\varrho$ within $\A_{rw}$, via $(rp,w,\T)$, either to an $\ell$-shallow collinearity, or to $\Omega(\ell^2)$ $\ell$-shallow co-circularities during $(\delta_1,\tau_0)$.
Since case (a) has been ruled out, $(qr,w,\T)$ is among the first $k+1$ such $(r,w)$-crossings to begin after any event that we charge. Hence, the above charging accounts for at most $O(k\ell^2N(n/\ell)+k\ell n^2\beta(n))$ quadruples $\varrho\in \TF^B$.
Otherwise, we end up with a subset of at most $3\ell$ points (perhaps including $p$ or $q$, or both) whose removal restores the Delaunayhood of $rw$ throughout $[\delta_1,\tau_0]$.

The similar second application of Theorem \ref{Thm:RedBlue} (over $(\tau_1,\delta_{rw})$) is done with respect to the reduced point set $P\setminus \{p'\}$ (where $p'$ denotes the point in $P\setminus\{p,q\}$ whose removal restores the Delaunayhood of $rw$ at time $\delta_{rw}$). It is necessary only if $\tau_1<\delta_{rw}$.

If at least one of the conditions (i), (ii) of that theorem holds, we charge $\varrho$ (via $(rp,w,\T)$) within $\A_{rw}$ either to an $(\ell+1)$-shallow collinearity, or to $\Omega(\ell^2)$ $(\ell+1)$-shallow co-circularities (which are $\ell$-shallow with respect to $P\setminus \{p'\}$).
By the choice of $\delta_{rw}$, $(qr,w,\T)$ is among the last $k+2$ such $(r,w)$-crossings to end after the event, so any $(\ell+1)$-shallow event in $\A_{rw}$ is charged by at most $O(k)$ quadruples $\varrho$.
Otherwise, we end up with a subset of at most $3\ell+1$ points (inclding $p'$, and perhaps also some of $p,q$) whose removal restores the Delaunayhood of $rw$ throughout $[\tau_1,\delta_{rw}]$.

\begin{figure}[htbp]
\begin{center}
\input{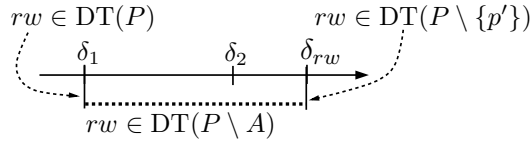}
\caption{\small Case (c): The edge $rw$ belongs to $\DT(P\setminus A)$ throughout the interval $[\delta_1,\delta_{rw}]$, which contains the last co-circularity $\delta_2$ of $p,q,r,w$. In addition, $rw$ belongs to $\DT(P)$ and $\DT(P\setminus \{p'\})$ at times $\delta_1$ and $\delta_{rw}$, respectively.}
\label{Fig:CrossRwB}
\vspace{-0.4cm}
\end{center}
\end{figure}

To conclude, we have a subset $A_{rw}$ of at most $6\ell+1$ points, {\it including $p'$}, and perhaps also some of $p,q$, whose removal restores the Delaunayhood of $rw$ throughout $[\delta_1,\delta_{rw}]$. To obtain the  crossing of $rw$ by $p$ or $q$ (which would occur in, respectively, $\DT((P\setminus A_{rw})\cup \{p\})$ or $\DT((P\setminus A_{rw})\cup \{q\})$), it suffices to show that $rw$ is hit by one of these two points during the interval $[\delta_1,\delta_{rw}]$. Notice that this interval contains $\delta_2\in (\delta_1,\t_{pq}]\subset (\delta_1,\xi_{wq}]$. See Figure \ref{Fig:CrossRwB}.
To do so, we distinguish between two possible sub-scenarios, depending on the precise order type of $\delta_2$, which is red-blue with respect to $pq$ and $rw$.

If the Delaunayhood of $rw$ is violated right {\it after} $\delta_2$ by $p\in \L_{rw}^-$ and $q\in \L_{rw}^+$, then,
since $\delta_2$ is the {\it last} co-circularity of $p,q,r,w$, Lemma \ref{Lemma:MustCross} implies that $rw$ is hit during $(\delta_2,\delta_{rw}]$ by at least one of $p,q$ (because $p'\neq p,q,r,w$), so we are done. (See Figure \ref{Fig:CrossRwHit} (left).)

Assume, then, that the Delaunayhood of $rw$ is violated right {\it before} $\delta_1$ by $p$ and $q$. Notice that this violation does not hold at time $\delta_1$. Hence, we can obtain the desired crossing of $rw$ in $(\delta_1,\delta_2)$ by applying the time-reversed variant of Lemma \ref{Lemma:MustCross} (for the point set $P=\{p,q,r,w\}$, backwards from $\delta_2$). 
The crucial observation is that $\delta_1$ and $\delta_2$ have different order types, which rules out the last case in Lemma \ref{Lemma:MustCross}. (See Figure \ref{Fig:CrossRwHit} (right).)

If $rw$ is hit during $(\delta_1,\delta_{rw}]$ by the point $p$, then, together with the crossing of $pw$ by $r$ (enforced in Part 1 by omitting $A_{pw}\setminus\{r\}$, where the subset $A_{pw}$ was obtained by applying Theorem \ref{Thm:RedBlue} in $\A_{pw}$), the triple $p,r,w$ now performs two Delaunay crosings within the triangulation $\DT\left(\left(P\setminus (A_{rw}\cup A_{pw})\right)\cup \{p,r\}\right)$.
Otherwise, if $rw$ is hit during $(\delta_1,\delta_{rw}]$ by $q$, the other triple $q,r,w$ performs two Delaunay crossings within the triangulation $\DT((P\setminus (A_{rw}\cup\{p\}))\cup \{q\})$, namely, the crossing of $qr$ by $w$ (prescribed by condition (B3)), and the just obtained crossing of $rw$ by $q$.

In both cases, a standard combination of Lemma \ref{Lemma:TwiceCollin} with the probabilistic argument of Clarkson and Shor implies that the overall number of the corresponding triples $(p,r,w)$ or $(q,r,w)$ in $P$ cannot exceed $O(\ell n^2)$. Since the quadruple $\varrho=(p,q,r,w)$ at hand is uniquely determined by each of its four sub-triples, this also bounds the overall number of such quadruples in $\TF^B$.

To conclude, we have established the following bound on the maximum possible cardinality of $\TF^B$:
\begin{equation}
T^B(n)=O\left(k\ell^2 N(n/\ell)+k^2N(n/k)+k\ell n^2\beta(n)\right).
\end{equation}

That is, we have expressed the maximum possible number of terminal quadruples of type B in terms of more elementary quantities which were introduced in Section \ref{Sec:Prelim}. Informally, here the system of our recurrences bottoms out, in the sense that no new quantities appear in the righ-hand side.

\subsection{Terminal quadruples of types C and D}\label{Subsec:TerminalCD}
We next establish near-quadratic recurrences for the maximum possible numbers $T^C(n)$ and $T^D(n)$ of terminal quadruples of types C and D, respectively, that can arise in an underlying set $P$ of $n$ moving points.
See Section \ref{Subsec:Stage4Left} for precise definitions of these two types of configurations. 

\begin{figure}[htbp]
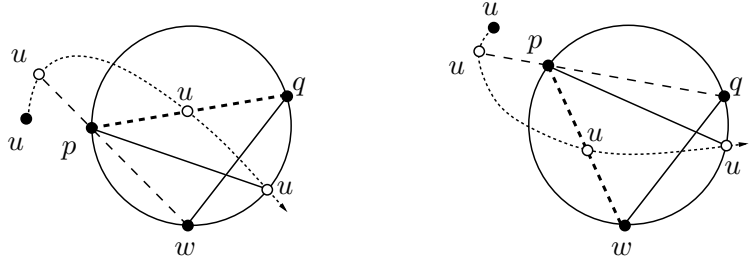

\begin{center}
\input{TypeC.pstex_t}\hspace{2cm}\input{TypeD.pstex_t}
\caption{\small Possible trajectories of $u$ in a terminal quadruple $\TQ=(w,q,u,p)$ of type C or D (resp., left and right). In both types, each of the unordered triples $p,u,q$ and $p,u,w$ is involved in a Delaunay crossing.}
\label{Fig:TypeCD}
\vspace{-0.6cm}
\end{center}
\end{figure}

Let $\TQ=(w,q,u,p)$ be a terminal quadruple of type C or D.
Notice that each of the (unordered) triples $u,p,w$ and $u,p,q$ is involved in a Delaunay crossing (see Figure \ref{Fig:TypeCD}).

Specifically, if $\TQ$ is of type C, we have a Delaunay crossing of $wu$ (or $uw$) by $p$ in $P\setminus \{q\}$, and a Delaunay crossing of $wq$ (or of $qw$) by $u$ in $P\setminus \{p\}$.
Similarly, if $\TQ$ is of type D, we have a Delaunay crossing of $qu$ (or of $uq$) by $p$ in $P\setminus \{p\}$. and a Delaunay crossing of $wp$ (or of $pw$) by $u$ in $P\setminus \{q\}$.

In both types, the four points of $\TQ$ are involved in a Delaunay co-circularity, right after which the Delaunayhood of $pu$ is violated by $w\in \L_{pu}^-$ and $q\in \L_{pu}^+$, and this is the last co-circularity of $w,q,u,p$.
We will use the above co-circularity to enforce a Delaunay crossing of $up$ by at least one of $w,q$.
As a result, one of the triples $u,p,w$ and $u,p,q$ will perform two single Delaunay crossings in a suitably refined subset of $P$, so our analysis will bottom out via Lemma \ref{Lemma:TwiceCollin}.

The desired crossing of $up$ can be enforced using exactly the same analysis as was used in Section \ref{Sec:ReduceToCrossings} to express the maximum possible number $N_E(n)$ of extremal Delaunay co-circularities in $P$ in terms of the maximum possible number $C(n/k)$ of Delaunay crossings that can arise in a subset (of $P$) of cardinality $n/k$. Nevertheless, we briefly review the argument of Section \ref{Sec:ReduceToCrossings} for the sake of completeness. 

Let $t_0$ denote the time of the above extremal Delaunay co-circularity of $w,q,u,p$.
If the edge $pu$ never re-enters $\DT(P)$ (leaving $\DT(P)$ at time $t_0$), then we can charge $\TQ$ to this last disappearance of $pu$ from $\DT(P)$, which occurs for at most $O(n^2)$ terminal quadruples $\TQ$ under consideration. Otherwise, let $t_1$ be the first time after $t_0$ when $up$ re-enters $\DT(P)$.
By Lemma \ref{Lemma:MustCross}, $pu$ is hit in $(t_0,t_1]$ by at least one of $w,q$. Namely, either $q$ crosses $up$ from $\L_{pu}^-$ to $\L_{pu}^+$, or $w$ crosses $pu$ in the opposite direction. Furthermore, this is the second and last collinearity of $p,q,u$ or $p,w,u$ (and, therefore, the {\it only} such collinearity of this order type to occur in $(t_0,t_1]$).

In both cases, we invoke Theorem \ref{Thm:RedBlue} to amplify the above second collinearity of $p,u,q$ or $p,u,w$ into an additional Delaunay crossing. 
Specifically, we fix a constant threshold $k>12$ and apply Theorem \ref{Thm:RedBlue} in $\A_{pu}$ over the interval
$(t_0,t_1)$. 

In cases (i) and (ii) of Theorem \ref{Thm:RedBlue}, we can charge $\TQ=(w,q,p,u)$ within $\A_{pu}$ either $\Omega(k^2)$ $k$-shallow co-circularities, or a $k$-shallow collinearity. Furthermore, each shallow event is charged at most $O(1)$ times, because it involves $p$ and $u$, and $t_0$ is the last disappearance of $pu$ from $\DT(P)$. Hence, the overall number of such terminal quadruples does not exceed $O\left(k^2N(n/k)+kn^2\beta(n)\right)$.

Finally, in case (iii) of Theorem \ref{Thm:RedBlue}, we end up with a subset $A$ of at most $3k$ points so that $pu$ belongs to $\DT(P\setminus A)$ throughout $[t_0,t_1]$. Thus, either $pu$ undergoes a single Delaunay crossing by $q$ in $(P\setminus A)\cup\{q\}$, or its reversed copy $up$ undergoes a single Delaunay crossing by $w$ in $(P\setminus A)\cup\{w\}$. 

Therefore, we can charge $\TQ$ to the corresponding triple $p,q,u$ or $p,w,u$ which performs two Delaunay crossings in a suitable subset of $P$. Lemma \ref{Lemma:TwiceCollin} together with the Clarkson-Shor argument imply that the overall number of such triples in $P$ cannot exceed $O(kn^2)$. Furthermore, each of them can be charged at most once, because $t_0$ is the last time when $pu$ disappears from $\DT(P)$ before being hit as above by $q$ or $w$.

To conclude, we have established the following recurrences for the above quantities $T^C(n)$ and $T^D(n)$:

\begin{equation}\label{Rec:TypeC}
T^C(n)=O\left(k^2N(n/k)+kn^2\beta(n)\right)
\end{equation}

and

\begin{equation}\label{Rec:TypeD}
T^D(n)=O\left(k^2N(n/k)+kn^2\beta(n)\right).
\end{equation}

\section{Proof of Theorem \ref{Thm:Balanced}}\label{Sec:Balanced}

Let $(pq,r,I=[t_0,t_1])$ be a $(p,r,k)$-chargeable Delaunay crossing, and let
$\I=[t_2,t_3]$ be the corresponding interval which certifies the $(p,r,k)$-chargeability of $(pq,r,I)$. In particular, at least $k$ counterclockwise $(q,r)$-crossings $(uq,r,I_u)$ occur within $\I$ (in the sense that $I_u\subseteq \I$).
In additon, the edge $pr$ belongs to $\DT(P)$ when $\I$ begins or ends, and there is a subset $A_0\subset P$ of $c_0=O(1)$ points whose removal restores the Delaunayhood of $pr$ throughout $\I$.

\begin{figure}[htbp]
\begin{center}
\input{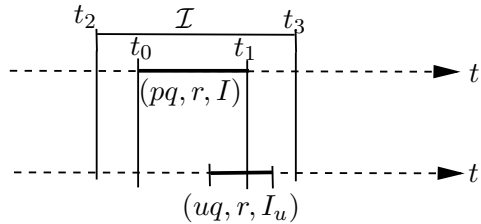}
\caption{\small The setup in the proof of Theorem \ref{Thm:Balanced}. The crossing $(pq,r,I=[t_0,t_1])$ is $(p,r,k)$-chargeable, for $\I=[t_2,t_3]$. We fix a counterclockwise $(q,r)$-crossing $(uq,r,I_u)$, which ends in $(t_1,t_3]$ (so $I_u\subseteq (t_0,t_3]$). The $(q,r)$-crossings $(pq,r,I)$ and $(uq,r,I_u)$ form a counterclockwise quadruple $(q,p,u,r)$.}
\label{Fig:ChargeabilityProof}
\end{center}
\vspace{-0.3cm}
\end{figure}

By Lemma \ref{Lemma: OrderOrdinaryCrossings}, each of the above $(q,r)$-crossings $(uq,r,I_u)$ occurs within one of the intervals $\I^+=(t_0,t_3]$ or $\I^-=[t_2,t_1)$. In particular, we have $I_u\subseteq (t_0,t_3]$ if and only if $r$ enters $\L^+_{uq}$ after entering $\L_{pq}^+$; see Figure \ref{Fig:ChargeabilityProof}. Without loss of generality, we assume that at least $\lceil k/2\rceil$ of these crossings occur within $(t_0,t_3]$.
Again, Lemma \ref{Lemma: OrderOrdinaryCrossings} implies that each such crossing must end within $(t_0,t_3]$.

\paragraph{Overview.} To establish Theorem \ref{Thm:Balanced}, we distribute the ``weight" of $(pq,r,I)$ over the above $\Omega(k)$ $(q,r)$-crossings $(uq,r,I_u)$ or, more precisely, over their respective arrangements $\A_{ur}$. 
(Recall that each counterclockwise $(q,r)$-crossing $(uq,r,I_u)$ is also a clockwise $(u,r)$-crossing.)
In what follows, we fix one of the first $\lceil k/2\rceil$ counterclockwise $(q,r)$-crossings that ends after time $t_1$ (and before $t_3$), and assume that its respective point $u$ does not belong to $A_0$. 
Our charging strategy is to make each such $u$ pay $\Theta(1/k)$ units of charge to $(pq,r,I)$, so that $(pq,r,I)$
receives a total of at least $1$ unit. The charging will be performed in one of two possible ways (depending on the structure of $\A_{ur}$ and on the motion of $p,q,u$, and $r$). 

We shall first try to charge $(uq,r,I_u)$ (rather than $(pq,r,I)$) to events within $\A_{ur}$ using the standard techniques of Section \ref{Sec:CountQuad} (involving Lemma \ref{Lemma:TwiceCollin} and Theorem \ref{Thm:RedBlue}). In case of success, $(uq,r,I_u)$ will be declared as {\it heavy} for $(pq,r,I)$ and will pay $\Theta(1/k)$ units of charge to $(pq,r,I)$. As we will show, the overall number of such crossings $(uq,r,I_u)$, that will be declared as heavy for {\it at least one} of their neighboring $(q,r)$-crossings, does not exceed $O(k^2 N(n/k)+kn^2\beta(n))$. 
Moreover, any crossing $(uq,r,I_u)$ will be charged (as heavy) by at most $\lceil k/2\rceil$ neighboring $(q,r)$-crossings $(pq,r,I)$, due to the $\lceil k/2\rceil$-proximity of the crossings $(pq,r,I)$ and $(uq,r,I_u)$, and will pay $\Theta(1/k)$ units of charge to each. 
Therefore, at most $O(k^2 N(n/k)+kn^2\beta(n))$ units of charge will be transferred in this fashion.

If the above strategy fails, we shall resort to a more subtle type of charging. In that case, we shall charge $(pq,r,I)$ (again within $\A_{ur}$) to $\Theta(k)$ $(4k)$-shallow co-circularities that involve $u,r$ and $p$ (together with some fourth point, not necessarily $q$), and each of these co-circularities will pay $\Theta(1/k^2)$ units of charge to $(pq,r,I)$. Moreover, we shall argue that each $(4k)$-shallow co-circularity can be charged in this latter manner by at most $O(1)$ crossings $(pq,r,I)$. Hence, at most $O(k^2N(n/k))$ units will be transferred in the second scheme. The theorem then follows from these two charging schemes.

Before proceeding with the above general strategy, we fix one such crossing $(uq,r,I_u)$ and establish several essential properties of it.

\begin{figure}[htbp]
\begin{center}
\input{Violated.pstex_t}\hspace{2cm}\input{ChargeabilityProof1.pstex_t}
\caption{\small Proof of Proposition \ref{Prop:c0}. Assuming $p\neq w$, the four points $w,u,r,p$ are involved in a red-blue co-circularity during the crossing $(uw,r,J_{uw})$. Since the Delaunayhood of $pr$ is then violated by $w$ and $u$, and $u$ is chosen outside $A_0$, the set $A_0$ must contain $w$.}
\label{Fig:Violated}
\vspace{-0.4cm}
\end{center}
\end{figure}

\begin{proposition}\label{Prop:c0}
With the above assumptions, and with $(uq,r,I_u)$ fixed, at most $c_0+1$ clockwise $(u,r)$-crossings $(uw,r,J_{uw})$ occur within $[t_0,t_3]$. 
\end{proposition}
\begin{proof}
Fix a clockwise $(u,r)$-crossing $(uw,r,J_{uw})$, such that $w\neq p$ and $J_{uw}\subset [t_0,t_3]$. Refer to Figure \ref{Fig:Violated}.

By Lemma \ref{Lemma:OnceCollin}, the points $w,u,r,p$ are involved during $J_{uw}$ in a co-circularity which is red-blue with respect to the edges $uw$ and $pr$. Hence, the Delaunayhood of $pr$ is violated by $u$ and $w$ either right before or right after this co-circularity. Since $[t_0,t_3]\subseteq \I$, the set $A_0$ must include at least one of the points $u,w$. Since, by assumption, $u\not\in A_0$, we must have $w\in A_0$, so there can be at most $c_0$ such crossings. Adding the possible crossing $(up,r,J_{up})$ yields the asserted bound. 
\end{proof}

Notice that the set $P$ induces a counterclockwise quadruple $\Q_u=(q,p,u,r)$ whose respective interval $\conn{I}{I_u}$ is contained in $[t_0,t_3]$.
The following proposition is stated in full generality and applies to {\it all} counterclockwise quadruples (i.e., not necessarily the ones that arise in the course of the present proof of Theorem \ref{Thm:Balanced}). It can be viewed simply as an extension of Lemma \ref{Lemma:ReturnsStillCross}.

\begin{proposition}\label{Prop:ReturnsAlsoCross}
Let $\sigma_u=(q,p,u,r)$ be a counterclockwise quadruple, with associated crossings $(pq,r,I)$ and $(uq,r,I_u)$. Suppose that the edge $rq$ is hit by the point $p$, and that this happens in the interval after $r$ enters $\L_{pq}^+$ and before $r$ enters $\L_{uq}^+$. Then $pr$ is also hit, during that same interval, by the point $u$.
\end{proposition}

\noindent{\bf Remarks.}
(1) Clearly, a symmetric statement holds if $rq$ is hit by $u$. Namely, in that case the edge $ru$ is hit by the point $p$. As a matter of fact, the proof of Proposition \ref{Prop:ReturnsAlsoCross} implies that the two scenarios coincide:
The edge $rq$ is hit by $p$ between the times when $r$ crosses $pq$ and $uq$ if and only if $rq$ is hit there by $u$ too.


\smallskip
\noindent(2) The reader might be tempted to use Lemma \ref{Lemma:TwiceCollin} in order to bound the number of such crossings $(uq,r,I_u)$, whose respective counterclockwise quadruples $\Q_u=(q,p,u,r)$ satisfy the conditions of Proposition \ref{Prop:ReturnsAlsoCross} (as was done, e.g., for clockwise Delaunay quadruples in case (a) of Section \ref{Subsec:Stage1Reg}). However, since we do not assume the edge $rq$ to be almost-Delaunay during $\conn{I}{I_u}$, the argument of Section \ref{Subsec:Stage1Reg} does not immediately apply to such instances.

\begin{proof}
Refer to Figure \ref{Fig:ReturnUCrossGeneral}. Notice that, according to Lemma \ref{Lemma:Crossing}, $p$ can hit $rq$ (as prescribed in the proposition) only during the gap between the intervals $I$ and $I_u$ of the two $(q,r)$-crossings of $\sigma_u$ (a gap that we therefore assume to exist).

\begin{figure}[htbp]
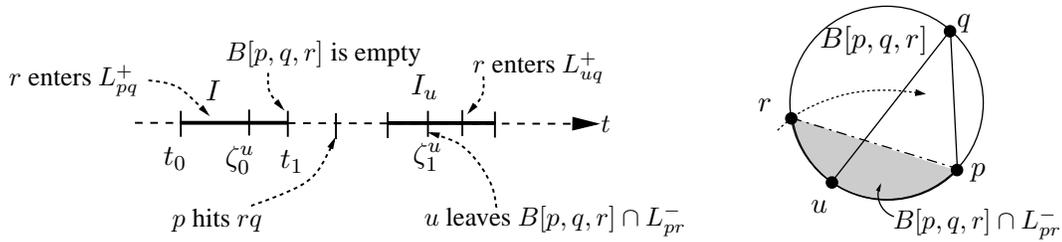

\begin{center}
\input{TimelineReturnU.pstex_t}\hspace{2cm}\input{Zeta1U.pstex_t}
\caption{\small Proof of Proposition \ref{Prop:ReturnsAlsoCross}. Left: The summary of events that are assumed to occur during $\conn{I}{I_u}$. Right: The point $u$ leaves the cap $B[p,q,r]\cap \L_{pr}^-$ at time $\zeta_1^u\in I_u\setminus I=I_u$.}
\label{Fig:ReturnUCrossGeneral}
\vspace{-0.4cm}
\end{center}
\end{figure} 

Since the points $p,q,r$ can be collinear at most twice, the halfplane $\L_{uq}^+$ contains $p$ when $r$ enters it during $I_u$. Therefore, and according to Lemma \ref{Lemma:OnceCollin}, the four points $p,q,u,r$ are involved  at some time $\zeta_1^u\in I_u\setminus I=I_u$ in a co-circularity, occurring before $r$ crosses $uq$; see Figure \ref{Fig:ReturnUCrossGeneral} (right). Right after this co-circularity the Delaunayhood of $uq$ is violated by $r\in \L_{uq}^-$ and $p\in \L_{uq}^+$. Note that at that very moment the point $u$ leaves the cap $B[p,q,r]\cap \L_{pr}^-$. 
Note also that, according to Lemma \ref{Lemma:OnceCollin}, the points $p,q,u,r$ are also involved in an earlier co-circularity which occurs at some time $\zeta_0^u\in I\setminus I_u=I$ (and before $p$ hits $rq$, which occurs between $I$ and $I_u$).
We distinguish between the following two scenarios.

\begin{figure}[htbp]
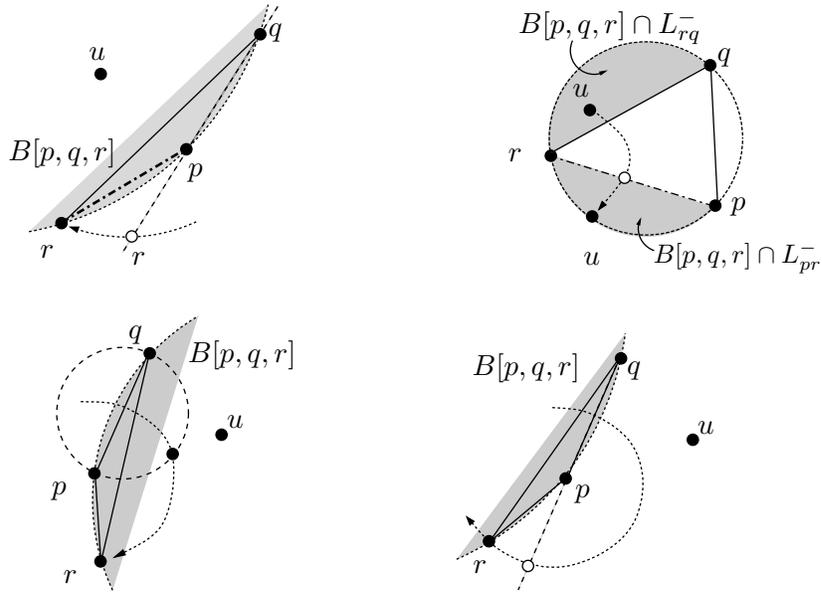

\begin{center}
\input{ReturnUCross.pstex_t}\hspace{3cm}\input{ReturnUCross3.pstex_t}\\
\vspace{0.5cm}
\input{ReturnUCross1.pstex_t}\hspace{3cm}\input{ReturnUCross2.pstex_t}
\caption{\small Proof of Proposition \ref{Prop:ReturnsAlsoCross}. Top: If $u$ lies in $\L_{pq}^-$ when $p$ hits $rq$ (top-left), then $u$ can exit the cap $B[p,q,r]\cap \L_{pr}^-$ only after crossing $pr$ (top-right). Bottom: The hypothetic scenario where $u$ lies in $\L_{pq}^+$ when $p$ hits $rq$. Right before $p$ hits $rq$, the disc $B[p,q,r]$ contains $u$, which must have entered it after $I$ (bottom-left). Right after that collinearity, $u$ lies outside $B[p,q,r]$, so it will have to re-enter $B[p,q,r]$ before $\zeta_1^u$ (bottom-right).}
\label{Fig:ReturnUCross}
\vspace{-0.4cm}
\end{center}
\end{figure} 

\noindent(i) If $u$ lies in $\L_{pq}^-$ when $p$ hits $rq$ (and $r$ re-enters $\L_{pq}^-$), then $u$ lies within the cap $B[p,q,r]\cap \L^-_{rq}$ right after this collinearity, as depicted in Figure \ref{Fig:ReturnUCross} (top-left). 
Right after this event and before $\zeta_1^u$, $u$ must move from this cap to the disjoint cap $B[p,q,r]\cap \L_{pr}^-$ (which it exits at time $\zeta_1^u$) either\footnote{Here we implicitly rely on the fact that the motion of $B[p,q,r]$ is continuous after the second collinearity of $p,q,r$.} through $pr$ (and through $rq$) or through the boundary of $B[p,q,r]$. See Figure \ref{Fig:ReturnUCross} (top-right). However, in the latter case $u$ would first have to leave its present cap through $\partial B[p,q,r]$, so the points $p,q,u,r$ would be co-circular at least twice during $(\zeta^u_0,\zeta^u_1)$, contradicting the assumption that any four points are co-circular at most three times. Hence, $u$ can enter $B[p,q,r]\cap \L_{pr}^-$ only through $pr$ and $rq$, as claimed in the proposition.

\noindent(ii) If $u$ lies in $\L_{pq}^+$ when $p$ hits $rq$, then $u$ lies within the disc $B[p,q,r]$ right before this event; see Figure \ref{Fig:ReturnUCross} (bottom-left).
By the definition of Delaunay crossings, the disc $B[p,q,r]$ contains no points of $P$ right after the end of $I$, as depicted in Figure \ref{Fig:ReturnUCross} (bottom-right). Hence, $u$ enters $B[p,q,r]$ at the end of $I$ and before $p$ hits $rq$. 
We also note that $u$ lies outside $B[p,q,r]$ right after the second collinearity of $p,q,r$, so $u$ must enter $B[p,q,r]$ (through its boundary) afterwards and before $\zeta^u_1$ (in order to exit it after $\zeta_1^u$). Similar to the preceding  scenario, we obtain four impossible co-circularities of $p,q,u,r$, showing that the present scenario cannot occur.
\end{proof}

\paragraph{Back to the proof of Theorem \ref{Thm:Balanced}.} With these preparations, we are finally ready to establish Theorem \ref{Thm:Balanced}. Recall that we have fixed a counterclockwise $(q,r)$-crossing $(uq,r,I_u)$ that ends in $(t_1,t_3]$, and which is among the first $\lceil k/2\rceil$ such $(q,r)$-crossings to end after $t_1$. Recall also that $u$ does not belong to the set $A_0$ (of size $c_0$, appearing in the definition of the $(p,r,k)$-chargeability of $(pq,r,I)$), and that the $(q,r)$-crossings $(pq,r,I)$ and $(uq,r,I_u)$ form a (not necessarily consecutive) counterclockwise $(q,r)$-quadruple $\Q_u=(q,p,u,r)$. 

We first claim that $r$ cannot cross $pq$ again between the times when it enters the halfplanes $\L_{pq}^+$ and  $\L_{uq}^+$ (during the two respective Delaunay crossings).
Indeed, otherwise a counterclockwise variant of Lemma \ref{Lemma:ReturnsStillCross} would imply that the edge $pr$ is hit by $u$ during the interval $\conn{I}{I_u}$. As the latter interval is contained in $[t_0,t_3]$, this is a clear contradiction to the assumed choice of $u$ outside $A_0$.
Similarly, $p$ cannot hit $rq$ between the times when $r$ enters the halfplanes $\L_{pq}^+,\L_{uq}^+$, for otherwise we would invoke Proposition \ref{Prop:ReturnsAlsoCross} to show that $pr$ is again hit by $u$ during $\conn{I}{I_u}\subseteq [t_0,t_3]$, and reach the same contradiction as above.

If the edge $pr$ is hit during $[t_1,t_3]$ by $q$ (which is the only remaining way in which $p,q,r$ can be collinear again), then the set $(P\setminus A_0)\cup \{q\}$ induces a Delaunay crossing of $pq$ by $r$, and a Delaunay crossing of $pr$ by $q$. A routine combination of Lemma \ref{Lemma:TwiceCollin} with the  probabilistic argument of Clarkson and Shor shows that this scenario happens for at most $O(n^2)$ Delaunay crossings $(pq,r,I)$. 

\begin{figure}[htbp]
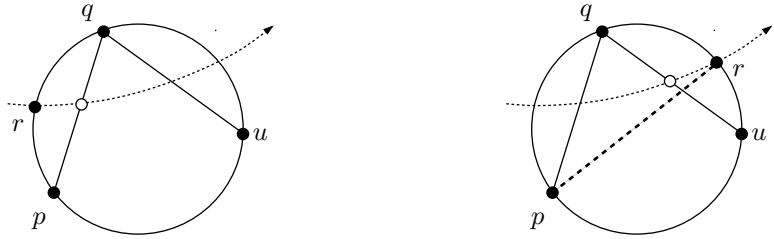

\begin{center}
\input{FirstU.pstex_t}\hspace{3cm}\input{SecondU.pstex_t}
\caption{\small The two co-circularities of $q,p,u,r$ which occur at times $\zeta_0^u\in I\setminus I_u$ (left) and $\zeta_1^u\in I_u\setminus I$ (right).}
\label{Fig:TwoCocircU}
\vspace{-0.4cm}
\end{center}
\end{figure}

It therefore suffices to focus on the scenarios where $r$ does not re-enter $\L_{pq}^-$ after $I$ and before it enters $\L_{uq}^+$ (through $uq$, during $I_u$).
As noted in Section \ref{Subseq:Quad} (see also the proof of Proposition \ref{Prop:ReturnsAlsoCross}), the four points $q,p,u,r$ are involved in co-circularities at some times $\zeta^u_0\in I\setminus I_u$ and $\zeta^u_1\in I_u\setminus I$; see Figure \ref{Fig:TwoCocircU}. Moreover, these are the only co-circularities of $p,q,u,r$ to occur during $I$ and $I_u$.

Consider the latter co-circularity, occurring at some time $\zeta^u_1\in I_u\setminus I$, which is red-blue with respect to the edges $pr,uq$. Since $r$ does not return to $\L_{pq}^-$, $p$ lies in $\L_{uq}^-$ when $r$ hits $uq$ during $I_u$. (See Figure \ref{Fig:TwoCocircU} (right).) Arguing as in Section \ref{Subsec:SingleProp} (see, e.g., the proofs of Lemmas \ref{Lemma:OnceCollin} and \ref{Lemma: OrderOrdinaryCrossings}), we can conclude that the Delaunayhood of $pr$ is violated right after time $\zeta^u_1$ by the points $u$ and $q$. 

We first argue that the above co-circularity at time $\zeta^u_1$ cannot be the {\it last} co-circularity of $q,p,u,r$.
Indeed, otherwise
Lemma \ref{Lemma:MustCross} (combined with the assumption that $q$ does not hit $pr$ during $[t_1,t_3]$) would imply that the edge $pr$ is hit by $u$ during the interval $(\zeta^u_1,t_3)$. However, in that case $u$ would belong to $A_0$, contrary to the choice of $u$.

To conclude, we can assume, from now on, that the co-circularity at time $\zeta^u_1$ is the {\it middle} co-circularity of the points $q,p,u,r$.
Hence, the preceding co-circularity, which occurs at time $\zeta^u_0\in I\setminus I_u$, must be the {\it first} co-circularity of these four points. 

To proceed, we distinguish between several topological scenarios, treating each in turn. In each of them, $(pq,r,I)$ receives $\Theta(1/k)$ units of charge (via $u$ alone, as reviewed in the beginning of this section). 
Recall that, with $(pq,r,I)$ fixed, $u$ and $(uq,r,I_u)$ can be chosen in $\Theta(k)$ possible ways. Hence, with an appropriate choice of the constants of proportionality, each $(p,r,k)$-chargeable crossing $(pq,r,I)$ will eventually receive at least 
one unit of charge.

\bigskip
\noindent{\bf Case (a).} The edge $ru$ is never Delaunay during $(-\infty,t_0]$. In this case, we classify the crossing $(uq,r,I_u)$
as {\it heavy} (for $(pq,r,I)$), and we make it pay $\Theta(1/k)$ units of charge to $(pq,r,I)$.

Notice that $(uq,r,I_u)$ is one of the first $c_0+2$ clockwise $(u,r)$-crossings (according to the standard order provided by Lemma \ref{Lemma: OrderOrdinaryCrossings}). Indeed, by Lemma \ref{Lemma:Crossing}, no such crossings begin before time $t_0$, when the edge $ru$ is not even Delaunay. In addition,
by Proposition \ref{Prop:c0}, at most $c_0+1$ clockwise $(u,r)$-crossings can begin after $t_0$ and before the beginning of $(uq,r,I_u)$, as each of them has to occur within the interval $[t_0,t_3]$.
In conclusion, the overall number of such crossings $(uq,r,I_u)$, that are classified as heavy for at least one of their neighboring $(q,r)$-crossings $(pq,r,I)$ (upon falling into case (a)), is at most $O(n^2)$.


\begin{figure}[htbp]
\begin{center}
\input{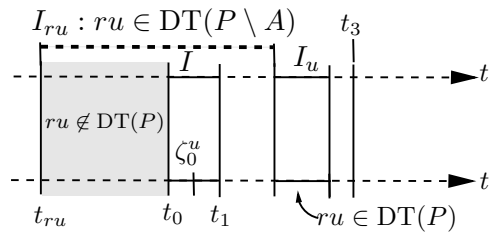}
\caption{\small Preparing for cases (b), (c), and (d): We pick the last time $t_{ru}$ in $(-\infty,t_0]$ when $ru$ is Delaunay and apply Theorem \ref{Thm:RedBlue} over the interval $I_{ru}$ (containing $\zeta_0^u$).}
\label{Fig:ChargeabilityB}
\vspace{-0.4cm}
\end{center}
\end{figure}

\bigskip
\noindent{\bf Preparing for cases (b), (c) and (d).} In each of the subsequent three cases, we assume that $ru$ appeared in $\DT(P)$ also before (or at) $t_0$.
Let $t_{ru}$ be the last time in $(-\infty,t_0]$ when $ru$ belongs to $\DT(P)$, and let $I_{ru}$ denote the subsequent interval that lasts from $t_{ru}$ to the beginning of $I_u$. Note that $I_{ru}$ contains $I\setminus I_u$, and therefore includes the time $\zeta_0^u$ of the first co-circularity of $p,q,u,r$. Refer to Figure \ref{Fig:ChargeabilityB}.


As a preparation, we apply Theorem \ref{Thm:RedBlue} in $\A_{ru}$ over $I_{ru}$ (with the same constant parameter $k$, and keeping in mind that $ru$ is Delaunay at both endpoints of $I_{ru}$), and then proceed depending on the outcome.

\medskip
\noindent{\bf Case (b).} If one of the Conditions (i), (ii) of Theorem \ref{Thm:RedBlue} is satisfied (i.e.,  $\A_{ru}$ contains either $\Omega(k^2)$ $k$-shallow co-circularities or a $k$-shallow collinearity, all of them occurring in $I_{ru}$), the crossing $(uq,r,I_u)$ is again classified as heavy for $(pq,r,I)$, and pays $\Theta(1/k)$ units of charge to it. 

We claim that the overall number of such crossings $(uq,r,I_u)$, that are classified as heavy for at least one of their neighbors $(pq,r,I)$ (within the present case (b)), is at most $O(k^2N(n/k)+kn^2\beta(k))$. 
To show this, we keep the crossing $(pq,r,I)$ fixed and charge $(uq,r,I_u)$ within $\A_{ru}$ either to $\Omega(k^2)$ $k$-shallow co-circularities, or to a $k$-shallow collinearity, which are assumed to occur during the respective interval $I_{ru}$.

We emphasize that the first endpoint $t_{ru}$ of $I_{ru}$ might depend on the choice of $(pq,r,I)$ from among those crossings that expect to receive $\Theta(1/k)$ units from $(uq,r,I_u)$. Furthermore, an event in $\A_{ru}$ might be charged by the same $(uq,r,I_u)$ in the context of several $(p,r,k)$-chargeable crossings $(pq,r,I)$ that charge $(uq,r,I_u)$ (for various values of $p$). Nevertheless, for each choice of an event in $\A_{ru}$ and each clockwise $(u,r)$-crossing $(uq,r,I_u)$,
all such episodes cause only one charging of this event by $(uq,r,I)$.


We next show that each event in $\A_{ru}$ is charged in the above manner by at most $O(1)$ crossings $(uq,r,I_u)$. 
Indeed, let $t^*$ be the time of a $k$-shallow event that we charge within $\A_{ru}$. 
Clearly, one can guess the points $u$ and $r$ of $(uq,r,I_u)$ in at most $O(1)$ ways, as they are involved in the event. Thus, it suffices to guess the third point $q$ of $(uq,r,I_u)$ (armed only with the knowledge of $t^*,r$ and $u$), which is done as follows.

Let $q$ be a potential third point, and let $(pq,r,I)$ be {\it any} $(p,r,k)$-chargeable crossing that receives $\Theta(1/k)$ units of charge from the corresponding crossing $(uq,r,I_u)$ (after the latter crossing is classified as heavy for $(pq,r,I)$, by the rule of case (b)).
By Lemma \ref{Lemma:Crossing}, no clockwise $(u,r)$-crossing $(uq,r,I_u)$ can begin during the respective interval $I_{ru}\cap (-\infty,t_0]$ (when $ru$ is not even Delaunay). Moreover, Proposition \ref{Prop:c0} implies that at most $c_0+1$ such $(u,r)$-crossings begin in the interval that lasts from $t_0$ to the beginning of $I_u$ (which is contained in $[t_0,t_3]$). Hence, $(uq,r,I_u)$ is among the first $c_0+2$ clockwise $(u,r)$-crossings to begin after $t^*$, so knowing $t^*,r$, and $u$ enables us to guess $(uq,r,I_u)$ in at most $O(1)$ ways (irrespective of the choice of $p$ and $(pq,r,I)$).

The number of $k$-shallow co-circularities in $\A_{ru}$, over all $r,u$, is at most $O(k^4N(n/k))$. Similarly, the number of $k$-shallow collinearities is $O(kn^2\beta(n))$. Each such event is charged by only $O(1)$ $(u,r)$-crossings $(uq,r,I_u)$ (which are declared as heavy in case (b), for at least one of their $(p,r,k)$-chargeable neighbors $(pq,r,I)$). Furthermore, each such crossing $(uq,r,I_u)$ charges either $\Omega(k^2)$ $k$-shallow co-circularities, or a $k$-shallow collinearity. All these considerations imply that the number of charging 
crossings $(uq,r,I_u)$ of this kind is $O(k^2N(n/k)+kn^2\beta(n))$, as claimed. 

Recall that, in the rest of the analysis, each of these $(u,r)$-crossings will pay $\Theta(1/k)$ units of charge to $O(k)$ ``neighboring" crossings $(pq,r,I)$, so these latter crossings will recieve in total $O(k^2N(n/k)+kn^2\beta(n))$ units of charge in this manner.

\medskip
\noindent{\bf Preparing for cases (c) and (d).}
Now suppose that Condition (iii) of Theorem \ref{Thm:RedBlue} holds. That is,
the Delaunayhood of $ru$ can be restored throughout $I_{ru}$ by removing a subset $A$ of cardinality at most $3k$. 
To handle this more difficult scenario, we first establish the following proposition.
\begin{proposition}\label{Prop:CrossUr}
With the above assumptions, the edge $ru$ is hit during $I_{ru}$ by at least one of the points $p,q$.
\end{proposition} 
\begin{proof}
The proof proceeds (essentially) along the same lines as in case (e) of Section \ref{Subsec:Stage1Reg}. (The main difference is that the quadruple $\Q_u$ under consideration is {\it counterclockwise}.)

We first get rid of the instances where $r$ crosses $\L_{uq}$ between the times when it enters the halfplanes $\L_{pq}^+$ and $\L_{uq}^+$ (in the respective intervals $I$ and $I_u$). 
Note that if $ru$ is hit there by $q$ then we are done (as it can happen only during the gap between $I$ and $I_u$, which is obviously covered by $I_{ru}$).

If $rq$ is hit by $u$, then a symmetric version of Proposition \ref{Prop:ReturnsAlsoCross} (see Remark (1) following the proposition), in which we switch the roles of $p$ and $u$ and reverse the direction of the time axis, implies that $p$ hits $uq$ between the times
when $r$ enters the halfplanes $\L_{pq}^+$ and $\L_{uq}^+$ (during the respective intervals $I$ and $I_u$). In particular, this latter collinearity of $u,r,p$ occurs after $t_0>t_{ru}$ and before $I_u$, and, therefore, also during $I_{ru}$. (As previously noted, this scenario is not only symmetric to the one assumed in Proposition \ref{Prop:ReturnsAlsoCross}, but, in fact, coincides with it.)

Finally, if $r$ hits $uq$, then a counterclockwise and time-reversed variant of Lemma \ref{Lemma:ReturnsStillCross} similarly implies that $ru$ is hit during $I_{ru}$ by $p$; see Figure \ref{Fig:ChargeabilityC} (left).
(As in the previous case, this collinearity occurs during $\conn{I}{I_u}$, between the times when $r$ enters the halfplanes $\L_{pq}^+$ and $\L_{uq}^+$.)

\begin{figure}[htbp]
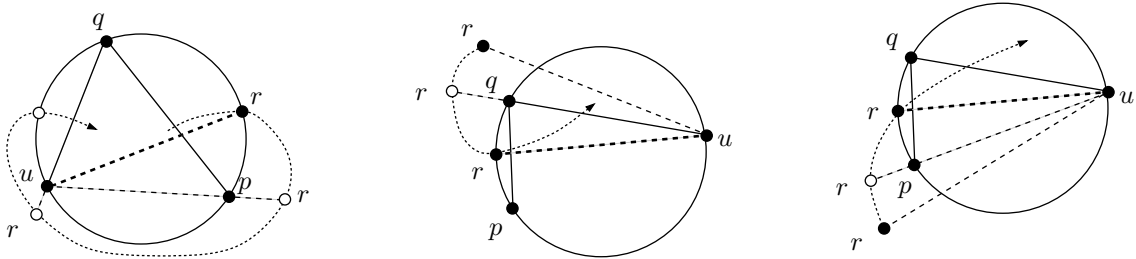

\begin{center}
\input{CrossRuReturn.pstex_t}\hspace{1.5cm}\input{CrossRu1.pstex_t}\hspace{1.5cm}\input{CrossRu.pstex_t}
\caption{\small Proof of Proposition \ref{Prop:CrossUr}: Arguing that $ru$ is hit, during $I_{ru}$, by at least one of $p$ or $q$. Left: The edge $rq$ is hit by $u$ between the times when $r$ crosses $pq$ and $uq$. Hence, the asserted crossing of $ur$ by $p$ follows from Proposition \ref{Prop:ReturnsAlsoCross}.
Center and right: The point $r$ remains in $\L_{uq}^-$ after entering $\L_{pq}^+$ and till the beginning of $I_u$. The Delaunayhood of $ru$ is violated, right before $\zeta_0^u$, by $p$ and $q$, so the asserted collinearity follows from Lemma \ref{Lemma:MustCross}.}
\label{Fig:ChargeabilityC}
\vspace{-0.4cm}
\end{center}
\end{figure} 

Let us then assume that $r$ remains in $\L_{uq}^-$ between the times when it enters the halfplanes $\L_{pq}^+$ and $\L_{uq}^+$. In particular, $u$ lies in $\L_{pq}^+$ when $r$ enters this halfplane, so the Delaunayhood of $ru$ is violated, right before time $\zeta^u_0$, by the points $p$ and $q$, as depicted in Figure \ref{Fig:ChargeabilityC} (center and right).
By (a time-reversal version of) Lemma \ref{Lemma:MustCross}, and since the co-circularity at time $\zeta^u_0$ is the {\it first} co-circularity of $q,p,u,r$, the edge $ru$ is hit during $I_{ru}$, and before $\zeta_0^u$, by at least one of the points $p,q$. Hence, the proposition holds also in this last remaining scenario.
\end{proof}

\medskip
\noindent{\bf Case (c).} If $ru$ is hit by $q$ during $I_{ru}$ then the triple $q,u,r$ defines two single Delaunay crossings within the triangulation $\DT((P\setminus A)\cup\{q\})$. 
In this case, the crossing $(uq,r,I_u)$ is again declared as a heavy and pays $\Theta(1/k)$ units of charge to $(pq,r,I)$.
A combination of Lemma \ref{Lemma:TwiceCollin} with the standard probabilistic argument of Clarkson and Shor yields an upper bound of $O(kn^2)$ on the overall number of such crossings $(uq,r,I_u)$, that are declared as heavy for at least one choice of $(pq,r,I)$ (upon falling into case (c)). 


\medskip
\noindent{\bf Case (d).} We can, therefore, assume that $ru$ is hit during $I_{ru}$ by $p$, so the reduced set $(P\setminus A)\cup \{p\}$ induces at least one Delaunay crossing of $ru$ by $p$. 
In this case, we say that the crossing $(uq,r,I_u)$ is {\it light} for $(pq,r,I)$, and distinguish between the following two subcases.

\medskip
\noindent{\bf Case (d1).} If at least one of the collinearities of $u,r,p$ that occur during $I_{ru}$ is $(4k)$-shallow, we 
directly charge $(pq,r,I)$ to it. In other words, in this case $(pq,r,I)$ receives $1$ unit of charge via $u$ alone, and it does not have to charge any other neighboring $(q,r)$-crossings.

We next argue that each $(4k)$-shallow collinearity, which occurs at some time $t^*$, is charged in the above manner by at most $O(1)$ $(p,r,k)$-chargeable crossings $(pq,r,I)$. 
Indeed, the points $p$ and $r$ of $(pq,r,I)$ can be guessed in $O(1)$ possible ways from among the three points involved in the charged collinearity, and their choice immediately determines the third point $u$ (which figures in the charging scenario of case (d1)).
The guessing of $q$, which is the last unknown point of $(pq,r,I)$, is done exactly as in case (b), and it requires only the knowledge of $t^*$, $r$ and $u$.
(As before, we use the property that $(uq,r,I_u)$ is among the first $c_0+2$ such clockwise $(u,r)$-crossings to begin after $t^*$.)
To conclude, the above charging accounts for at most $O(kn^2\beta(n))$ crossings $(pq,r,I)$.


\medskip
\noindent{\bf Case (d2).} It thus remains to handle the scenario where all collinearities of $u,p,r$ that occur during $I_{ru}$ are $(4k)$-deep. 

 
We first argue that $\A_{ru}$ contains at least $k$ $(4k)$-shallow co-circularities, each occurring within the previously defined interval $I_{ru}$ and involving $p,u,r$ and some fourth point of $P$. 
Indeed, the open disc $B[u,r,p]$ contains no points of $P\setminus A$ when the above crossing of $ru$ by $p$ begins, within the reduced triangulation $\DT((P\setminus A)\cup \{p\})$. (If $ru$ undergoes more than one crossing by $p$ within $\DT((P\setminus A)\cup \{p\})$, we consider the first such crossing.) Since the corresponding collinearity of $u,r,p$ is not $(4k)$-shallow (and the cardinality of $A$ is at most $3k$), the disc $B[u,r,p]$ ``swallows" at least $k$ points of $P\setminus A$ before $ru$ is hit by $p$, which can enter $B[u,r,p]$ only through its boundary. Since at the beginning of the process $B[u,r,p]$ contains only (at most $3k$) points of $A$, the first $k$ points that $B[u,r,p]$ ``swallows" form with $u,r$ and $p$ $k$ co-circularities, all of which are $(4k)$-shallow.

Each of the above $(4k)$-shallow co-circularities pays $\Theta(1/k^2)$ units of charge to $(pq,r,I)$.
Therefore, $(pq,r,I)$ still receives at least $\Theta(1/k)$ units of charge via $(uq,r,I_u)$. 
To complete our analysis, we argue, almost exactly as in the previous case (d1), that each $(4k)$-shallow co-circularity, which occurs at some fixed time $t^*$, is charged in this manner by at most $O(1)$ crossings $(pq,r,I)$. 
Indeed, the points $p,r$ and $u$ can be chosen in at most $O(1)$ possible ways from among the four points that are co-circular at time $t^*$. Moreover, the knowledge of $t^*$, $r$ and $u$ enables us to guess the last unknown point  $q$ of $(pq,r,I)$ in at most $c_0+2$ possible ways, as was done in cases (b) and (d1).

To conclude, in case (d2) the crossing $(pq,r,I)$ receives a total $\Theta(1/k)$ units of charge from $\Theta(k)$ $(4k)$-shallow co-circularities within $\A_{ru}$ (each involving $p,r$ and $u$), where each co-circularity is charged by at most $O(1)$ crossings.

\paragraph{Wrap up.} To finish the proof of Theorem \ref{Thm:Balanced}, it remains to check that all the $(p,r,k)$-chargeable crossings $(pq,r,I)$ (over all possible $p,r\in P$) receive a total of at most $O(k^2N(n/k)+kn^2\beta(n))$ units of charge from neighboring heavy $(q,r)$-crossings $(uq,r,I_u)$ and from $(4k)$-shallow collinearities and co-circularities in appropriate arrangements $\A_{ru}$. 

Indeed, the overall number of crossings $(uq,r,I_u)$ that are classified as heavy (upon falling into one of the cases (a)--(c)), for at least one of their neighbors $(pq,r,I)$, is at most $O(k^2N(n/k)+kn^2\beta(n))$. Moreover, a heavy crossing $(uq,r,I_u)$ pays $\Theta(1/k)$ units of charge to $(pq,r,I)$ only if these crossings are $\lceil k/2\rceil$-consecutive (as $(q,r)$-crossings), so it pays at most $O(1)$ units of charge in total.

Furthermore, we have shown that any $(4k)$-shallow co-circularity or collinearity is charged, through the mechanism of case (d), by $O(1)$ crossings $(pq,r,I)$.
Namely, in case (d1) each $(4k)$-shallow collinearity pays $1$ unit of charge to each of the $O(1)$ possible charging crossing $(pq,r,I)$, so the total charge paid by these collinearities is $O(kn^2\beta(n))$.
In contrast, in case (d2) each $(4k)$-shallow co-circularity pays each time only $\Theta(1/k^2)$ units of charge, so the total charge paid by these co-circularities is $O\left(\frac{1}{k^2}k^4 N(n/k)\right)=O\left(k^2N(n/k)\right)$.

Finally, each $(p,r,k)$-chargeable crossing $(pq,r,I)$ charges $\lceil k/2\rceil$ neighboring $(q,r)$-crossings $(uq,r,I_u)$. Except for case (d1), where $(pq,r,I)$ receives via $(uq,r,I_u)$ one unit of charge (from a $(4k)$-shallow collinearity of $u,r$ and $p$), $(pq,r,I)$ recieves each time $\Theta(1/k)$ units of charge, either directly from $(uq,r,I_u)$ (when that last crossing is heavy), or from certain $(4k)$-shallow events within the corresponding arrangement $\A_{ru}$ (when $(uq,r,I_u)$ is light).
In either case, $(pq,r,I)$ receives one at least one unit of charge, and the proof of Theorem \ref{Thm:Balanced} is now complete. $\Box$


\paragraph{Remark.}
It is instructive to compare the arguments used in cases (b) and (d) of the above analysis. Notice that both of them proceed by charging events that occur in $\A_{ru}$ during $I_{ru}$.

In case (b), each $k$-shallow event under consideration is only known to involve $r$ and $u$ (but not necessarily $p$ or $q$). This information appears to be sufficient for guessing $q$ and $(uq,r,I_u)$, but not necessarily $p$ and $(pq,r,I)$. Hence, we cannot directly charge $(pq,r,I)$ to such events in $\A_{ru}$, so the charging is performed indirectly, via the crossing $(uq,r,I_u)$, which is then classified as heavy for $(pq,r,I)$.
(Note, though, that the same crossing $(uq,r,I_u)$ can be heavy for $\Omega(k)$ neighboring $(q,r)$-crossings $(pq,r,I)$. This is compensated by the fact that $u$ and $(uq,r,I_u)$ can be chosen in $\Theta(k)$ possible ways.)

In case (d), the $(4k)$-shallow events under consideration are more restricted and involve {\it three} fixed points $u,r,p$. As in case (b), the knowledge $u,r$, and the time $t^*$, of each event, enables us to guess $q$ and $(uq,r,I_u)$ in $O(1)$ possible ways.
However, since the point $p$ is now also involved in the event, we can now guess it too in $O(1)$ possible ways.
This enables direct charging of such events by $(pq,r,I)$.



\appendix

\section{On Co-circularities and Collinearities of Points Moving at Unit Speeds}\label{AppSec:UnitSpeeds}

\begin{lemma}
Let $P$ be a finite collection of points in the plane, each moving along some straight line at unit speed.
Then (i) any four points of $P$ can be co-circular at most three times, and (ii) no triple of points can be collinear more than twice.
\end{lemma}

\begin{proof}
To see (i), we note that each co-circularity of a quadruple $\{p_i=(x_i(t),y_i(t))\mid 1\leq i\leq 4\}$ (in $P$) occurs at a time $t$ when the following determinant is equal to zero (see, e.g., \cite{Ed2,EM}):

\[\small D(t)= \left| \begin{array}{cccc}
1 & 1 & 1 & 1\\
x_1(t) & x_2(t) & x_3(t) & x_4(t)\\
y_1(t) & y_2(t) & y_3(t) & y_4(t)\\
x_1^2(t)+y^2_1(t) & x_2^2(t)+y_2^2(t)  & x_3^2(t)+y_3^2(t) & x_4^2(t)+y_4^2(t)\end{array} \right|\] 

Since each $p_i$ is moving along some line in $\reals^2$, its respective location $(x_i(t),y_i(t))$ can be represented as $(x_i+u_it,y_i+v_it)$, where $(x_i,y_i)$ is the location of $p_i$ at the time $t=0$.
Furthermore, since each $p_i$ is moving at unit speed, we obtain $u_i^2+v_i^2=1$.

Substituting $x_i(t)=x_i+u_it$ and $y_i(t)=y_i+v_it$ into the previous expression for $D(t)$, and cancelling the equal terms $(u_i^2+v_i^2)t^2=t^2$ in the bottom row of the determinant, we can replace the equation $D(t)=0$ with its cubic equivalent, with at most {\it three solutions}.

To see (ii), we note that each collinearity of a triple $\{p_i(t),\mid 1\leq i\leq 3\}$ occurs at a time $t$ when the following determinant is equal to zero:

\[\small F(t)= \left| \begin{array}{ccc}
1 & 1 & 1\\
x_1(t) & x_2(t) & x_3(t)\\
y_1(t) & y_2(t) & y_3(t) \end{array} \right|\] 

Substituting $x_i(t)=x_i+u_it$ and $y_i(t)=y_i+v_it$, for $1\leq i\leq 3$, we get that the equation $F(t)=0$ is quadratic (for any choice of $u_i$ and $v_i$), with at most {\it two solutions}.
\end{proof}

\section{The General Position Assumption}\label{AppSec:GenPos}
In our analysis we assume that no five points can become co-circular during the motion, no four points can become collinear, no two points can coincide, and no two events of either a co-circularity of four points or of collinearity of three points can occur simultaneously.
In addition, we assume that in every co-circularity event involving some four points $a,b,p,q\in P$, each of the points, say $a$, crosses the circumcircle of the other three points $b,p,q$; that is, it lies outside the circle right before the event and inside right afterwards, or vice versa. Similarly, we assume that in every collinearity event involving some triple of points of $P$, each of the points crosses the line through the remaining two points.
Degeneracies in the point trajectories of the above kinds can be handled, both algorithmically and combinatorially, by any of the standard symbolic perturbation techniques, such as simulation of simplicity \cite{EM}; for combinatorial purposes, a sufficiently small generic perturbation of the motions will get rid of any such degeneracy, without decreasing the number of topological changes in the diagram.

\section{Proof of Theorem \ref{Thm:RedBlue}}\label{Append:RedBlue}
In this section we establish Theorem \ref{Thm:RedBlue}.
Without loss of generality, we assume that the edge $pq$ is Delaunay at time $t_0$. (If $pq$ is Delaunay at time $t_1$ then we can argue in a fully symmetrical fashion.)

\begin{figure}[htbp]
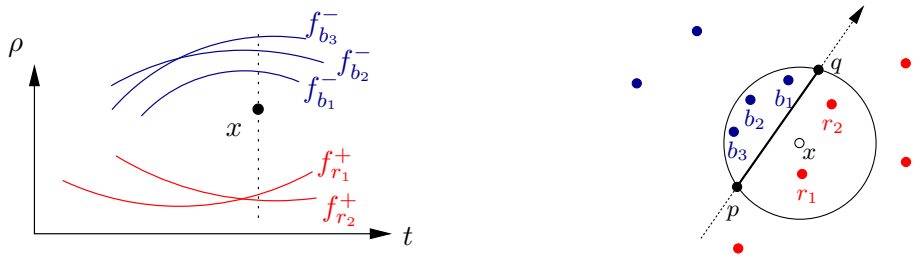

\begin{center}
\input{RedBlueLevels.pstex_t}\hspace{3cm}\input{RedBlueLevels1.pstex_t}
\caption{\small Left: The point $x=(t,\rho)$ lies below three blue functions and above two red functions, so its blue and red levels are $3$ and $2$, respectively. Right: The circumdisc centered at (signed) distance $\rho$ from $\L_{pq}$ and touching $p$ and $q$ at time $t$ contains the three corresponding blue points and two red points.}
\label{Fig:RedBlueLevels}
\end{center}
\end{figure}

Consider the portion of the red-blue arrangement associated with $pq$ within the time interval $(t_0,t_1)$. As above, refer to the parametric plane in which this arrangement is represented as the $t\rho$-plane, where $t$ is the time axis and $\rho$ measures signed distances  from $\L_{pq}$.
We define the {\it red} (resp., {\it blue}) {\it level} of a point $x=(t,\rho)$ in this parametric $\reals^2$ as the number of red (resp., blue) functions that lie below (resp., above) $x$ (in the $\rho$-direction). See Figure \ref{Fig:RedBlueLevels}.
It is easily checked that the level of a co-circularity event at time $t$, with circumcenter at distance $\rho$ from $\L_{pq}$, is the sum of the red and the blue levels of $(t,\rho)$.

We distinguish between the following (possibly overlapping) cases:

\smallskip
\noindent {\bf (a)} $p$ and $q$ participate in a $k$-shallow collinearity with a third point $r$ at some moment during $I$. That is, Condition (i) is satisfied. (Note that here we do not care whether $r$ crosses $pq$ or $\L_{pq}\setminus pq$.)

Suppose that this does not happen. That is, each time when a point $r\in P$ changes its color from red to blue or vice versa, the number of points on each side of $\L_{pq}$ is larger than $k$. Hence, either the number of points on each side of $\L_{pq}$ is always larger than $k$ (during $(t_0,t_1)$), or the sets of red and blue points remain fixed throughout $(t_0,t_1)$ (no crossing takes place), and the size of one of them is at most $k$. More concretely, either one of the sets contains fewer than $k$ points at the start of $I$, and then no crossing can ever occur during $I$, or both sets contain at least $k$ points at the start of $I$, and this property is maintained during $I$, by assumption. In the latter case Condition (iii) trivially holds, since removal of all points in $P\cap \L_{pq}^+$ or in $P\cap \L_{pq}^-$ guarantees that $pq$ is a hull edge throughout $(t_0,t_1)$, and thus belongs to the Delaunay triangulation. Hence, we may assume that the number of red points, and the number of blue points, are always both larger than $k$ during $(t_0,t_1)$.

\begin{figure}[htbp]
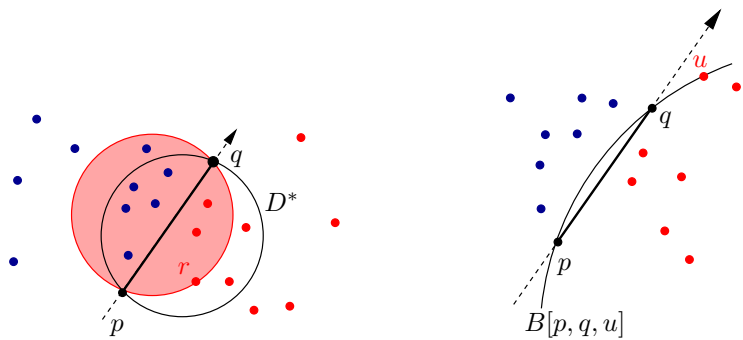

\begin{center}
\input{DeepDeep.pstex_t} \hspace{2cm} \input{CrossOutside.pstex_t}
\caption{\small Left: Case (b). The disc $D^*$ contains at least $\lceil k/3 \rceil=5$ red points, and at least $\lceil k/3 \rceil$ blue points. If $r$ lies at red level at most $\lceil k/3 \rceil$, it belongs to $D^*$. Hence, the circumdisc $B[p,q,r]$ contains at least $\lceil k/3 \rceil$ blue points, so the blue level of $f^+_r$ is at least $\lceil k/3 \rceil$. Right: Case (c). The setup right after time $t'$ when $u$ crosses $\L_{pq}\setminus pq$. $B[p,q,u]$ contains at least $k$ red points and no blue points.}
\label{Fig:DeepDisc}
\end{center}
\end{figure} 

\smallskip
\noindent {\bf (b)} At some moment $t_0\leq t^*\leq t_1$ there is a disc $D^*$ that touches $p$ and $q$, and contains at least $\lceil k/3\rceil$ red points and at least $\lceil k/3\rceil$ blue points.
In particular, for each of the $\lceil k/3\rceil$ shallowest red functions $f_r^+$ at time $t^*$, its respective red point $r$ belongs to $D^*$. and similarly for the $\lceil k/3\rceil$ shallowest blue functions.
See Figure \ref{Fig:DeepDisc} (left). Before we use the existence of $D^*$ we first conduct the following structural analysis.

Let $f_r^+$ be a red function which is defined at time $t_0$, and whose red level is then at most $\lfloor k/6\rfloor$. (Recall that, at time $t_0$, the blue level of any red function is $0$ since $pq$ belongs to $\DT(P)$.) We claim that either $f_r^+$ is defined and continuous throughout $(t_0,t_1)$ and its red level is always at most $\lceil k/3\rceil$, or $r$ participates in at least $\lceil k/6\rceil$ red-red and/or red-blue co-circularities, all of which are $\lceil k/3\rceil$-shallow. 

Indeed, the circumdisc $B[p,q,r]$ contains at most $\lfloor k/6\rfloor$ red points (and no blue points) at time $t_0$, and it moves continuously as long as $r$ remains in $\L_{pq}^+$. By the time at which either (the graph of) $f_r^+$ reaches red level $\lceil k/3\rceil$ or $r$ hits $\L_{pq}$, this disc ``swallows" either at least $\lceil k/6\rceil$ red points (either in the former case or in the latter case when $r$ crosses $\L_{pq}\setminus pq$) or at least $\lceil k/6\rceil$ blue points (in the latter case when $r$ crosses $pq$). (Recall that, by assumption, the number of red points and the number of blue points is always larger than $k$ during $I$.) We thus obtain at least $\lceil k/6\rceil$ $\lceil k/3\rceil$-shallow red-red or red-blue co-circularities involving $p,q,r,$ and a fourth (red or blue) point. 

To recap, if at least $\lfloor k/12\rfloor$ red functions, which at time $t_0$ are among the $\lceil k/6\rceil$ shallowest red functions, reach red level at least $\lceil k/3\rceil+1$, or have a discontinuity at $\rho=-\infty$ or $+\infty$ (at a crossing of $\L_{pq}$ by the corresponding point), then we encounter $\Omega(k^2)$ co-circularities (involving $p$ and $q$) which are $k$-shallow, so Condition (ii) holds.

Hence, we may assume that at least $\lceil k/12 \rceil$ red functions $f_r^+$ that are among the $\lceil k/6\rceil$ shallowest red functions at time $t_0$, are defined throughout $(t_0,t_1)$, and their red level always remains at most $\lceil k/3\rceil$.
Fix any such red function $f_r^+$.
Clearly, the red point $r$ that defines $f_r^+$ belongs to $D^*$ at time $t^*$, and the circumdisc $B[p,q,r]$ contains at least $\lceil k/3\rceil$ blue points. See Figure \ref{Fig:DeepDisc} (left).
This implies that the blue level of $f_r^+$ reaches $\lceil k/3\rceil$ so (since the blue level was $0$ at time $t_0$) $r$ participates in at least $\lfloor k/6\rfloor$ $\lceil k/3\rceil$-shallow co-circularities during $(t_0,t^*)$. Repeating this argument for each of the remaining $\lceil k/12\rceil$ such red functions, we conclude that Condition (ii) is again satisfied.

\smallskip
\noindent {\bf(c)} Suppose that neither of the two cases (a), (b) holds. 
Let $A_R$ (resp., $A_B$) be the subset of all points $u$ whose red (resp., blue) functions $f_u^+$ (resp., $f^-_u$) appear at red (resp., blue) level at most $\lceil k/3 \rceil$ at some moment during $(t_0,t_1)$. 

Since the situation in (b) does not occur, we can restore the Delaunayhood of $pq$, throughout the entire interval $(t_0,t_1)$, by removing all points in $A_R\cup A_B$. To see this, suppose that $pq$ is not Delaunay (in $\DT(P\setminus (A_R\cup A_B))$) at some time $t_0< t^*< t_1$. This is witnessed by a disc $D^*$ whose boundary passes through $p$ and $q$ and which contains a red point $r\not\in A_R$ and a blue point $b\not\in A_B$. Since the red level of $f_r^+$ is greater than $\lceil k/3\rceil$ at time $t^*$, $D^*$ must also contain the $\lceil k/3\rceil$ red points corresponding to the $\lceil k/3\rceil$ shallowest red functions at time $t^*$, and, symmetrically, also the $\lceil k/3\rceil$ blue points corresponding to the $\lceil k/3 \rceil$ shallowest blue functions at time $t^*$. But then the disc $D^*$ satisfies the conditions of Case (b), contrary to assumption.

Let $A_R^o$ (resp., $A_B^o$) be the set of $k$ points whose red (resp., blue) functions are shallowest at time $t_0$.
It remains to consider the case where at least $k$ points $u$ in $A_R\cup A_B$ belong to neither of $A_R^o,A_B^o$, for otherwise Condition (iii) is trivially satisfied, with a removed set of size at most $3k$.
Fix such a point $u$ and consider the first time $t^*\in (t_0,t_1)$ when its red function $f_u^+$ has red level at most $\lceil k/3\rceil$, or its blue function $f_u^-$ has blue level at most $\lceil k/3\rceil$. Without loss of generality, suppose that at time $t^*$ the red function $f_u^+$ has red level at most $\lceil k/3\rceil$.
We claim that $u$ does not cross $pq$ during $(t_0,t^*]$. Indeed, if there were such a crossing from $\L_{pq}^-$ to $\L_{pq}^+$ then the blue function $f_u^-$ would tend to $\infty$ right before the crossing, and its blue level would then be $0$ even before $t^*$, contrary to the choice of $t^*$. Similarly, if the crossing were from $\L_{pq}^+$ to $\L_{pq}^-$ then the red level of $f_u^+$ would be $0$ just before the crossing, again contradicting the choice of $t^*$. 

First, assume that $u$ does not cross $\L_{pq}$ during $(t_0,t^*)$, so the graph of $f_u^+$ is continuous during this time interval. Hence, the motion of the circumdisc $B[p,q,u]$ is also continuous.
Since $u\not\in A_R^o$, at time $t_0$ the circumdisc $B[p,q,u]$ contains at least $k$ red points and no blue points. At time $t^*$,
$B[p,q,u]$ contains $\lceil k/3\rceil$ red points and fewer than $\lceil k/3\rceil$ blue points (otherwise Case (b) would occur). 
Hence, we encounter at least $\lfloor k/3\rfloor$ $k$-shallow co-circularities during $(t_0,t^*)$, each involving $p,q,u$ and some other point of $P$.

Now, suppose $u$ crosses $\L_{pq}\setminus pq$ during $(t_0,t^*)$, and consider the last time $t'$ when this happens.
We can use exactly the same argument as in the ``continuous" case but now starting from $t'$. Indeed, $f_u^+$ is continuous during $(t',t^*]$ and, right after $t'$, the circumdisc $B[p,q,u]$ contains (all the red points and thus) at least $k$ red points, and no blue points. See Figure \ref{Fig:DeepDisc} (right).

Repeating this argument for all such points $u\in A_R\cup A_B\setminus(A_R^o\cup A_B^o)$, we get $\Omega(k^2)$ $k$-shallow co-circularities which occur during $(t_0,t_1)$ and involve $p$ and $q$. Hence, Condition (ii) is again satisfied. $\Box$

\section{The number of double Delaunay crossings}\label{Subsec:Double}
In this subsection we show that any set $P$ of $n$ points moving as above in $\reals^2$ admits at most $O(n^2)$ double Delaunay crossings.
Since double Delaunay crossings are not possible if no ordered triple of points can be collinear more than once
(i.e., if for any $p,q,r$ the third point $r$ can hit the segment $pq$ at most once), we may assume throughout this subsection that no triple of points in $P$ can be collinear more than twice.

Without loss of generality, we only bound the number of such double Delaunay crossings $(pq,r,I)$ whose point $r$ crosses through $pq$ from $\L_{pq}^-$ to $\L_{pq}^+$ during the first collinearity of $p,q,r$ (and then returns back to $\L_{pq}^-$ during the second collinearity).
Indeed, if the crossing $(pq,r,I)$ does not satisfy the above condition then they are satisfied by $(qp,r,I)$.
Our goal is to show that (on average) a point $r$ of $P$ is involved in only few Delaunay crossings of edges that share the same endpoint $p$.

The following theorem provides certain structural properties of two double crossings that share the same crossing point ($r$) and one endpoint ($p$) of the crossed edges.

\begin{figure}[htbp]
\begin{center}
\input{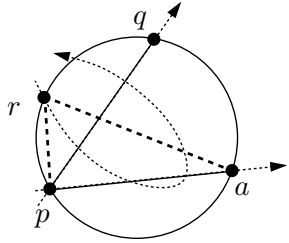}
\caption{\small The trace of $r$ according to Theorem \ref{Thm:OrderSpecialCrossings}. The four points $p,q,a,r$ are involved during $I$ in two co-circularities, which are red-blue with respect to the edges $pq$ and $ra$.}
\label{Fig:StayDelaunay}
\end{center}
\end{figure} 

\begin{theorem}\label{Thm:OrderSpecialCrossings}
Let $(pq,r,I)$ and $(pa,r,J)$ be two double Delaunay crossings of $p$-edges (that is, edges incident to $p$) $pq,pa$ by the same point $r$. 
Assume that
the first collinearity of $p,q,r$ occurs before the first collinearity of $p,a,r$.
Then the following properties hold (with the conventions assumed above):\\
\indent(i) $a$ lies in $\L_{pq}^+$ at both times when $r$ hits $pq$.\\
\indent (ii) $q$ lies in $\L_{pa}^-$ at both times when $r$ hits $pa$.\\
\indent (iii) The points $p,q,a,r$ are involved during $I\setminus J$ in two co-circularities, both of them red-blue with respect to $pq$ and occurring when $r\in \L^-_{pq}$ and $a\in \L^+_{pq}$. \\
\indent (iv) One of the two co-circularities in (iii) occurs before the beginning of $J$; right before it the Delaunayhood of $ra$ is violated by $p$ and $q$. A symmetric such co-circularity occurs after the end of $J$;
right after it the Delaunayhood of $ra$ is again violated by $p$ and $q$.
In particular, $J\subset I$.
\end{theorem}
The schematic description of the motion of $r$ during $I$, according to the above theorem, is depicted in Figure \ref{Fig:StayDelaunay} (right).
Clearly, a suitable variant of Theorem \ref{Thm:OrderSpecialCrossings} exists also for similar pairs of double crossings of incoming $p$-edges $qp,ap$ that are oriented {\it towards} $p$ (again, by the same point $r$).
\begin{proof}
We first establish Part (ii) of the theorem.
The crucial observation is that the first collinearity of $p,a,r$ occurs when $r$ lies in $\L_{pq}^+$ (i.e., during the interval between the two collinearities of $p,q,r$). 
Indeed, otherwise the point $a$ must lie in $\L_{pq}^+=\L_{pr}^+$ at both collinearities of $p,a,r$, and $q$ must lie in $\L_{pa}^+$ at both collinearities of $p,a,r$. 
We shall prove that, in this hypothetical setup, the points $p,q,a,r$ are involved in two co-circularities during $I$ which are red-blue with respect to $pq$, and in a symmetric pair of co-circularities during $J$, both of them red-blue with respect to $pa$. That will clearly contradict the assumption that any four points can be co-circular at most three times.

Indeed, in the above situation the point $a$ lies in the cap $B[p,q,r]\cap \L_{pq}^+$ shortly before the first collinearity of $p,q,r$, and shortly after their second collinearity. 
Since $B[p,q,r]$ contains no points at the beginning of $I$, the point $a$ must have entered this cap before the first collinearity of $p,q,r$. Moreover, $a$ can enter this cap only through the boundary of $B[p,q,r]$, for otherwise it would hit $pq$ during $I$, and no point of $P\setminus\{p,q,r\}$ can hit $pq$ during its Delaunay crossing by $r$. This argument gives us the first of the promised two red-blue co-circularities that $p,q,a,r$ define with respect to $pq$. The second such co-circularity is symmetric to the first one, and occurs when $a$ leaves the cap $B[p,q,r]\cap \L_{pq}^+$ (and after $r$ returns to $\L_{pq}^-$ through $pq$). See Figure \ref{Fig:DoubleFour} (left).
The other pair of co-circularities, both red-blue with respect to $pa$, is obtained by applying a fully symmetric argument to the cap $B[p,a,r]\cap \L_{pa}^+$ and the point $r$. See Figure \ref{Fig:DoubleFour} (center). (For example, we can switch the roles of $q$ and $a$ by reversing the direction of the time axis.)
Finally, all four co-circularities are distinct, because the same co-circularity cannot be red-blue with respect to two edges $pq,pa$ with a common endpoint. 

\begin{figure}[htbp]
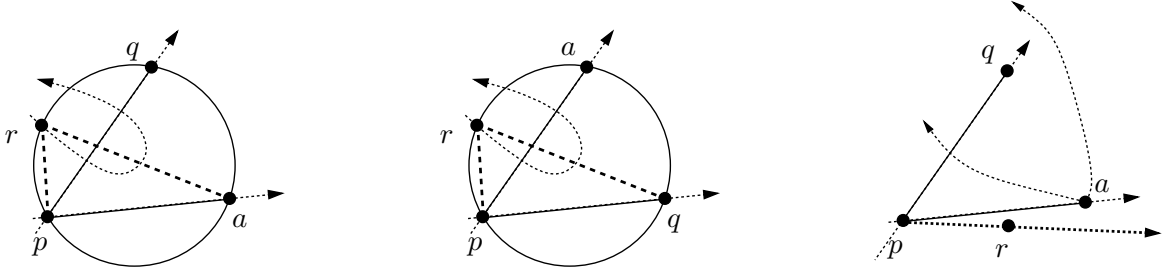

\begin{center}
\input{DoubleFourCocircs1.pstex_t}\hspace{2cm}\input{DoubleFourCocircs.pstex_t}\hspace{2cm}\input{CannotReturn.pstex_t}
\caption{\small Proof of Theorem \ref{Thm:OrderSpecialCrossings}. Left and center: The hypothetical case where $r$ first hits $pa$ within $\L_{pq}^-$, after twice hitting $pq$. The points $p,q,a,r$ are involved in a pair of co-circularities during $I$, and in a symmetric pair of co-circularities during $J$. Right: The hypothetical traces of $a$ if it enters $\L_{pq}^+$ before $r$ (and before the second collinearity of $p,a,r$ occurs).}
\label{Fig:DoubleFour}
\end{center}
\end{figure}

Hence, we can assume, from now on, that the first time when $r$ hits $pa$ occurs when both points lie in $\L_{pq}^+$.
To complete the proof of Part (ii), it suffices to show that the points $a$ and $r$ still remain in $\L_{pq}^+$ during the second collinearity of the triple $p,a,r$.
Indeed, otherwise $a$ must lie in $\L_{pq}^-$ when $r$ hits $pq$ for the second time, because, untill it crosses $pa$ again, $a$ lies in $\L_{pr}^-$ which coincides with $\L_{pq}^-$ at the second crossing of $pq$ by $r$. See Figure \ref{Fig:DoubleFour} (right).
That is, $a$ must cross $\L_{pq}$ from $\L_{pq}^+$ to $\L_{pq}^-$ while $r$ still remains in $\L_{pq}^+$, and before $r$ hits the edges $pq,pa$ for the second time. In particular, the above collinearity of $p,q,a$ must occur during $I\cap J$. Clearly, the point $a$ can potentially cross $\L_{pq}$ in three ways.
If $a$ crosses $\L_{pq}$ within $pq$, this contradicts the definition of $I$ as the interval of the Delaunay crossing of $pq$ by $r$. If $a$ hits $\L_{pq}\setminus pq$ within the ray emanating from $q$ then (at that very moment) $q$ hits $pa$, which contradicts the definition of $J$. Finally, $a$ cannot hit $\L_{pq}\setminus pq$ within the outer ray emanating from $p$ before an additional (and forbidden) collinearity of $p,a,r$ takes place.  This establishes part (ii), and the analysis given above immediately implies part (i) two.

Part (i) follows immediately from Part (ii), because $a$ lies in $\L_{pr}^+$ during both collinearities of $p,q,r$.

Parts (iii) and (iv) follow from Parts (i) and (ii). Indeed, recall that the open disc $B[p,q,r]$ contains no points of $P$ at the beginning of $I$. Right before $r$ hits $pq$ for the first time, the right cap $B[p,q,r]\cap \L_{pq}^+$ of this disc contains $a$. Clearly, $a$ first enters this cap through the corresponding portion of $\partial B[p,q,r]$. This determines the first red-blue co-circularity with respect to $pq$, right before which the Delaunayhood of $ra$ is violated by $p$ and $q$. The symmetric such co-circularity occurs during $I$ when the point $a$ leaves the cap $B[p,q,r]\cap \L_{pq}^+$, after the second collinearity of $p,q,r$. Clearly, the Delaunayhood of $ra$ is violated right after that co-circularity by $p$ and $q$. By Lemma \ref{Lemma:Crossing}, neither of these co-circularities can occur during $J$, because $ra$ remains Delaunay throughout $J$. Hence, the former one occurs, according to the previously established Parts (i) and (ii), before $J$, and the latter one occurs after $J$. This establishes parts (iii) and (iv), and completes the proof.
\end{proof}

\begin{theorem}\label{AppThm:SpecialCrossings}
Let $P$ be a set of $n$ points, whose motion in $\reals^2$ respects the following conventions: (i) any four points can be co-circular at most three times, and (ii) no three points can be collinear more than twice. Then $P$ admits at most $O(n^2)$ double Delaunay crossings.
\end{theorem}
\begin{proof}
We fix a pair of points $p,r$ in $P$. Our strategy is to show that, for an average such pair, there is at most a constant number of double Delaunay crossings of $p$-edges by $r$.
Indeed, let $(pq_1,r,I_1),(pq_2,r,I_2),$ $\ldots,(pq_k,r,I_k)$ be the complete list of such double Delaunay crossings of $p$-edges by $r$, and assume that $r$ hits the edges $pq_1,pq_2,\ldots,p_{q_k}$, for the first time, in this same order.
By Theorem \ref{Thm:OrderSpecialCrossings}, the respective intervals of the above double crossings form a nested sequence $I_1\supset I_2\supset \ldots \supset I_k$.

\begin{figure}[htbp]
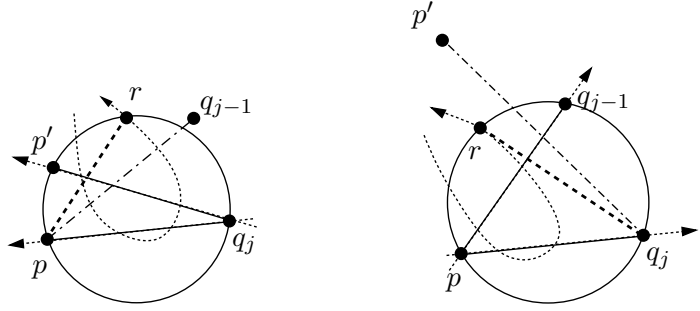

\begin{center}
\input{QuadraticDouble1.pstex_t}\hspace{2cm}\input{QuadraticDouble2.pstex_t}
\caption{\small Proof of Theorem \ref{AppThm:SpecialCrossings}. Left: If the double crossing $(p'q_j,r,I')$ ends before the end of $I_{j-1}$ then the second co-circularity of $q_j,p,p',r$ occurs during $I_{j-1}$. Right: If the double crossing $(p'q_j,r,I')$ ends after $I_{j-1}$ then the second co-circularity of $p,q_{j-1},q_j,r$ occurs during $I'$.}
\label{Fig:QuadDouble}
\end{center}
\end{figure} 

Clearly, the first crossing $(pq_1,r,I_1)$ can be uniquely charged to the pair $p,r$.
Now assume that $k>1$. We show that each of the additional double Delaunay crossings $(pq_j,r,I_j)$, for $2\leq j\leq k$, 
can be uniquely charged to the corresponding pair $q_j,r$.
Specifically, we show that no double Delaunay crossing of incoming $q_j$-edges $p'q_j$ (that is, $p$-edges that are oriented towards $p$), by $r$, can end after $I_j$. In other words, $(pq_j,r,I_j)$ is the ``last" such double crossing.

Indeed, fix $2\leq j\leq k$ as above. We first show that no double crossing of the form $(p'q_j,r,I')$ can end during the interval which lasts from the end of $I_j$ and to the end of $I_{j-1}$. Indeed, suppose to the contrary that such a situation occurs, and apply a suitable variant of Theorem \ref{Thm:OrderSpecialCrossings} to the double Delaunay crossings of $q_j$-edges $p'q_j$ and $pq_j$ by $r$. 
By Part (iv) of that theorem, $I_j$ is contained in $I'$, and the four points $q_j,p,p',r$ are involved in a red-blue co-circularity with respect to $p'q_j$ during the second portion of $I'\setminus I_j$. See Figure \ref{Fig:QuadDouble} (left). Right after that co-circularity, the Delaunayhood of $pr$ is violated by $q_j$ and $p'$. If $I'$ ends before the end of $I_{j-1}$, the above co-circularity must occur during $I_{j-1}$ (as $I_{j-1}\supset I_j$), which contradicts Lemma \ref{Lemma:Crossing} (applied to the crossing of $pq_{j-1}$ by $r$).

It remains to show that no double Delaunay crossing $(p'q_j,r,I')$, as above, can end after the end of $I_{j-1}$. Indeed, by Part (iv) of Theorem \ref{Thm:OrderSpecialCrossings} (now applied to the double crossings of the $p$-edges $pq_{j-1}$ and of $pq_j$, by $r$), the points $p,q_{j-1},q_j,r$ are involved in a co-circularity during the second portion of $I_{j-1}\setminus I_j $. Right after this co-circularity, the Delaunayhood of $q_jr$ is violated by $p$ and $q_{j-1}$. 
If the interval $I'$ (which contains $I_j$) ends after the end of $I_{j-1}$, the aforementioned co-circularity must occur during $I'$; see Figure \ref{Fig:QuadDouble} (right). However, this is another contradiction to Lemma \ref{Lemma:Crossing} (now applied to the crossing of $p'q_j$ by $r$, which takes place during $I'$).

We have shown that every double Delaunay crossing can be uniquely charged to an (ordered) pair of points of $P$, so their number is $O(n^2)$, as asserted.
\end{proof}

\section{Proof of Lemma \ref{Lemma:Incremental}}\label{AppendSec:Incremental}
Assume with no loss of generality that $r$ lies in $\L_{pq}^-$.
Clearly, it is sufficient to establish only the Delaunayhood of the edge $rq$; the Delaunayhood of $pr$ follows in a fully symmetrical manner. 

The crucial observation is that the cap $B[p,q,r]\cap \L_{pq}^-$ has $Q$-empty interior (or, else, $pq$ would be Delaunay also in $Q\cup \{r\}$). That is, in terms of the static red-blue arrangement of $pq$, the corresponding blue function $f_r^+$ of $r$ coincides with the blue upper envelope $E^-$.

Assume for a contradiction that $rq$ is not Delaunay in $Q\cup \{r\}$.
We now consider the static red-blue arrangement of $rq$. Let $x\in Q\cap \L_{rq}^+$ be the point whose function $f_x^+$ (all functions in this argument are from the red-blue arrangement of $rq$) coincides with the red lower envelope $E^+$ (again, with respect to $rq$). In particular, we have $f^+_x\leq f^+_p$ (as is easily checked, $p\in \L_{rq}^+$, when $r\in \L_{pq}^-$). Clearly, $x$ cannot be equal to $p$, for then the disc $B[p,q,r]$ would have $Q$-empty interior. Indeed, we argued that $B[p,q,r]\cap \L_{pq}^-$ is $Q$-empty, and a similar argument shows that $B[p,q,r]\cap \L_{rq}^+$ would also have to be empty if $x$ and $p$ coincide, from which the emptiness of the whole interior follows. It follows that $pq$ is Delaunay in $Q\cup \{r\}$,
contradicting the definition of a Delaunay crossing. See Figure \ref{Fig:StayDelaunay} (left).
Moreover, $x$ cannot lie in $\L_{pq}^-$, for it would then have to lie in $B[p,q,r]\cap \L_{pq}^-$ (because $f_x^+<f_p^+)$, which is impossible since this portion of $B[p,q,r]$ is $Q$-empty. Thus, $p\in \L_{xq}^-$.

\begin{figure}[htbp]
\begin{center}
\input{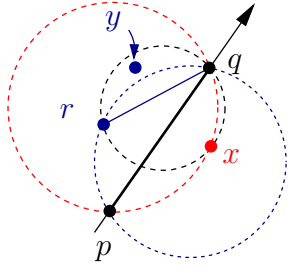}
\caption{\small Left: Proof of Lemma \ref{Lemma:Incremental}.}
\label{Fig:StayDelaunay}
\end{center}
\end{figure}

Since $rq$ is not Delaunay, the disc $B=B[q,r,x]$ contains another point $y\in Q\cap\L_{rq}^-$, which is easily seen to lie in $\L_{pq}^-$ and in $\L_{xq}^-$.
We can move $B$ so that its boundary continues to touch $x$ and $q$ and its portion within $\L_{xq}^-$ expands, until its boundary touches $p$, $q$ and $x$, and its interior contains $y$. This implies that $pq$ does not belong to $\DT(Q)$, which contradicts the definition of a Delaunay crossing. $\Box$

\end{document}